\documentclass[12pt]{article}

\usepackage{amsmath}
\usepackage{amssymb}
\usepackage{amsfonts}
\usepackage{amsthm}
\usepackage{appendix}
\usepackage{natbib}

\newcommand{\SE}{{\rm I\kern-.21em E}}

\newcommand{\SP}{{\rm I\kern-.21em P}}
\newcommand{\SR}{{\rm I\kern-.21em R}}
\newcommand{\SN}{{\rm I\kern-.21em N}}

\usepackage{geometry}
\usepackage{xcolor}

\DeclareMathOperator*{\argmin}{argmin}

\DeclareMathOperator*{\argmax}{argmax}

\newcounter{result}

\newtheorem{theorem}[result]{\textbf{Theorem}}

\newtheorem{proposition}[result]{\textbf{Proposition}}

\newtheorem{example}{\textbf{Example}}
\newtheorem{remark}{\textbf{Remark}}

\newtheorem{lemma}{\textbf{Lemma}}

\newcommand{\SONE}{{\rm 1\kern-.25em I}}
\newcommand{\SC}{{\rm\kern.26em\vrule width.03em height1.4ex
depth-.05ex\kern-.29em C}}
\newcommand{\SQ}{{\rm\kern.27em\vrule width.03em height1.4ex
depth-.05ex\kern-.30em Q}}
\newcommand{\SZ}{{\rm\kern.26em\vrule width.03em height0.5ex depth0ex\kern.03em
        \vrule width.03em height1.47ex depth-1ex\kern-.35em Z}}

\newcounter{llst}

\newenvironment{abet}{\begin{list}{\rm (\alph{llst})}{\usecounter{llst}
\setlength{\itemindent}{0em} \setlength{\leftmargin}{3em}
\setlength{\labelwidth}{2em}
\setlength{\labelsep}{1em}}}{\end{list}}

\begin{document}
\newgeometry{top=2cm,bottom=2cm} 
\title{\Large{Anonymous and Strategy-Proof Voting under\\ Subjective Expected Utility Preferences\thanks{The author thanks  Edi Karni and Yves Sprumont for useful exchanges and stimulating conversations. Thoughtful comments and suggestions by Juan Moreno-Ternero, Alessandro Pavan, Sudipta Sarangi,  Charles Sprenger, William Thomson, and John Weymark are gratefully acknowledged. 
}}}
\author{Eric Bahel\thanks{Address for correspondence: Department of Economics,
 Virginia Polytechnic Institute and State University,
~\mbox{Blacksburg}, VA 24061-0316, USA. Email:
erbahel@vt.edu}\\
\textit{\small Virginia Tech }\\ {\small and} \\ \textit{\small National Science Foundation\thanks{\textit{Any opinions, findings, interpretations, conclusions or recommendations expressed in this
paper are those of its author and do not necessarily reflect the views of the National Science Foundation.}
}}}

\date{\small July 2024}
 \maketitle
\begin{abstract} We study three  axioms in the  model of constrained social choice under uncertainty  where (i) agents have subjective expected utility preferences  over acts and (ii) different states of nature have (possibly) different sets of available outcomes. \textit{Anonymity} says that  agents' names or labels  should never play a role in the mechanism used to select the social act. \textit{Strategy-proofness} requires  that reporting one's true preferences  be a (weakly) dominant strategy for each agent in the associated direct revelation game.  \textit{Range-unanimity} essentially says that a  feasible act must be selected by society whenever it is reported as every  voter's favorite act  within the range of the  mechanism. We first show that every social choice function satisfying these three axioms can be factored as a product of voting rules that are either constant or binary (always yielding one of two pre-specified outcomes in each state). We describe four  basic types of binary factors: three of these types are novel to this literature and exploit the voters' subjective beliefs. Our characterization result then states that a social choice function is anonymous, strategy-proof and range-unanimous if and only if every binary factor (in its canonical factorization) is of one of these four basic types.

\end{abstract}

 \emph{JEL classification}:  D71, D81, D82, R53

\medskip\indent  \textbf{Keywords}: {\small social act, subjective expected utility, anonymity, strategy-proofness.}
\medskip
\restoregeometry

\newpage
\section{Introduction}

Real-life social choice problems typically involve uncertainty. For instance, voters in a democratic country  must elect their leaders without knowing in advance whether the successful candidates will have to deal with a financial crisis,  a pandemic,  or some other uncertain event. Since there is no objective way of assessing the likelihood of these random events,  voters must form their own subjective beliefs in  order to decide which candidates they want to vote for. Other applications of our work include committee voting  under uncertainty. For instance, members of a city council may vote to select the site of a new public facility (such as an airport) under conditions of uncertainty about the viability of the respective candidate sites (with the uncertainty resolved only after teams of engineers have completed feasibility studies). Likewise, shareholders often vote between  merger or acquisition options for their company; and they may have different assessments of the likelihood of random events (such as a nationalization or an armed conflict) that can render some investment options infeasible or unprofitable in the future. 

\subsection{Problem and main axioms}\label{SubsectProblem}
The present paper examines collective decision-making under uncertainty. We consider a  group whose members must choose a course of action, without knowing which one of the possible states of nature will  materialize in the future.  The set of available outcomes  may vary depending on the realized state of nature; and each course of action (that the group may select) yields, for every possible state of nature, an outcome available in this state. 
Following \cite{savage1972foundations}, we model uncertain prospects as \textit{acts}, which are mappings from the set of  relevant states of nature into the set of conceivable outcomes. Since these acts are of interest to every member of society, we   refer to them as social acts. 

 We assume that agents' preferences over acts are of \cite{savage1972foundations}'s Subjective Expected Utility (SEU) type. Hence, each agent must report a state-independent valuation function (expressing her tastes over outcomes) and a probability measure (giving her subjective belief about the likelihood of every event).\footnote{The interested reader can consult \cite{karni1993subjective} for an extension of Savage's subjective expected utility theory that includes state-dependent valuations. }
 A voting mechanism must then select a social act at any reported preference profile. Given this context of uncertainty, our main objective  is to study and describe voting mechanisms  satisfying two well-known axioms in social choice theory. \textit{Anonymity} says  that voters' identities (or social characteristics) should bear no role in the collective decision process. \textit{Strategy-proofness} is an incentive-compatibility property saying that no individual voter should ever have incentives to misreport her valuation function or her subjective belief. Since not all rankings over acts can be generated by SEU preferences, note that the Gibbard-Satterthwaite \citep{gibbard1973manipulation,satterthwaite1975strategy} theorem does not apply; and this opens the door to the possibility of anonymous mechanisms that are strategy-proof. As explained in Subsection \ref{Sectliterature}, the current work is also motivated by an impossibility result stemming from  \cite{bahel2020strategyproof}: there exists no anonymous mechanism satisfying strategy-proofness and \textit{unanimity} (\textit{i.e.}, an act should be selected by society whenever it is the favorite feasible act of all voters). It hence follows that democratic and strategy-proof social choice imposes the weakening of the unanimity axiom.

\subsection{Illustration: voting on a group activity}\label{ToyEx}
To illustrate the framework and the mechanisms proposed in this paper, we offer the following problem as a  toy example of collective decision-making under uncertainty. Suppose that a group of ten high school friends decide to reunite in their  hometown to  spend an entire day together. The reunion will take place in a few months, at some agreed-upon date; and the friends must collectively decide   on the main activity that they will carry out during the gathering. The set of candidate activities is described as follows: \textit{amusement park (P), beach (B), movie marathon (M), skydiving (D), indoor volleyball (V)}. Importantly, it is not known in advance what the weather will be on reunion day in their hometown; and some of the candidate activities may only be available  under specific weather conditions. More precisely, assume that the five possible types of weather are \textit{cloudy (c), rainy (r), snowy (o),  sunny (s), and windy (w)}; and the  activities available in these  states of nature are  $X_c=\{B, M, P,V\}$, $X_r=\{M, V\}=X_o=X_w$, $X_s=\{B, D,  M, P,V\}$.

The group must collectively select \textbf{today} a feasible course of action, which is a contingent plan specifying, for each state of nature, an activity available in this state. For example, observe that the course of action $f=(M,M,M,M,M)$, which recommends a movie marathon regardless of the state of nature, is feasible; and so is the course of action $g=(P,V,V,V,B)$ recommending the amusement park in case of cloudy skies, volleyball in case of rain or snow or wind gusts, and the beach in the event of a sunny reunion day. 
The collective decision process must be based on the friends' rankings of such contingent plans; and under SEU preferences the ranking of each  group member can be described by the combination of (i) her valuations of the respective activities and (ii) her subjective beliefs about the likelihood of the respective states of nature. 
The collective decision process should be democratic (i.e., anonymous) and immune to manipulations by any of the friends (which is guaranteed by strategy-proofness). This example will be  recalled (both in Subsection \ref{SubMainresults} and Section \ref{SectChar}) to illustrate our  results.

\subsection{Main contribution and results}\label{SubMainresults}
To our main axioms of anonymity and strategy-proofness, we add the natural and minimal requirement of range-unanimity, which says that an act should be chosen by the voting mechanism whenever it is reported as the favorite of every voter among acts  yielding in each state  an outcome within the range of outcomes selected (by the mechanism) for this state. Range-unanimity is much weaker than the standard unanimity axiom ---see \cite{bahel2020strategyproof}, which is in turn weaker than requiring (\textit{ex post} or \textit{ex ante}) Pareto optimality of the social act  selected. Specifically, the main contribution of the present work is the full  characterization of the class of anonymous, strategy-proof and range-unanimous voting mechanisms under SEU preferences.

Our first result, stated in Proposition \ref{Propbinarydecomp}, shows that any voting mechanism satisfying these three axioms can be expressed as a product of  sub-mechanisms (or factors), each of which is \textbf{(i)} defined on a distinct cell of some partition of the state space and  \textbf{(ii)} either constant or \textit{binary} (\textit{i.e.}, the range of outcomes selected by the factor, as one varies the reported  preference profile, must be a pair). It is crucial to note that the result of Proposition  \ref{Propbinarydecomp} implies a useful separability property: each factor in the product is a voting mechanism on its own, since the sub-act  it selects (on the associated cell of the partition of the state space) may vary \textbf{only} with the agents' \textbf{conditional beliefs} on that cell and their valuations of the outcomes used. Moreover, the factors all inherit from the original mechanism our properties of anonymity, strategy-proofness and range-unanimity.

Exploiting this factorization result, one can thus focus on the class of binary voting mechanisms (which select in every state of nature one of two pre-specified outcomes, say, $a$ and $b$). We offer below a description of four sub-classes of factors  constituting the building blocks of our characterization result. 

\subsubsection{Simple factors}\label{IntroSimpleFactor}
The first and most straightforward of these building blocks are  the so-called \textit{simple factors}, which make no use of the voters' beliefs. A simple factor is implemented by asking each agent which of the two outcomes she values more, either  $a$ or $b$. 
If the reported number of  supporters of $a$ is greater than some exogenously chosen threshold, then the factor selects the outcome $a$ in every state of its cell of the partition. Otherwise, the outcome $b$ is selected in every state of the  cell. For instance, recalling our group activity example of Subsection \ref{ToyEx},  a quasi-dictatorial factor can be implemented on the  subset of states $\{c,s\}$ as follows: ask each voter to choose between $B$ and $P$. In any of these two states of nature, the group will go to the beach if at least \textbf{six} of the ten friends prefer $B$ to $P$ (otherwise, they will play go to the amusement park).

\smallskip
Unlike the simple factors described above, the following three types of factors do make use of the voters' subjective beliefs.

\subsubsection{Quasi-dictatorial factors}\label{IntroQuasiDictFAct}
The second building block, dubbed \textit{quasi-dictatorial factor}, is implemented using an exogenous menu of (at least three) non-nested events whose union covers the associated cell of the partition.This factor selects $a$ ($b$) in every state of the cell if the number of agents who prefer $a$ to $b$ is greater (lower) than 1. If \textbf{exactly one agent}  reports preferring $a$ to $b$, then the unique supporter of $a$ is asked to choose the event in the menu that she finds most likely ($a$ is then selected in that event while $b$ is chosen in the complementary event).

 Recalling our example of Subsection \ref{ToyEx}, one can define 
 the following quasi-dictatorial factor on the subset of states $\{r,o,w\}$. Ask each group member to choose between movie marathon ($M$) or volleyball ($V$). If the number of supporters of $M$ is greater (lower) than one, then the group will watch movies (play volleyball) in each of the three states rainy, snowy, windy. In the case where there is exactly one supporter of $M$, this person is asked to report which of the three states $r,o,w$ she finds most likely ($M$ is then selected in this state, and $V$ is selected on each of the other two states).

Remarkably, even a quasi-dictatorial mechanism satisfies (strategy-proofness, range-unanimity and) anonymity: this is because the quasi-dictator (the unique supporter of $a$, when such an isolated agent exists) is chosen endogenously: each voter in fact assumes this role at different profiles of reported valuations.

\subsubsection{Dyadic factors}\label{IntroDyadFactor}
Our third  building block, which we refer to as \textit{dyadic factor}, requires \textbf{(i)} a partition of the associated cell into two events, say, $E $ and $F$, \textbf{(ii)} two integer thresholds $\underline{k}\leq \bar{k}$, and \textbf{(iii)} a non-decreasing function $H:\SN \times \SN \rightarrow \{0,1\}$, whose two arguments are natural numbers. If the reported number of supporters of $a$ is lower (greater) than $\underline{k}$ ($\bar{k}$) then the dyadic factor selects the outcome $b$ ($a$) in every state of the cell. Whenever the number of supporters of $a$ is between $\underline{k}$ and $\bar{k}$ (inclusively), the agents are asked to vote between \textbf{(I)} having $a$ on the event $E$ and $b$ on the event $F$, and \textbf{(II) }having  $b$ on the event $E$ and $a$ on the event $F$. The integer $m_a$ ($m_b$) then records the number of votes in favor of option (I) among supporters of $a$ ($b$); and the mechanism indeed selects option (I) if $H(m_a, m_b)=1$ ---otherwise option (II) is selected.\footnote{It may be worth noting that the map $H$ is akin to (but requires more constraints than) the voting by committees schemes of \cite{barbera1991voting} and \cite{barbera2005voting}.} It is shown that, if $\underline{k}< \bar{k}$, then our axioms require linear isoquants for the function $H$ (i.e., every vote, whether coming from a supporter of $a$ or from a supporter of $b$, must be given a weight of 1). This requirement does not apply when $\underline{k}= \bar{k}$, which means that the designer may then use weighted-majority voting rules,  assigning for instance a higher (lower) weight to the votes of supporters of $a$ ($b$).

For illustration, recalling our group activity example, one can implement the following dyadic factor on the subset of states $\{c,s\}$.  Ask each of the ten friends to choose between the beach ($B$) or the amusement park ($P$). If at least seven (at most three) people prefer $B$ to $P$, then the group will go the beach (park) in the event of either a sunny or a cloudy day. In the middle case (where four, five, or six agents prefer the beach) each of the ten friends is asked to vote between \textbf{[I]} the sub-act  yielding $B$ ($P$) in the cloudy (sunny) state   and \textbf{[II]} the sub-act  yielding $P$ ($B$) in the cloudy (sunny) state, with  option I selected if it gets at least six favorable votes (otherwise option II is selected). 

\subsubsection{Iso-filtering factors}\label{IntroIsoFiltFactor}
The fourth and most complex building block, called \textit{iso-filtering factor}, includes a sequence of votes (between pairs of sub-acts) that are contingent on the number $k$ of supporters of $a$ vs $b$, with specific constraints (described in Section \ref{Binarysection}) linking these sub-acts as $k$ varies between some exogenous thresholds $\underline{k}$ and $ \bar{k}$.\footnote{The pair of thresholds $\underline{k}, \bar{k}$ (used in dyadic and iso-filtering factors) is slightly reminiscent of the ``consent quotas'' introduced by \cite{samet2003between} in the context of group identification problems, where society uses a vote to determine whether individuals belong to a certain group. The resulting consent rules have recently been generalized by \cite{cho2023reaching}, whose framework allows for different levels of group membership.}   

For instance, in the group activity application of Subsection \ref{ToyEx}, an iso-filtering factor can be defined on the  subset of states $\{r,o,w\}$ as follows. Ask every group member  to express support for either $M$ or $V$. If the number of supporters of $M$ is at most  three (at least six), then the group will play volleyball  (watch movies) in each of the three states rainy, snowy, windy. If exactly four members support $M$ then each of the ten friends is asked to vote between  \textbf{[A]} the sub-act yielding $M$ ($V$) in state $r$ (in states $o$ and $w$) and  \textbf{[B]} the sub-act yielding $V$ ($M$) in states  $r$ and $w$ (in state $o$), with option  selected if and only if  at least one supporter of $V$  votes for it. In the case where exactly five members support $M$ then each of the ten friends is asked to vote between \textbf{[C]} the sub-act yielding $M$ ($V$) in states $r$ and $o$ (in state $w$) and \textbf{[D]} the sub-act yielding $V$ ($M$) in state $r$ (in states $o$ and $w$), with option C selected if and only if at least one supporter of $M$  votes for it.

Note that, as the numbers $k$ of supporters of $M$ increases from 4 to 5, state $o$ is ``filtered out'' of its binary status, since the range of outcomes it  yields when $k=5$ is $\{M\}$,  as opposed to $\{M,V\}$ when $k=4$. This is where the name ``filtering factor'' comes from. Moreover, as explained in Section \ref{Binarysection}, strategy-proofness requires in such a situation that the constant outcome selected (when $k=5$) be precisely the one whose number of supporters is increasing (which is $M$ in this example). This specific filtering constraint explains our use of the term ``iso-filtering'' to qualify this fourth factor.



\medskip
Our main result, which is stated in Theorem \ref{ThmChar}, asserts that \textit{a voting mechanism is anonymous, strategy-proof and range-unanimous if and only if every non-constant factor in its canonical factorization is a simple, quasi-dictatorial,  dyadic, or iso-filtering building block.} 
The proofs of the respective results are available in the appendix of Section \ref{appendix}.

\subsection{Related Literature}\label{Sectliterature}

As pervasive as uncertainty may be in real-life group decisions, it has not received much attention within the social choice literature. Regardless, the present work connects to a few strands of the literature on social choice problems. First, a closely related work  is that of \cite{bahel2020strategyproof}, whose main contribution is to identify strategy-proof social choice functions that are non-dictatorial and exploit the agents' beliefs (with sufficiently many states, these mechanisms can potentially use the preferences of all voters in the collective decision process). However, the main drawback of \cite{bahel2020strategyproof}'s  mechanisms is that they are not implementable in a democratic manner (since none of them are anonymous). Indeed, the definition of these mechanisms, called \textit{locally bilateral top selections}, requires the exogenous assignment of  ``control rights'' over states  to specific agents, who then dictate the social outcome in states assigned to them. By contrast, the strategy-proof voting mechanisms identified in the present work are implementable in a democratic society, since they satisfy anonymity and treat all voters the same way.

Second, a few authors, such as \cite{gibbard1977manipulation}, \cite{hylland1980strategy}, \cite{dutta2007strategy, dutta2008strategy},  have examined social choice under \textit{risk} (rather than uncertainty), that is, in the context where society must choose between lotteries (with  objective and publicly known probabilities). These papers do not deal with the notion of subjective beliefs and their solutions do not apply to the present framework. The distinguished mechanisms in these works are the so-called \textit{random dictatorships}, which select every agent's favorite outcome with some exogenous probability.  These random dictatorships are strategy-proof, but typically not anonymous (only the random dictatorship giving each agent the same exogenous probability of dictatorship is anonymous).

 Third, there exists a strand of literature focusing on the elicitation of an expert's beliefs about the likelihood of uncertain events in which she has no stake. 
The earliest elicitation mechanisms for this problem are the proper scoring rules \citep{mccarthy1956measures, savage1971elicitation}, which offer a trade-off between accuracy and incentive-compatibility. Other methods include \cite{finetti1974theory}'s promissory notes and \cite{ducharme1973intrasubject}'s reservation probability mechanism, of which \cite{karni2009mechanism} proposed a variant achieving both accuracy and strategy-proofness. The main differences between these works and the present paper lies in the fact that (i) our model features multiple agents (as opposed to a single expert), and (ii) these agents do have stakes ---expressed by their valuation functions-- in the outcomes generated by the social act.\footnote{In a work tangentially related to this strand, \cite{bahel2021strategy} examine  belief elicitation for a group of agents (who all agree on the ranking of two outcomes: collective success and collective failure) when the range of events where success may occur is exogenously constrained.} 

Fourth, a number of papers have linked group decision problems repeated over multiple time periods in order to improve the incentive-compatibility of  single-period mechanisms. This approach includes for instance \cite{casella2005storable}, \cite{jackson2007overcoming}, \cite{casella2008simple}, and \cite{hortala2012qualitative}. Although the mechanisms proposed in this strand are anonymous and Bayesian incentive-compatible, they are not strategy-proof   and require the designer to have statistical information about the agents preferences (which is neither needed nor available in the present analysis).


\section{The model}\label{SectionModel}
There is a finite set of agents $N=\left\{ 1,...,n\right\} $ with $n\geq 2,$
a finite set of states of nature $\Omega $ with $\left\vert \Omega
\right\vert \geq 2,$ and a finite set of outcomes $X$ with $\left\vert
X\right\vert \geq 2.$ Outcomes should be interpreted as public alternatives
(such as policies or allocations) that are of interest to all agents.
We call \textit{event} any subset   of $\Omega $. A  \textit{social act} (often  referred to simply as ``an act'') is a map $f\in
X^{\Omega }$.

\subsection{Preferences }\label{Prefs}
Each  agent $i$'s preference ordering $\succcurlyeq _{i}$ over acts
is assumed to be of the \textit{subjective expected utility} type: there exist a
valuation function $v_{i}:X\rightarrow \mathbb{R}$ (defined over outcomes) and a subjective
probability measure $p_{i}: 2^{\Omega}\rightarrow [0,1]$ (defined over events)   such that for all $f,g\in
X^{\Omega }$,
\begin{equation}\label{EqdefSEU}
f\succcurlyeq_{i}g\Leftrightarrow \mathbf{E}^{p_i}_{v_i}(f)\geq \mathbf{E}^{p_i}_{v_i}(g),
\end{equation}%
where   $\mathbf{E}^{p_i}_{v_i}(h):=\sum\limits_{\omega \in \Omega }p_{i}(\omega
)v_{i}(h(\omega ))$ for all $h\in X^{\Omega}$.

Note that we 
often omit curly brackets to alleviate notation: we will occasionally write $\omega $ instead of $\left\{ \omega \right\}$, $i$ instead of $\{i\}$, and so on.  Since the set of
acts is finite, neither the valuation function $v_{i}$ nor the subjective
probability measure $p_{i}$ representing the preference ordering $%
\succcurlyeq _{i}$ are determined uniquely; but this will not affect our analysis. We call $\mathbf{E}^{p_i}_{v_i}(h)$ the subjective expected utility associated with the act $h$ (under the valuation function $v_i$ and the belief $p_i$). 

For simplicity in the exposition, we rule out indifference between distinct acts.\footnote{Allowing for indifference further  complicates the analysis without bringing any important insights. The discussion of Section \ref{Concl} describes some tie-breaking rules that can be used to extend our results to the case where indifference between distinct acts is possible.} This implies that for any $(v_{i},p_{i})$ representing $\succcurlyeq_{i}$, (a) $v_{i}$ is injective and (b) $p_{i}$ is injective: for all $%
A,B\subseteq \Omega ,$ $p_{i}(A)=p_{i}(B)\Rightarrow
A=B.$ Because $p_{i}(\emptyset )=0$, it follows from (ii) that $%
p_{i}(\omega )>0$ for all $\omega \in \Omega .$ We further assume, without
loss of generality, that $v_{i}$ is normalized: $\min_{X}v_{i}=0<%
\max_{X}v_{i}=1.$ We denote by $\mathcal{V}$ the set of normalized injective
valuation functions $v_{i}$ and by $\mathcal{P}$ the set of (necessarily
positive) injective measures $p_{i}$, which we call \textit{beliefs. }The
domain of preferences is  the set of  pairs $(v_i,p_i)$ that generate a strict
ordering of the set of acts, that is,
\begin{equation}\label{Eqdomain}
\mathcal{D}=\left\{ (v_{i},p_{i})\in \mathcal{V}\times \mathcal{P}%
:\mathbf{E}^{p_i}_{v_i}(f) \neq \mathbf{E}^{p_i}_{v_i}(g), \forall f,g\in
X^{\Omega }  \text{ s.t. } f\neq g\right\} .
\end{equation}

\subsection{Feasibility constraint and social choice functions}\label{Feasibility}
In every state of nature $\omega \in \Omega$, a subset $%
X_{\omega }\subseteq X$ of outcomes are available. 
An act is \textit{feasible }if and only if it belongs to the set $%
\times _{\omega \in \Omega }X_{\omega }$, that is to say,\footnote{
Obviously, our model of constrained social choice includes as a particular case the ``unconstrained model'' where $X_{\omega}=X$ for all $\omega\in \Omega$, that is to say, the case where all outcomes in $X$ are always available, regardless of which state of nature materializes.} a feasible act yields in any given state of nature an outcome that is available in this state.

 A \textit{social choice function} (or SCF) is a mapping $\varphi :%
\mathcal{D}^{N}\rightarrow \times _{\omega \in \Omega }X_{\omega }$ defined 
from the set  $\mathcal{D}^{N}$ (where $\mathcal{D}$ comes from (\ref{Eqdomain})) to the set of feasible acts $\times _{\omega \in
\Omega }X_{\omega }$.  
We denote the ordered list $((v_{1},p_{1}),$ $%
...,(v_{n},p_{n}))\in \mathcal{D}^{N}$ by $(v,p).$ 

A SCF $\varphi $ may in principle choose different acts for profiles $%
(v,p)$ and $(v^{\prime },p^{\prime })$ that represent the same profile of
preferences $(\succcurlyeq _{1},...,\succcurlyeq _{n})$; but of course the
requirement of strategy-proofness (defined below) will rule this out. Hence, with a
slight abuse of terminology, we  may call every $(v,p)\in \mathcal{D}^{N}$
a \textit{preference profile}. We call $v=(v_{1},...,v_{n})\in \mathcal{V}%
^{N}$ a \textit{valuation profile} and $p=(p_{1},...,p_{n})\in \mathcal{P}%
^{N}$ a \textit{belief profile. }For every preference profile $(v,p)\in
\mathcal{D}^{N}$ and every $\omega \in \Omega $, we denote by $\varphi
(v,p;\omega )$ the feasible outcome chosen by the act $\varphi (v,p)$ in state $%
\omega $. Finally, for any given state $\omega\in \Omega$, let   $R^\varphi_\omega: =\{\varphi
(v,p;\omega ),   (v,p)\in \mathcal{D}^{N}\}$  denote  the set of outcomes selected by $\varphi$ in $\omega$; and write $R^\varphi: =\{f\in X^\Omega ~\vert~ f(\omega)\in R^\varphi_\omega, \forall \omega\in \Omega\}$.

\subsection{Social planner and basic axioms}\label{planner}
Consider a social planner in charge of designing a social choice function (or voting mechanism) that will pick a social act for the group at  any possible preference profile reported by its members. 

In a democratic society, the respective votes of distinct agents should be treated in the same way, regardless of their names, labels, or social characteristics. This fairness requirement is formalized by the anonymity axiom, which is defined as follows. 
Let $\Pi(N)$  denote the set of permutations of $N$. Given  any  $\sigma\in \Pi(N)$ and any preference profile  $(v,p)\in \mathcal{D}^N$, let  $(\sigma v, \sigma p)   \in \mathcal{D}^N$ be the preference profile defined by $(\sigma v)_i=v_{\sigma(i)}$ and $(\sigma p)_i=p_{\sigma(i)}$, for all $i\in N$. In other words,  $(\sigma v, \sigma p)$ obtains  by swapping the respective preferences reported in the profile $(v,p)$ as suggested by the permutation $\sigma$. 

A SCF $\varphi $ is \textit{anonymous} if, for all $ \sigma \in \Pi(N)$ and all $ (v,p)\in \mathcal{D}^{N}$, we have 
\begin{equation}\label{Eqanonym}
\varphi (\sigma v, \sigma p) =\varphi(v,p).
\end{equation}

Under anonymity, only the set of reported preferences (rather than the agents' names or characteristics) matters in determining the social act. This means that questions of the type ``which agent reported that s/he finds state 1 more likely than state 2?'') become irrelevant.  Indeed, 
remark from (\ref{Eqanonym}) that, given such a SCF, agents may well report their preferences anonymously; and still society will  be able to use the mechanism to select a social act at each and every (anonymous) preference profile that may be reported. Anonymity is the foundation of  democratic social choice.

Moreover, since preferences are private information, it is important that the SCF
used by the social planner induce  agents to report their preferences
truthfully. This incentive-compatibility requirement  is  captured  in our  model by the strategy-proofness axiom, which is formally defined as follows.

\medskip

Write $v_{-i}$ and $p_{-i}$ for the valuation
and belief sub-profiles obtained by deleting $v_{i}$ from $v$ and $p_{i}$
from $p$, respectively. A SCF $\varphi $ is \textit{strategy-proof} if, for
all $i\in N,$ all $(v,p)\in \mathcal{D}^{N},$ and all $(v_{i}^{\prime
},p_{i}^{\prime })\in \mathcal{D},$%
\begin{equation}\label{defSP}
\mathbf{E}_{v_{i}}^{p_{i}}(\varphi (v,p))\geq \mathbf{E}_{v_{i}}^{p_{i}}(\varphi
((v_{i}^{\prime },v_{-i}),(p_{i}^{\prime },p_{-i}))).
\end{equation}%
The well-known axiom of strategy-proofness means that distorting one's preference ---by
misrepresenting one's valuation function or one's beliefs-- is never
profitable, given the  fixed preferences reported by the other agents.
At every
profile $(v,p)$ and for every agent $i$, the truthful report $(v_i,p_i)$ is a weakly
dominant strategy in the preference revelation game generated by $\varphi $
at $(v,p).$



We now introduce a third axiom, which can be viewed as a minimal efficiency and consistency requirement. Given a valuation function $v_i\in \mathcal{V}$ and a subset of outcomes $A\subseteq X$, let $\tau_A(v_i):=\argmax\limits_{x\in A} v_i(x)$ be the (unique) maximizer of $v_i$ on $A$.

 We say that a SCF $\varphi $ is \textit{range-unanimous} if, for  all $f\in R^\varphi$ and all $(v,p)\in \mathcal{D}^N$, we have:
 \begin{equation}\label{Eqweakunanim}
\left[ f(\omega)=\tau_{R^\varphi_\omega}(v_i), \forall i\in N, \forall \omega \in \Omega \right]\Rightarrow [\varphi(v,p) =f ]
\end{equation}
 
In  words, range-unanimity means that  an act $f$ should be selected by the SCF $\varphi$ whenever it is every single agent's favorite (feasible) act within $R^\varphi$.  Moreover, one can  see from (\ref{Eqweakunanim}) that range-unanimity is weaker than   
 unanimity [\textit{i.e.}, the requirement that an act $f\in  \times_{\omega\in \Omega} X_\omega$   should be selected whenever all agents report $f$ as their favorite feasible act].
 
Remark  that unanimity is a very mild requirement: it does not impose any restriction on the selected  act at preference profiles where the agents do not all report the same favorite (feasible) act. As already pointed out,  our axiom range-unanimity is even milder, since it imposes restrictions on the social choice derived from $\varphi$ only if the common favorite act belongs to $R^\varphi$. It is important to note that range-unanimity is a natural requirement precisely because the designer does not exogenously impose the set of acts on which she will apply the restricted unanimity condition: rather, that restricted set of acts ($R^\varphi$) is endogenously determined by the mechanism $\varphi$ itself.

Our objective in this work is to describe the structure of the class of SCF satisfying anonymity, strategy-proofness and range-unanimity.



\section{Factorization and binary factors}
\label{binary factors} 

The objective of this section is to show that any SCF satisfying our three axioms can be decomposed into ``sub-mechanisms'' satisfying some specific properties (such as having a limited range). This decomposition will then be used to characterize the family of SCFs satisfying our three axioms.

We first introduce  additional notation and terminology.
Given a belief $p_i\in \mathcal{P}_i$ and a nonempty event $F\in 2^\Omega$, we write $p_{i}$ to refer to the associated \textit{conditional belief} on $F$, that is to say, $p_{i/F}(E):=\frac{p_i(E\cap F)}{p_i(F)}$, for all $E\subseteq F$. Furthermore, given any belief profile $p\in \mathcal{P}^N$, let $p_{/F}:=(p_{i/F})_{i\in N}$ be the associated profile of conditional beliefs. We use the shorthand $\mathcal{P}_{/F}$ to refer to the set of conditional beliefs on $F$.

Given a partition of $\Omega$ into nonempty events $\Omega_1,\ldots, \Omega_K$ (where $K\geq 2$) and an act $f\in X^\Omega$, we use the notation $f_{\Omega_k}=(f(\omega))_{\omega\in \Omega_k}$ to refer to the sub-act induced by $f$ on the event $\Omega_k$ (for all $k=1,\ldots,K$); and we will often  write $f=f_{\Omega_1}\oplus\ldots \oplus f_{\Omega_K}$.
We will also write $\varphi_{\Omega_k}(v,p)=(\varphi(v,p))_{\Omega_k}$, 
for any given SCF $\varphi$ and any $(v,p)\in \mathcal{D}^N$.

\medskip
 We say that a SCF $\varphi :\mathcal{D}^{N}\rightarrow \times_{\omega \in \Omega }X_{\omega }$ admits  a \textit{non-trivial factorization} if there exist an integer $K\geq 2$, a partition of $\Omega$ into nonempty events $\Omega_1,\ldots, \Omega_K$, and $K$ functions $\varphi_k: \mathcal{V}^N\times (\mathcal{P}_{/\Omega_k})^N\rightarrow \times _{\omega \in \Omega_k}X_{\omega }$ (for $k=1,\ldots,K$) such that 
 \begin{equation}\label{EqDecomp}
     \varphi(v,p)= \varphi_{1} (v,p_{/\Omega_1})\oplus\ldots \oplus \varphi_{K} (v,p_{/\Omega_K}), ~\forall (v,p)\in \mathcal{D}^N.
 \end{equation}

It is important to remark from (\ref{EqDecomp}) that the sub-act selected by $\varphi$ on each event $\Omega_k$ may vary only with (the valuations and)   the conditional beliefs  $p_{/\Omega_K}$: the rest of the information contained in the  belief profile $p$ is irrelevant to the choice of this sub-act on $\Omega_k$. Hence, each function $\varphi_{\Omega_k}$ is itself a \textit{bona fide} SCF, defined using the reduced state space $\Omega_k$ (for $k=1,\ldots,K$). If (\ref{EqDecomp}) holds, we call each of these sub-SCFs  $\varphi_{k}$  a  \textit{factor}  of  $\varphi$.  It is also worth noting from (\ref{EqDecomp}) that each factor must always choose a feasible sub-act, that is to say, 
$\varphi_{k}(v,p_{/\Omega_k})\in \times_{\omega \in \Omega_k} X_{\omega }$ for all $(v,p)\in \mathcal{D}^N$.


Of course, a SCF  $\varphi$ may in general admit multiple factorizations.  
However, as explained below,   one particular factorization always stands out.

A non-trivial  factorization [as described in (\ref{EqDecomp})] is called \textit{canonical} if each of  the factors $\varphi_{k}$ admits only a trivial factorization; and in that case we  refer to each $\varphi_{k}$ as  a \textit{prime factor} of $\varphi$.
  
  A SCF $\varphi$ that does not admit a non-trivial factorization  has a \textit{trivial} canonical factorization exhibiting a unique prime factor (which is $\varphi$ itself). 
Using this observation and the definition given by  (\ref{EqDecomp}), it is easy to see that every SCF $\varphi :\mathcal{D}^{N}\rightarrow \times_{\omega \in \Omega }X_{\omega }$ admits a unique canonical factorization into  prime factors.

We say that a factor $\varphi_{\Omega_k}$ of a SCF  $\varphi_{\Omega_k}$ is (I) 
\textit{binary} if there exist distinct $a,b\in X$ such that $\varphi_{\Omega_k} (v,p_{/\Omega_k})\in \{a,b\}^{\Omega_k}$, for all $(v,p)\in \mathcal{D}^N$; (II) \textit{constant} if there exists $c\in X$ such that $\varphi_{\Omega_k} (v,p_{/\Omega_k})\in \{c\}^{\Omega_k}$, for all $(v,p)\in \mathcal{D}^N$. Finally, a \textit{non-constant factor} is simply a factor $\varphi_k$ that is not constant.

\medskip


\medskip We are now ready to state  an important consequence of our  axioms. The following lemma and proposition essentially state that, together, our three properties lead to a canonical factorization into factors that are either binary or constant.

\begin{lemma}\label{Lembinarydecomp}
    Consider a SCF $\varphi :\mathcal{D}^{N}\rightarrow \times_{\omega \in \Omega }X_{\omega }$ that is anonymous, strategy-proof and range-unanimous. Moreover, assume that  $|R^\varphi_\omega|\geq 2$, for all $\omega\in \Omega$ (there is no state where the outcome selected by $\varphi$ is independent of the reported preferences).    
Then $\varphi$ has a factorization into binary factors.
\end{lemma}

Allowing then the possibility of states $\omega$ where the social outcome selected is independent of the reported preferences (i.e., $|R^\varphi_\omega|=1$), one easily gets the following result as a corollary to Lemma \ref{Lembinarydecomp} (the proof of Proposition \ref{Propbinarydecomp}  is omitted).

\begin{proposition}\label{Propbinarydecomp} \textbf{\textit{Factorization}}\\
    Consider a SCF $\varphi :\mathcal{D}^{N}\rightarrow \times_{\omega \in \Omega }X_{\omega }$ that is anonymous, strategy-proof and range-unanimous. Then  $\varphi$ admits a factorization where every factor is either binary or constant. Moreover, each  factor of $\varphi$ is itself anonymous, range-unanimous and strategy-proof.
\end{proposition}

Note that the sub-act chosen by a constant factor (on the associated event) does not vary at all with the preference profile. However, as we shall see, the structure of binary factors can be quite complex; and in general  these sub-SCFs offer a wide range of possibilities in terms of the sub-acts that society may  select as the reported preference profile varies. 

Exploiting the result of Proposition \ref{Propbinarydecomp}, we may now turn to the analysis of binary social choice functions: this is the objective of the next section.



\section{Decomposition of a binary SCF}\label{Binarysection}

In this section,  we consider the special case of a \textbf{binary} social choice function (BSCF), that is, a SCF  which, in every state $\omega \in \Omega$, selects  one of two prespecified outcomes. Recall that  $N$ is the set of agents, $X$ is the set of outcomes, and $\Omega$  is the set of states of nature. We fix once and for all the two prespecified outcomes  $a,b\in \cap_{\omega\in \Tilde{\Omega}}X_{\omega}$ (with $a\neq b$) used in this section.

\subsection{Preliminaries}
For all nonempty $\tilde{\Omega}\in 2^\Omega$,  let $\mathcal{P}(\tilde{\Omega})$ be the collection of (injective) probability measures defined on $\tilde{\Omega}$, and denote by $\mathcal{D}(\tilde{\Omega})$  the set containing all  pairs $(v_i,p_i)\in \mathcal{V}\times \mathcal{P}(\tilde{\Omega})$ generating  a strict SEU ranking over the set  $X^{\tilde{\Omega}}$.
 
\subsubsection{Valuation sections and vote between pairs of events}
Let us introduce some additional notation. For all $v\in \mathcal{V}^N$, define
\begin{align}\label{EqNa}
  N^v_a &:= \{i\in N: v_i(a)>v_i(b)\},~ n^v_a:=| N^v_a|; \\ \nonumber
  N^v_b &:= \{i\in N: v_i(a)<v_i(b)\}, ~ n^v_b:=| N^v_b|.
\end{align}
Since every $v_i\in \mathcal{V}$ is injective, note from (\ref{EqNa}) that $N^v_b=N\setminus N^v_a$ always holds (for any $v\in\mathcal{V}^N$); and hence $n^v_a+n^v_b=n$. For any  $k=0,1,\ldots, n$,  denote by $\mathcal{V}_k^N:=\{v\in \mathcal{V}^N: n^v_a=k\}$ the set of valuation profiles where exactly $k$ agents  prefer $a$ to $b$. It is easy to see that we have the disjoint union $\mathcal{V}^N=\bigcup\limits_{k=0}^n  \mathcal{V}^N_k$. Given any $k\in \{0,1,\ldots,n\}$, we refer to $\mathcal{V}^N_k$ as a \textit{section of valuation profiles} (or simply a valuation section).

Considering any $E,F\in 2^{\tilde{\Omega}}$, $S\in 2^N$, and $p\in \mathcal{P}(\tilde{\Omega})$, we will use the shorthand $$\eta_p(S, E|F):= |\{i\in S: p_i(E)>p_i(F)\}|.$$
In words, for any disjoint $E$ and $F$,  the integer $\eta_p(S, E|F)$ records the number of voters in $S$ who prefer having their top outcome in the event $E$ (and the other outcome in $F$) to having their  top outcome in  $F$ (and the other outcome in $E$). 

Moreover, for any binary sub-act $f\in \{a,b\}^{\tilde{\Omega}}$,  define $f^{x}:=\{\omega\in \tilde{\Omega}: f(\omega)=x\}$ (for  $x=a,b $). Since a binary sub-act  $f\in \{a,b\}^{\tilde{\Omega}}$ is fully characterized by  the event  $f^a$, it  will often be convenient to use the notation $[f^a]_a^{\Tilde{\Omega}}$  to refer to any binary sub-act $f$.

\subsubsection{Dipartitions and filters}
Let $\Tilde{\Omega}\in 2^\Omega$ be such that  $|\tilde{\Omega}|\geq 3$.
We call  \textit{dipartition} of $\Tilde{\Omega}$ any ordered collection of the form 
 \[\mathcal{C}=\{(E_1,F_1), \ldots,(E_{M},F_{M}), (G_a, G_b)\},\]
 where  $M\geq 1$ is an integer and  $E_1,F_1,\ldots, E_{M},F_{M}, G_a,G_b\in 2^{\Tilde{\Omega}}$ are  pairwise disjoint events  satisfying $E_1\cup F_1\cup\ldots\cup E_{M}\cup F_{M}\cup  G_a \cup G_b=\Tilde{\Omega}$.   Note that we require $E_m,F_m\neq \emptyset$ for all $m=1,\ldots, M$; and $G_a\cup G_b\neq \emptyset$ (we allow  at most one of the events $G_a,G_b$ to be empty). 

 The intuition for a dipartition is as follows: for a fixed valuation section $\mathcal{V}_k^N$, the mechanism designer can use each pair of events $(E_m,F_m)$, with  $m=1,\ldots,M$, to offer voters a choice between two sub-acts ($aE_m\oplus bF_m$ and $bE_m\oplus aF_m$), whereas the outcome $a$ ($b$) will invariably be selected in the event $G_a$ ($G_b$), regardless of the voters' beliefs. 

 Letting $\mathcal{C}=\{(E_1,F_1), \ldots,(E_{M},F_{M}), (G_a, G_b)\}$ be a dipartition of $\Tilde{\Omega}$, define the binary relation $R_{\mathcal{C}}$ on $\tilde{\Omega}$ as follows: $\forall \omega,\omega'\in \tilde{\Omega}$,  we have $\omega R_{\mathcal{C}} \omega'$ if and only if there exists $m\in \{1,\ldots,M\}$ such that $\omega, \omega' \in E_m\cup F_m$.

 
\noindent
A sequence of dipartitions of $\tilde{\Omega}$, defined by $\mathcal{C}^k=\{(E_1^k,F_1^k), \ldots,(E_{M_k}^k,F_{M_k}^k), (G^k_a, G^k_b)\}$ (with $k=\underline{k},\ldots,\Bar{k}$),  will be called a \textit{filter} if \textbf{(i)} the transitive closure of its union relation $\bigcup\limits_{k=\underline{k}  }^{\bar{k}}R_{\mathcal{C}^k}$ has full graph $\tilde{\Omega}\times \tilde{\Omega}$ and \textbf{(ii)} for all $k= \underline{k},\ldots, \Bar{k}-1$, $m=1,\ldots,M_{k}$ and $m'=1,\ldots,M_{k+1}$, we have
\begin{align}
 \label{EqFilterdef1}
     [(E_{m}^{k},F_{m}^{k}) \notin \mathcal{C}^{k+1} ] &\Rightarrow [F^{k}_m\subseteq G^{k+1}_a];\\ \label{EqFilterdef2}
     [(E_{m'}^{k+1},F_{m'}^{k+1}) \notin \mathcal{C}^{k} ] &\Rightarrow [E^{k+1}_{m'}\subseteq G^{k}_b].
\end{align}

In other words, a sequence of dipartitions $\{\mathcal{C}^k\}_{k=\underline{k},\ldots,\Bar{k}}$ is a filter if (I)  for all states $\omega,\omega' \in \tilde{\Omega}$, there exist a sequence of states $\{\omega_t\}_{t=1,\ldots,T}$ and a sequence of  integers $\{k_t\}_{t=1,\ldots,T+1}\in \{\underline{k},\ldots,\bar{k}\}^{T+1}$ s.t. $\omega R_{\mathcal{C}^{k_1}} \omega_1 R_{\mathcal{C}^{k_2}} \ldots R_{\mathcal{C}^{k_T}} \omega_{T}R_{\mathcal{C}^{k_{T+1}}}\omega'$; and (II) a pair of events $(E,F)$ that is removed from the voters' menu  [as the number of supporters of some outcome increases] must have at least one event that is ``filtered out'' of its binary status.\footnote{ Remark that (\ref{EqFilterdef1}) and (\ref{EqFilterdef2}) essentially describe the same requirement: these two conditions are mirror images of one another as  the roles of $a$ and $b$ in the analysis are reversed. This is why we can summarize them both in the statement (II).} 

\smallskip
 The following example illustrates  the notion of filter and its two properties (as well as their respective violations).

\begin{example}\label{Exdpart1} Suppose that $\Tilde{\Omega}=\{\omega_1,\omega_2, \ldots,\omega_7\}$. Consider the following sequence of dipartitions of $\Tilde{\Omega}$ (where $\underline{k}=1$ and $\bar{k}=3$): \begin{align*}
    \mathcal{C}^1= &\{(\omega_1, \omega_4),(\omega_2,\omega_3),(\emptyset, \omega_5\omega_6\omega_7)\};\\\mathcal{C}^2=&\{(\omega_5\omega_6,\omega_1 ),(\omega_2,\omega_3)(\omega_4, \omega_7)\};\\
    \mathcal{C}^3= &\{(\omega_7, \omega_2\omega_5), (\omega_1\omega_3\omega_4\omega_6,\emptyset)\}.
\end{align*} 
It is not difficult to check that $\{\mathcal{C}^1, \mathcal{C}^2, \mathcal{C}^3\}$ is a filter. 
Indeed, note  that we have (A) $\omega R_{\mathcal{C}^1} \omega' \Leftrightarrow [\omega,\omega'\in \{\omega_1,\omega_4\} \mbox{ or } \omega,\omega'\in \{\omega_2,\omega_3\}] $; (B) $\omega R_{\mathcal{C}^2} \omega'$  if and only if  $ [\omega,\omega'\in \{\omega_1,\omega_5,\omega_6 \} \mbox{ or } \omega,\omega'\in \{\omega_2,\omega_3\}]$; and (C) $\omega R_{\mathcal{C}^2} \omega'$  if and only if $ \omega,\omega'\in \{\omega_2,\omega_5, \omega_7\}$. Combining (A)-(C), we conclude that the transitive closure of $R_{\mathcal{C}^1}\cup R_{\mathcal{C}^2}\cup R_{\mathcal{C}^3}$ has full graph. Remark in particular that we \textbf{do not have} $\omega_1 (R_{\mathcal{C}^1}\cup R_{\mathcal{C}^2}\cup R_{\mathcal{C}^3})\omega_7$. However, $\omega_1$ and $\omega_7$ are indeed connected by the transitive closure of this union relation: $\omega_1 R_{\mathcal{C}^2} \omega_5 R_{\mathcal{C}^3} \omega_7$. Moreover, remark that  (\ref{EqFilterdef1}) and (\ref{EqFilterdef2}) are also satisfied: for instance, $(\omega_1,\omega_4)\in \mathcal{C}^1\setminus \mathcal{C}^2$, $(\omega_5\omega_6,\omega_1)\in \mathcal{C}^2\setminus \mathcal{C}^1$; and we indeed have both $\omega_4\in G^2_a$ and $\omega_5\omega_6\subseteq G_b^1$.

As a first counter-example, observe that the sequence of dipartitions $\{\mathcal{C}^1, \mathcal{C}^2\}$ is not a filter, since the   states $\omega_1$ and $\omega_7$ (in particular) are not connected via the transitive closure of $R_{\mathcal{C}^1}\cup R_{\mathcal{C}^2}$.

As a second counter-example,  the sequence of dipartitions $\{\mathcal{C}^1, \mathcal{C}^2, \bar{\mathcal{C}}^3\}$, with $\bar{\mathcal{C}}^3=\{(\omega_7, \omega_3\omega_5), (\omega_1\omega_2\omega_4\omega_6,\emptyset)\}$ is not a filter because the  $(\omega_2,\omega_3)\in \mathcal{C}^2\setminus \bar{\mathcal{C}}^3$, \textbf{but} $\omega_3\notin \bar{G}^3_a=\omega_1\omega_2\omega_4\omega_6$.
\end{example}

\subsection{Building blocks}\label{SubBuilding Blocks}

We fix  $\tilde{\Omega}\in 2^\Omega$ for the remainder of Section \ref{Binarysection}; and we  may thus use the shorthand $\tilde{\mathcal{D}}:=\mathcal{D}(\tilde{\Omega})$.
To further alleviate  notation, we will also write  $[f^a]_a$ (instead of the more cumbersome $[f^a]^{\Tilde{\Omega}}_a$) to denote any act $f\in \{a,b\}^{\Tilde{\Omega}}$. Recall from (\ref{EqNa}) that $N_a^v$ is the set of players who value $a$ more than $b$; and $n_a^v=|N^v_a|$ gives the number of such players.

The following lines define four distinct types of  $\tilde{\Omega}$-BSCF. These types will constitute the building blocks of our characterization result.

\subsubsection{Simple factors}\label{SimpleFactor}
As suggested by its name, this first type is easy to define:  the chosen act in fact does not depend  the voters'  beliefs (it may vary only with the valuation profile).
For the formal definition of a \textit{simple factor}, pick a quota $\bar{k}\in \{1,\ldots, n\}$.  For any $(v,p)\in \tilde{\mathcal{D}}^N$, select the social act 
\begin{equation}\label{EqSimpleFact}
  \varphi (v,p)=\left \{ \begin{array}{cl}
                             [\emptyset]_a & \mbox{ if } n^v_a<\bar{k}; \\ \\
\left[\tilde{\Omega}\right]_a &\mbox{ if }  n^v_a\geq \bar{k}.  \end{array} \right.
  \end{equation}
  
In words, (\ref{EqSimpleFact}) means that the sub-act selected on $\tilde{\Omega}$ is always constant, either $b\tilde{\Omega}$ or $a\tilde{\Omega}$, with the latter sub-act selected if and only if the number of supporter of $a$ vs $b$ is at least equal to the exogenously chosen quota $\bar{k}$. 
It is easy to see from  (\ref{EqSimpleFact})  that a simple factor is an anonymous, range-unanimous and strategy-proof mechanism defined on $\tilde{\Omega}$. 

Recalling our group activity application, the reader may refer to Subsection \ref{IntroSimpleFactor} for an example of simple factor defined on $\tilde{\Omega}=\{c,s\}$, with a quota of $\bar{k}=6$ needed for the beach to be chosen by the group over the amusement park.

\subsubsection{Quasi-dictatorial factors}\label{QDictFactor}

Assume for this building block that $|\tilde{\Omega}|\geq 3$. For the formal definition of a \textit{quasi-dictatorial factor}, fix a collection $\mathcal{E}$ of at least three non-nested events covering  $\tilde{\Omega}$.  If $n_a^v=0$ then select  the constant act $b  \tilde{\Omega}$; if $n_a^v\geq 2$ then select the constant act $a \tilde{\Omega}$; otherwise we have $n_a^v=1$ and the unique supporter of $a$ is asked to choose her best event in  the menu $\mathcal{E}$  ($a$ is chosen in that best event, while $b$ is chosen in the complementary event). This means that, for all $(v,p)\in (\mathcal{D}(\tilde{\Omega}))^N$,
\begin{equation}\label{EqQDictFact}
  \varphi (v,p)=\left \{ \begin{array}{cl}
                             [\emptyset]_a & \mbox{ if } n^v_a=0; \\
                             \left[\argmax\limits_{E\in \mathcal{E}}p_i(E)\right]_a &\mbox{ if } N^v_a=\{i\};\\
 \left[\tilde{\Omega}\right]_a &\mbox{ if } n^v_a\geq 2.  \end{array} \right.
  \end{equation}

Remark from (\ref{EqQDictFact}) that a quasi-dictatorial factor is anonymous! Indeed, even though a single player dictates society's choice between the events in $\mathcal{E}$, this quasi-dictator (the unique supporter of $a$, when such an isolated player exists) is chosen endogenously. Note in fact that every agent $i$ gets to assume this role of quasi-dictator as we vary the valuation profile $v$. Using (\ref{EqQDictFact}), it is not difficult to check that a quasi-dictatorial factor is strategy-proof and range-unanimous.

For illustration, see Subsection \ref{IntroQuasiDictFAct} where  the collection of non-nested events $\mathcal{E}=\{\{r\}, \{o\}, \{w\}\}$ is used to define a quasi-dictatorial factor  on $\tilde{\Omega}=\{r,o,w\}$.

\subsubsection{Dyadic factors}\label{DyadFactor}
Assume that $|\tilde{\Omega}|\geq 2$; and let $\{E,F\}$ be a  partition of  $\tilde{\Omega}$ (with $E,F\neq \emptyset$).  Pick two thresholds  $\underline{k}, \overline{k}\in \{1,\ldots, n-1\}$ such that $\underline{k}\leq \overline{k}$. Then select an exogenous non-decreasing function $H:\SN \times \SN \rightarrow \{0,1\}$, with $H$ having linear isoquants  if  $\underline{k}<\overline{k}$, that is to say, $H(m_1,m_2)=H(m'_1,m'_2)$ for all $(m_1,m_2),(m'_1,m'_2)$ such that $m_1+m_2=m_1'+m_2'$.

The  associated \textit{dyadic factor} can be described follows.
If  $n^v_a<\underline{k}$ ($n^v_a>\bar{k}$) then   the outcome $b$ ($a$) is selected in every state of the cell. Whenever  $n^v_a\in \{\underline{k}, \ldots, \bar{k}\}$, the agents are  asked to vote between the sub-acts $\left[E\right]_a$ and $\left[F\right]_a$. Note that each agent would like their preferred outcome (either $a$ or $b$) to be selected in the event they find more likely (either $E$ or $F$); and this rule guides their vote as they choose between these two sub-acts. In other words, the  numbers of supporters of the respective outcomes $a$ and $b$ who prefer the first sub-act are precisely $\eta_p(N^v_a, E|F)$ and $\eta_p(N^v_b, F|E)$). The sub-act $\left[E\right]_a$ will then be selected if and only if $H(\eta_p(N^v_a, E|F), \eta_p(N^v_b, F|E))=1$.

Summarizing the above description, the formal definition of this dyadic factor is as follows:  
 for all $(v,p)\in \tilde{\mathcal{D}}^N$, 
\begin{equation}\label{EqQDyadFact}
  \varphi (v,p)=\left \{ \begin{array}{cl}
                           [\emptyset]_a & \mbox{ if } n^v_a\leq  \underline{k}-1; \\
                             \left[E\right]_a  &\mbox{ if } \underline{k}\leq  n^v_a\leq \overline{k}  \mbox{ and } H(\eta_p(N^v_a, E|F), \eta_p(N^v_b, F|E) )=1;\\
                             \left[F\right]_a &\mbox{ if } \underline{k}\leq  n^v_a\leq \overline{k}  \mbox{ and } H(\eta_p(N^v_a, E|F), \eta_p(N^v_b, F|E) )=0;\\
                             \left[\tilde{\Omega}\right]_a & \mbox{ if } n^v_a\geq  \overline{k}  +1.
                             \end{array} \right.
  \end{equation}
Note from (\ref{EqQDyadFact}) that a dyadic factor is anonymous and range-unanimous. Moreover, its strategy-proofness is guaranteed by the  facts that $H$ is non-decreasing and (if $\underline{k}<\overline{k}$) $H$ exhibits linear isoquants ---a violation of either of these conditions would cause the mechanism defined in (\ref{EqQDyadFact}) to fail strategy-proofness.

Interestingly, the requirement of linear isoquants means that, whenever $\underline{k}<\overline{k}$, the associated dyadic factor gives the same weight to all votes (whether they come from supporters of $a$ or from supporters of $b$). The reader may refer to Subsection \ref{IntroDyadFactor} for such an example of dyadic factor with equal weights for both groups of supporters. 

On the other hand, if $\underline{k}=\overline{k}$, this requirement of linear isoquants is not needed (and  the fact that $H$ is non-decreasing suffices for strategy-proofness). Thus, the designer may  use a weighted-majority  rule (whenever $ n^v_a=\underline{k}=\bar{k}$), assigning for instance a higher (lower) weight to the votes of supporters of $a$ ($b$), in order to choose between the two sub-acts  $\left[E\right]_a$ and  $\left[E\right]_b$. In the most extreme version of this weight inequality, when $n^v_a=\underline{k}=\bar{k}$, the designer may well choose to select $\left[E\right]_a$ over $\left[E\right]_b$ if and only if $\eta_p(N^v_a, E|F)\geq \bar{t}$, where  $\bar{t}\in \{1,\ldots,\bar{k}\}$ is some exogenously selected threshold ---in this particular case the function $H$ is constant with respect to its second argument.

\subsubsection{Filtering factors}\label{FilteringFactor} The fourth type of binary factor requires $n\geq 3$ and $|\Tilde{\Omega}|\geq 3$. 
Since this is the most complex of our four building blocks, it may be useful to start with an example involving only three states.
\begin{example}\label{ExSimplefiltering}
    Suppose that $\tilde{\Omega}=\{\omega_1,\omega_2, \omega_3\}$, $N=\{1,\ldots,n\}$, with $n\geq 3$.
\medskip\noindent
Let us then define the SCF $\varphi$ as follows:  for all  $(v,p)$, 
\begin{equation}\label{EqFiltEx}
  \varphi (v,p)=\left \{ \begin{array}{cl}
                            [\emptyset]_a& \mbox{ if } n^v_a=0; \\
                            \left[\omega_1\right]_a &\mbox{ if } n^v_a=1 \mbox{ and } \eta_p(N_b^v, \omega_2|\omega_1)\geq 1;\\
\left[\omega_2\right]_a  &\mbox{ if } n^v_a=1 \mbox{ and } \eta_p(N_b^v, \omega_2|\omega_1)=0;\\
  \left[\omega_1 \omega_2\right]_a  & \mbox{ if }  n^v_a=2 \mbox{ and } \mu_a^v(p; \omega_2,\omega_3)\geq 1;\\
 \left[\omega_1 \omega_3\right]_a & \mbox{ if }  n^v_a=2 \mbox{ and } \mu_a^v(p; \omega_2,\omega_3)=0 ;
                              \\ \left[\tilde{\Omega}\right]_a & \mbox{ if }  n^v_a\geq 3.
                             \end{array} \right.
  \end{equation}
   First, note that $\varphi$ will  select a given  outcome $x$ (either $a$ or $b$) in all states of $\Tilde{\Omega}$ whenever $x$ is unanimously preferred (to the other outcome). In fact the constant act $a\tilde{\Omega}$ is selected as soon as  there are three (or more) supporters of $a$. 
Interestingly, note that the selected act when the number of supporters of $a$ is either $1$ or $2$ can be described by using the filter $\{\mathcal{C}^1, \mathcal{C}^2\}$ such that $\mathcal{C}^1=\{(\omega_2, \omega_1), (\emptyset,\omega_3)\}$, $\mathcal{C}^2=\{(\omega_3, \omega_2), (\omega_1,\emptyset)\}$. Indeed, remark from (\ref{EqFiltEx}) that, if $n_a^v=1$, then $b$ is selected in $\omega_3$ and  the two supporters of $b$ (and only them) are asked to vote between  two acts, $a\omega_2 \oplus b \omega_1$  and $a\omega_1 \oplus b \omega_2 $, with the former act chosen if and only if  at least one supporter of $b$ prefers it. A similar choice between $a\omega_2 \oplus b \omega_3$  and $a\omega_3 \oplus b \omega_2$ is offered to the two supporters of $a$ whenever $n_a^v=2$ (with the outcome $a$ selected in $\omega_1$).
\end{example}

One can check that the BSCF described in Example \ref{ExSimplefiltering} is strategy-proof, range-unanimous and  anonymous. Moreover, this BSCF  is prime (it does not admit any non-trivial factors), even though it is  not of one of the three  previous types (simple, quasi-dictatorial, and dyadic factors). In fact this BSCF is of the fourth (and final) type of buidling block, which we formally define in the following lines.

\medskip
To  implement a \textit{filtering factor}, pick two thresholds $\underline{k},  \bar{k}\in \{1,\ldots,n-1\}$ such that $\underline{k}<\bar{k}$;
 a filter of $\tilde{\Omega}$,  denoted by \[\mathcal{C}^k=\{(E_1^k,F_1^k), \ldots,(E_{M_k}^k,F_{M_k}^k), (G^k_a, G^k_b)\}\] (with $k\in \{\underline{k},\ldots, \Bar{k}\}$);
    and  quotas   $\tilde{t}^k_m\in \{1,\ldots,k+1\}$, $ \hat{t}^k_m\in   \{1,\ldots, n-k+1\}$, for any $k=\underline{k},\ldots,\bar{k}$ and $m=1,\ldots,M_{k}$.
    
Select the constant act $a\tilde{\Omega}=[\tilde{\Omega}]_a$ ($b\tilde{\Omega}=[\emptyset]_a$) whenever $n_a^v\geq \bar{k}+1$ ($n_a^v\leq \underline{k}-1$).
Recall from the definition of a filter that, for each $k\in \{\underline{k},\ldots, \Bar{k}\}$, the collection $\{E_{1}^k\cup F_{1}^k, \ldots, E_{M_k}^k\cup F_{M_k}^k,   G^k_a\cup  G^k_b\}$ is a partition of $\tilde{\Omega}$ into $M_k+1$ cells.
 In each of the $M_k$ cells of the form $ E_{m}^k\cup F_{m}^k$,  offer voters a choice between the two sub-acts $aE_m^k\oplus b F_m^k$ and $aF_m^k\oplus b E_m^k$ whenever $n_a^v=k$, with the latter sub-act selected if and only if [the number of supporters of $a$ who prefer it to the former is at least $\tilde{t}^k_m$, or the number of supporters of $b$ who prefer it to the former is at least $\hat{t}^k_m$]. 
This means that $aF_m^k\oplus b E_m^k$ will be selected in this cell if [$\eta_p(N^v_a, F^k_m|E^k_m)\geq \tilde{t}^k_m$ or $\eta_p(N^v_b, E^k_m|F^k_m)\geq \hat{t}^k_m$]; and otherwise the sub-act $aE_m^k\oplus b F_m^k$ is selected in this cell.
 Moreover, in the cell $G^k_a\cup G^k_b$ the designer will select the sub-act $aG^k_a\oplus bG^k_b$ regardless of the reported preferences (whenever $n_a^v=k$).
Taking the concatenation of these $M_k+1$ sub-acts, the subset of $\tilde{\Omega}$ where $a$ is selected is hence   
   $ Z^k(v,p):=G^k_a \cup (\bigcup\limits_{m\in   \bar{L}^k(v,p)} E^k_m )\cup (\bigcup\limits_{m\in  L^k(v,p)} F^k_m )$, where
$L^k(v,p):=\{m\in \{1,\ldots, M_k\}:  \eta_p(N^v_a, E^k_m|F^k_m)\geq \hat{t}^k_m     \mbox{ or } \eta_p(N^v_b, F^k_m|E^k_m)\geq \tilde{t}^k_m$\} and $\bar{L}^k(v,p)=\{1,\ldots,M_k\}\setminus L^k(v,p)$.

  To summarize the previous explanations, the associated filtering factor $\varphi$ is defined as follows:
 for all $(v,p)\in \tilde{\mathcal{D}}^N$,
\begin{equation}\label{EqTriad}
  \varphi (v,p)=\left \{ \begin{array}{cl}
                              [\emptyset]_a & \mbox{ if } n^v_a\leq \underline{k}-1; \\
                                                        \left[Z^k(v,p)\right]_a&\mbox{ if } \underline{k}\leq  n^v_a\leq \bar{k};\\
                             \left[\tilde{\Omega}\right]_a & \mbox{ if } n^v_a\geq \bar{k}+1.
                             \end{array} \right.
\end{equation}

As an illustration, remark that the BSCF $\varphi$ of Example \ref{ExSimplefiltering} is the filtering factor associated with the thresholds $\underline{k}=1$, $\bar{k}=2$, the filter $\mathcal{C}^1=\{(\omega_2, \omega_1), (\emptyset,\omega_3)\}$, $\mathcal{C}^2=\{(\omega_3, \omega_2), (\omega_1,\emptyset)\}$, and the voting quotas $\hat{t}^1_1=1, \tilde{t}^1_1=2$, $\hat{t}^2_1=1, \tilde{t}^2_2=2$.

One can check that the  full graph assumption  (for the transitive closure of the union relation of the associated filter) guarantees that a filtering factor  admits only a trivial factorization (it is hence a prime factor). 
Moreover, remark 
from (\ref{EqTriad}) that a filtering factor is anonymous and range-unanimous; but it is not necessarily strategy-proof (as explained in the next example).

\begin{example}\label{Exdpart3}  Suppose that $\Tilde{\Omega}=\{\omega_1,\omega_2, \ldots,\omega_7\}$ and $N=\{1,2,\ldots,10\}$. Let $\varphi$  be the filtering factor  
defined by the filter 
\begin{align}\label{EqExFilter}
    \mathcal{C}^1=& \{(\omega_1\omega_2\omega_3, \omega_4),(\emptyset, \omega_5\omega_6\omega_7)\}\\ \nonumber
    \mathcal{C}^2= &\{(\omega_6, \omega_1), (\omega_5,\omega_2), (\omega_3\omega_4, \omega_7)\} \\ \nonumber
 \mathcal{C}^3= &\{(\omega_7,\omega_5\omega_6), (\omega_1\omega_2\omega_3\omega_4, \emptyset)\}
\end{align}
and the voting quotas $\tilde{t}^k_m=1,\hat{t}^k_m=2$ for all $k\in \{1,2,3\}$, $m\in \{1,\ldots,M_k\}$. At any preference profile $(v,p)$ where $N^v_a=\{9,10\}$, $p_1(\omega_6)>1/2$, and  $p_1(\omega)$, $p_i(\omega_1)v_i(a)+p_i(\omega_6)v_i(b)<p_i(\omega_6)v_i(a)+p_i(\omega_1)v_i(b)$ but   $p_i(\omega_1)v_i(a)+p_i(\omega_6)v_i(b)<p_i(\omega_6)v_i(a)+p_i(\omega_1)v_i(b)$ for all $i\neq 1$, note from (\ref{EqTriad}) that $\varphi(v,p;\omega_6)=a$. On the other hand, picking $v'_1$ such that $v'_1(a)>v'_1(b)$ and writing $v'=(v'_1,v_2,\ldots,v_{10})$, remark that $n_a^{v'}=3$; and it hence comes from (\ref{EqTriad}) that $\varphi(v',p;\omega_6)=b$. Hence, voter 1 can manipulate $\varphi$ at $(v,p)$ ---and $\varphi$ is thus not strategy-proof.
\end{example}

Consider a filtering factor $\varphi$ defined by $\mathcal{C}^k=\{(E_1^k,F_1^k), \ldots,(E_{M_k}^k,F_{M_k}^k), (G^k_a, G^k_b)\}$ (which is a filter,  with $k= \underline{k},\ldots, \Bar{k}$) and voting quotas   $\tilde{t}^k_m\in \{1,\ldots,k\}$, $ \hat{t}^k_m\in   \{1,\ldots, n-k\}$ for any $k=\underline{k},\ldots,\bar{k}$ and $m=1,\ldots,M_{k}$. We call $\varphi$ an \textit{iso-filtering factor} if, for all $k= \underline{k},\ldots, \Bar{k}-1$, $m=1,\ldots,M_{k}$ and $m'=1,\ldots,M_{k+1}$, we have
\begin{align}
      \label{EqTriadFilter2}
      [E^k_{m}\cap  F_{m'}^{k+1}\neq \emptyset] &\Rightarrow  [\hat{t}^{k}_{m}=\tilde{t}^{k+1}_{m'}=1];\\
\label{EqTriadFilter0}
  [(E_{m'}^{k+1},F_{m'}^{k+1}) \in \mathcal{C}^{k} ] &\Rightarrow  [\hat{t}^{k}_{m}=\tilde{t}^{k}_m=1=\hat{t}^{k+1}_{m'}=\tilde{t}^{k+1}_{m'}].
\end{align}

Together, (\ref{EqTriadFilter2}) and (\ref{EqTriadFilter0})  mean that any state $\omega$ that is not filtered out of its binary status (as we move from a valuation section $\mathcal{V}_k^N$ to the next one $\mathcal{V}_{k+1}^N$) must be associated with the quotas $\hat{t}^{k}_{m}=\tilde{t}^{k+1}_{m'}=1$, \textit{i.e.}, a single favorable vote from a supporter of $b$  triggers the selection of this outcome in $\omega$ in section $\mathcal{V}_k^N$.

The following example illustrates the notion of  iso-filtering factor.
\begin{example}\label{Exdpart4}  
Suppose that $\Tilde{\Omega}=\{\omega_1,\omega_2, \ldots,\omega_7\}$ and $N=\{1,2,\ldots,10\}$.

\noindent
(i)- It is easy to see that the filtering factor $\varphi$ of Example \ref{Exdpart3} is not iso-filtering ---since it violates (\ref{EqTriadFilter2}).

\noindent
(ii)- Let $\varphi'$  be the filtering factor  
defined by the filter introduced in (\ref{EqExFilter}) and the voting quotas $\hat{t}^k_m=1,\tilde{t}^k_m=1$ for all $k\in \{1,2,3\}$, $m\in \{1,\ldots,M_k\}$. Then one can check that deviations (from truth-telling) similar to the one described in  Example \ref{Exdpart3} are no longer profitable; and $\varphi'$  satisfies (\ref{EqTriadFilter2}). Moreover, (\ref{EqTriadFilter2}) and (\ref{EqTriadFilter0}) are trivially satisfied, since $\hat{t}^k_m=1$ and $\tilde{t}^k_m=1$ always hold. Hence,  $\varphi'$ is an iso-filtering factor.
\end{example}

The next result provides a  characterization of the set of filtering factors that satisfy strategy-proofness. 

\begin{proposition}{\label{PropTriadic}}\textbf{Strategy-proof filtering}
\newline
A filtering factor $\varphi$ is strategy-proof if and only if it is iso-filtering.
\end{proposition}

 As an illustration, Proposition \ref{PropTriadic} says that the   filtering factor $\varphi'$ described in Example \ref{Exdpart4}-(ii) is strategy-proof.
Since every filtering factor is anonymous and range-unanimous, remark that the set of iso-filtering factors also corresponds to the class of filtering factors satisfying our three axioms.

For any $p\in (\mathcal{P}(\tilde{\Omega}))^N$ and any nonempty $C\in 2^{\tilde{\Omega}}$, let $p_{/C}=(p_{i/C})_{i\in N}$ denote the  conditional belief profile induced by $p$ on the event $C$.

We are now ready to characterize the set of BSCF that are anonymous, range-unanimous and strategy-proof.

\begin{proposition}{\label{LemmaDec}}\textbf{BSCF Factorization}
\newline
\noindent Suppose that $\varphi: (\mathcal{D}(\tilde{\Omega}))^N\rightarrow \{a,b\}^{\tilde{\Omega}}$ is an anonymous, range-unanimous and strategyproof  $\tilde{\Omega}$-BSCF. Then there exist (i) an integer $T\geq 1$,
(ii) a partition $\{\Omega_1, \ldots, \Omega_T\}$ of $\tilde{\Omega}$, and (ii) $T$ sub-BSCFs $\varphi_t: (\mathcal{D}(\Omega_t))^N\rightarrow \{a,b\}^{\Omega_t}$, for $t=1,\ldots, T$, such that:
\[\varphi(v,p)= \varphi_1(v,p_{/\Omega_1}) \oplus \ldots \oplus \varphi_T(v,p_{/\Omega_T}),  ~\forall (v,p)\in (\mathcal{D}(\tilde{\Omega}))^N ; \]
and each $\varphi_t$  is  a simple, a quasi-dictatorial, a  dyadic, or an iso-filtering factor.
\end{proposition}

\medskip
As emphasized in its definition, note that the existence of an iso-filtering factor requires at least three voters and three states of nature ($n\geq 3$  and $|\Tilde{\Omega}|\geq 3$).  It is also worth noting that the existence of a quasi-dictatorial factor requires at least three states  $(|\Tilde{\Omega}|\geq 3)$.

\section{Characterization: Main result}\label{SectChar}

This section fully describes  the class of anonymous, range-unanimous and strategy-proof social choice functions. Let us return to our original state space $\Omega$; and  consider once again  the full set of  outcomes $X$, with $X_\omega\subseteq X$ for all $\omega 
\in \Omega$.

Remark that the following result  follows from the combination of Proposition \ref{Propbinarydecomp} and Proposition \ref{LemmaDec}. 

\begin{theorem}{\label{ThmChar}}\textbf{Characterization result}
\newline
\noindent Consider a constrained SCF  $\varphi: \mathcal{D}^N\rightarrow \times_{\omega \in \Omega}X_\omega$. Then $\varphi$ is anonymous, strategy-proof and range-unanimous if and only if  any non-constant  factor in its canonical factorization is simple, quasi-dictatorial,  dyadic, or iso-filtering. 
\end{theorem}
As mentioned before, the canonical factorization of a SCF $\varphi$ may include an iso-filtering factor only if we have at least three voters and three states of nature. 
Quasi-dictatorial factors require at least two voters and three states. Finally, observe that simple and dyadic factors require at least two voters and two states.

\medskip
Recall now the high school reunion example of Section \ref{ToyEx}, where the set of states of nature is $\Omega=\{c,r,o,s,w\}$ and the respective sets of outcomes available in these states are $X_c=\{B, M, P,V\}$, $X_r=\{M, V\}=X_o=X_w$, $X_s=\{B, D,  M, P,V\}$. Let us illustrate Theorem \ref{ThmChar} by providing two examples of factorable SCF satisfying our three axioms.

First, consider the partition of the state space into two cells $\Omega_1=\{c,s\}$ and   $\Omega_2=\{r,o,w\}$.  Let $\varphi_1$ be the dyadic factor (defined on $\Omega_1$)  asking each of the ten friends to express  support for either the beach ($B$) or the amusement park ($P$). If at least seven (at most three) people prefer $B$ to $P$, then the group will go the beach (park) in the event of either a sunny or a cloudy day. In the middle case (where four, five, or six agents prefer the beach) each of the ten friends is asked to vote between the two sub-acts \textbf{(I) }$Bc\oplus Ps$  and \textbf{(II) } $Pc\oplus Bs$, with  option (I) selected if it gets at least six favorable votes ---otherwise option (II) is selected.  For the other cell of the partition, let $\varphi_2$ be the quasi-dictatorial factor asking each group member to express support for either a movie marathon ($M$) or volleyball ($V$). If the number of supporters of $M$ is greater (lower) than one, then the group will watch movies (play volleyball) in each of the three states rainy, snowy, windy. In the case where there is exactly one supporter of $M$, this person is asked to choose her favorite among the three  sub-acts $Mr\oplus V(ow)$, $Mo\oplus V(rw)$, $Mw\oplus V(rw)$. Letting then $\varphi=\varphi_1\oplus \varphi_2$, remark that  $\varphi $ belongs to the class of mechanisms described by Theorem \ref{ThmChar}.

A second factorable SCF can be described as follows. Partition the state space into three cells $\Omega'_1=\{s\}$, $\Omega'_2=\{c\}$, $\Omega'_3=\{r,o,w\}$. Let $\varphi'_1$ be the constant factor always selecting skydiving ($D$) in the event of a sunny day. Moreover, consider the simple factor $\varphi'_2$ asking each of the friends to express support for either the beach or skydiving, with the former activity carried out by the group in case of a cloudy day if at least four members support it. Finally, let  $\varphi'_3$ be the iso-filtering factor
 asking every agent to express support for either $V$ or $M$ and selecting a sub-act on $\Omega'_3$ as follows. If the number of supporters of $V$ is at most four (at least seven), then the group will play volleyball  (watch movies) in each of the three states rainy, snowy, windy. If exactly five agents support $V$ then each of the ten friends is asked to vote between the two sub-acts \textbf{(I)} $Vo\oplus M(rw)$ and \textbf{(II)} $Vr\oplus M(ow)$, with option I selected if  (at least \textit{one} supporter of $M$  votes for it or at least \textit{two} supporters of $V$ vote for it). In the case where exactly six agents support $V$ then each of the ten friends is asked to vote between the two sub-acts \textbf{(III)}  $V(or)\oplus Mw$ and \textbf{(IV)} $V(ow)\oplus Mr$, with option III selected if and only if (at least \textit{one} supporter of $V$  votes for it or at least \textit{three} supporters of $V$ vote for it).  Defining then the SCF $\varphi'=\varphi'_1\oplus \varphi'_2\oplus \varphi'_3$, observe that $\varphi'$  belongs to the class of mechanisms described by Theorem \ref{ThmChar}.

\smallskip
Conversely, it is important  to note from Theorem \ref{ThmChar}  that all anonymous, strategy-proof and range-unanimous SCF must  have a similar structure: they can always be decomposed into prime (constant or binary) factors, each of which resembles one of the five factors $\varphi_1,\varphi_2, \varphi'_1$, $\varphi'_2, \varphi'_3$ described in the previous paragraphs.

\section{Discussion}\label{Concl}
Our three axioms are logically independent. To see this, note first that the locally bilateral top selections of \cite{bahel2020strategyproof} satisfy strategy-proofness and range-unanimity, but they are not anonymous (as explained in  Subsection \ref{Sectliterature}). Second, remark that any  SCF maximizing some weighted sum (with fixed and positive weights) of the agents' subjective expected utility is anonymous and range-unanimous, but obviously not strategy-proof (in particular, if $|X|\geq 3$, such a rule cannot be decomposed into binary or constant factors). Third, we provide a simple example of SCF that is strategy-proof and anonymous, but not range-unanimous. Suppose that there are five voters, three states of nature ($\omega_1,\omega_2,\omega_3$) and four outcomes ($a,b,c,d$), with $X_{\omega_1}=\{a,b\}$ and $X_{\omega_2}=X_{\omega_3}=\{a,c,d\}$. Use simple-majority voting between the two acts $f=a\omega_1\oplus c\omega_2 \oplus d\omega_3$ and  $g=b\omega_1\oplus d\omega_2 \oplus c\omega_3$. Note that this voting mechanism is not range-unanimous because the outcome $c$, which is available within the range of outcomes selected in $\omega_2$ ($\omega_3$) by this SCF, is not chosen in both of these states when all voters value the sub-act $c \omega_2 \omega_3$ above
 all else (this is the case at preference profiles $(v,p)$ such that $v_i(c)\approx 1$ and $p_i(\omega_2)\approx p_i(\omega_3)\approx 1/2$, for all voters $i=1,\ldots,5$).

To simplify the exposition, our analysis has set aside indifference between distinct acts. Allowing for indifference, one would need to couple each of the mechanisms identified in Theorem \ref{ThmChar} with a tie-breaking rule specifying (i) which outcome to choose whenever a voter must pick one of  two indifferent outcomes and (ii) which event an agent will vote for whenever she must choose between two events she finds equally likely to occur. An example of tie-breaking rule  is as follows. Pick an arbitrary linear order of the set of outcomes, and break ties between distinct outcomes as suggested by the selected order. Likewise, pick an arbitrary linear order of the set of states and, when faced with two indifferent events, select the event whose lowest state (according to the arbitrary order) is lower if the two lowest  states are different. Otherwise, select either the event containing a single state or (if both  have at least two states) the event whose second-lowest state is lower,  and so on and so forth. It is not difficult to check that this  tie-breaking rule  is strategy-proof and anonymous. The description  of the full class of strategy-proof and anonymous tie-breaking rules is an open question.

It is worth noting that the mechanisms identified in our characterization of Theorem \ref{ThmChar} are \textit{group strategy-proof}, that is to say, no group of voters can improve the  situation of each of its members  via a coordinated deviation from truth-telling. 
The interested reader can consult for instance the works of \cite{barbera2001introduction} and \cite{serizawa1999strategy}, which offer  a study of group strategy-proof mechanisms for social choice in a deterministic world. 

The assumption  ---made in Subsection \ref{Feasibility}-- that the set of feasible acts is a product set ($\times_{\omega\in \Omega} X_{\omega}$) is useful for tractability. Note that, in the case where it is not a product set, our analysis  still applies if one can extract from the set of of feasible acts a degenerate rectangle (exhibiting at least one face with distinct vertices).

\section{Conclusion}\label{Concl2}
In the model of constrained collective choice where society must choose between uncertain prospects, our work has established that a SCF satisfies anonymity, strategy-proofness and range-unanimity if and only if every non-constant factor in its canonical decomposition is simple, dyadic, quasi-dictatorial, or iso-filtering. Simple factors are akin to quota voting [see for instance \cite{levin1995introduction}] between two constant acts. 
To the best of our knowledge,  quasi-dictatorial, dyadic, and iso-filtering factors are novel to this literature; and each of these three types of mechanism exploits both the valuations and the subjective beliefs of all voters in a strategy-proof and anonymous way. 

 It was important for exhaustiveness and classification purposes to explore  the full class of mechanisms satisfying strategy-proofness, anonymity and range unanimty.  The question of choosing between the different building blocks identified is an interesting one: future works may motivate and add new axioms (to our three basic requirements) in order to  characterize  some subclasses (for instance, voting rules whose factors are dyadic) within the family described in Theorem \ref{ThmChar}.


	\bibliographystyle{apalike}
	\bibliography{references}

\newpage
\section{Appendix: proofs of the results}\label{appendix}

\subsection{Preliminaries}

We introduce some notation and terminology to be used in the proofs. For any event $A\in 2^\Omega$, let $\overline{A}:=\Omega\setminus A$ denote the complementary event of $A$. Given a nonempty event $A$. we say that a SEU preference  $(v_i,p_i)\in \mathcal{D}$ is $A$-\textit{lexicographic} if, for all $f,g\in X^\Omega$, we have
\begin{equation}\label{Eqdeflexico}
    [\sum_{\omega \in A}p_i(\omega)v_i(f(\omega))>\sum_{\omega \in  A}p_i(\omega)v_i(g(\omega))]\Rightarrow [\mathbf{E}_{v_i}^{p_i}(f)> \mathbf{E}_{v_i}^{p_i}(g)]
\end{equation}
In essence, under a lexicographic preference  $(v_i,p_i)\in \mathcal{D}$,  an act $f$ will be preferred to another act $g$ whenever the sub-act $f_{A}$ is preferred to the sub-act $g_{A}$. 
By extension, we say that a preference profile $(v,p)\in \mathcal{D}^N$ is $A$-lexicographic if $(v_i,p_i)$ is $A$-lexicographic for all $i\in N$.


   For any $\theta\in (0,1)$ and $A\in 2^\Omega\setminus \{\emptyset\}$, define the belief $p_i^{\theta,\Omega'}$ as follows: $\forall \omega \in \Omega$,
\begin{equation}\label{Eqdefcond}
p_i^{\theta,A}(\omega):=\left\{\begin{array}{cc}
                               \frac{ \theta p_i(\omega)}{p_i(A)} & \mbox{ if } \omega \in A; \\
                              \frac{(1- \theta) p_i(\omega)}{p_i(\overline{A})}
                             & \mbox{ if } \omega \in \overline{A}.
                            \end{array} \right.
                    \end{equation}
In words, (\ref{Eqdefcond}) means that the belief $p_i^{\theta,A}$ gives an aggregate probability of $\theta$ ($1-\theta)$ to $A$ ($\overline{A}$) while preserving $p_i$'s conditional probability distributions on  $A$ and $\overline{A}$. The following remarks will be useful.

\begin{remark}\label{remarklexico}
Consider a nonempty event $A\in 2^{\Omega}$; and fix an arbitrary preference $(v_i,p_i)\in \mathcal{D}$. Then $(v_i,p^{\theta,A}_i)$ is $A$-lexicographic for $\theta$ sufficiently close to 1; and $(v_i,p^{\theta,A}_i)$ is $\overline{A}$-lexicographic for $\theta$ sufficiently close to 0.
\end{remark}

\begin{remark}\label{remarkSPlexico}
Let $\varphi:\mathcal{D}^N\rightarrow\times_{\omega \in \Omega }X_{\omega }$ be a strategy-proof SCF. Fix  $A\in 2^{\Omega}\setminus\{\emptyset\}$, $i\in N$, $(v,p)\in \mathcal{D}^N$ and $\theta',\theta''\in (0,1)$  s.t. $(v_i,p_i^{\theta,A})$ is $\overline{A}$-lexicographic and  $(v_i,p_i^{\theta'',A})$ is $\overline{A}$-lexicographic. Then the following implication holds:
\[[\varphi_{A}(v,(p^{\theta,A}_i,p_{-i}))=\varphi_{A}(v,(p_i^{\theta'',A},p_{-i}))]\Rightarrow [\varphi(v,p)=\varphi(v,(p^{\theta,A}_i, p_{-i}))=\varphi(v,(p^{\theta'',A}_i, p_{-i}))].\] 
\end{remark}

\medskip

Finally, for any given  act $f\in X^{\Omega}$,  nonempty event $A\in 2^\Omega$, and  outcome $x\in X$, define $f^x_{A}:=\{\omega\in A: f(\omega)=x\}$. Likewise, given a SCF $\varphi:\mathcal{D}^N\rightarrow \times_{\omega \in \Omega }X_{\omega }$, we  often write $\varphi^x_{A}(v,p)$ to refer to the (possibly empty) event where the sub-act $\varphi_{A}(v,p)$ yields the outcome $x$.

\subsection{Proof of Lemma \ref{Lembinarydecomp}}
Let  $\varphi :\mathcal{D}^{N}\rightarrow \times_{\omega \in \Omega }X_{\omega }$  be an anonymous, strategy-proof and range-unanimous SCF. Furthermore, assume as in the statement  of the lemma that  $|R^\varphi_\omega|\geq 2$, for all $\omega\in \Omega$.

The proof proceeds in three steps.

\medskip\noindent\textbf{Step 1.} For all $\omega\in \Omega$, we have $|R^\varphi_\omega|=2$.

\smallskip\noindent
Let $\omega \in \Omega$ be an arbitrary state of nature. We proceed by contradiction. Assume  that $|R^\varphi_\omega|\geq 3$. In other words, there exist distinct $a,b,c\in X_\omega$ and $(w,q)$, $(w',q'), (w'',q'')\in \mathcal{D}^N$ such that 
\begin{equation}\label{Eqcontradbinary}
    \varphi (w,q)=a, \varphi (w',q')=b, \varphi (w'',q'')=c.
\end{equation}

In the following lines we construct a deterministic social choice function [in the sense of Gibbard (1973) and Satterthwaite (1975)] defined over profiles of linear orderings of the set $\{a,b,c\}$.  Let $L^{ab}$ be the linear ordering (defined over $\{a,b,c\}$) such that $a$ is preferred to $b$, and $b$ is preferred to $c$. In the same spirit, the other five  linear orderings of $\{a,b,c\}$ can be denoted by $L^{ac}, L^{ba}, L^{bc}, L^{ca}, L^{cb}$.  Write then $\mathcal{L}=\{L^{ab},L^{ac}, L^{ba}, L^{bc}, L^{ca}, L^{cb}\}$.

Fix now six valuation functions $v_k^{ab},v_k^{ac},v_k^{ba},v_k^{bc},v_k^{ca},v_k^{cb}\in \mathcal{V}$ such that, for all distinct $\bar{x}, \bar{y}\in \{a,b,c\}$, we have
\begin{equation}\label{Eqvalabc}
   1=v_k^{\bar{x}\bar{y}}(\bar{x})> v_k^{\bar{x}\bar{y}}(\bar{y})>v_k^{\bar{x}\bar{y}}(x)> \min_{x\in \{a,b,c\}}v_k^{\bar{x}\bar{y}}(x)> \max_{x\in X\setminus \{a,b,c\}}v_k^{\bar{x}\bar{y}}(x).
\end{equation}
For instance, under the valuation function $v_k^{ab}$, agent $k$'s top outcome is $a$, then  $b$ has the second-highest valuation and  $c$ the third-highest (any outcome not in $\{a,b,c\}$ has a lower valuation than each of the three outcomes  $a,b,c$). For all distinct $\bar{x}, \bar{y}\in \{a,b,c\}$, let  $v^{\bar{x}\bar{y}}$ be the valuation profile where each agent $i\in N$ has the valuation function $v_k^{\bar{x}\bar{y}}$.

Next, for any distinct $\bar{x}, \bar{y}\in \{a,b,c\}$, define
\begin{eqnarray}\label{Eqdeltaminval}
    \delta^{\bar{x}\bar{y}}:=\min_{\tiny \begin{array}{cc}
        x,y\in X \\
        x\neq y 
    \end{array}}|v^{\bar{x}\bar{y}}_k(x) - v^{\bar{x}\bar{y}}_k(y)|.\end{eqnarray}
Recalling that each valuation function in  $\mathcal{V}$ is injective, observe from (\ref{Eqdeltaminval}) that each $ \delta^{\bar{x}\bar{y}}$ is positive. Hence, defining 
$\delta:=\min\limits_{\tiny \begin{array}{cc}
        \bar{x}, \bar{y}\in \{a,b,c\}  \\
       \bar{x}\neq \bar{y} 
    \end{array}}\delta^{\bar{x}\bar{y}}$; we conclude that $\delta>0$.

Fix  $\varepsilon\in (0, \frac{\delta}{1+\delta})$; and pick  a belief profile  $p^{\varepsilon}\in \mathcal{P}^N$ that is symmetric (i.e., $p^{\varepsilon}_i=p^{\varepsilon}_j$ for all $i,j\in N$) and satisfies
\begin{equation}\label{Eqbeliefdelta}
    p^{\varepsilon}_i(\omega)= 1-\varepsilon   ~\hbox{ [i.e.,    $p^{\varepsilon}_i(\omega)\in (\frac{1}{1+\delta}, 1)$]} .
\end{equation}
Combining (\ref{EqdefSEU}), (\ref{Eqbeliefdelta}) and the definition of $\delta$, remark  that the profile $(v,p^\varepsilon)$ is $\{\omega\}$-lexicographic for any $v\in \{v_k^{ab},v_k^{ac},v_k^{ba},v_k^{bc},v_k^{ca},v_k^{cb}\}^N$. 
In particular, \textbf{at the preference profile }$(v^{ab},p^{\varepsilon})$,  every agent $i\in N$ prefers the act $\varphi(w,q)$ [which yields $i$'s top outcome in state $\omega$, by  (\ref{Eqcontradbinary})] to any act $f$ such  $f(\omega)\neq a=\varphi(w,q)$. Therefore, repeated application of the strategy-proofness axiom [defined in (\ref{defSP})] yields $\varphi(v^{ab},p^{\varepsilon})=a$. The same line of reasoning allows to write\begin{eqnarray}\label{EqArrayabc}
\varphi(v^{ac},p^{\varepsilon})=a; ~\varphi(v^{bc},p^{\varepsilon})=b; ~
\varphi(v^{cb},p^{\varepsilon})=c.
\end{eqnarray}

Combining (\ref{Eqvalabc}), (\ref{Eqbeliefdelta}) and (\ref{EqArrayabc}), observe that, \textbf{at any preference profile} $(v,p^\varepsilon)$ \textbf{such that} $v\in \{v_k^{ab},v_k^{ac},v_k^{ba},v_k^{bc},v_k^{ca},v_k^{cb}\}^N$, every agent  $i\in N$ prefers each of the three acts  $\varphi(v^{ac},p^{\varepsilon})$,$\varphi(v^{bc},p^{\varepsilon}), \varphi(v^{cb},p^{\varepsilon})$ above to any act $f$ such  $f(\omega)\notin \{a,b,c\}$. Thus, repeatedly applying strategy-proofness once again allows to write
\begin{equation}\label{Eqabcrange}
  \left[v\in \{v_k^{ab},v_k^{ac},v_k^{ba},v_k^{bc},v_k^{ca},v_k^{cb}\}^N\right] \Rightarrow \left[\varphi(v,p^\varepsilon;\omega) \in \{a,b,c\}\right]~
\end{equation}

Let then $\rho_{GS}: \mathcal{L}^N \rightarrow \{a,b,c\}$ be the deterministic choice function defined as follows: for any $L=(L^{\bar{x}_1\bar{y}_1}, \ldots,L^{\bar{x}_n\bar{y}_n})\in \mathcal{L}^N$ (where $\bar{x}_i, \bar{y}_i\in \{a,b,c\}$ are distinct for each $i\in N$),  
\begin{equation}\label{eqdefrhoabc}
    \rho_{GS}(L)=\varphi((v_k^{\bar{x}_1\bar{y}_1}, \ldots,v_k^{\bar{x}_n\bar{y}_n}), p^\varepsilon; \omega).
\end{equation}
    Note from (\ref{Eqabcrange})-(\ref{eqdefrhoabc}) that $ \rho_{GS}$ is well-defined; and by (\ref{EqArrayabc}) it has full range $\{a,b,c\}$. Moreover, it is not difficult to check that  $\rho_{GS}$ is strategy-proof and anonymous (since $\varphi$ is strategy-proof and anonymous, and $p^\varepsilon$ is $\{\omega\}$-lexicographic and symmetric). But  this contradicts the Gibbard-Satterthwaite theorem, which requires  that $\rho_{GS}$ be dictatorial rather than anonymous (recall that $n\geq 2$ and $|\{a,b,c\}|=3$).

\medskip\noindent\textbf{
Step 2.} The SCF $\varphi$ admits a ``binary pre-factorization'', that is, there exist distinct pairs $(a_1,b_1),\ldots, (a_K,b_K)\in X^2$ and  functions $\varphi_{1}:\mathcal{D}^N\rightarrow \{a_1,b_1\}^{\Omega_1}, \ldots,  \varphi_{k}: \mathcal{D}^N \rightarrow \{a_K,b_K\}^{\Omega_K}$ (where $K\geq 1$ and $\{\Omega_1,\ldots,\Omega_k\}$ is a partition of $\Omega$) such that $\varphi(v,p)= \varphi_{1} (v,p)\oplus\ldots \oplus \varphi_{k} (v,p), ~\forall (v,p)\in \mathcal{D}^N$.

\medskip\noindent
To complete Step 2, consider the partition of $\Omega$ into nonempty cells $\Omega_1,\ldots, \Omega_K$ (where $K\geq 1$)   such that, for all $k=1,\ldots, K$,
\begin{equation}\label{EqdefOmegaK}
    [\omega,\omega'\in \Omega_k] \Leftrightarrow [R^\varphi_\omega=R^\varphi_{\omega'}].
\end{equation}
Combining the result of Step 1 with (\ref{EqdefOmegaK}), we may  write $R^\varphi_\omega=\{a_k,b_k\}$, for all $k=1,\ldots,K$ and all $\omega\in \Omega_k$.
The desired result thus obtains by letting  $\varphi_{k}:\mathcal{D}^N\rightarrow \{a_k,b_k\}$  be the function defined by $ \varphi_{k}(v,p)=(\varphi(v,p))_{\Omega_k}$. Note that, for each  $k=1,\ldots,K$, the function $\varphi_{k}$ is well-defined, since it comes from (\ref{EqdefOmegaK}) that $(\varphi(v,p))_{\Omega_k}\in \{a_k,b_k\}^{\Omega_k}$ for all $\omega \in \Omega_k$. This concludes Step 2.

\medskip
 In the trivial case where $K=1$, remark that, together, Step 1 and Step 2 suffice to conclude the proof of Lemma \ref{Lembinarydecomp} (since we have $p=p_{/\Omega_1}$ in that case). 
\textbf{Suppose  then in what follows that} $K\geq 2$. 

\medskip
For all $(v,p)\in \mathcal{D}^N$ and $k\in\{1,\ldots,K\}$, let  $I_k(v,p)$ be the collection of agents $i\in N$ whose preferences $(v_i,p_i$) are $\Omega_k$-lexicographic; and write  $I_{-k}(v,p)$ to denote the collection of agents $i\in N$ whose preferences  $(v_i,p_i)$  are $\overline{\Omega_k}$-lexicographic. In addition, define  $l_{k}(v,p):=|I_k(v,p)|$ and $l_{-k}(v,p):=|I_{-k}(v,p)|$.

Moreover,  
for any nonempty $\Omega'\in 2^\Omega$, 
denote by $\mathcal{P}(\Omega')$  the set of injective probability measures defined over $\Omega'$. Write $\mathcal{D}(\Omega')$ to refer to the collection of profiles $(v,q)\in \mathcal{V}^N\times (\mathcal{P}(\Omega'))^N$ that induce a strict ranking of the set of sub-acts $X^{\Omega'}$. 

Combining the strategy-proofness of $\varphi$ with (\ref{Eqdeflexico}), observe that we must have: $\varphi_{k}(v,p)=\varphi_{k}(v,p')$ whenever the conditions $l_k(v,p)=l_k(v,p')=n$ and $p_{/\Omega_k}=p'_{/\Omega_k}$ hold. Hence,   there exists a strategy-proof and anonymous sub-SCF $\varphi^*_{k}:(\mathcal{D}(\Omega_k))^N\rightarrow \{a_k,b_k\}^{\Omega_k}$ (for each $k=1,\ldots,K$) such that
\begin{eqnarray}\label{EqlexSep1}
\varphi_{\Omega_k}(v,p)=&  \varphi^*_{k}(v,p_{/\Omega_k}) & \hbox{ if }  l_{k}(v,p)=n. 
\end{eqnarray}

Define then the SCF  $\varphi^*$ as  follows:   $\varphi^*(v,p)=\varphi^*_{1}(v,p_{/\Omega_1})\oplus \ldots\oplus \varphi^*_{K}(v,p_{/\Omega_K})$, for all $(v,p)\in \mathcal{D}^N$.
Since $K\geq 2$ and the desired result holds when there exists only one cell in the partition of the state space, we may assume by induction that, for each $k=1,\ldots,K$,
\begin{eqnarray}\label{EqlexSep2}
\varphi_{\overline{\Omega_{k}}}(v,p)= \varphi^*_{\overline{\Omega_{k}}}(v,p)=\oplus_{k'\neq k}\varphi^*_{k'}(v,p_{/\Omega_{k'}}) & \hbox{ if }  l_{-k}(v,p)=n. 
\end{eqnarray}

\bigskip\noindent\textbf{
Step 3.} For any $k=1,\ldots, K$, we have $\varphi_{k} (v,p)=\varphi^*_k(v,p_{/\Omega_k})$, for all $(v,p) \in \mathcal{D}^N$.

\medskip\noindent
For all $m=0,1,\ldots n-1$, define $$_m\mathcal{D}^N:=\{(v,p)\in \mathcal{D}^N: |\{j\in N\setminus1: (v_j,p_j)=(v_1,p_1)\}|=m\}.$$
In words, $_m\mathcal{D}^N$ is the set of preference profiles such that agent 1 has the same preferences over $X^
{\Omega}$ as exactly $m$ other agents.   Note that we have $\mathcal{D}^N= \cup_{m=0}^{n-1} ~_m\mathcal{D}^N $. 

We will use backward induction over $m$ to prove that 
\begin{equation}\label{EqIdwithLexk}
    \varphi_{k}(v,p)=\varphi^*_k(v,p_{/\Omega_k}), \forall (v,p)\in \mathcal{D}^N, \forall k\in \{1,\ldots,K\}.
\end{equation}

First, note that (\ref{EqIdwithLexk}) holds by range-unanimity for all $(v,p)\in \!_{n-1}\mathcal{D}^N$. \textbf{Fix now } $m\in \{0,\ldots,n-2\}$ and assume by backward induction that
\begin{equation}\label{IndHypom}
    \varphi(v,p)=\varphi^*(v,p), \forall (v,p)\in \!_{m+1}\mathcal{D}^N.
\end{equation}

Given any  $(w,q)\in \mathcal{D}^{n-m}$, one can form the preference profile 
\begin{equation}\label{Eqpsiclaim3}
\psi(w,q)=((w_1,q_1),(w_2,q_2),\overbrace{(w_1,q_1),\ldots,(w_1,q_1)}^{m \mbox{ times}}, (w_{-12},q_{-12}))   \in \mathcal{D}^N.
\end{equation}


Define then the SCF $^m \dot{\varphi}:\mathcal{D}^{n-m} \rightarrow \times_{k=1}^K\{a,b\}^{\Omega_k}$ as follows: 
\small{ \begin{align}\label{EqDotdPhi}
 ^m\dot{\varphi} (w,q)=\varphi(\psi(w,q)), 
\end{align}}
for all $(w,q)\in \mathcal{D}^{n-m}$ and all $k\in \{1,\ldots,K\}$.
Let  $J_v^p:=\{j\in N: (v_j,p_j)=(v_1,p_1)\}$. We prove a few claims below.

\bigskip\noindent
\textbf{Claim 1.} For all $(v,p) \in \!_{m}\mathcal{D}^N$, we have $\varphi(v,p)=\! ^m \dot{\varphi}((v_1,p_1), (v_{-J_v^p}, p_{-J_v^p}))$. 

\smallskip\noindent
This  claim easily  follows from (\ref{EqDotdPhi}), the anonymity of $\varphi$ the definitions of $\!_{m}\mathcal{D}^N$ and $J_v^p$.

\bigskip\noindent
\textbf{Claim 2.} $^m\dot{\varphi}$ is strategy-proof (for all $m=0,\ldots, n-1$).

\smallskip\noindent Fix $m\in \{0,\ldots,n-1\}$ and $(w,q) \in \mathcal{D}^{n-m}$.  For any  $i=2,\ldots,n-m$,  combining (\ref{EqDotdPhi}) with anonymity and strategy-proofness of $\varphi$ yields: 
for all $ (w'_i,q'_i) \in \mathcal{D}$,
\small{ \begin{align}\label{EqSPdirectdot}
 \mathbf{E}_{w_i}^{q_i}(^m\dot{\varphi}((w,q)) =&\mathbf{E}_{w_i}^{q_i}   (\varphi((w_1,q_1),\ldots,(w_1,q_1), (w_{2},q_{2}),\ldots, \! (w_{n-m},q_{n-m})))  \\ \nonumber
 \geq &\mathbf{E}_{w_i}^{q_i}   (\varphi((w_1,q_1), \ldots,(w_1,q_1),(v_{2},p_{2}),\ldots,(v'_i,p'_i),\ldots, \! (v_{n-m},p_{n-m})))  \\ \nonumber
 = &\mathbf{E}_{w_i}^{q_i}   (^m\dot{\varphi}((w_1,q_1), (w_{2},q_{2}),\ldots,(w'_i,q'_i),\ldots, \! (q_{n-m},q_{n-m})))  .
\end{align}}
Moreover, given any $(w'_1,q'_1)\in \mathcal{D}$, combining (\ref{EqDotdPhi}) and the anonymity of $\varphi$ with repeated application (for all agents $i=1,\ldots,m+1$) of  strategy-proofness of $\varphi$  allows to write
\small{\begin{align} \label{EqSPSuccDot}
 \mathbf{E}_{w_1}^{q_1}(^m\dot{\varphi}((w,q)) =&\mathbf{E}_{w_i}^{q_i}   (\varphi((w_1,q_1),\ldots,(w_1,q_1), (w_{2},q_{2}),\ldots, \! (w_{n-m},q_{n-m})))  \\ \nonumber
 \geq &\mathbf{E}_{w_1}^{q_1}   (\varphi((w'_1,q'_1), \ldots,(w_1,q_1),(v_{2},p_{2}),\ldots,(v'_i,p'_i),\ldots, \! (v_{n-m},p_{n-m})))  \\ \nonumber \vdots&  \\ \nonumber
 \geq&\mathbf{E}_{w_1}^{q_1}   (^m\dot{\varphi}((w'_1,q'_1),\ldots,(w'_1,q'_1),  (w_{2},q_{2}),\ldots, \! (q_{n-m},q_{n-m}))) 
 \\ \nonumber
 = &\mathbf{E}_{w_1}^{q_1}   (^m\dot{\varphi}((w'_1,q'_1), (w_{2},q_{2}),\ldots, \! (q_{n-m},q_{n-m})))  .
\end{align}}
Using (\ref{EqSPdirectdot})-( \ref{EqSPSuccDot}), we conclude that  $^m\dot{\varphi}$ is strategy-proof.

\medskip Let us introduce some additional notation before stating the next claim. Recalling Remark \ref{remarklexico}, note that, for any given $(v,p)\in \mathcal {D}^N$ and $i\in N$, there exist 
$^1_{v_i}p_i$ and $^0_{v_i}p_i \in \mathcal{P}$ such that   
\begin{align}\label{EqLex1pi}
  (w_i,~\!^1_{v_i}p_i) \mbox{ is } \Omega_1\mbox{-lexicographic},  \\ \label{EqLex0pi} (w_i,~\!^0_{v_i}p_i)  \mbox{ is } \overline{\Omega_1}\mbox{-lexicographic},
\end{align} for any  valuation function $w_i$ obtained either by taking $w_i=v_j$ (for some $j\in N$) or by taking $w_i(x)=v_i(h(x))$ for any $x\in X$ (where $h$ is a permutation of the set $X$).\footnote{The existence of $^1_{v_i}p_i$ and $^0_{v_i}p_i $ follows from Remark \ref{remarklexico}, since the set of possible transformations (of $v_i$ into $w_i$) described above is finite for any given $(v,p)\in \mathcal{D}$.} 

Let $\mathbf{B}:=\left\{\{a_1,b_1\}, \ldots, \{a_k,b_k\}\right\}$;  and  define the following set of preference profiles
\begin{equation}\label{EqbarDN}
    _{m}\mathcal{\bar{D}}^N:=\left\{(v,p)\in \!_{m}\mathcal{D}^N:  \left\{ \begin{array}{ll}
    2\notin J^p_v,\\
         p_{1/\overline{\Omega_1}}=p_{2/\overline{\Omega_1}}, \mbox{ and } \\
       (v_1(x)-v_1(y))(v_2(x)-v_2(y))>0, \forall \{x,y\}\in \mathbf{B}\setminus \{\{a_1,b_1\}\} 
    \end{array}  \right. \right\}.
\end{equation}
Remark from (\ref{EqbarDN}) that (i) $ _{m}\mathcal{\bar{D}}^N$ is nonempty,\footnote{To see why $ _{m}\mathcal{\bar{D}}^N$ is nonempty, remark that it contains for example  all profiles $(v,p)$ such that $  p_{1/\bar{\Omega}_1}=p_{2/\bar{\Omega}_1}$ and $v_1(a_1)=1=v_2(b_1)>v_1(b_1)=v_2(a_1)>v_1(x)=v_2(x)$  for all $x\in X\setminus \{a_1,b_1\}$.} and (ii) given any $(v,p)\in \!_{m}\mathcal{D}^N$ and any $k=2,\ldots,K$,  agent 1 and agent 2 have the same ranking of all sub-acts in the set $\{a_k,b_k\}^{\Omega_k}$. 

Since  the proofs of the remaining claims involve only variations of the first and second arguments of $^m\dot{\varphi}$, it will often  be convenient to\textbf{ drop the rest of the profile} by writing for instance $^m\dot{\varphi}(w'_1,w'_{2}), (q'_1,q'_{2}))$ instead of the full $^m\dot{\varphi}((w'_1,w'_{2}, w_{-12}), (q'_1,q'_{2}, q'_{-12}))$.

\bigskip\noindent
\textbf{Claim 3.} $^m\dot{\varphi}((v_1,v_{2}), (~\!^1_{v_1}p_1,~\!_{v_{2}}\!\!^0 p_{2}))= \!^m\dot{\varphi}(v_1,v_{2}),(~\!^0_{v_1}p_1,~\!^0_{v_{2}}p_{2}))$, for all $(v,p) \in \!_{m}\mathcal{\bar{D}}^N$. 

\medskip\noindent First, remark that for any given $k=2,\ldots, K$ and any $(v,p)\in \!_{m}\mathcal{\bar{D}}^N$ such that $(v_2,p_2)$ is $\Omega_k$-lexicographic and $c_k=\argmax\limits_{x\in \{a_k,b_k\}}v_1(x)$, we have
\begin{equation}\label{Eqmaxprobk1}
p_1\left(\left(~\!^m\dot{\varphi}((v_1,v_{2}),(p_1,p_{2})\right)^{c_k}_{\Omega_k} \right)\leq p_1 \left(\left(\varphi^{*}_k(v,p)\right)^{c_k}\right).
\end{equation}

Indeed, since $((\underbrace{v_2,\ldots,v_2}_{J_v^p}, v_{-J_v^p}),(\underbrace{p_2,\ldots,p_2}_{J_v^p}, p_{-J_v^p})\in \!_{m+1}\mathcal{\bar{D}}^N$ [because $(v,p)\in \!_{m}\mathcal{\bar{D}}^N$], note from (\ref{IndHypom}) that $(~\!^m\dot{\varphi}\left((v_2, v_2),(p_2,p_2)\right))_{\Omega_k}^{c_k}= \left(\varphi^*_k ((v_2,\ldots,v_2, v_{-J_v^p}),(p_2\ldots,p_2, p_{-J_v^p})\right)^{c_k}=(\varphi^*_k((v,p))^{c_k}$ ---where the last equality comes from strategy-proofness of $\varphi^*_k$ and the facts that $\varphi_{\Omega_k}(v,p),\varphi^{*}_k(v,p)\in \{a_k,b_k\}^{\Omega_k}$, $ p_{1/\bar{\Omega}_1}=p_{2/\bar{\Omega}_1}$ and $(v_1(a_k)-v_1(b_k))(v_2(a_k)-v_2(b_k))>0$. The result stated in (\ref{Eqmaxprobk1}) then follows by applying strategy-proofness of $~\!^m\dot{\varphi}$ to agent 1, given that $(v_2,p_2)$ is $\Omega_k$-lexicographic and  $v_2(c_k)=\max\{v_2(a_k, v_2(b_k)\}$. Moreover, using (\ref{Eqmaxprobk1}) and applying strategy-proofness of $~\!^m\dot{\varphi}$ to agent 2 in a similar way, we get 
\begin{equation}\label{Eqmaxprobk2}
p_2\left(\left(~\!^m\dot{\varphi}((v_1,v_{2}),(p_1,p_{2})\right)^{c_k}_{\Omega_k} \right)\leq p_2\left(\left(\varphi^{*}_k(v,p)\right)^{c_k} \right), 
\end{equation}
for all  $ k=2,\ldots,K$ and  $(v,p)\in \!_{m}\mathcal{\bar{D}}^N$ such that $c_k=\argmax\limits_{x\in \{a_k,b_k\}}v_1(x)$.

\medskip
Second, fix $(v,p) \in \!_{m}\mathcal{\bar{D}}^N$; and write $c_k=\argmax\limits_{x\in \{a_k,b_k\}}v_1(x)=\argmax\limits_{x\in \{a_k,b_k\}}v_2(x)$ for all $k=2,\ldots,K$ ---recall from (\ref{EqbarDN}) that the agents 1 and 2 have the same rankings of the pair $\{a_k,b_k\}$ for any $k\neq 1$. It thus comes from (\ref{Eqmaxprobk2}) that 
\begin{equation}\label{Eq3C3}
p_2\left(\left(~\!^m\dot{\varphi}((v_1,v_{2}), (~\!^1_{v_1}p_1,~\!_{v_{2}}\!\!^0 p_{2})\right)^{c_k}_{\Omega_k} \right)\leq p_2\left(\left(\varphi^{*}_k(v,p)\right)^{c_k} \right), ~ \forall k\in \{2,\ldots,K\}.
\end{equation}
Since $~\!^m\dot{\varphi}((v_1,v_{2}), (~\!^1_{v_1}p_1,~\!_{v_{2}}\!\!^0 p_{2}))_{\Omega_k}\in \{a_k,b_k\}^{\Omega_k}$ (for all $k\neq 1$) and $(v_2, ~\!_{v_{2}}\!\!^0 p_{2})$ is $\bar{\Omega}_1$-lexicographic, the combination of  strategy-proofness of $~\!^m\dot{\varphi}$ with (\ref{Eqmaxprobk2}) implies that 
\[p_2\left(\left(~\!^m\dot{\varphi}((v_1,v_{2}), (~\!^1_{v_1}p_1,~\!_{v_{2}}\!\!^0 p_{2})\right)^{c_k}_{\Omega_k} \right)= p_2\left(\left(\varphi^{*}_k(v,p)\right)^{c_k} \right),\]  for all $k\in \{2,\ldots,K\}$. Since $p_2$ is injective, and $(v_1,p_1)$ and $(v_2,p_2)$ induce the same ranking of all acts in $\{a_k,b_k\}^{\Omega_k}$ (for all $k\neq1$), we have thus shown  that
\begin{align}\label{EqC3ComplBind}
(~\!^m\dot{\varphi}((v_1,v_{2}), (~\!^1_{v_1}p_1,~\!_{v_{2}}\!\!^0 p_{2})))_{\Omega_k}= & (~\!^m\dot{\varphi}((v_1,v_{2}),(~\!^0_{v_1}p_1,~\!^0_{v_{2}}p_{2})))_{\Omega_k} \\ \nonumber
=&   \varphi^*_k ((v_2,\ldots,v_2, v_{-J_v^p}),(p_2\ldots,p_2, p_{-J_v^p}),\\ \nonumber
 =& \varphi^*_k(v,p),~ \forall (v,p) \in \!_{m}\mathcal{\bar{D}}^N, \forall k\neq 1.
\end{align}

Finally, combining (\ref{EqC3ComplBind}) with the fact that the preferences of each agent $i=1,2$ over sub-acts $\{a_1,b_1\}^{\Omega_1}$ are the same from $(v_1,v_{2}), (~\!^0_{v_1}p_1,~\!_{v_{2}}\!\!^0 p_{2})$ to $(v_1,v_{2}), (~\!^1_{v_1}p_1,~\!_{v_{2}}\!\!^0 p_{2})$, we can use 
strategy-proofness of $~^m\dot{\varphi}$ to write $^m\dot{\varphi}((v_1,v_{2}), (~\!^1_{v_1}p_1,~\!_{v_{2}}\!\!^0 p_{2}))= ~\!^m\dot{\varphi}((v_1,v_{2}),(~\!^0_{v_1}p_1,~\!^0_{v_{2}}p_{2}))$, for all $ (v,p) \in \!_{m}\mathcal{\bar{D}}^N$.

\bigskip\noindent
\textbf{Claim 4.} $^m\dot{\varphi}((v_1,v_{2}), (~\!^0_{v_1}p_1,~\!_{v_{2}}\!\!^1 p_{2}))= ~\!^m\dot{\varphi}((v_1,v_{2}),(~\!^0_{v_1}p_1,~\!^0_{v_{2}}p_{2}))$, for all $(v,p) \in \!_{m}\mathcal{\bar{D}}^N$. 

\smallskip\noindent
Switching the roles of agent 1 and agent 2, an argument similar to the proof of Claim 3 yields the desired result.

\bigskip\noindent
\textbf{Claim 5.} $^m\dot{\varphi}((v_1,v_{2}), (~\!^1_{v_1}p_1,~\!_{v_{2}}\!\!^1 p_{2}))= \!^m\dot{\varphi}(v_1,v_{2}),(~\!^1_{v_1}p_1,~\!^0_{v_{2}}p_{2}))= \!^m\dot{\varphi}(v_1,v_{2}),(~\!^0_{v_1}p_1,~\!^1_{v_{2}}p_{2}))$, for all $(v,p) \in \!_{m}\mathcal{\bar{D}}^N$.

\smallskip\noindent
Note   that, since  $(v_1, ~\!^1_{v_1}p_1)$ and $(v_2, ~\!^1_{v_2}p_2)$ are both $\Omega_1$-lexicographic, strategy-proofness of $^m\dot{\varphi}$ gives: for all $(v,p)\in \mathcal{D}^N$,
\begin{align}
\label{Eqgvsf1}
\mbox{If } ~ \!^m\dot{\varphi}((v_1,v_{2}), (~\!^1 _{v_1}p_1,~\!_{v_{2}}\!\!^1p_2))\neq 
 ~\!^m\dot{\varphi}((v_1,v_{2}), (~\!^0_{v_1}p_1,~\!_{v_{2}}\!\!^1p_2)) \mbox{ then}  \nonumber  \\  
 [p_1 (~\!^
 m\dot{\varphi}_{\Omega_1}^{a_1}((v_1,v_{2}), (~\!^1_{v_1}p_1,~\!_{v_{2}}\!\!^1 p_{2})))-p_1 (~\!^
 m\dot{\varphi}_{\Omega_1}^{a_1}((v_1,v_{2}), (~\!^0_{v_1}p_1,~\!_{v_{2}}\!\!^1 p_{2})))](v_1(a_1)-v_1(b_1))>0,
\end{align}
and 
\begin{align}
\label{Eqgvsf2}
\mbox{if } ^m\dot{\varphi}((v_1,v_{2}), (~\!^1_{v_1}p_1,~\!_{v_{2}}\!\!^1p_2))\neq  \!^m\dot{\varphi}((v_1,v_{2}), (~\!^1_{v_1}p_1,~\!_{v_{2}}\!\!^0p_2)) \mbox{ then}  \nonumber  \\  
 [p_2 (~\!^m\dot{\varphi}_{\Omega_1}^{a_1}((v_1,v_{2}), (~\!^0_{v_1}p_1,~\!_{v_{2}}\!\!^1 p_{2})))-p_2(~\!^m\dot{\varphi}_{\Omega_1}^{a_1}((v_1,v_{2}), (~\!^0_{v_1}p_1,~\!_{v_{2}}\!\!^1 p_{2})))](v_2(a_1)-v_2(b_1))>0.
\end{align}


Next,  for any $v_i\in \mathcal{V}$, rank the outcomes in $X$ so as to have 
\begin{equation}\label{EqrankX}
  v_i(x^{v_i}_1)>v_i(x^{v_i}_2)>\ldots\ >v_i(x^{v_i}_{|X|}).  
\end{equation}
 In other words, for all $r=1,\ldots,|X|$, the outcome $x^{v_i}_r$ is the $r$th highest-valued in  $X$ under the valuation function $v_i$. Moreover, we use the shorthand $d_{v_i}:=|\{x\in X: \min\{v_i(a_1),v_i(b_1)\}<v_i(x)<\max\{v_i(a_1),v_i(b_1)\} \}|$.
For all $(v,p) \in \!_{m}\mathcal{\bar{D}}^N$ and $i\in $, one can then define a new valuation function $\Tilde{v}_i\in \mathcal{V}$ as follows:
\begin{equation}\label{Eqdeftildevi}
 \tilde{v}_i(x)=   \left\{\begin{array}{cl} v_i\left(x^{v_i}_{|X|-r+1}\right)& \mbox{if } x\neq a_1,b_1 \mbox{ and } x=x^{v_i}_r; \\
       
         v_i\left(x^{v_i}_{|X|-r-d_{v_i}}\right)& \mbox{if } v_i(x)=\max\{v_i(a_1), v_i(b_1)\}  \mbox{ and } x=x^{v_i}_r;
         \\
        v_i\left(x^{v_i}_{|X|-r+d_{v_i}+2}\right)& \mbox{if } v_i(x)=\min\{v_i(a_1), v_i(b_1)\}  \mbox{ and } x=x^{v_i}_r. 
    \end{array}\right.
\end{equation}
Combining (\ref{Eqdeftildevi}) and (\ref{EqbarDN}), observe that the valuation  functions $v_i$ and $\tilde{v}_i$ disagree on the ranking on any pair of outcomes  $\{x,y\}\in \mathbf{B}\setminus\{\{a_1,b_1\}\}$; and they agree on the ranking of the pair $\{a_1,b_1\}$.   
Moreover, (\ref{Eqdeftildevi}) and (\ref{EqbarDN}) imply that  $\Tilde{v}_i(x)=v_i(h(x))$ (given any $x\in X$ and $i=1,2$), for some permutation $h$ of the set $X$. It thus comes from (\ref{EqLex1pi})-(\ref{EqLex0pi}) that, for any $(v,p) \in ~\!_{m}\mathcal{\bar{D}}^N$, $(\Tilde{v}_i,~\!^1_{v_i}p_i)$ is $\Omega_1$-lexicographic and $(\Tilde{v}_i,~\!^0_{v_i}p_i)$ is $\overline{\Omega_1}$-lexicographic (where $i=1,2$). Finally, notice that $(v,p) \in ~\!_{m}\mathcal{\bar{D}}^N$ implies $\left((\Tilde{v}_i,p_i)_{i\in 2\cup J_v^p}, (v_i,p_i)_{i\in N\setminus (2\cup J_v^p)}\right)\in ~\!_{m}\mathcal{\bar{D}}^N$.

\medskip
From this point on, we complete two tasks to prove Claim 5.

\medskip
\textbf{Task 1.}  Suppose  there exists $(u,\pi)\in \!_m\mathcal{\Bar{D}}^N$ such that $^m\dot{\varphi}((\Tilde{u}_1,u_2),(~\!^1_{u_1}\pi_1,~ \!^0_{u_2}\pi_2))= $ $ \!^m\dot{\varphi}((\Tilde{u}_1,u_2),(~\!^0_{u_1}\pi_1, ~\!^1_{u_2}\pi_2)) \neq ~\!^m\dot{\varphi}((\Tilde{u}_1,u_2),(~\!^1_{u_1}\pi_1,~ \!^1_{u_2}\pi_2))$ and, in case $K\geq 3$, $\pi_i(\Omega_{k+1}\cup\ldots \cup \Omega_K)< \min\limits_{\tiny\begin{array}{cc} E,F\subseteq \Omega_k \mbox{ s.t. } E\neq F\\
w_i=u_i,\Tilde{u}_i
    \end{array}} |(w_i(a_k)-w_i(b_k))(\pi_i(E)-\pi_i(F))|$, for any $i=1,2$ and $k=2,\ldots,K-1$. 
Then we show that these properties contradict the strategy-proofness of $~\!^m\dot{\varphi}$.

\medskip Suppose that  $(u,\pi)\in \!_m\mathcal{\Bar{D}}^N$ and  $f= ~\!^m\dot{\varphi}((\Tilde{u}_1,u_2),(~\!^1_{u_1}\pi_1,~ \!^0_{u_2}\pi_2))= ~\!^m\dot{\varphi}((\Tilde{u}_1,u_2),(~\!^0_{u_1}\pi_1, ~\!^1_{u_2}\pi_2)) $  $\neq ~\!^m\dot{\varphi}((\Tilde{u}_1,u_2),(~\!^1_{u_1}\pi_1,~ \!^1_{u_2}\pi_2))=g$. 
Note first that, recalling (\ref{EqLex1pi})-(\ref{EqLex0pi}), one is allowed to write $~\!^1_{u_i}\pi_i= ~\!^1_{\tilde{u}_i}\pi_i$ and $~\!^0_{u_i}\pi_i= ~\!^0_{\tilde{u}_i}p_i$ (for $i=1,2$). 
Then recalling (\ref{Eqgvsf1})-(\ref{Eqgvsf2}) gives the two inequalities
\begin{align*}
  [\pi_1(g^{a_1}_{\Omega_1})-\pi_1(f^{a_1}_{\Omega_1})](\Tilde{u}_1(a_1)-\Tilde{u}_1(b_1))>0, ~[\pi_2(g^{a_1}_{\Omega_1})-\pi_2(f^{a_1}_{\Omega_1})](u_2(a_1)-(u_2(b_1))>0;  
\end{align*}
and it hence follows (by strategy-proofness of $^m\dot{\varphi}$) that
\begin{align}\label{Eq1Task0}
  \sum_{k=2}^K[\pi_1(g^{a_k}_{\Omega_k})-\pi_1(f^{a_k}_{\Omega_k})](\Tilde{u}_1(a_k)-\Tilde{u}_1(b_k))<0 ;\\ \label{Eq2Task0}
    \sum_{k=2}^K[\pi_2(g^{a_k}_{\Omega_k})-\pi_2(f^{a_k}_{\Omega_k})]((u_2(a_k)-u_2(b_k))<0.
\end{align}
But note that (\ref{Eq1Task0}) and  (\ref{Eq2Task0}), which mean that $f_{\overline{\Omega_1}}$ is preferred to  $g_{\overline{\Omega_1}}$ at both $(\Tilde{u}_1,\pi_1)$ and $(u_2,\pi_2)$,    contradict one another. Indeed,  letting $k^*=\min\{k\in \{2,\ldots, K:  f^{a_k}_{\Omega_k} \neq g^{a_k}_{\Omega_k} \}\}$ and recalling  $\pi_1(\Omega_{k^*+1}\cup\ldots \cup \Omega_K)< \min\limits_{\tiny\begin{array}{cc} E,F\subseteq \Omega_{k^*} \\\mbox{ s.t. } E\neq F
    \end{array}} |(\tilde{u}_1(a_k)-\tilde{u}_1(b_k))(\pi_1(E)-\pi_1(F))|$, note from (\ref{Eq1Task0}) that we must have 
    \begin{align}\label{Eq1.1Task0}[\pi_1(g^{a_{k^*}}_{\Omega_{k^*}})-\pi_1(f^{a_{k^*}}_{\Omega_{k^*}})](\Tilde{u}_1(a_{k^*})-\Tilde{u}_1(b_{k^*}))<0.
    \end{align}
    
Since $\pi_{1/\overline{\Omega_1}}=\pi_{2/\overline{\Omega_1}}$ [because $(u,\pi)\in \!_m\mathcal{\Bar{D}}^N$] and $(\tilde{u}_1(a_k)-\tilde{u}_1(b_k))(u_2(a_k)-u_2(b_k))<0$ [from (\ref{EqbarDN}) and (\ref{Eqdeftildevi})], the  inequality in (\ref{Eq1.1Task0}) implies $[\pi_2(g^{a_{k^*}}_{\Omega_{k^*}})-\pi_2(f^{a_{k^*}}_{\Omega_{k^*}})]((u_2(a_{k^*})-u_2(b_{k^*}))>0$, which contradicts (\ref{Eq2Task0}) ---since  $\pi_2(\Omega_{k^*+1}\cup\ldots \cup \Omega_K)< \min\limits_{\tiny\begin{array}{cc} E,F\subseteq \Omega_{k^*} \\\mbox{ s.t. } E\neq F
    \end{array}} |((u_2(a_k)-u_2(b_k))((\pi_2(E)-\pi_2(F))|$.

\medskip
\textbf{Task 2.} Assuming by contradiction that Claim 5 does not hold, we now prove the existence of a profile $(u,\pi)\in \!_m\mathcal{\Bar{D}}^N$ such as described in Task 1.

\smallskip Fix then $(v',p')\in \!_m\mathcal{\Bar{D}}^N$ such that $~\!^m\dot{\varphi}((v'_1,v'_{2}),(~\!^1p'_1,~\!^0p'_{2})) \neq ~\!^m\dot{\varphi}((v'_1,v'_{2}),(~\!^1p'_1,~\!^1p'_{2}))$.
Given  any $\varepsilon> 0$, one can pick a belief $p^{\varepsilon}\in \mathcal{P}^N$ such that \textbf{(I)} $p^{\varepsilon}_i=p'_i$ for all $i\notin 2\cup J^{p'}_{v'}$\; \textbf{(II)}  $p^{\varepsilon}_{j/\Omega_k}=p'_{j/\Omega_k}$ for all $j=1,2$  and  $k=1,\ldots,K$; \textbf{(III)} $p^{\varepsilon}_j=p^{\varepsilon}_1$ for all $j\in J^{p'}_{v'}$; \textbf{(IV)} If $K\geq 3$ then
$p^{\varepsilon}_i(\Omega_{k+1}\cup\ldots \cup \Omega_K)< \varepsilon p^{\varepsilon}_i(\Omega_k)$ for all $i=1,2$ and  $k=2,\ldots,K-1$.  Note in particular that $p^{\varepsilon}$ preserves the conditional belief of all agents over each $\Omega_k$ (for $k=1,\ldots,K$).


Combining (\ref{EqbarDN}),  $(v',p')\in \!_m\mathcal{\Bar{D}}^N$ and the properties (I)-(III) above, remark  that $(v',p^\varepsilon) \in ~\!_m\mathcal{\Bar{D}}^N$ for all  $ \varepsilon \geq 0$.
Pick  $\varepsilon^*$ such that \textbf{(V)} $0<\varepsilon^*< \min\limits_{\tiny\begin{array}{cc} E,F\subseteq \Omega_k \mbox{ s.t. } E\neq F \\
       k=2,\ldots,K \\
       i=1,2\\
       w=v'_i,\tilde{v}'_i
    \end{array}} |(w_i(a_k)-w_i(b_k)) \frac{|p'_i(E)-p'_i(F)|}{p'(\Omega_k)} $.  Defining then $(v,p):=(v', p^{\varepsilon^*})$, it comes from  (IV)-(V)  that, if $k\geq 3$, we have 
    \begin{equation}\label{EqC5eps}
  p_i(\Omega_{k+1}\cup\ldots \cup \Omega_K)< \min\limits_{\tiny\begin{array}{cc} E,F\subseteq \Omega_k \\\mbox{ s.t. } E\neq F  \\
   w=v_i,\tilde{v}_i
    \end{array}} |(v_i(a_k)-v_i(b_k)) (p_i(E)-p_i(F))|, ~\forall i=1,2, \forall k=2,\ldots,K-1.    \end{equation}

Recalling (\ref{EqC3ComplBind}) ---and the facts that $v=v'$ and $p_{i/\Omega_k}=p'_{i/\Omega_k}$ (for $i=1,2$ and $k=2,\ldots,K$), we may write the following: for all $k=2,\ldots,K$
\begin{equation}\label{Eq2Task1}
  (~\!^m\dot{\varphi}((v_1,v_{2}),(~\!^1p_1,\!^0p_{2})))_{\Omega_k} =\varphi^*_k(v,p)=\varphi^*_k(v',p') = (~\!^m\dot{\varphi}((v'_1,v'_{2}),(~\!^1p'_1,~\!^0p'_{2})))_{\Omega_k}.
\end{equation}
Since  $((v'_1,v'_{2}),(~\!^0p'_1,~\!^1p'_{2}))$ and $((v_1,v_{2}),(~\!^0p_1,~\!^1p_{2}))$ induce the same ranking of all sub-acts in $\{a_1,b_1\}^{\Omega_1}$ for every $i=1,2$, combining (\ref{Eq2Task1}) with strategy-proofness of $~\!^m\dot{\varphi}$ yields
\begin{equation}\label{Eq3Task1}
  ~\!^m\dot{\varphi}((v_1,v_{2}),(~\!^1p_1,\!^0p_{2}))=~\!^m\dot{\varphi}((v'_1,v'_{2}),(~\!^1p'_1,~\!^0p'_{2})).
\end{equation}
Moreover, observe that strategy-proofness of $~\!^m\dot{\varphi}$ implies 
 \begin{equation}\label{Eq4Task1}
   (~\!^m\dot{\varphi}((v_1,v_{2}),(~\!^1p_1,\!^1p_{2})))_{\Omega_1}=(~\!^m\dot{\varphi}((v'_1,v'_{2}),(~\!^1p'_1,~\!^1p'_{2})))_{\Omega_1},
\end{equation}
since $((v'_1,v'_{2}),(~\!^1p'_1,~\!^1p'_{2}))$  and $((v_1,v_{2}),(~\!^1p_1,~\!^1p_{2}))$ $\Omega_1$-lexicographic and induce the same ranking of all sub-acts in $\{a_1,b_1\}^{\Omega_1}$ for every $i=1,2$.

\medskip
 Together, (\ref{Eq3Task1})-(\ref{Eq4Task1}) and the assumption that $ ~\!^m\dot{\varphi}((v'_1,v'_{2}),(~\!^1p_1,~\!^0p_{2})) \neq ~ \!^m\dot{\varphi}((v_1,v_{2}),(~\!^1p_1,~\!1p_{2}))$ imply that  $ ~\!^m\dot{\varphi}((v_1,v_{2}),(~\!^1p_1,\!^0p_{2}))\neq ~\!^m\dot{\varphi}((v_1,v_{2}),(~\!^1p_1,~\!^1p_{2}))$. Combining this finding with Claims 3-4, we thus get
 \begin{equation}\label{Eq5Task1}
   \!^m\dot{\varphi}((v_1,v_{2}),(~\!^0p_1,\!^1p_{2}))\neq ~\!^m\dot{\varphi}((v_1,v_{2}),(~\!^1p_1,~\!^1p_{2})).
\end{equation}

\textbf{To fix ideas, assume without loss of generality that} $v_1(a_1)>v_1(b_1)$. Since $(v,p)\in \!_m\mathcal{\Bar{D}}^N$, this also means that $v_2(a_1)>v_2(b_1)$.
For any act $h\in \times_{k=1}^K\{a_1,b_1\}^{\Omega_k}$ such that $h^{a_1}\neq \emptyset$, define  
\begin{align}
\label{EqdefPh} 
 \mathcal{P}_h:=&\{\Tilde{p}_1\in \mathcal{P}: \Tilde{p}_{1/\overline{\Omega_1}} =p_{2/\overline{\Omega_1}}  \mbox{ and } \Tilde{p}_1(\omega)>\Tilde{p}_1(\Omega_1\setminus h^{a_1}_{\Omega_1}), \forall \omega \in h^{a_1}_{\Omega_1}\};    \\ \label{EqdefEh}
 \mathcal{E}_h:=&\{h^{a_1}_{\Omega_1}\}\cup \{E\in 2^{\Omega_1}: h^{a_1}_{\Omega_1}\setminus E\neq \emptyset\}.
\end{align}

Remark from  (\ref{EqdefPh})-(\ref{EqdefEh}) that $\Tilde{p}_1(h^{a_1}_{\Omega_1})=\max\limits_{E\in \mathcal{E}_h}\Tilde{p}_1(E)$, for all $h\in \times_{k=1}^K\{a_1,b_1\}^{\Omega_k}$ and $\Tilde{p}_1\in \mathcal{P}_h$. Hence, letting $f=\!^m\dot{\varphi}((v_1,v_{2}),(~\!^0p_1,~\!^1p_{2}))$ and $\Tilde{f}=\!^m\dot{\varphi}((\tilde{v}_1,\tilde{v}_{2}),(~\!^0p_1,~\!^1p_{2}))$, it comes from (\ref{Eqgvsf1}) and strategy-proofness of $^m\dot{\varphi}$ that 
\begin{align}\label{Eq1PhwarrantyC5}
  \!^m\dot{\varphi}((v_1,v_{2}),(~\!^0p'_1,~\!^1p_{2})) =   \!^m\dot{\varphi}((v_1,v_{2}),(~\!^1p'_1,~\!^1p_{2})), \forall p'_1\in \mathcal{P}_f;
  \\ \nonumber
   \!^m\dot{\varphi}((\tilde{v}_1,\tilde{v}_{2}),(~\!^0p'_1,~\!^1p_{2})) =   \!^m\dot{\varphi}((\tilde{v}_1,\tilde{v}_{2}),(~\!^1p'_1,~\!^1p_{2})), \forall p'_1\in \mathcal{P}_{\tilde{f}}.
\end{align}
We recursively construct a sequence of beliefs  $\{\!_tq\}_{t\in \SN}$   and  a sequence  of acts $\{\!_tf\}_{t\in \SN}$   as follows. For all $t\in \SN$ and $i=1,2$, write $w^t_i=v_i$ if $t$ is even; and write $w^t_i=\tilde{v}_i$ if $t$ is odd. Let $_0q_1:=p_1$,  $_0f:=f$; and  pick $_1q_1\in \mathcal{P}_f$.  Moreover, for all $t\geq 1$, define 
\begin{align}\label{EqdefftC5}
   _{t}f:=  ~\!^m\dot{\varphi}((w^{t}_1,w^{t}_2),(~\!_{t}\!^0q_1,~\!^1p_{2}))\\  \nonumber
    _{t}\tilde{f}:= ~ \!^m\dot{\varphi}((w^{t}_1,w^{t}_2),(~\!_{t}\!^1q_1,~\!^1p_{2}));
\end{align}
and if $_{t}f=~\! _{t}\tilde{f}$ then let $_{t+1}q_1= ~\!_{t}q_1 $, otherwise pick  $_{t+1}q_1\in \mathcal{P}_{ _{t}f}$.

Since $_1q_1\in \mathcal{P}_f$,  notice from (\ref{Eq5Task1}) and  (\ref{Eq1PhwarrantyC5}) that $_1q_1\neq ~\!_0q_1 = p_1$; and we thus have either $_{1}f=~\! _{1}\tilde{f}$ [in which case $_{2}q_1= ~\!_{1}q_1$] or it comes from the combination of (\ref{Eqgvsf2}) and $v_2(a_1)>v_2(b_1)$ that $p_2(_1f^{a_1}_{\Omega_1})<p_2(_{1}\tilde{f}^{a_1}_{\Omega_1})=p_2(_0f^{a_1}_{\Omega_1})$ where the last equality follows from strategy-proofness of $ ~\!^m\dot{\varphi}$ and the fact that $_1q_1\in \mathcal{P}_f$.
Repeating this argument for all $t\geq 2$ allows to see  that the sequence $\{\!_tq\}_{t\in \SN}$ must have a \textit{stationary point}, that is to say, there exists $\bar{t}\geq 1$ s.t. $\!_{\bar{t}+1}q=~\!_{\bar{t}}q$. Otherwise,  combining (\ref{EqdefftC5}) and (\ref{Eqgvsf2}) yields: for any $t\in \SN\setminus \{0\}$,
\begin{align*}
   p_2(_tf^{a_1}_{\Omega_1})<p_2(_{t-1}f^{a_1}_{\Omega_1})<\ldots<p_2(_0f^{a_1}_{\Omega_1}),
\end{align*} in contradiction with the finiteness of the set of acts $X^\Omega$.

Letting then $t^*$ be the smallest stationary point of 
$\{\!_tq\}_{t\in \SN}$, note from (\ref{EqdefftC5}) and (\ref{Eq1PhwarrantyC5}) that we must have 
\begin{align}\label{Eqpstationary1}
   ~\!^m\dot{\varphi}((\Tilde{v}_1,\Tilde{v}_2),(~\!_{t^*}^0q_1,\!^1p_{2})) =&  ~  \!^m\dot{\varphi}((v_1,v_2),(~\!_{t^*}^0q_1,\!^1p_{2})) \overbrace{=}^{\mbox{\tiny Claims 3-4}}     ~\!^m\dot{\varphi}_{\Omega_1}((v_1,v_2),(~\!_{t^*}^1q_1,\!^0p_{2})) 
     \\ \label{Eqpstationary2}
     \!^m\dot{\varphi}((v_1,v_2),(~\!_{t^*}^0q_1,\!^1p_{2}))=& ~ \!^m\dot{\varphi}((v_1,v_2),(~\!_{t^*}^1q_1,\!^1p_{2}))&
\end{align}

Moreover, since  $(v_i(a_1)-v_i(b_1))(\Tilde{v}_i(a_1)-\Tilde{v}_i(b_1))>0$ [by (\ref{Eqdeftildevi})] and $(\tilde{v}_i,\!^1p_i)$ and $(v_i,\!^1p_i)$ are both $\Omega_1$-lexicographic (for $i=1,2$), remark that 
\begin{align}\label{Eqpstationary3}
  \!^m\dot{\varphi}_{\Omega_1}((\tilde{v}_1,\tilde{v}_2),(~\!_{t^*}^1q_1,\!^1p_{2}))&=  ~\!^m\dot{\varphi}_{\Omega_1}((\tilde{v}_1,v_2),(~\!_{t^*}^1q_1,\!^1p_{2})) \\ \label{Eqpstationary4}
     \!^m\dot{\varphi}_{\Omega_1}((\tilde{v}_1,v_2),(~\!_{t^*}^0q_1,\!^1p_{2})) &\underbrace{=}_{\tiny \mbox{from } (\ref{Eqpstationary1})- (\ref{Eqpstationary2}) }  ~\!^m\dot{\varphi}_{\Omega_1}((\tilde{v}_1,v_2),(~\!_{t^*}^1q_1,\!^1p_{2})) \\
\label{Eqpstationary5}
     \!^m\dot{\varphi}_{\Omega_1}((\tilde{v}_1,v_2),(~\!_{t^*}^1q_1,\!^0p_{2})) &=   ~\!^m\dot{\varphi}_{\Omega_1}((v_1,v_2),(~\!_{t^*}^1q_1,\!^0p_{2})) 
\end{align}
 
Together, (\ref{Eqpstationary3})-(\ref{Eqpstationary5}) and strategy-proofness of $^m\dot{\varphi}$ give
\begin{equation}\label{Eqpstationary6}
   \!^m\dot{\varphi}((\tilde{v}_1,v_2),(~\!_{t^*}^1q_1,\!^0p_{2}))=   \!^m\dot{\varphi}((\tilde{v}_1,v_2),(~\!_{t^*}^0q_1,\!^1p_{2}))=  ~\!^m\dot{\varphi}_((\tilde{v}_1,v_2),(~\!_{t^*}^1q_1,\!^1p_{2})) .
\end{equation}

Finally, note from (\ref{EqdefftC5}) and  $_{t+1}q_1\in \{~\!_{t}q_1\}\cup \mathcal{P}_{_tf}$ that (A) $_tq_{1/\overline{\Omega_1}}=p_{2/\overline{\Omega_1}}$, for all $t\geq 0$; and (B)
$ _tq_1(\Omega_{k+1}\cup\ldots \cup \Omega_K)< \min\limits_{\tiny\begin{array}{cc} E,F\subseteq \Omega_k \\\mbox{ s.t. } E\neq F  \\
   w_1=v_1,\tilde{v}_1
    \end{array}} |(w_1(a_k)-w_1(b_k)) (_tq_1(E)-_tq_1(F))|, \forall k=2,\ldots,K-1$ [from (\ref{EqC5eps}) and (\ref{EqdefPh})]. Combining these two properties (A)-(B) with (\ref{Eqpstationary6}), we conclude that $(u_1,u_2),(\pi_1,\pi_{2}))=(v^{t^*}_1,v^{t^*}_2),(_{t^*}q_1,p_{2}))$ is as described in Task 1.

Hence, we have shown that, for all $(v,p)\in \mathcal{\Bar{D}}^N$,
\begin{align}\label{EqcombC3-5}
    ^m\dot{\varphi}((v_1,v_{2}), (~\!^0_{v_1}p_1,~\!_{v_{2}}\!\!^0 p_{2}))= & ~\!^m\dot{\varphi}((v_1,v_{2}),(~\!^1_{v_1}p_1,~\!^0_{v_{2}}p_{2})), \\ \nonumber
    =& ~ ^m\dot{\varphi}((v_1,v_{2}), (~\!^0_{v_1}p_1,~\!_{v_{2}}\!\!^1 p_{2})) \\ \nonumber
    =& ~ ^m\dot{\varphi}((v_1,v_{2}), (~\!^1_{v_1}p_1,~\!_{v_{2}}\!\!^1 p_{2})).
\end{align}
Moreover, from   (\ref{IndHypom}) and (\ref{EqbarDN}), remark that  
\begin{align*}
^m\dot{\varphi}_{\overline{\Omega_1}}((v_1,v_{2}), (~\!^0_{v_1}p_1,~\!_{v_{2}}\!\!^0 p_{2}))= &~^m\dot{\varphi}_{\overline{\Omega_1}}((v_2,v_{2}), (~\!^0_{v_2}p_2,~\!_{v_{2}}\!\!^0 p_{2})) \\ \nonumber
= &~\varphi^*_{\overline{\Omega_1}}(v,p).   
\end{align*}
and it thus follows from strategy-proofness of $^m\dot{\varphi}$ and $\varphi^*$ that 
\begin{equation}\label{EqlastC5}
 ^m\dot{\varphi}((v_1,v_{2}), (~\!^0_{v_1}p_1,~\!_{v_{2}}\!\!^0 p_{2}))=  \varphi^*(v,p).
\end{equation}
Finally, the combination of (\ref{EqcombC3-5}), (\ref{EqlastC5}) and Remark 2 guarantees the result of Claim 5.

\medskip
The following claim generalizes Claim 5. For any $k=1,\ldots,K$, define the set of valuation profiles
\begin{equation}\label{EqbarDNab}
    _{m}\mathcal{\bar{D}}^N_{k}:=\left\{(v,p)\in \!_{m}\mathcal{D}^N:  \left\{ \begin{array}{ll}
    2\notin J^p_v,\\
         p_{1/\overline{\Omega_k}}=p_{2/\overline{\Omega_k}}, \mbox{ and } \\
       (v_1(x)-v_1(y))(v_2(x)-v_2(y))>
       
    0, \forall \{x,y\}\in \mathbf{B}\setminus \{\{a_k,b_k\}\} 
    \end{array}  \right. \right\}.
\end{equation}

\noindent
\textbf{Claim 6.} $^m\dot{\varphi}((v_1,v_{2}), (~\!^1_{v_1}p_1,~\!_{v_{2}}\!\!^1 p_{2}))= \!^m\dot{\varphi}(v_1,v_{2}),(~\!^1_{v_1}p_1,~\!^0_{v_{2}}p_{2}))= \!^m\dot{\varphi}(v_1,v_{2}),(~\!^0_{v_1}p_1,~\!^1_{v_{2}}p_{2}))$, for all $(v,p) \in ~\!_{m}\mathcal{\bar{D}}^N_k$ and all $k\in \{1,\ldots,k\}$.

\medskip\noindent
Fixing $k\in \{1,\ldots,k\}$, the proof of Claim 6 is identical to that of Claim 5 (and it is omitted). 

\medskip\noindent
Let us make a few useful  observations before stating the next claims.
First, it is not difficult to check that, for all $v\in \mathcal{D}^N$, there exists $k_1,k_2\in \{1,\ldots,K\}$ such that 
\begin{align}\label{EqadjacencyQuasi}
    [\min\{v_1(a_{k_1}),v_1(b_{k_1})\}<v_1(x)<\max\{v_1(a_{k_1}),v_1(b_{k_1})\}]&\Rightarrow [\{a_{k_1},x\}, \{b_{k_1},x\}\notin  \mathbf{B} ];  \\ \nonumber
 [\min\{v_2(a_{k_2}),v_2(b_{k_2})\}<v_2(x)<\max\{v_2(a_{k_2}),v_2(b_{k_2})\}]&\Rightarrow [\{a_{k_2},x\}, \{b_{k_2},x\}\notin  \mathbf{B} ].
\end{align}
In other words, for any valuation function $v_i\in \mathcal{V}$, we can always find $\{a_k,b_k\}\in \mathbf{B}$ such that no  outcome $x$ valued between $v_i(a_k)$ and $v_i(b_k)$  shares a cell $\Omega_{l}$ ($l\neq k$) ---in our partition of the state space-- with either $a_k$ or $b_k$. 

Second, note that, for all $(w,q)\in ~ \!_m\mathcal{D}^N$, we have 
\begin{align}\label{Eq1C6}
    ^m\dot{\varphi}_{\Omega_1}((w_1,w_2),(~\!^1q_1,~\!^1q_2))= \varphi^*_1(w,q_{/\Omega_1}).   
\end{align}

To see why (\ref{Eq1C6}) holds, fix $(w,q)\in  ~\!_m\mathcal{D}^N$ and assume without loss that $2\notin J_w^q$  (anonymity of $\varphi$ allows this, since $m\leq n-2$ and hence $|J^q_w|\leq n-1$). 

Pick a preference profile  $(w',q')\in ~\!_m\mathcal{D}^N$ such that \textbf{(I)} $w'_2(x)=w_1(x)$ for all $x\neq a_{k_1},b_{k_1}$; \textbf{(II)} $(w'_2(a_{k_1})-w'_2(b_{k_1}))(w_2(a_{k_1})-w_2(b_{k_1}))>0$; \textbf{(III)} $w'_i=w_i$ for  $i\in N\setminus 2$; and  \textbf{(IV)} $q'_{2/\overline{\Omega_1}}=q_{1/\overline{\Omega_1}}$, $q'_{2/\Omega_1}=q_{2/\Omega_1}$, and $q'_{i}=q_i$ for $i\in N\setminus 2$. 

\medskip
Note from (I)-(IV) above  that $(w',q')\in ~\!_m\mathcal{\bar{D}}^N_{k_1}$. Hence,  combining Claim 6 with strategy-proofness of $~\!^m\dot{\varphi}$  gives
\begin{align}\nonumber
   ^m\dot{\varphi}_{\Omega_1}((w_1,w_2),(~\!^1_{w_1}q_1,~\!^1_{w_2}q_2)) &= ~\!^m\dot{\varphi}_{\Omega_1}((w'_1,w'_2),~\!^1_{w'_1}q'_1,~\!^1_{w'_2}q'_2))\\ \nonumber
   &  \underbrace{=}_{\tiny \mbox{Claim 6}}\varphi_1^*(w',q'_{/\Omega_1}) \\ \nonumber
  & \underbrace{=}_{\tiny \mbox{(II)-(IV)}}\varphi_1^*(w,q_{/\Omega_1}),
\end{align}
since $((w_1,w_2),(~\!^1q_1,~\!^1q_2))$ and $((w'_1,w'_2),(q'_1,q'_2))$, which are both $\Omega_1$-lexicographic, yield (for each agent) the same ranking of all acts in $\{a_1,b_1\}^{\Omega_1}$. This proves (\ref{Eq1C6}).

\medskip
Given any  valuation function $v_i\in \mathcal{V}$, two distinct outcomes $a,b$ will be called $v_i$-\textbf{adjacent} if there exists no outcome  $x\in X$ such that $\min \{v_i(a),v_i(b)\}<v_i(x)<\max\{v_i(a),v_i(b)\}$. The next claim can then be stated as follows.

\bigskip\noindent
\textbf{Claim 7.} $\varphi(v,p)=\varphi^*(v,p_{})$, for all $(v,p) \in \!_{m}\mathcal{D}^N$ such that $2\notin J_v^p$ and $a_k,b_k$ are $v_i$-adjacent for all $i=1,2$ (where $k\in\{1,\ldots,K\})$.

\medskip\noindent
We show the claim for  $k=1$ without loss of generality. Fix $(v,p) \in \!_{m}\mathcal{D}^N$ such that $a_1,b_1$ are $v_i$-adjacent for $i=1,2$; and if $k\geq 3$ then
\begin{equation}
\label{Eq0C7}
p_1(\Omega_{\delta(k+1)}\cup\ldots \cup \Omega_{\delta(K)})< \min\limits_{\tiny\begin{array}{cc} E,F\subseteq \Omega_{\delta(k)}\\ \mbox{ s.t. } E\neq F
    \end{array}} |(v_i(a_\delta(k))-v_i(b_{\delta(k)}))(p_i(E)-p_i(F))|, ~\forall k\in \{2,\ldots,K\},
\end{equation}
where $\delta$ is an arbitrary permutation of the set $\{2,\ldots,K\}$.

Exploiting Claim 6 and using an argument similar to the proofs of Claims 3-4,  we get
\begin{align}\label{Eq3C6}
   ^m\dot{\varphi}((v_1,v_2),(~\!^1p_1,~\!^1p_2)) &=  ~\!^m\dot{\varphi}((v_1,v_2),(~\!^0p_1,~\!^1p_2)) = ~\! ^m\dot{\varphi}((v_1,v_2),(~\!^1p_1,~\!^0p_2)).
   \end{align}
We show next that 
\begin{align}\label{Eq4C6}
   ^m\dot{\varphi}((v_1,v_2),(~\!^0p_1,~\!^0p_2)) &=  ~\!^m\dot{\varphi}((v_1,v_2),(~\!^0p_1,~\!^1p_2)).
   \end{align}
Pick a preference  $(v'_1,p'_1)\in \mathcal{D}$ such that \textbf{(I)} $(v'_1(a_1)-v'_1(b_1))= (v_2(a_1)-v'_2(b_1))<0$ and $\{v'_1(a_1),v'_1(b_1)\}=\{v_1(a_1),v_1(b_1)\}$; \textbf{(II)} $v'_1(x)=v_1(x)$ for all    $x\neq a_1,b_1$; \textbf{(III)}  $p'_{1/\overline{\Omega_1}}=p_{1/\overline{\Omega_1}}$ and $p'_{1/\Omega_1}=p_{2/\Omega_1}$.

Remark from (I)-(III) [and the fact that $a_1,b_1$ are  $v_1$-adjacent] that $a_1,b_1$ are also $v'_1$-adjacent; and $((v'_1,v_{-1}), (p'_1,p_{-1})) \in \!_{m}\mathcal{D}^N$.  Hence, it comes from (\ref{Eq3C6}) that  
\begin{equation}\label{Eq4.1C6}
    ^m\dot{\varphi}((v'_1,v_2),(~\!^1p'_1,~\!^0p_2)) =  ~\!^m\dot{\varphi}((v'_1,v_2),(~\!^0p'_1,~\!^1p_2)).
\end{equation}

Letting then $f= ~\!^m\dot{\varphi}((v'_1,v_2),(~\!^0p'_1,~\!^0p_2))$, $g=~\!^m\dot{\varphi}((v'_1,v_2),(~\!^0p'_1,~\!^1p_2))=$ $  ~\!^m\dot{\varphi}((v'_1,v_2),(~\!^1p'_1,~\!^0p_2)) $, and assuming by contradiction that $f\neq g$, remark that an argument similar to the proof of (\ref{Eqgvsf1})-(\ref{Eqgvsf2}) gives
\begin{align*}
   [p'_1(g_{\Omega_1}^{a_1}) -p'_1(fg_{\Omega_1}^{a_1})](v'_1(a_1)-v'_2(b_1))>0;            \\
  [p_2(g_{\Omega_1}^{a_1}) -p_2(fg_{\Omega_1}^{a_1})](v_2(a_1)-v_2(b_1))>0 .
\end{align*}
But note that, together, these two inequalities contradict the combination of (I) and (III), which says that [$p'_{1/\Omega_1}=q_{2/\Omega_1}$ and $(v'_1(a_1)-v'_2(b_1))(v_2(a_1)-v_2(b_1))<0 $]. Hence, we  have $f=g$; and it follows that
\begin{align}\label{Eq5C6}
   ~\! ^m\dot{\varphi}_{\overline{\Omega_1}}((v'_1,v_2),(~\!^0p'_1,~\!^0p_2))=&    ~\!^m\dot{\varphi}_{\overline{\Omega_1}}((v'_1,v_2),(~\!^0p'_1,~\!^1p_2)) 
   \\ \nonumber
 = & ~\!^m\dot{\varphi}_{\overline{\Omega_1}}((v_1,v_2),(~\!^0p_1,~\!^1p_2)),   
\end{align}
where the last equality obtains  by combining strategy-proofness of $^m\dot{\varphi}$ with (\ref{Eq0C7}) and the fact that $(v_1,~\!^0 p_1)$ and $(v'_1,~\!^0p'_1)$  are both $\overline{\Omega_1}$-lexicographic and induce exactly the same ranking of all sub-acts in $\{a_k,b_k\}^{\Omega_k}$ (for each $k=2,\ldots,K$).\footnote{The $v_1$-adjacency of $a_1,b_1$  implies that one can permute (if necessary) the valuations of $a_1$ and $b_1$, without affecting the relative position of $a_1$ ($b_1$) with regard to any other outcome $x\in X\setminus \{a_1,b_1\}$. This is precisely how $v_1'$ is constructed from $v_1$.} For much the same reasons, one can also write the equality
\begin{equation}\label{Eq6C6}
    \!^m\dot{\varphi}_{\overline{\Omega_1}}((v'_1,v_2),(~\!^0p'_1,~\!^0p_2)) = ~\!^m\dot{\varphi}_{\overline{\Omega_1}}((v_1,v_2),(~\!^0p_1,~\!^0p_2)).
\end{equation}

Combining  (\ref{Eq4.1C6})-(\ref{Eq6C6}), we get $~ \!^m\dot{\varphi}_{\overline{\Omega_1}}((v_1,v_2),(~\!^0p_1,~\!^0p_2))=\!^m\dot{\varphi}_{\overline{\Omega_1}}((v_1,v_2),(~\!^0p_1,~\!^1p_2))$; and one may then use Remark 2 to see that (\ref{Eq4C6}) indeed holds.

We have thus shown that, for any $(v,p) \in \!_{m}\mathcal{D}^N$ and any permutation $\delta$ of $\{2,\ldots,K\}$ such that $a_1,b_1$ are $v_i$-adjacent for $i=1,2$, and $p_1$ meets (\ref{Eq0C7}), we have
\begin{align*}
    ~\!^m\dot{\varphi}((v_1,v_2),(~\!^0p_1,~\!^0p_2))&=~\!^m\dot{\varphi}((v_1,v_2),(~\!^1p_1,~\!^0p_2))
    \\ & =~\!^m\dot{\varphi}((v_1,v_2),(~\!^0p_1,~\!^1p_2))\\
    &=~\!^m\dot{\varphi}((v_1,v_2),(~\!^1p_1,~\!^1p_2));
\end{align*}

  Hence, combining (\ref{Eq1C6}),  repeated application of Remark 2, and (\ref{EqlexSep2})   
 allows to conclude the proof of Claim 7.

\bigskip\noindent
\textbf{Claim 8.} $\varphi(v,p)=\varphi^*(v,p_{})$, for all $(v,p) \in ~\!_{m}\mathcal{D}^N$.

\medskip\noindent
Fix $(v,p) \in \!_{m}\mathcal{D}^N$ such that 
\begin{equation}
\label{Eq0C8}
p_1(\Omega_{\delta(k+1)}\cup\ldots \cup \Omega_{\delta(K)})< \min\limits_{\tiny\begin{array}{cc} E,F\subseteq \Omega_{\delta(k)}\\ \mbox{ s.t. } E\neq F
    \end{array}} |(v_i(a_\delta(k))-v_i(b_{\delta(k)}))(p_i(E)-p_i(F))|, ~\forall k\in \{2,\ldots,K\},
\end{equation}
where $\delta$ is an arbitrary permutation of the set $\{2,\ldots,K\}$. Using the anonymity of $\varphi$, we may focus on the case where $2\notin J^p_v$ (since $m\leq n-2$). Moreover, recalling (\ref{EqadjacencyQuasi}),  assume without loss of generality that $k_2=1$, that is to say, \begin{eqnarray}\label{EqadjacencyQuasi1}
    [\min\{v_2(a_{1}),v_2(b_{1})\}<v_2(x)<\max\{v_2(a_{1}),v_2(b_{1})\}]&\Rightarrow [\{a_{1},x\}, \{b_{1},x\}\notin  \mathbf{B} ].
\end{eqnarray}

First, exploiting Claim 7 and using an argument similar to the proofs of Claims 3-4, one can write
\begin{align}\label{Eq3C8}
   ^m\dot{\varphi}((v_1,v_2),(~\!^1p_1,~\!^1p_2)) &=  ~\!^m\dot{\varphi}((v_1,v_2),(~\!^0p_1,~\!^1p_2)) = ~\! ^m\dot{\varphi}((v_1,v_2),(~\!^1p_1,~\!^0p_2));
   \end{align}
Repeating the procedure used to show (\ref{Eq4C6}), we get
\begin{align}\label{Eq4C8}
   ^m\dot{\varphi}((v_1,v_2),(~\!^0p_1,~\!^0p_2)) &=  ~\!^m\dot{\varphi}((v_1,v_2),(~\!^0p_1,~\!^1p_2)).
   \end{align}

Combining  (\ref{Eq3C8})-(\ref{Eq4C8}) gives $~ \!^m\dot{\varphi}_{\overline{\Omega_1}}((v_1,v_2),(~\!^0p_1,~\!^0p_2))=\!^m\dot{\varphi}_{\overline{\Omega_1}}((v_1,v_2),(~\!^0p_1,~\!^1p_2))$; and one may then use Remark 2 to see that
\begin{align*}
    ~\!^m\dot{\varphi}((v_1,v_2),(~\!^0p_1,~\!^0p_2))&=~\!^m\dot{\varphi}((v_1,v_2),(~\!^1p_1,~\!^0p_2))
    \\ & =~\!^m\dot{\varphi}((v_1,v_2),(~\!^0p_1,~\!^1p_2))\\
    &=~\!^m\dot{\varphi}((v_1,v_2),(~\!^1p_1,~\!^1p_2));
\end{align*}

  Hence, the combination of (\ref{Eq1C6}),  repeated application of Remark 2, and (\ref{EqlexSep2})   
 gives the desired result. This concludes the proof of Lemma \ref{Lembinarydecomp}.

\endproof

\section{Proof of Proposition \ref{LemmaDec}}



\proof We use induction over the number of states of nature. Note first that the statement of Proposition \ref{LemmaDec}  easily holds for any $\tilde{\Omega}\in 2^\Omega$ such that $|\tilde{\Omega}|=1$: in this case note that the decomposition gives only a simple sub-SCF (no beliefs are involved since there is only one state of nature).

\bigskip
Next, consider a \textbf{fixed} SCF $\varphi: (\mathcal{D}(\tilde{\Omega}))^N\rightarrow \{a,b\}^{\tilde{\Omega}}$ such that $|\tilde{\Omega}| =S\geq 2$; and assume by induction that the statement of Proposition \ref{LemmaDec} holds for any strategyproof, anonymous, range-unanimous and  binary SCF $\hat{\varphi}: (\mathcal{D}(\hat{\Omega}))^N\rightarrow \{a,b\}^{\hat{\Omega}}$ such that $|\hat{\Omega}|\in \{1,\ldots, S-1\}$.
Moreover, suppose that $\varphi$ is strategyproof, anonymous, range-unanimous and  binary.
Since $\tilde{\Omega}$ is fixed, we use the shorthand  notation $\tilde{\mathcal{P}}:=\mathcal{P}(\tilde{\Omega})$ and $\tilde{\mathcal{D}}:=\mathcal{D}(\tilde{\Omega})$ without possible confusion. Given any $v_i\in \mathcal{V}$, write $\mathcal{\tilde{P}}_{v_i}=\{p_i\in \mathcal{\tilde{P}}: (v_i,p_i)\in \mathcal{D}\}$. We also  use the shorthand $\mathcal{\tilde{P}}_v=\times_{i\in N}\mathcal{\tilde{P}}_{v_i}$, for any $v\in \mathcal{V}^N$.
Finally, for all $E\in 2^{\tilde{\Omega}}$, define the complementary event $\overline{E}:=\tilde{\Omega}\setminus E$. With a slight abuse of notation, we often write $\overline{\omega}$ instead of $\overline{\{\omega\}}$ (for any $\omega\in \tilde{\Omega}$).


\begin{lemma}{\label{LemmaOrdinal}}\textbf{Ordinality}
\newline
\noindent Since $\varphi$ is binary (choosing either $a$ or $b$ in every state $\omega\in\tilde{\Omega} $), strategy-proofness of $\varphi$ implies that the act selected by $\varphi$ does not change if the reported valuations change in a way that preserves each agent's relative ranking of $a$ and $b$. More precisely, for all $ v,w\in \mathcal{V}^N$ and  $p\in \mathcal{\tilde{P}}_v\cap \mathcal{\tilde{P}}_w $, we have
\begin{equation*}
  \left[ (v_i(a)-v_i(b))(w_i(a)-w_i(b))>0, \forall i\in N\right]\Rightarrow [ \varphi(v,p)= \varphi(w,p)].
\end{equation*}
\end{lemma}

\medskip
The proof of Lemma \ref{LemmaOrdinal} is omitted. It is a direct consequence of the fact that  an agent $i$'s ranking of all binary acts $f,g\in \{a,b\}^{\tilde{\Omega}}$ remains exactly the same from   $(v_i,p_i)$ to $(w_i,p_i)$, provided that $(v_i(a)-v_i(b))(w_i(a)-w_i(b))>0$.

\bigskip
For any tuple $x\in \SR^N$ and any subset $T\in 2^N$, write $x_T= (x_i)_{i\in T}$.
  We may now state the following  results.

\begin{lemma}{\label{LemmaSections}}\textbf{Sectional Allocation}
\newline
\noindent For each $k\in \{1,\ldots,n-1\}$, there  exists a map $\alpha^k$ ($k=1,\ldots, n-1$) such that $\alpha^k: \tilde{\mathcal{P}}^{k}\times \tilde{\mathcal{P}}^{n-k} \rightarrow 2^{\tilde{\Omega}}$ and, for all $k\in \{1,\ldots, n-1\}, v\in \mathcal{V}_k^N,  p\in \tilde{\mathcal{P}}^N$, we have
\begin{align*}
 \varphi(v,p)= a  (\alpha^k(p_{N^v_a}, p_{N^v_b})) \oplus ~b  (\overline{\alpha^k(p_{N^v_a}, p_{N^v_b})}).
\end{align*}
\end{lemma}
\proof  Pick any $k\in \{1,\ldots, n-1\}$. Then fix a profile $\bar{u}^{k}\in \mathcal{V}^N$ such that
\begin{align}\label{EqNaproof}
 \bar{u}^{k}_i(a)=1> \bar{u}^{k}_i(b)=0,& ~~\mbox{ for } i=1,\ldots, k;   \\\nonumber
  \bar{u}^{k}_i(b)=1> \bar{u}^{k}_i(a)=0,& ~~\mbox{ for } i=k+1,\ldots, n.
\end{align}
Remark from (\ref{EqNaproof}) that $N^{\bar{u}^{k}}_a= \{1,\ldots, k\}$ and $N^{\bar{u}^{k}}_b= \{k+1,\ldots,n\}$.

Next, define the  maps $\alpha^k$ as follows:  $\forall q\in \tilde{\mathcal{P}}^N$,
\begin{equation}\label{EqDefSection1}
  \alpha^k(q_{\{1,\ldots,k\}}, q_{\{k+1,\ldots, n\}}):=\varphi^a(\bar{u}^{k},q).
\end{equation}
Observe from the combination of Lemma \ref{LemmaSections} and (\ref{EqDefSection1}) that
\begin{equation}\label{EqDefSection2}
 \overline{\alpha^k(q_{\{1,\ldots,k\}}, q_{\{k+1,\ldots, n\}})}=\varphi^b(\bar{u}^{k},q).
\end{equation}

Fix now $(v,p)\in \tilde{\mathcal{D}}$ such that $v\in \mathcal{V}^N_k$; and define  $(w,q)\in \tilde{\mathcal{D}}$ as follows:
\begin{equation}\label{Eq0Lem2proof}
I.~ w_{\{1,\ldots,k\}}=v_{N_a^v}, w_{\{k+1,\ldots,n\}}=v_{N_b^v} ; ~
 II.~ q_{\{1,\ldots,k\}}=p_{N_k^v}, q_{\{k+1,\ldots,n\}}=p_{N_b^v}.\end{equation}
Since $\varphi$ is anonymous, remark that we must have $\varphi(v,p)= \varphi(w,q)$. Moreover, given that $N^{\bar{u}^{k}}_a= N^{w}_a=\{1,\ldots, k\}$, Lemma \ref{LemmaOrdinal} implies that
\begin{equation}\label{EqLem2proof}\varphi(w,q)= \varphi(\bar{u}^{k},q)=a  (\alpha^k(q_{\{1,\ldots,k\}}, q_{\{k+1,\ldots, n\}}) \oplus ~b  (\overline{\alpha^k(q_{\{1,\ldots,k\}}, q_{\{k+1,\ldots, n\}}}),\end{equation}
where the last equality stems from (\ref{EqDefSection1})-(\ref{EqDefSection2}).
Finally, combining (\ref{Eq0Lem2proof}-II) and (\ref{EqLem2proof}), one can write $\varphi(v,p)= a  (\alpha^k(p_{N^v_a}, p_{N^v_b})) \oplus ~b  (\overline{\alpha^k(p_{N^v_a}, p_{N^v_b})})$. \endproof

Given any  $i,j\in N$, denote by $\sigma_{ij}$ the permutation of $N$ defined by $\sigma_{ij}(i)=j$, $\sigma_{ij}(j)=i$ and $\sigma_{ij}(i')=i'$, for all $i'\in N\setminus \{i,j\}$. By extension, given a set $M\in 2^N$ such that $i,j\in M$ and a belief profile $p\in \tilde{\mathcal{P}}^M$, we write $\sigma_{ij}p$ to denote the belief profile obtained from $p$ by  swapping the beliefs of agent $i$ and agent $j$ (every other agent's belief remaining the same).

\begin{lemma}{\label{LemmaSemiAnon}}\textbf{Side-Anonymity}
\newline
\noindent For all $k\in \{1,\ldots, n-1\},   (p,q)\in \tilde{\mathcal{P}}^{k}\times \tilde{\mathcal{P}}^{n-k}$, we have
\begin{align*}
I. & ~\alpha^k(\sigma_{ij}p,q)= \alpha^k(p,q), \forall  i,j\in \{1,\ldots,k\};
\\II. & ~ \alpha^k(p,\sigma_{ij}q)= \alpha^k(p,q),  \forall i,j\in \{k+1,\ldots,n\}.
\end{align*}
\end{lemma}

\smallskip\noindent The proof of Lemma {\ref{LemmaSemiAnon}} is omitted: it easily obtains by combining Lemma \ref{LemmaSections} with the anonymity of $\varphi$.

\begin{lemma}{\label{LemmaSemiSP}}\textbf{Probability maximizers and minimizers}
\newline
\noindent For all $k\in \{1,\ldots, n-1\},   (p,q)\in \tilde{\mathcal{P}}^{k}\times \tilde{\mathcal{P}}^{n-k}$, we have
\begin{align*}
I. & ~  p_i(\alpha^k(p,q)) =\max\limits_{p'_i \in \mathcal{\tilde{P}}} ~p_i(\alpha^k((p_i,p_{-i}),q), ~\forall i\in \{1,\ldots, k\};
\\II. & ~   q_i(\alpha^k(p,q))= \min\limits_{q'_i \in \mathcal{\tilde{P}}} ~q_i(p, (q_i, q_{-i})), ~ \forall i\in\{k+1, \ldots,n\}.
\end{align*}
\end{lemma}

\smallskip\noindent The proof of Lemma \ref{LemmaSemiSP} is omitted: it easily obtains by combining Lemma \ref{LemmaSections} with the strategy-proofness of $\varphi$. Recall that, when comparing two  binary and distinct acts $f,g\in \{a,b\}^{\tilde{\Omega}}$, each agent $i\in N^v_a$ ($i\in N^v_b$) always prefers the act that maximizes (minimizes) her probability of obtaining the outcome $a$.

\bigskip
We call every $\alpha^k$  a \emph{sectional allocation rule} (or SAR, for short).  Exploiting Lemmas \ref{LemmaSections}-\ref{LemmaSemiSP}, one can derive further properties of the SARs $\alpha^k$ ($k=0,\ldots, n$) associated with $\varphi$.

Since $\varphi$ is unanimous, note that $\alpha^0$ and $\alpha^n$ are completely determined:
\[\alpha^0(p,q)=\emptyset~ \mbox{ and } ~\alpha^n(p,q)=\Omega, \]
for all $  (p,q)\in \tilde{\mathcal{P}}^{k}\times \tilde{\mathcal{P}}^{n-k}$.

We  now study the SARs $\alpha^k$ such that $k\in \{1,\ldots, n-1\}$. Let then  

\begin{equation}\label{EqdefsectRange}
\mathcal{E}^k:=\{E\in 2^\Omega: \alpha^k(p,q)=E, \mbox{ for some }  (p,q)\in \tilde{\mathcal{P}}^k\times \tilde{\mathcal{P}}^{n-k}\}
\end{equation}
denote the \emph{range} of $\alpha^k$, for any $k\in \{1,\ldots, n-1\}$.
We prove below that each $\mathcal{E}^k$ is \emph{a collection of non-nested events}, that is to say, no two distinct events in $\mathcal{E}^k$ are comparable with respect to set inclusion.

\begin{lemma}\label{LemmaClut}\textbf{Non-nestedness}

\noindent Fix $k\in \{1,\ldots, n-1\}$ and let $E,E'\in \mathcal{E}^k$ be such that $E\neq E'$.  Then we have $E\setminus  E'\neq \emptyset$ and $E' \setminus E \neq \emptyset$.
\end{lemma}

\proof Fix $k\in \{1,\ldots, n-1\}$  and suppose there exist $(p,q),(p',q')\in  \tilde{\mathcal{P}}^k\times \tilde{\mathcal{P}}^{n-k}$ satisfying $E=\alpha^k(p,q)$ and $F=\alpha^k(p',q')$. Suppose by contradiction that $E\subsetneq E'$. Assume without loss of generality that $E'$ is \textit{maximal} (i.e., there exists no $E''\in \mathcal{E}^k$ such that $E'\subsetneq E''$) and $E$ is \textit{minimal} (i.e., there exists no $E''\in \mathcal{E}^k$ such that $E''\subsetneq E$). 

Pick then $(\bar{p}, \bar{q})\in  \tilde{\mathcal{P}}^k\times \tilde{\mathcal{P}}^{n-k}$ such that for all  $i\in \{1,\ldots,k\}$,
\begin{equation}\label{EqLexENN}
  \left\{ \begin{array}{l}
                          \bar{p}_i(\omega)> \bar{p}_i(\tilde{\Omega}\setminus E),   \forall \omega\in E;  \\
                              \bar{p}_i(\omega)> \bar{p}_i(\tilde{\Omega}\setminus E),  \forall \omega\in E'\setminus E;
                             \end{array} \right.
\end{equation}
and, for all $i\in \{k+1,n\}$,
\begin{equation}\label{EqLexEprNN}
  \left\{\begin{array}{l}
                          \bar{q}_i(\omega)> \bar{q}_i(E'),   \forall \omega\in \tilde{\Omega}\setminus E';  \\
                              \bar{q}_i(\omega)> \bar{q}_i(E),  \forall \omega\in E'\setminus E.
                             \end{array} \right.
\end{equation}

 From (\ref{EqLexENN}) and the fact that $E'$ is maximal, remark that each $i\in \{1,\ldots,k\}$ finds $E'$ more likely than any other event in $\mathcal{E}^k$; and hence repeatedly applying Lemma   \ref{LemmaSemiSP}-I gives
 \begin{equation}\label{Eq1NN}
 \alpha^k(\bar{p}, q')=E'.
 \end{equation}

 Likewise, the combination of  (\ref{EqLexEprNN}), Lemma  \ref{LemmaSemiSP}-II, and the minimality of $E$ yields
 \begin{equation}\label{Eq2NN}
 \alpha^k(p, \bar{q})=E.
 \end{equation}

Next, note that we must have $\alpha^k(\bar{p}, (q'_k+1,\ldots, q'_{n-1}, \Bar{q}_n))=E'$. Indeed, on the one hand, assuming that there exists $\omega\in (\tilde{\Omega}\setminus E')\cap \alpha^k(\bar{p}, (q'_k+1,\ldots, q'_{n-1}, \Bar{q}_n))$ would lead to $\Bar{q}_n(\alpha^k(\bar{p}, (q'_k+1,\ldots, q'_{n-1}, \Bar{q}_n)))>\bar{q}_n(E')$ [by (\ref{EqLexEprNN})]; and this would violate Lemma \ref{LemmaSemiSP}-II. On the other hand, writing $F=\alpha^k(\bar{p}, (q'_k+1,\ldots, q'_{n-1}, \Bar{q}_n)))\subsetneq E'$ would  lead to $q'_n(F)<q'_n(E') \overbrace{=}^{\mbox{ by } (\ref{Eq1NN})} q'_n(\alpha^k(\bar{p}, q'))$, in violation of Lemma \ref{LemmaSemiSP}-II.
Repeating this procedure for all $i=n, n-1,\ldots,k+1$, we get
\begin{equation}\label{Eq3NN}
    \alpha^k(\Bar{p},\bar{q})=E'.
\end{equation}

Similar to the procedure described in the previous paragraph, one can Combine (\ref{EqLexENN}), (\ref{Eq2NN}), and  repeated application of Lemma \ref{LemmaSemiSP}-I for $i=1,\ldots,k$ to get $\alpha^k(\Bar{p},\bar{q})=E$. But note that this last result contradicts (\ref{Eq3NN}). \endproof

\begin{lemma}\label{LemmaEUnanimous}\textbf{Event-unanimity}
\newline
\noindent Fix $k\in \{1,\ldots, n-1\}$ and suppose that $E\in \mathcal{E}^k$. Then for any $E$-dominant $\pi_1\in \mathcal{\tilde{P}}$ and any $\overline{E}$-dominant $\pi_2 \in \mathcal{\tilde{P}}$, we have $\alpha^k((\underbrace{\pi_1,\ldots,\pi_1}_{k~ times}), (\underbrace{\pi_2,\ldots,\pi_2}_{n-k~ times}))=E$.
\end{lemma}
\proof Pick any $(\pi_1,\pi_2)\in \mathcal{\tilde{P}}^2 $  such that (i) $\pi_1$ is $E$-dominant and  (ii) $\pi_2$ is $\overline{E}$-dominant. Considering the belief profile $(p,q)=(\underbrace{\pi_1,\ldots,\pi_1}_{k~ times}), (\underbrace{\pi_2,\ldots,\pi_2}_{n-k~ times}))\in \mathcal{\tilde{P}}^k\times \mathcal{\tilde{P}}^{n-k}$, observe from Lemma \ref{LemmaSemiSP} and Lemma \ref{LemmaClut} that 
\begin{equation}\label{EqproofEUn}
  E=\argmax_{E'\in \mathcal{E}^k} p_i(E') =\argmin_{E'\in \mathcal{E}^k}p_j(E'), ~\forall i\in \{1,\ldots,k\}, \forall j\in \{k+1,\ldots,n\}.
\end{equation}
Hence, at the belief profile $(p,q)$, all agents $i\in N$ unanimously find the event $E$ to be the most desirable in the range of $\alpha^k$. Since $E\in \mathcal{E}^k$, note that there exists $(p',q')\in \mathcal{\tilde{P}}^k\times \mathcal{\tilde{P}}^{n-k}$ such that $\alpha^k(p',q')=E$.   Combining Lemma \ref{LemmaSemiSP} and (\ref{EqproofEUn}), we thus have
\begin{eqnarray*}\label{EqundyadE1}
  E&=& \alpha_{\Omega_1}^k(p',q')=\alpha^k((p_1,p'_{-1}), q')))= \alpha^k_{\Omega_1}((p_1,p_2,p'_3\ldots,p'_k), q'))=\ldots= \alpha^k_{\Omega_1}(p, q')\\
   &=& \alpha^k_{\Omega_1}(p, (q_{k+1},q'_{-(k+2)}))= \alpha^k_{\Omega_1}(p, (q_{k+1},q_{k+2}, q'_{k+3},\ldots, q'_n)))=\ldots=\alpha^k_{\Omega_1}(p,q).
\end{eqnarray*}
This completes the proof.
 \endproof

\bigskip
For any nonempty set of events $\mathcal{A}\subseteq 2^{\Omega}\setminus \{\emptyset\}$, we call
\textit{decomposition }of $\mathcal{A}$  any collection $\left\{ \mathcal{A}%
_{1},...,\mathcal{A}_{L}\right\} $ such that%
\begin{align} \label{Eqdecomp1}
 \mathcal{A}_{l} \subseteq &~ 2^{\Omega}\setminus \{\emptyset\}\text{ for }l=1,...,L; \\ \label{Eqdecomp2}
A_{l}\cap A_{l^{\prime }}=&~\varnothing \text{ for all }%
A_{l}\in \mathcal{A}_{l},\text{ }A_{l^{\prime }}\in \mathcal{A}_{l^{\prime
}},\text{ }l\neq l^{\prime }; \\ \label{Eqdecomp3}
\mathcal{A} =&\left\{
\bigcup\limits_{l=1}^{L}A_{l}:(A_{1},...,A_{L})\in \mathcal{A}_{1}\times
...\times \mathcal{A}_{L}\right\} .
\end{align}%
A decomposition $\left\{ \mathcal{A}%
_{1},...,\mathcal{A}_{L}\right\}$ of $\mathcal{A}$ is called \textit{maximal }if there is no other
decomposition $\left\{ \mathcal{A}_{1}^{\prime },...,\mathcal{A}_{L^{\prime}}^{\prime }\right\} $ of $\mathcal{A}$ such that $L^{\prime }>L.$ \cite{barbera2005voting} prove that every set $\mathcal{A}\subseteq
2^{\Omega}\setminus \{\emptyset\}$ has a unique maximal decomposition ---see also \cite{svensson2008strategy}. The sets $\mathcal{A}_{1},...,\mathcal{A}_{L}$ are the
\textit{components }of $\mathcal{A}.$ Given any component $\mathcal{A}_l$ in the decomposition $\{\mathcal{A}_{1},...,\mathcal{A}_{L}\}$ of $\mathcal{A}$, we say that $\mathcal{A}_l$ is
\begin{itemize}
 \item  \emph{trivial} if $|\mathcal{A}_l| = 1$;
  \item \emph{dyadic} if $|\mathcal{A}_l|=2$;
  \item \emph{rich} if $|\mathcal{A}_l| \geq 3$.
\end{itemize}
Hence, each component of the maximal decomposition is  trivial, dyadic, or rich. Moreover,  a collection of events $\mathcal{A}$ will be called \emph{indecomposable} if its maximal decomposition has a single component (which must then be $\{\mathcal{A}\}$). A collection of events $\mathcal{A}$ will be called \emph{richly decomposable} if  its maximal decomposition exhibits at least one rich component.  Note in particular that if $\mathcal{A}$ is an indecomposable collection of events such that $|\mathcal{A}|\geq 3$, then $\mathcal{A}$ is richly decomposable.

Finally, call a state $\omega \in \tilde{\Omega}$ \emph{simple} if  $|\{\varphi(\bar{u}^{k},p;\omega): p\in \mathcal{\tilde{P}}_{\bar{u}^{k}}\}|=1, \forall k\in \{1,\ldots, n\} $, with $\bar{u}^{k}\in \mathcal{V}^k$ introduced in (\ref{EqNaproof}). In other words, given Lemma \ref{LemmaOrdinal}, a state is simple if,  for each valuation section $\mathcal{V}^N_k$, the selected outcome is constant (\emph{i.e.}, independent of the belief profile.)

 For any $\Omega_1\in 2^{\tilde{\Omega}}$, denote by $\varphi_{\Omega_1}$ the mapping selecting (at any preference profile) the restriction to $\Omega_1$ of the act selected by $\varphi(v,p)$ to $\Omega_1$, that is to say, $\varphi_{\Omega_1}(v,p)$ denotes the sub-act of $\varphi(v,p)$ that is selected on $\Omega_1$, for all $(v,p)\in \mathcal{\tilde{D}}^N$. Likewise, we define $\alpha^k_{\Omega_1}(p_{\{1,\ldots,k\}}, p_{\{k+1,\ldots,n\}}):=\alpha^k (p_{\{1,\ldots,k\}}, p_{\{k+1,\ldots,n\}})\cap \Omega_1$, for all $k=1,\ldots,n$ and $(p_{\{1,\ldots,k\}}, p_{\{k+1,\ldots,n\}})\in \mathcal{\tilde{P}}^{\{1,\ldots,k\}}\times \mathcal{\tilde{P}}^{\{k+1,\ldots,n\}}$.

\begin{lemma}\label{LemmaDomalloc}  \textbf{Dominant-belief allocation}
\newline
\noindent Fix $k\in \{1,\ldots, n-1\}$ and $\hat{\Omega}\in 2^{\tilde{\Omega}}$. Consider a map $\gamma:(\mathcal{P}(\hat{\Omega}))^k\times (\mathcal{P}(\hat{\Omega}))^{n-k} \rightarrow 2^{\hat{\Omega}}$ satisfying,  for all $(p,q)\in (\mathcal{P}(\hat{\Omega}))^k\times (\mathcal{P}(\hat{\Omega}))^{n-k}$, (A)  $ p_i(\gamma(p,q)) =\max\limits_{p'_i \in \mathcal{P}(\hat{\Omega})} ~p_i(\gamma((p'_i,p_{-i}),q)$, for all $i\in \{1,\ldots, k\}$;
        (B) $q_j(\gamma((p,q)) =\min\limits_{q'_i \in \mathcal{P}(\hat{\Omega})} ~q_j(\gamma(p,(q'_j,q_{-j})$, for all $j\in \{1,\ldots, n-k\}$.
              \newline\noindent Then the following properties hold: for all $\Omega_1\in 2^{\hat{\Omega}_1}$, $p,p' (\mathcal{P}(\hat{\Omega}))^k $, $q,q'\in (\mathcal{P}(\hat{\Omega}))^{n-k}$,
\newline\noindent
\textbf{(I)}-~ if $p,p'$ are $\Omega_1$-dominant and $p_{/\Omega_1}=p'_{/\Omega_1}$, then we have $\gamma_{\Omega_1}(p,q)=\gamma_{\Omega_1}(p',q)$;
\newline
\noindent
\textbf{(II)}- if $q,q'$ are $\Omega_1$-dominant and $q_{/\Omega_1}=q'_{/\Omega_1}$, then we have $\gamma_{\Omega_1}(p,q)=\gamma_{\Omega_1}(p,q')$.

\end{lemma}

\proof  Let $\hat{\Omega}, \gamma, \Omega_1$ be as in the statement of Lemma \ref{LemmaDomalloc}.

\noindent
(I) Fix $p,p'\in (\mathcal{P}(\hat{\Omega}))^k$ such that $p$ and $p'$ are $\Omega_1$-dominant; and let $q\in (\mathcal{P}(\hat{\Omega}))^{n-k}$.
To see why $\gamma_{\Omega_1}(p,q)=\gamma_{\Omega_1}(p',q)$ holds,  define $p^0=p$, $p^i=(p'_i, p^{i-1}_{-i})$, for  $i=1,\ldots,k$ (so that $p^k=p'$). Assuming  $\gamma_{\Omega_1}(p^1,q)\neq \gamma_{\Omega_1}(p^0,q)$ would mean  that  $p'_1(\gamma_{\Omega_1}(p^1,q))>p'_1((\gamma_{\Omega_1}(p^0,q))$, because the reverse inequality would lead to $p'_1(\gamma(p^1, q))<p'_1(\gamma(p^0,q))$ (since $p'_1$ is $\Omega_1$-dominant), in contradiction with assumption (I). Given that $p_{/\Omega_1}=p'_{/\Omega_1}$, remark that  $p'_1(\gamma_{\Omega_1}(p^1,q))>p'_1((\gamma_{\Omega_1}(p^0,q))$ is equivalent to $p_1(\gamma_{\Omega_1}(p^1,q))>p_1(\gamma_{\Omega_1}(p^0,q))$ [since $\gamma_{\Omega_1}(p^1,q), \gamma_{\Omega_1}(p^0,q)\in 2^{\Omega_1}$]. Given that $p_1$ is $\Omega_1$-dominant,  the previous inequality implies $p_1(\gamma(p^1, q))>p_1(\gamma(p^0, q))$, which contradicts assumption (I).
 Hence, we must have instead $\gamma_{\Omega_1}(p^1,q)= \gamma_{\Omega_1}(p^0,q)$. Repeating this argument $k$ times, one can write $\gamma_{\Omega_1}(p,q)=\gamma_{\Omega_1}(p^0,q)= \gamma_{\Omega_1}(p^1,q) =\ldots=\gamma_{\Omega_1}(p^{k-1},q)=\gamma_{\Omega_1}(p^k,q)=\gamma_{\Omega_1}(p',q)$.

 \medskip\noindent (II) Fix now $p\in (\mathcal{P}(\hat{\Omega}))^k $ and   $q,q'\in (\mathcal{P}(\hat{\Omega}))^{n-k}$ such that $q$ and $q'$ are $\Omega_1$-dominant. Define $q^0=q$, $q^i=(q'_i, q^{i-1}_{-i})$, for  $i=1,\ldots,n-k$ (so that $q^{n-k}=q'$). Assuming  $\gamma_{\Omega_1}(p,q^1)\neq \gamma_{\Omega_1}(p,q^0)$ would mean  that  $q'_{1}(\gamma_{\Omega_1}(p,q^{1}))<q'_{1}((\gamma_{\Omega_1}(p,q^0))$, because the reverse inequality would lead to $q'_{1}(\gamma(p,q^{1}))>q'_{1}((\gamma(p,q^0))$ (since $q'_{1}$ is $\Omega_1$-dominant), in contradiction with assumption (II). Given that $q_{/\Omega_1}=q'_{/\Omega_1}$, remark that  $q'_{1}(\gamma_{\Omega_1}(p,q^{1}))<q'_{1}((\gamma_{\Omega_1}(p,q^0))$ is equivalent to $q_{1}(\gamma_{\Omega_1}(p,q^{1}))<q_{1}((\gamma_{\Omega_1}(p,q^0))$ [since $\gamma_{\Omega_1}(p,q^{1}), \gamma_{\Omega_1}(p,q^{0})\in 2^{\Omega_1}$]. Given that $q_1$ is $\Omega_1$-dominant,  the previous inequality implies $q_1(\gamma(p,q^{1}))<q_1(\gamma(p,q^{0}))$, which contradicts assumption (II).
 Hence, we must have instead $\gamma_{\Omega_1}(p,q^1)= \gamma_{\Omega_1}(p,q^0)$. Repeating this argument $n-k$ times, one can write $\gamma_{\Omega_1}(p,q)=\gamma_{\Omega_1}(p,q^0)= \gamma_{\Omega_1}(p,q^1) =\ldots=\gamma_{\Omega_1}(p,q^{n-k-1})=\gamma_{\Omega_1}(p,q^{n-k})=\gamma_{\Omega_1}(p,q')$.
  \endproof

\begin{lemma}\label{LemmaInvariancePres}  \textbf{Monotonic sectional invariance}\\
\noindent Let $k\in \{1,\ldots,n-1\}$ and $\omega\in \tilde{\Omega}$.  Assume that $\omega\in \alpha^k(p,q)$, for all $(p,q)\in \mathcal{\tilde{P}}^{k}\times\mathcal{\tilde{P}}^{n-k}$.
Then it also holds that $\omega\in \alpha^{k+1}(p,q)$, for all $(p,q)\in  \mathcal{\tilde{P}}^{k+1}\times\mathcal{\tilde{P}}^{n-k-1}$.
\end{lemma}
\proof Fix $k\in \{1,\ldots,n-1\}$ and $\omega\in \tilde{\Omega}$ such that $\omega\in \alpha^k(p,q)$, for all $(p,q)\in \mathcal{\tilde{P}}^{k}\times \mathcal{\tilde{P}}^{n-k}$.
Note first that the result trivially holds if $k=n-1$, since  range-unanimity gives  $\alpha^n(p,q)=\tilde{\Omega}$, for all $(p,q)\in  \mathcal{\tilde{P}}^{n}$. Assume then that $1\leq k\leq n-2$ (i.e., $n\geq 3$); and suppose by contradiction that there exists a pair $(p,q)\in \mathcal{\tilde{P}}^{k+1}\times \mathcal{\tilde{P}}^{n-k-1}$ such that $\omega\notin \alpha^{k+1}(p,q)$.
Pick  $(\bar{v},\bar{p})\in \mathcal{\tilde{D}}^N$  such that
\[\mbox{(I) } N^{\bar{v}}_a=\{1,\ldots,k\},  \mbox{ (II) }  \bar{p}_{k+1}(\omega)>1/2,       \mbox{ (III) } \bar{p}_i=\left\{\begin{array}{cl}
                                                                                                                                                         p_i & \mbox{if } i=1,\ldots,k ;\\
                                                                                                                                                         q_{i-k-1} & \mbox{if }  i=k+2,\ldots,n .
                                                                                                                                                       \end{array} \right.  \]
Note from (I) and Lemma \ref{LemmaSections} that $\varphi^a(\bar{v},\bar{p})=\alpha^k(\bar{p}_{\{1,\ldots,k\}},\bar{p}_{\{k+1,\ldots,n\}})$.                                                                                                                                                       Combining this equality with $\omega\in \alpha^k(\bar{p}_{\{1,\ldots,k\}},\bar{p}_{\{k+1,\ldots,n\}})$ thus gives $\bar{p}_{k+1}(\varphi^a(\bar{v},\bar{p}))\geq \bar{p}_{k+1}(\omega)> 1/2$, where the strict inequality comes from (II).

But $\bar{p}_{k+1}(\varphi^a(\bar{v},\bar{p}))> 1/2$ leads to a manipulation of $\varphi$ at $\varphi^a(\bar{v},\bar{p})$ by agent $k+1$. Agent $k+1$ can report $(v'_{k+1}, p'_{k+1})$ such that (IV) $v'_{k+1}(a)>v'_{k+1}(b)$ and (V) $p'_{k+1}=q_1$ to induce the choice of the act $\varphi((v'_{k+1},\bar{v}_{-(k+1)}), (p'_{k+1},\bar{p}_{-k+1}))$. Define $p'=(p'_{k+1}, \bar{p}_{-(k+1)})$. Combining (I) and (IV) gives $N^{(v'_{k+1},\bar{v}_{-(k+1)})}_a=\{1,\ldots,k+1\}$;  and  combining (III) and (V) yields $(p'_{\{1,\ldots,k+1\}}, p'_{\{k+2,\ldots,n\}}=(p,q)$. It thus comes from Lemma \ref{LemmaSections} that  $\varphi^a((v'_{k+1},v_{-(k+1)}),p')=\alpha^{k+1}(p,q)$; and since $\omega\notin \alpha^{k+1}(p,q)$, it comes from (II) that $\bar{p}_{k+1}(\varphi^a((v'_{k+1},v_{-(k+1)}),p'))\leq 1-\bar{p}_{k+1}(\omega)<1/2$.

Combining the respective results of the previous two paragraphs, we have
\[\bar{p}_{k+1}(\varphi^a((v'_{k+1},v_{-(k+1)}),p'))<\bar{p}_{k+1}(\varphi^a(\bar{v}, \bar{p})).\]
                                                                 Given that $\varphi$ is binary, this inequality means that agent 2 benefits from misreporting $(v'_{k+1}, p'_{k+1})$ at the profile $(\bar{v}, \bar{p})$, which contradicts the strategy-proofness of $\varphi$. \endproof

\begin{lemma}\label{LemmaSepSimpleDict} ~ \textbf{Constant and dictatorial separation}\\
\noindent Let $j,k\in \{1,\ldots,n\}$, $\omega_1\in \tilde{\Omega} $, $\Omega_1\in 2^{\tilde{\Omega}}\setminus \{\emptyset, \tilde{\Omega}\}$; and assume that $ \{A_1,\ldots, A_L\}\subset 2^{\Omega_1}$ is a collection of non-nested events covering $\Omega_1$ (with $L\geq 2$).

\smallskip\noindent \textbf{I.}  Suppose that, for all $v\in\mathcal{V}_k^N$, $|\{\varphi(v,p;\omega_1): p\in \mathcal{\tilde{P}}_v\}|=1$. Then we have
\[ \varphi_{\overline{\omega_1}}(v,p)=\varphi_{\overline{\omega_1}}(v,q), \mbox{ for all } v\in \mathcal{V}_k^N \mbox{ and }  p,q\in \mathcal{\tilde{P}}_v \mbox{ s.t. } p_{/\overline{\omega_1}}= q_{/\overline{\omega_1}}.\]

\smallskip\noindent \textbf{II.}  Suppose  $\varphi^a_{\Omega_1}(v,p)=\argmax\limits_{l=1,\ldots,L}p_j(A_l)$, for all $(v,p)\in \mathcal{\tilde{D}}^N$ s.t. $N^v_a=\{j\}$. Then we have
\[ \varphi_{\overline{\Omega_1}}(v,p)=\varphi_{\overline{\Omega_1}}(v,q), \mbox{ for all } (v,p)\in \mathcal{\tilde{D}}^N \mbox{ s.t. } N^v_a=\{j\} \mbox{ and } p_{/\overline{\Omega_1}}= q_{/\overline{\Omega_1}}.\]
\end{lemma}
\proof Let $j,k\in \{1,\ldots,n\}$, $\omega_1\in \tilde{\Omega} $, $\Omega_1\in 2^{\tilde{\Omega}}\setminus \{\emptyset, \tilde{\Omega}\}$ and $ \{A_1,\ldots, A_L\}\subset 2^{\Omega_1}$ be as in the statement of Lemma \ref{LemmaSepSimpleDict}.

\smallskip\noindent \textbf{I.} Suppose that  $|\{\varphi(v,p;\omega_1): p\in \mathcal{\tilde{P}}^N\}|=1$ for all $v\in\mathcal{V}_k^N$, \emph{i.e.}, for all $v\in \mathcal{V}_k^N$, the outcome chosen in $\omega_1$ is the same, regardless of $p\in \mathcal{\tilde{P}}_v$. Assume without loss of generality that
\begin{equation}\label{EqSepI}\varphi(v,p,\omega_1)= a, ~ \forall v\in \mathcal{V}^N_k, \forall p\in \mathcal{\tilde{P}}_v.
\end{equation}

Next, let us fix $ v\in \mathcal{V}_k^N$ and $p,q\in \mathcal{\tilde{P}}_v$ such that $p_{/\overline{\omega_1}}= q_{/\overline{\omega_1}}$.
Define $p^0=p$ and let $p^i=(q_i, p^{i-1}_{-i})$ for all $i=1,\ldots,n$ [so that $p^n=q$]. Moreover, define $~^if:=\varphi(v,p^i)$ for all $i\in \{0,1,\ldots,n\}$.  Recalling from (\ref{EqSepI}) that $~^if(\omega_1)=~^{i-1}f(\omega_1)=a$, one can use strategy-proofness of $\varphi$ for each agent $i\in \{1,\ldots,n\}$, respectively at $(v,p^{i-1})$ and $(v,p^{i})$, to write
\begin{align}\label{EqConstsep1}
  p_i(\omega_1)v_i(a) + \sum_{\omega\neq \omega_1} p_i(\omega)v_i(~^{i-1}f(\omega))&\geq p_i(\omega_1)v_i(a) + \sum_{\omega\neq \omega_1} p_i(\omega)v_i(f_{i}(\omega)) \\
  \label{EqConstsep2}
   q_i(\omega_1)v_i(a) + \sum_{\omega\neq \omega_1} q_i(\omega)v_i(~^{i-1}f(\omega))&\leq q_i(\omega_1)v_i(a) + \sum_{\omega\neq \omega_1} q_i(\omega)v_i(f_{i}(\omega))
\end{align}
Since $p_{i/\overline{\omega_1}}= q_{i/\overline{\omega_1}}$, note that (\ref{EqConstsep2}) can be rewritten as
\[\frac{1}{p_i(\overline{\omega_1})}\sum_{\omega\neq \omega_1} p_i(\omega)v_i(~^{i-1}f(\omega))\leq   \frac{1}{p_i(\overline{\omega_1})} \sum_{\omega\neq \omega_1} p_i(\omega)v_i(f_{i}(\omega)).\]
That is to say,
\begin{align}\label{EqConstsep3}
  \sum_{\omega\neq \omega_1} p_i(\omega)v_i(~^{i-1}f(\omega))&\leq   \sum_{\omega\neq \omega_1} p_i(\omega)v_i(f_{i}(\omega)).
\end{align}
Combining (\ref{EqConstsep3}) and (\ref{EqConstsep1}) thus gives
\[p_i(\omega_1)v_i(a) + \sum_{\omega\neq \omega_1} p_i(\omega)v_i(~^{i-1}f(\omega))= p_i(\omega_1)v_i(a) + \sum_{\omega\neq \omega_1} p_i(\omega)v_i(f_{i}(\omega)),\]
which means that $~^if\sim_{(v_i,p_i)} ~^{i-1}f$. Since $(v_i,p_i)\in \mathcal{\tilde{D}}$ induces a strict ranking of $X^{\tilde{\Omega}}$ (no indifference between distinct acts), we  conclude that $~^if=~^{i-1}f$ for all $i\in \{1,\ldots,n\}$. That is to say, $~^n f=~^{n-1}f=\ldots=~^1 f=~^0 f$ and hence $\varphi_{\overline{\omega_1}}(v,p)=\varphi_{\overline{\omega_1}}(v,q)$.

\bigskip\noindent \textbf{II.} Suppose  $\varphi^a_{\Omega_1}(v,p)=\argmax\limits_{l=1,\ldots,L}p_j(A_l)$, for all $(v,p)\in \mathcal{\tilde{D}}^N$ s.t. $N^v_a=\{j\}$, \textit{i.e.}, whenever  agent $\{j\}$ is the only one  preferring $a$ to $b$, the subset of $\Omega_1$ where the outcome a is selected is  the event   $j$ finds most likely within $\{A_1,\ldots,A_L\}$.
We proceed in two steps.

\smallskip\noindent
\emph{Claim 1}:  $\varphi_{\overline{\Omega_1}}(v,p)=\varphi_{\overline{\Omega_1}}(v,q)$  for all  $v\in \mathcal{V}$ s.t. $N^v_a=\{j\}$ and any  $p,q\in \mathcal{\tilde{P}}_v$ satisfying [$p_{/\overline{\Omega_1}}= q_{/\overline{\Omega_1}}$ and $\argmax\limits_{l=1,\ldots,L}p_j(A_l)= \argmax\limits_{l=1,\ldots,L}q_j(A_l)$].

\smallskip\noindent The proof of Claim 1 will not be given explicitly: it is a variant of the argument given in  \textbf{I} above. Indeed, fixing  $v\in \mathcal{V}_k^N, p,q$ as in the statement of Claim 1, one must define once again $~^if$ ($i=0,\ldots, n$).  Notice that we have $~^{i-1} f_{\Omega_1}=\varphi_{\Omega_1}(v,(p_i,p^{i-1}_{-i}))=\varphi_{\Omega_1}(v,(q_i,p^{i-1}_{-i}))=\varphi_{\Omega_1}(v,(p_i,p^{i}_{-i}))=~^i f_{\Omega_1}$ for  $i=1,\ldots,n$, since (i) the selected sub-act in $\Omega_1$ is independent of the belief of any $i\neq j$ and (ii) $\argmax\limits_{l=1,\ldots,L}p_j(A_l)= \argmax\limits_{l=1,\ldots,L}q_j(A_l)$. It then suffices to replace  $\omega_1$ with $\Omega_1$ in (\ref{EqConstsep1})-(\ref{EqConstsep2})  to obtain in the same manner $\varphi_{\overline{\Omega_1}}(v,p)=~^0 f=\ldots=~^n f=\varphi_{\overline{\Omega_1}}(v,q)$, given that $p_{/\overline{\Omega_1}}= q_{/\overline{\Omega_1}}$.

\medskip\noindent
\emph{Claim 2}: For any  $v\in \mathcal{V}^N$ s.t. $N^v_a=\{j\}$ and any $p,q\in \mathcal{\tilde{P}}_v$ s.t.$p_{/\overline{\Omega_1}}= q_{/\overline{\Omega_1}}$, there exist $p',q'\in \mathcal{\tilde{P}}_v$ satisfying (i)  $\varphi_{\overline{\Omega_1}}(v,p')=\varphi_{\overline{\Omega_1}}(v,q')$  (ii) $\argmax\limits_{l=1,\ldots,L}p_j(A_l)= \argmax\limits_{l=1,\ldots,L}p_j'(A_l)$, (iii) $\argmax\limits_{l=1,\ldots,L}q_j(A_l)= \argmax\limits_{l=1,\ldots,L}q'_j(A_l)$.

\medskip
To see why Claim 2 holds, fix $v\in \mathcal{V}^N$ s.t. $N^v_a=\{j\}$ and any $p,q\in \mathcal{\tilde{P}}_v$ s.t.$p_{/\overline{\Omega_1}}= q_{/\overline{\Omega_1}}$. Recalling (\ref{Eqdefcond}), note that we can choose $\theta\in (0,1)$ small enough to guarantee that both $p^{\theta,\Omega_1}$ and $q^{\theta,\Omega_1}$ are $\overline{\Omega_1}$-dominant. Letting $p'=p^{\theta,\Omega_1}$ and $q'=q^{\theta,\Omega_1}$, it is easy to see from (\ref{Eqdefcond}) that   ii) $\argmax\limits_{l=1,\ldots,L}p_j(A_l)= \argmax\limits_{l=1,\ldots,L}p_j'(A_l)$, (iii) $\argmax\limits_{l=1,\ldots,L}q_j(A_l)= \argmax\limits_{l=1,\ldots,L}q'_j(A_l)$.
We show next  that $\varphi_{\overline{\Omega_1}}(v,p')=\varphi_{\overline{\Omega_1}}(v,q')$ must hold. Let $p'^0=p'$ and, for all $i=1,\ldots,n$, define $p'^{i}=(q'_i,p'^{i-1}_{-i})$ (so that $p'^n=q'$). Moreover, observe that strategyproofness of $\varphi$  yields $\varphi_{\overline{\Omega_1}}(v,p'^i)=\varphi_{\overline{\Omega_1}}(v,p'^{i-1})$ for all $i=,1\ldots,n$.
Indeed, letting $i\in \{1,\ldots,n\}$, $f= \varphi(v,p'^{i-1})$, $g=\varphi(v,p'^{i})$, and assuming $f_{\overline{\Omega_1}} \neq g_{\overline{\Omega_1}}$, one can use the strategyproofness of $\varphi$ at $(v,p'^{i-1})$  to write
\begin{equation}\label{EqDictSep1}
  \sum_{\omega\in \overline{\Omega_1}}p'_i(\omega) v_i(f(\omega))+ \sum_{\omega\in \overline{\Omega_1}}p'_i(\omega) v_i(f(\omega))>  \sum_{\omega\in \overline{\Omega_1}}p'_i (\omega) v_i(g(\omega))+ \sum_{\omega\in \overline{\Omega_1}}p'_i(\omega) v_i(g(\omega))\end{equation}
Given that the belief $p'_i$ is $\overline{\Omega_1}$-dominant, note that (\ref{EqDictSep1}) implies
\begin{equation}\label{EqDictSep2}
 \sum_{\omega\in \overline{\Omega_1}}p'_i(\omega) v_i(f(\omega))>  \sum_{\omega\in \overline{\Omega_1}}p'_i(\omega) v_i(g(\omega)).
\end{equation}
Since $p'_{i/\overline{\Omega_1}}= q'_{i/\overline{\Omega_1}}$, remark that (\ref{EqDictSep2}) is equivalent to
\begin{equation}\label{EqDictSep3}
 \sum_{\omega\in \overline{\Omega_1}}q'_i(\omega) v_i(f(\omega))>  \sum_{\omega\in \overline{\Omega_1}}q'_i(\omega) v_i(g(\omega)).
 \end{equation}
Given that $q'_i$ is $\overline{\Omega_1}$-dominant, it thus comes from (\ref{EqDictSep3}) that
\begin{equation}\label{EqDictSep4}
 \sum_{\omega\in \overline{\Omega_1}}q'_i(\omega) v_i(f(\omega))+ \sum_{\omega\in \overline{\Omega_1}}q'_i(\omega) v_i(f(\omega))>  \sum_{\omega\in \overline{\Omega_1}}q'_i(\omega) v_i(g(\omega))+ \sum_{\omega\in \overline{\Omega_1}}q'_i (\omega) v_i(g(\omega)),
 \end{equation}
which contradicts the strategyproofness of $\varphi$ at $(v,p'^{i})$. Therefore, we must have
$\varphi_{\overline{\Omega_1}}(v,p')=\varphi_{\overline{\Omega_1}}(v,p'^0)=\varphi_{\overline{\Omega_1}}(v,p'^{1})=\ldots=\varphi_{\overline{\Omega_1}}(v,p'^n)= \varphi_{\overline{\Omega_1}}(v,q')$, and Claim 2 is shown.

\bigskip Finally, combining Claim 1 and Claim 2, it is  easy to see that $\varphi_{\overline{\Omega_1}}(v,p)=\varphi_{\overline{\Omega_1}}(v,q)$, for all $(v,p)\in \mathcal{\tilde{D}}^N$ s.t. $N^v_a=\{j\}$ and  $p_{/\overline{\Omega_1}}= q_{/\overline{\Omega_1}}$.
\endproof

To ease on notation, we will often write $I^{m'}_m=\{m,\ldots, m'\}$, for any integers $m,m'\in \SN$ such that $m\leq m'$. We can then state the following lemma.

\begin{lemma}\label{LemmaDyad}  \textbf{Dyadic separation}\\
\noindent Fix $k\in I_1^{n-1}$ and $\hat{\Omega}\in 2^{\tilde{\Omega}}$. Consider a map $\gamma:(\mathcal{P}(\hat{\Omega}))^k\times (\mathcal{P}(\hat{\Omega}))^{n-k} \rightarrow 2^{\hat{\Omega}}$ such that, for all $(p,q)\in (\mathcal{P}(\hat{\Omega}))^k\times (\mathcal{P}(\hat{\Omega}))^{n-k}$, we have:   $\mathbf{(I)}$  $ p_i(\gamma(p,q)) =\max\limits_{p'_i \in \mathcal{P}(\hat{\Omega})} ~p_i(\gamma((p'_i,p_{-i}),q),$ $\forall i\in I^k_1$;
          $\mathbf{(II)}$  $q_j(\gamma((p,q)) =\min\limits_{q'_i \in \mathcal{P}(\hat{\Omega})} ~q_j(\gamma(p,(q'_j,q_{-j}) \forall j\in I^{n-k}_1$;   $\mathbf{(III)}$  $\gamma(\sigma_{ij}p,q)= \gamma(p,q)$, $\forall i,j\in I^k_1$;  $\mathbf{(IV)}$  $\gamma(p,\sigma_{i,j}q)=\gamma(p,q)$, $\forall i,j\in I_1^{n-k}$.

\noindent Suppose that the maximal decomposition of the range of $\gamma$ has $M$ dyadic components (given by $\{E_1,F_1\}, \ldots, \{E_M,F_M\}$, where $M\geq 1$) and no rich component.
Then for every $m\in \{1,\ldots, M\}$ there exists a non-decreasing function $H_m: \SN \times \SN \rightarrow \{0,1\}$ s.t.   $H_m(0,0)=0$, $H_m(k,n-k)=1$  and, for all $(p,q)\in \mathcal{\tilde{P}}^k\times \mathcal{\tilde{P}}^{n-k}$,
\[[\gamma_{E_m\cup F_m}(p,q)=E_m]  \mbox{\textbf{ iff} }
[H_m(\eta_{p}(I^{k}_1, E_m|F_m),  \eta_{q}(I^{1}_{n-k}, F_m|E_m))=1].\]
\end{lemma}
\proof Let $k\in I_1^{n-1}$ and $\hat{\Omega}\in 2^{\tilde{\Omega}}$. We  prove Lemma \ref{LemmaDyad} by induction over the number $M$ of dyadic components exhibited by the range of $\gamma$.

\noindent
\underline{\textbf{Case 1}}:
Consider  a map $\gamma: (\mathcal{P}(\hat{\Omega}))^k\times (\mathcal{P}(\hat{\Omega}))^{n-k} \rightarrow 2^{\hat{\Omega}}$ whose range $\mathcal{R}=\{\gamma (E), E\in 2^{\hat{\Omega}}\}$ has a maximal decomposition with a \textbf{unique dyadic component} $\{E_1,F_1\}$, and \textbf{no rich component} (a trivial component may or may not exist). Assume also that $\gamma$ satisfies (I)-(IV) as in the statement of Lemma \ref{LemmaDyad}. Write $\Omega_1=E_1\cup F_1$.

Recall that, for any $C\subset \hat{\Omega}$, a belief $p_i\in \mathcal{P}(\hat{\Omega})$ is called  $C$-dominant if we have $\min\limits_{\tiny\begin{array}{c}
                                                                                                                   E,F\subseteq C \\
                                                                                                                   E\neq F
                                                                                                                 \end{array}} |p_i(E)-p_i(F)|>p_i(\hat{\Omega}\setminus C)$.
We prove the claim below.

\medskip\noindent
\emph{Claim 1}: There exists a non-decreasing function $H: \SN \times \SN \rightarrow \{0,1\}$   s.t.  $H(0,0)=0$, $H(k,n-k)=1$ and, for all  $\Omega_1\mbox{-dominant } (p,q)\in \mathcal{\tilde{P}}^k\times \mathcal{\tilde{P}}^{n-k}$,
\[[\gamma_{\Omega_1}(p,q)=E] \mbox{ \textbf{iff} }
[H(\eta_{(p,q)}(I^{k}_1, E_1|F_1),  \eta_{(p,q)}(\eta_{q}(I^{n-k}_{1}, F_1|E_1))=1].\]

\noindent
 For any $C\in \{E_1,F_1\}$, define the set
\[ \Lambda^C_{\Omega_1}:=\{(p,q)\in \mathcal{\tilde{P}}^k\times \mathcal{\tilde{P}}^{n-k}: (p,q) \mbox{ is } \Omega_1\mbox{-dominant} \mbox{ and } \gamma_{\Omega_1}(p,q)=C\}.\]
Obviously, for any $\Omega_1$-dominant $(p,q)\in \mathcal{\tilde{P}}^k\times \mathcal{\tilde{P}}^{n-k}$,
we have $(p,q)\in  \Lambda^{E_1}_{\Omega_1}\cup  \Lambda^{F_1}_{\Omega_1}$, since $\{E_1,F_1\}$ is the range of $\gamma_{\Omega_1}$.
We proceed in two steps to prove Claim 1:  we show first that $\Lambda^{E_1}_{\Omega_1}, \Lambda^{F_1}_{\Omega_1}\neq \emptyset$; and then we prove the existence of $H$.

\medskip\noindent\emph{Step 1}: $\Lambda^{E_1}_{\Omega_1}, \Lambda^{F_1}_{\Omega_1}\neq \emptyset$.

\smallskip\noindent
Since $\{E_1,F_1\}$ is a component of the maximal decomposition of $\mathcal{R}$ (the range of $\gamma$), there exists $(p',q')\in \mathcal{\tilde{P}}^k\times \mathcal{\tilde{P}}^{n-k}$ such that $\gamma_{\Omega_1}(p', q')=E_1$.
Fix now  an $\Omega_1$-dominant profile $(p,q)\in \mathcal{\tilde{P}}^k\times \mathcal{\tilde{P}}^{n-k}$ s.t.
\begin{equation}\label{EqDyadUnan}
  E_1=\argmax\{p_i(E_1), p_i(F_1)\}=\argmin \{q_j(E_1), q_j(F_1)\}, \forall i\in I_1^k, \forall j\in I_{1}^{n-k}.
\end{equation}
Equation (\ref{EqDyadUnan}) guarantees that all agents find the event $E_1$ to be the most desirable in the range of $\gamma_{\Omega_1}$ (recall that agents in $I_{1}^{n-k}$ seek to minimize the probability of the event where $a$ is selected  ).
Define now $(p^0,q^0)=(p',q')$ and $p^{i}=(p_{i}, p_{i}^{i-1})$, $q^{j}=(q_{j}, q_{j}^{j-1})$, for all $i\in I_{1}^{k}$ and $j\in I_{1}^{n-k}$.
Then, since $(p,q)$ is $\Omega_1$-dominant, the combination  of Equation (\ref{EqDyadUnan}) and assumptions (I)-(II)  gives:
\begin{eqnarray}\label{EqHknk}
 \nonumber E_1&=& \gamma_{\Omega_1}((p_1,p'_{-1}), q'))= \gamma_{\Omega_1}((p_1,p_2,p'_3\ldots,p'_k), q'))=\ldots= \gamma_{\Omega_1}(p, q')\\
   &=& \gamma_{\Omega_1}(p, (q_{1},q'_{-1}))= \gamma_{\Omega_1}(p, (q_{1},q_{2}, q'_{3},\ldots, q'_{n-k}))=\ldots=\gamma_{\Omega_1}(p,q).
\end{eqnarray}


Of course, the analogous of (\ref{EqHknk}) obtains in a similar way, \emph{i.e.}, given  any $\Omega_1$-dominant profile $(p,q)\in  \mathcal{\tilde{P}}^k\times \mathcal{\tilde{P}}^{n-k}$ s.t. $ E_1=\argmax\{p_i(E_1), p_i(F_1)\}=\argmin \{q_j(E_1), q_j(F_1)\}$ for all $i\in I_{1}^{k}, \forall j\in I_{1}^{n-k}$, we have
\begin{eqnarray}\label{EqH00} F_1=\gamma_{\Omega_1}(p,q).
\end{eqnarray}

We have thus shown that $\Lambda^{E_1}_{\Omega_1}, \Lambda^{F_1}_{\Omega_1}\neq \emptyset$, which means that the restriction of $\gamma_{\Omega_1}$ to the set of $\Omega_1$-dominant beliefs has full range $\{E_1,F_1\}$.

\medskip\noindent\emph{Step 2}: Existence of $H$.

\smallskip\noindent First, note that assumption (I) yields  $\gamma_{\Omega_1}(p,q)=\gamma_{\Omega_1}(p'_{i},p_{-i}),q)$, for any $\Omega_1$-dominant $(p,q)\in   \mathcal{\tilde{P}}^k\times \mathcal{\tilde{P}}^{n-k}$ and  $\Omega_1$-dominant $p'_i\in \mathcal{\tilde{P}}$ s.t. $(p_i(E_1)-p_i(F_1))(p'_i(E_1)-p'_i(F_1))>0$. Likewise, assumption (II) yields $\gamma_{\Omega_1}(p,q)=\gamma_{\Omega_1}(p,(q'_i,q_{-i}))$,
for any $\Omega_1$-dominant $(p,q)\in   \mathcal{\tilde{P}}^k\times \mathcal{\tilde{P}}^{n-k}$ and $\Omega_1$-dominant $p'_i\in \mathcal{\tilde{P}}$ such that $(p_i(E_1)-p_i(F_1))(p'_i(E_1)-p'_i(F_1))>0$. Applying these two properties repeatedly (as many times as there are agents $i$ s.t. $p_i\neq p'_i$ or $q_i\neq q'_i$), it hence follows that $\gamma_{\Omega_1}(p,q)= \gamma_{\Omega_1}(p',q')$, for any $(p,q),(p',q')\in  \mathcal{\tilde{P}}^k\times \mathcal{\tilde{P}}^{n-k}$ s.t. [$(p_i(E_1)-p_i(F_1))(p'_i(E_1)-p'_i(F_1))>0, \forall i\in I_1^k$] and [$(q_j(E_1)-q_i(F_1))(q'_i(E_1)-q'_i(F_1))>0, \forall j\in I_1^{n-k}$].

Second, note  from assumption (III) that, for any $i,j\in I_{1}^{k}$, we have  $\gamma_{\Omega_1}(p,q)=\gamma_{\Omega_1}(\sigma_{ij}p,q)$. Likewise, it comes from assumption (IV) that $\gamma_{\Omega_1}(p,q)=\gamma_{\Omega_1}(p,\sigma_{ij}q)$, for any $i,j\in I_{1}^{n-k}$.
Combining these observations with the result of the previous paragraph, one can  claim that, for any $\Omega_1$-dominant $(p,q),(p',q')\in  \mathcal{\tilde{P}}^k\times \mathcal{\tilde{P}}^{n-k}$,
  \begin{align}\label{EqdefH0} \nonumber
  [(\eta_{(p,q)}(I_{1}^{k}, E_1|F_1), \eta_{(p,q)}(I_{1}^{n-k}, F_1|E_1))= (\eta_{(p',q')}(I_{1}^{k}, E_1|F_1), \eta_{(p',q')}(I_{1}^{n-k}, F_1|E_1))] \Rightarrow   \\ [\gamma_{\Omega_1}(p,q)= \gamma_{\Omega_1}(p',q')].
\end{align}

Define then $H$ as follows. For any $(t,t')\in \{1,\ldots,k\}\times\{1,\ldots,n-k\}$, let $H(t,t')=1$  [respectively, $H(t,t')=0$] if there exists a profile $(p,q)\in \Lambda^{E_1}_{\Omega_1}$ [$(p,q)\in \Lambda^{F_1}_{\Omega_1}$] such that
$(\eta_{(p,q)}(I_{1}^{k}, E_1|F_1), \eta_{(p,q)}(I_{1}^{n-k}, F_1|E_1))=(t,t')$. For any $(t,t')\in \SN\times \SN$ s.t. $t>k$ or $t'>n-k$, let $H(t,t')=1$.
Recalling (\ref{EqdefH0}), we have thus shown that there exists a function $H: \SN \times \SN \rightarrow \{0,1\}$   s.t., for all  $\Omega_1$-dominant profiles $(p,q)\in \mathcal{\tilde{P}}^k\times \mathcal{\tilde{P}}^{n-k}$,
\begin{equation}\label{EqdefH}
  [\gamma_{\Omega_1}(p,q)=E_1] \mbox{ \textbf{iff} }
[H(\eta_{(p,q)}(I_{1}^{k}, E_1|F_1),  \eta_{(p,q)}(I_{1}^{n-k}, F_1|E_1))=1].
\end{equation}
Given that $\gamma_{\Omega_1}$ and $H$ are both binary,  (\ref{EqdefH}) also means that,  for all  $\Omega_1$-dominant profiles $(p,q)\in \mathcal{\tilde{P}}^k\times \mathcal{\tilde{P}}^{n-k}$,
\[ [\gamma_{\Omega_1}(p,q)=F_1] \mbox{ \textbf{iff} }
[H(\eta_{(p,q)}(I_{1}^{k}, E_1|F_1),  \eta_{(p,q)}(I_{1}^{n-k}, F_1|E_1))=0].\]
Moreover, combining (\ref{EqdefH}) respectively with (\ref{EqHknk}) and (\ref{EqH00}), one can write $H(k,n-k)=1$ and $H(0,0)=0$.

Since $\eta_{(p,q)}(I_{1}^{k}, E_1|F_1)\leq k$ and $\eta_{(p,q)}(I_{1}^{n-k}, F_1|E_1)\leq n-k$,  it now remains to argue that the function $H$ is  non-decreasing on  $\{0,1,\ldots,k\}\times \{0,1,\ldots, n-k\}$. Fix $(t,t')\in \{0,1,\ldots, k-1\}\times\{0,1,\ldots,n-k\}$  and let $(p,q)\in \mathcal{\tilde{P}}^k\times \mathcal{\tilde{P}}^{n-k}$ be an $\Omega_1$-dominant profile  such that $\eta_{(p,q)}(I_{1}^{k}, E_1|F_1)=t$ and $\eta_{(p,q)}(I_{1}^{n-k},F_1|E_1)=t'$.
Given that $\eta_{(p,q)}(I_{1}^{k}, E_1|F_1)=t\leq k-1$, there exists $i\in I_{1}^{k}$ such that $p_i(E_1)<p_i(F_1)$.
Hence, letting $p'_i\in \mathcal{\tilde{P}}$ be an $\Omega_1$-dominant belief s.t. $p'_i(E_1)>p'_I(F_1)$, it comes that $\eta_{(p',q)}(I_{1}^{k}, E_1|F_1)=t+1$ and $\eta_{(p',q)}(I_{1}^{n-k}, F_1|E_1)=t'$, where $p':=(p'_i,p_{-i})$. Note then from assumption (I) that $p'_i(\gamma(p',q))\geq p'_i(\gamma(p,q))$; and because $p'_i$ is $\Omega_1$-dominant, we conclude that $p'_i(\gamma_{\Omega_1}(p',q))\geq p'_i(\gamma_{\Omega_1}(p,q))$. Since $p'_i(E_1)>p'_I(F_1)$, the previous inequality means that $[\gamma(p,q)=E_1]\Rightarrow [\gamma(p',q)=E_1]$. Recalling (\ref{EqdefH}), we have thus shown that $[H(t,t')=1]\Rightarrow [H'(t+1,t')=1]$; and hence $H$ is non-decreasing in its first argument.

Exploiting assumption (II), one can use a similar procedure to show that  $H$ is non-decreasing in its second argument as well. This concludes the proof of Claim 1.

  \medskip
\noindent
\underline{\textbf{Case 2}}:
 Consider now a map  $\gamma: (\mathcal{P}(\hat{\Omega}))^k\times (\mathcal{P}(\hat{\Omega}))^{n-k} \rightarrow 2^{\hat{\Omega}}$ whose range $\mathcal{R}=\{\gamma (E), E\in 2^{\hat{\Omega}}\}$ has a maximal decomposition  $\{E_1,F_1\}, \ldots, \{E_M,F_M\}$ , with $M\geq 2$ dyadic components (and no rich component). Suppose also that $\gamma$ satisfies (I)-(IV) as in the statement of Lemma \ref{LemmaDyad}. 
 
 \smallskip\noindent To simplify the exposition, assume without loss that $\mathcal{R}$ has no trivial component, that is to say,\footnote{If $\mathcal{R}$ has dyadic components $\{E_1,F_1\}, \ldots, \{E_M,F_M\}$ \textbf{and} a trivial component $\{E_0\}$, then it follows that $E_0\subset \gamma(p,q), \forall (p,q)\in (\mathcal{P}(\hat{\Omega}))^k\times (\mathcal{P}(\hat{\Omega}))^{n-k}$. One can then define $\underline{\Omega}=\hat{\Omega}\setminus E_0$ and a map $\underline{\gamma}: (\mathcal{P}(\underline{\Omega}))^k\times (\mathcal{P}(\underline{\Omega}))^{n-k} \rightarrow 2^{\underline{\Omega}}$ such that $\gamma(p,q)= \underline{\gamma}(p_{/\underline{\Omega}},q_{/\underline{\Omega}})\setminus E_0$, for all $(p,q)\in (\mathcal{P}(\hat{\Omega}))^k\times (\mathcal{P}(\hat{\Omega}))^{n-k}$. Note that this map $\underline{\gamma}$ inherits  the properties (I)-(IV) of $\gamma$;  and its range can be decomposed in exactly $M$ dyadic components $\{E_1,F_1\}, \ldots, \{E_M,F_M\}$ (with no trivial or rich component).}
  \begin{equation}\label{EqRangedyad}
  \mathcal{R}=\{C_1\cup\ldots\cup C_M,~ C_i\in \{E_m,F_m\} \mbox{ for } m=1,.\ldots,M \}.
    \end{equation}

  Given Claim 1, we may  assume \textbf{by induction} that any map $\gamma': (\mathcal{P}(\bar{\Omega}))^k\times (\mathcal{P}(\bar{\Omega}))^{n-k} \rightarrow 2^{\bar{\Omega}}$ (with $\bar{\Omega}\subseteq \tilde{\Omega}$), meeting (I)-(IV) and whose range has a maximal decomposition with $M'$ dyadic components (and no rich component), satisfies the statement of Lemma \ref{LemmaDyad} whenever $M'\leq M-1$.

  Write $\Omega_m=E_m\cup F_m$, for all $m=1,\ldots,M$; and let us use the shorthand notation $\Omega_{-1}=\Omega_2\cup \ldots\cup\Omega_M$, i.e.,  $\Omega_1\cup \Omega_{-1}=\hat{\Omega}$.
 Next, we show the claim below.

\medskip\noindent
\emph{Claim 2}: There exists a non-decreasing function $H_1: \SN \times \SN \rightarrow \{0,1\}$   s.t.  $H_1(0,0)=0$, $H_1(k,n-k)=1$ and, for all  $\Omega_1\mbox{-dominant } (p,q)\in (\mathcal{P}(\hat{\Omega}))^k\times (\mathcal{P}(\hat{\Omega}))^{n-k}$,
\[[\gamma_{\Omega_1}(p,q)=E_1] \mbox{ \textbf{iff} }
[H_1(\eta_{(p,q)}(I^{k}_1, E_1|F_1),  \eta_{(p,q)}(I^{n}_{k+1}, F_1|E_1))=1].\]

\noindent First, remark from Lemma \ref{LemmaDomalloc} that, for all $\Omega_1$-dominant $(p,q),(p',q')\in  (\mathcal{P}(\hat{\Omega}))^k\times (\mathcal{P}(\hat{\Omega}))^{n-k}$ such that
$p_{/\Omega_1}=p'_{/\Omega_1}$ and $q_{/\Omega_1}=q'_{/\Omega_1}$, we have $\gamma(p,q)=\gamma(p',q')$. In other words, there exists a map $\gamma':(\mathcal{P}(\Omega_1))^{k}\times (\mathcal{P}(\Omega_1))^{n-k}$ such that
\begin{equation}\label{Eqgammapdyad}
  \gamma_{\Omega_1}(p,q)= \gamma'(p_{/\Omega_1}, q_{/\Omega_1}), \mbox{ for any } \Omega_1\mbox{-dominant } (p,q),\in (\mathcal{P}(\hat{\Omega}))^k\times (\mathcal{P}(\hat{\Omega}))^{n-k}.
 \end{equation}
Using (\ref{Eqgammapdyad}) and the fact that $\gamma$ meets (I)-(IV), it is not difficult to check that $\gamma'$ satisfies  (I)-(IV) as well. Moreover, note that the maximal decomposition of the range of $\gamma'$ has $\{E_1,F_1\}$ as its unique (and dyadic) component. It thus comes from our induction hypothesis that there exists a non-decreasing function $H_1: \SN \times \SN \rightarrow \{0,1\}$   s.t.  $H_1(0,0)=0$, $H_1(k,n-k)=1$ and, for all  $\Omega_1\mbox{-dominant } (p,q)\in (\mathcal{P}(\hat{\Omega}))^k\times (\mathcal{P}(\hat{\Omega}))^{n-k}$,
\[[\gamma'(p_{/\Omega_1}, q_{/\Omega_1})=E_1] \mbox{ \textbf{iff} }
[H_1(\eta_{(p,q)}(I^{k}_1, E_1|F_1),  \eta_{(p,q)}(I^{n}_{k+1}, F_1|E_1))=1].\]
Recalling (\ref{Eqgammapdyad})   allows to conclude the proof of Claim 2.

\bigskip
\medskip\noindent
\emph{Claim 3}: For  $m=2,\ldots,M$,  there exists a non-decreasing function $H_m: \SN \times \SN \rightarrow \{0,1\}$     s.t.  $H_m(0,0)=0$, $H_m(k,n-k)=1$ and, for all  $\Omega_{-1}\mbox{-dominant } (p,q)\in(\mathcal{P}(\hat{\Omega}))^k\times (\mathcal{P}(\hat{\Omega}))^{n-k}$,
\[[\gamma_{\Omega_m}(p,q)=E_m] \mbox{ \textbf{iff} }
[H_m(\eta_{p}(I^{k}_1, E_m|F_m),  \eta_{q}(I^{n-k}_{1}, F_m|E_m))=1].\]

\noindent The proof parallels that of Claim 2. Observe from Lemma \ref{LemmaDomalloc} that, for any $\Omega_{-1}$-dominant $(p,q),(p',q')\in  \mathcal{\tilde{P}}^k\times \mathcal{\tilde{P}}^{n-k}$ such that
$p_{/\Omega_{-1}}=p'_{/\Omega_{-1}}$ and $q_{/\Omega_{-1}}=q'_{/\Omega_{-1}}$, we have $\gamma(p,q)=\gamma(p',q')$. In other words, there exists a map $\gamma'':(\mathcal{P}(\Omega_{-1}))^{k}\times (\mathcal{P}(\Omega_{-1}))^{n-k}$ such that
\begin{equation}\label{Eqgammapdyad3}
  \gamma(p,q)= \gamma''(p_{/\Omega_{-1}}, q_{/\Omega_{-1}}), \mbox{ for any }\Omega_{-1}\mbox{-dominant } (p,q),\in  (\mathcal{P}(\hat{\Omega}))^k\times (\mathcal{P}(\hat{\Omega}))^{n-k}.
 \end{equation}
Using (\ref{Eqgammapdyad3}) and the fact that $\gamma$ meets (I)-(IV), it is not difficult to check that $\gamma''$ satisfies  (I)-(IV) as well. In addition, since $ \Omega_{-1}=\Omega_{2}\cup\ldots \cup\Omega_{M}$, one can see that $\{E_2,F_2\}, \ldots, \{E_M,F_M\}$ are the only components of the maximal decomposition of the range of $\gamma''$. It thus follows from our induction hypothesis that there exist $M-1$ non-decreasing functions $H_m: \SN \times \SN \rightarrow \{0,1\}$    such that, for each $m\in \{2,\ldots,M\}$, we have $H_m(0,0)=0$, $H_m(k,n-k)=m$ and, for all  $\Omega_{-1}\mbox{-dominant } (p,q)\in (\mathcal{P}(\hat{\Omega}))^k\times (\mathcal{P}(\hat{\Omega}))^{n-k}$,
\[[\gamma''_{\Omega_m}(p,q)=E_m] \mbox{ \textbf{iff} }
[H_m(\eta_{p}(I^{k}_1, E_m|F_m),  \eta_{q}(I^{n-k}_{1}, F_m|E_m))=1].\]
Recalling (\ref{Eqgammapdyad3})   allows to conclude the proof of Claim 3.

\bigskip
\medskip\noindent
\emph{Claim 4}: Let $(p,q)\in (\mathcal{P}(\hat{\Omega}))^k\times (\mathcal{P}(\hat{\Omega}))^{n-k}$ be such  that $(\eta_{p}(I^{k}_{1}, E_1|F_1), \eta_{q}(I^{n-k}_{1}, F_1|E_1))=(k,n-k)$.
Then for  $m=2,\ldots,M$, we have
\[[\gamma_{\Omega_m}(p,q)=E_m] \mbox{ \textbf{iff} }
[H_m(\eta_{p}(I^{k}_1, E_m|F_m),  \eta_{q}(I^{n-k}_{1}, F_m|E_m))=1].\]

\smallskip\noindent
We proceed in two steps to prove Claim 4.

\medskip\noindent\emph{Step 1}: Suppose that  $(p,q)\in (\mathcal{P}(\hat{\Omega}))^k\times (\mathcal{P}(\hat{\Omega}))^{n-k}$ is  $\Omega_{-1}$-dominant and satisfies
 \[(\eta_{p}(E_1|F_1), I^{k}_{1}, \eta_{q}(I^{n-k}_{1}, F_1|E_1)  )=(k,n-k).\] Then  $\gamma_{\Omega_1}(p,q)=E_1$ holds.

\medskip\noindent
Fix  an $\Omega_{-1}$-dominant  $(p,q)\in (\mathcal{P}(\hat{\Omega}))^k\times (\mathcal{P}(\hat{\Omega}))^{n-k}$  such that $\eta_{p}(I^{k}_{1}, E_1|F_1)=k$ and $\eta_{q}(I^{n-k}_{1}, F_1|E_1)=n-k$, i.e., each of  the $n$ agents prefers $E_1$ (to  $F_1$) in the cell $\Omega_1$. Let $C=\bigcup\limits_{m=2}^M C_m=\gamma_{\Omega_{-1}}(p,q)$ be the sub-event chosen on $\Omega_{-1}$ at $(p,q)$, where $C_m\in \{E_m, F_m\}$ for $m=2,\ldots,M$.
Relabeling events if necessary, we will assume without loss that $C_m=E_m$ for all $m=2,\ldots, M$.
Recalling (\ref{EqRangedyad}), note that there exists a unique event $E\in \mathcal{R}$ such that $E=E_1\cup C= \bigcup\limits_{m=1}^M E_m$. We will show that $\gamma(p,q)=E$.

Pick  an   $\Omega_{-1}$-dominant belief profile
$(p',q')\in (\mathcal{P}(\hat{\Omega}))^k\times (\mathcal{P}(\hat{\Omega}))^{n-k}$ such that
 \begin{align}\label{EqOmega1comp0}
 E_m=& \argmax\{p'_i(E_m),p'_i(F_m)\} \\
 \label{EqOmega1comp1} =& \argmin\{q'_i(E_m),q'_i(F_m)\}, ~\forall (i,j,m)\in I_1^k\times I_1^{n-k}\times I_1^M; \\
 (A)~ \label{EqOmega1cond} p_{/\Omega_1}=&p'_{/\Omega_1}; ~(B)~ q_{/\Omega_{1}}=q'_{/\Omega_{1}}.
  \end{align}
Let $p^0,p^1,\ldots,p^k$  be the sequence of  belief profiles defined by $p^0=p$ and $p^{i}=(p'_{i}, p^{i-1}_{-i})$ for all $i=1,\ldots, k$ (so that $p^k=p'$).
Likewise, define  $q^0,q^1,\ldots,q^{n-k}$ as follows: $q^0=q$ and $q^{i}=(q'_{i}, q^{i-1}_{-i})$ for all $i=1,\ldots, n-k$ (so that $q^{n-k}=q'$).

Remark from (\ref{EqOmega1comp0}) that, for all  $m\in I_2^M$,  $\eta_{p^1}(I_1^k, E_m|F_m)\geq \eta_{p^0}(I_1^k, E_m|F_m).$ Given that $H_m$ is non-decreasing, the previous inequality  implies \[H_m(\eta_{p^1}(I_1^k, E_m|F_m), \eta_{q}(I_1^{n-k}, F_m|E_m)) \geq H_m(\eta_{p}(I_1^k, E_m|F_m), \eta_{q}(I_1^{n-k}, F_m|E_m)), ~\forall   m\in I_1^M.\]
 Since  $H_m(\eta_{p}(I_1^k, E_m|F_m), \eta_{q}(I_1^{n-k}, F_m|E_m))=1$ by Claim 3 [recall that $\gamma_{\Omega_{-1}}(p,q)=E_2\cup \ldots\cup E_M$ and $(p,q)$ is $\Omega_{-1}$-dominant], the last inequality allows to write as well   
 \[H_m(\eta_{p^1}(I_1^k, E_m|F_m), \eta_{q}(I_1^{n-k}, F_m|E_m))=1.\] It thus comes from Claim 3 that $\gamma_{\Omega_{-1}}(p^1,q)=E_2\cup \ldots\cup E_M=C= \gamma_{\Omega_{-1}}(p,q)$. Combining this equality with (I) and (\ref{EqOmega1cond})-A  then gives $\gamma(p^1,q)=\gamma(p,q)$, since $p^1_j=p_{j}$ for all $j\neq 1$.

Repeating the procedure described in the previous paragraph $k$ times, it comes that $\gamma(p^k,q)=\ldots= \gamma(p^1,q)=\gamma(p,q)$. In addition, once can see from (\ref{EqOmega1comp1}) that, for all  $m\in I_2^M$,
 $\eta_{q^1}(I_1^{n-k}, F_m|E_m)\geq \eta_{q}(I_1^{n-k}, F_m|E_m)$.
Given that $H_m$ is non-decreasing, the previous inequality  implies \[H_m(\eta_{p^k}(I_1^k, E_m|F_m), \eta_{q^1}(I_1^{n-k}, F_m|E_m)) \geq H_m(\eta_{p^k}(I_1^k, E_m|F_m), \eta_{q}(I_1^{n-k}, F_m|E_m)), ~\forall   m\in I_1^M.\]

Observing that $H_m(\eta_{p^k}(I_1^k, E_m|F_m), \eta_{q}(I_1^{n-k}, F_m|E_m))=1$ from  $\gamma(p^k,q)=\gamma(p,q)$ and Claim 3, we get from the previous inequality   $H_m(\eta_{p^k}(I_1^k, E_m|F_m), \eta_{q^1}(I_1^{n-k}, F_m|E_m))=1$ as well.
It thus comes from Claim 3 that $\gamma_{\Omega_{-1}}(p^k,q^1)=E_2\cup \ldots\cup E_M= \gamma_{\Omega_{-1}}(p^k,q)$. Combining $\gamma_{\Omega_{-1}}(p^k,q^1)=\gamma_{\Omega_{-1}}(p^k,q)$ with (II) and  (\ref{EqOmega1cond})-B then gives $\gamma(p^k,q^1)=\gamma(p^k,q)=\gamma(p,q)$.  Repeating this argument $n-k$ times, one can write $\gamma(p',q')=\gamma(p^k,q^{n-k})=\ldots=\gamma(p^k,q^1)=\gamma(p,q)$.

It now remains to see from (\ref{EqOmega1comp0})-(\ref{EqOmega1comp1})  that, for all $m=1,\ldots,M$
 \begin{equation}\label{EqOmega1comp3}\eta_{p'}(I_1^k, E_m|F_m)=k ; ~\eta_{q'}(I_1^{n-k}, F_m|E_m)=n-k.
 \end{equation}
 Since (\ref{EqOmega1comp3}) and (\ref{EqRangedyad}) mean that, at the profile $(p',q')$ all agents unanimously view $E=E_1\cup\ldots\cup E_M$ as the best event in the range $\mathcal{R}$ of $\gamma$, note that the combination (I)-(II) requires $\gamma(p',q')=E$. Indeed, since there exists $(p'',q'')\in \mathcal{\tilde{P}}^k\times \mathcal{\tilde{P}}^{n-k}$ such that $\gamma(p'',q'')=E$, writing $\gamma(p',q')\neq E$ would violate either (I) or (II) as we move from   $(p',q')$ to $(p'',q'')$ by changing the belief of one agent at a time (the argument is standard). Thus, we have $\gamma(p,q)=\gamma(p',q')=E_1\cup\ldots\cup E_M=E$; and this concludes Step 1.

For all $\theta \in (0,1)$ and $\pi\in \mathcal{P}(\hat{\Omega})$, define the belief $\pi^\theta(\omega)$ as follows: for all $\omega\in \hat{\Omega}$,  \begin{equation}\label{EqDyadcon}
  \pi^\theta(\omega)=\left\{\begin{array}{cl}
                                                                        \theta  \frac{  \pi(\omega)}{\pi(\Omega_1)} & \mbox{ if } \omega \in \Omega_1; \\
                                                                          (1-\theta) \frac{\pi(\omega)}{1-\pi(\Omega_1)} & \mbox{ if } \omega \in \Omega_{-1}.
                                                                         \end{array} \right.
                                                                         \end{equation}
We now proceed to the second step of the proof of Claim 4.

\medskip\noindent\emph{Step 2}: Let $(p,q)\in (\mathcal{P}(\hat{\Omega}))^k\times (\mathcal{P}(\hat{\Omega}))^{n-k}$  be such that  $(\eta_{p}(I^{k}_{1}, E_1|F_1), \eta_{q}(I^{n-k}_{1}, F_1|E_1))=(k,n-k)$. Then
for  $m=2,\ldots,M$, we have
\[[\gamma_{\Omega_m}(p,q)=E_m] \mbox{ \textbf{iff} }
[H_m(\eta_{p}(I^{k}_1, E_m|F_m),  \eta_{q}(I^{n-k}_{1}, F_m|E_m))=1].\]

\medskip\noindent Fix  $(p,q)\in (\mathcal{P}(\hat{\Omega}))^k\times (\mathcal{P}(\hat{\Omega}))^{n-k}$ s.t. $(\eta_{p,q)}(I^{k}_{1}, E_1|F_1), \eta_{(p,q)}(I^{1}_{n-k}, F_1|E_1))=(k,n-k)$. Note from (\ref{EqDyadcon}) that
$(p^\theta, q^\theta)$ is $\Omega_{1}$-dominant [$\Omega_{-1}$-dominant] if $\theta$ is close enough to one [zero]. Fix then $\theta, \theta'\in (0,1)$ such  that  $(p^\theta, q^\theta)$ is $\Omega_{-1}$-dominant and  $(p^{\theta'}, q^{\theta'})$ is $\Omega_{1}$-dominant.
Recalling Step 1, note that $\gamma_{\Omega_1}(p^\theta, q^\theta)=E_1$. Given that $p^{\theta'}_1$ is $\Omega_1$-dominant and $p_1(E_1)>p_1(F_1)$ [i.e., $p^{\theta'}_1(E_1)>p^{\theta'}_1(F_1)$], remark from the combination of assumption (I) and [$\gamma_{\Omega_1}(p^\theta, q^\theta)=E_1$] that we must have $\gamma_{\Omega_1}((p^{\theta'}_1,p_{-1}^\theta), q^\theta)= \gamma_{\Omega_1}(p^\theta, q^\theta)=E_1$.
Next, using $\gamma_{\Omega_1}((p^{\theta'}_1,p_{-1}^\theta), q^\theta)= \gamma_{\Omega_1}(p^\theta, q^\theta)$ and recalling from (\ref{EqDyadcon}) that $p^{\theta'}_1$ and $p_1^\theta$ have the same conditional distribution over $\Omega_{-1}$, remark from (I) that $\gamma_{-\Omega_1}((p_1^{\theta'},p_{-1}^\theta), q^\theta)=\gamma_{-\Omega_1}(p^\theta, q^\theta)$, and hence $\gamma((p_1^{\theta'},p_{-1}^\theta), q^\theta)=\gamma(p^\theta, q^\theta)$, must hold.

But then note that  $\gamma((p_1^{\theta'},p_{-1}^\theta), q^\theta)=\gamma(p^\theta, q^\theta)$ implies $\gamma((p_1,p_{-1}^\theta), q^\theta)=\gamma(p^\theta, q^\theta)$.
To show this implication, suppose first that   $\gamma_{\Omega_1}((p_1,p_{-1}^\theta), q^\theta)=\gamma_{\Omega_1}(p^\theta, q^\theta)$. Recalling that $p_1$ and $p_1^\theta$ have the same conditional distribution over $\Omega_{-1}$, one can combine $\gamma_{\Omega_1}((p_1,p_{-1}^\theta), q^\theta)=\gamma_{\Omega_1}(p^\theta, q^\theta)$ and (I) to conclude that  $\gamma_{\Omega_{-1}}((p_1,p_{-1}^\theta), q^\theta) =\gamma_{\Omega_{-1}}(p^\theta, q^\theta)$; and it hence follows in this case  that  $\gamma((p_1,p_{-1}^\theta), q^\theta)=\gamma(p^\theta, q^\theta)$.

Suppose next that $\gamma_{\Omega_1}((p_1,p_{-1}^\theta), q^\theta)\neq \gamma_{\Omega_1}(p^\theta, q^\theta)=\gamma_{\Omega_1}(p_1^{\theta'},p_{-1}^\theta), q^\theta)=E_1$. Combining this difference with assumption (I) yields \textbf{(a)} $p_1^{\theta'}(\gamma_{\Omega_1}((p_1,p_{-1}^\theta), q^\theta))<p_1^{\theta'}(E_1)$, since $p_1^{\theta'}$ is $\Omega_1$-dominant.
Moreover, given that  $p_1^{\theta}$ is $\Omega_{-1}$-dominant, assumption (I) yields the inequality
 \textbf{(b)} $p_1^{\theta}(\gamma_{\Omega_{-1}}((p_1,p_{-1}^\theta), q^\theta)) \leq p_1^{\theta}(\gamma_{\Omega_{-1}}((p_1^\theta, q^\theta))$. Recalling from (\ref{EqDyadcon}) that $p^{\theta'}_1$ and $p_1$ [$p^{\theta}_1$ and $p_1$] have the same conditional distribution over $\Omega_{1}$ [$\Omega_{-1}$], we may rewrite the inequalities (a)-(b) above as
\begin{eqnarray}\label{Eq1lexClaim4}
  p_1(\gamma_{\Omega_1}((p_1,p_{-1}^\theta), q^\theta))<&p_1(\gamma_{\Omega_1}(p^\theta, q^\theta)) \\  \label{Eq2lexClaim4}
 p_1(\gamma_{\Omega_{-1}}((p_1,p_{-1}^\theta), q^\theta))\leq &p_1(\gamma_{\Omega_{-1}}(p^\theta, q^\theta))   
\end{eqnarray}
The sum of (\ref{Eq1lexClaim4}) and (\ref{Eq2lexClaim4}) thus gives 
 $p_1(\gamma((p_1,p_{-1}^\theta), q^\theta))<p_1(\gamma(p^\theta, q^\theta))$.  But note that this inequality contradicts assumption (I).

We have thus shown that $\gamma((p_1,p_{-1}^\theta), q^\theta)=\gamma(p^\theta, q^\theta)$. Repeating this argument $k$ times, we get
\begin{equation}\label{EqDyadtheta1}
  \gamma(p^\theta, q^\theta)= \gamma((p_1,p_{-1}^\theta), q^\theta)=\ldots =\gamma((p_1,\ldots, p_k), q^\theta)=\gamma(p, q^\theta).
\end{equation}
Moreover, exploiting (II) and applying  a similar argument $n-k$ times to $q^\theta$, one can write as well
\begin{equation}\label{EqDyadtheta2}\gamma(p, q^\theta)= \gamma(p,(q_1,q^\theta_{-1}))=\ldots=\gamma(p,(q_1,\ldots, q^{n-k})).
\end{equation}
Combining (\ref{EqDyadtheta1}) and (\ref{EqDyadtheta2}) then gives $\gamma(p,q)=  \gamma(p^\theta, q^\theta)$. Since $(p^\theta, q^\theta)$ is $\Omega_{-1}$-dominant, using this equality in  Claim 3 then  yields: for all $m=2,\ldots,M$,
\[[\gamma_{\Omega_m}(p,q)=E_m] \mbox{ \textbf{iff} }
[H_m(\eta_{p^\theta}(I^{k}_1, E_m|F_m),  \eta_{q^\theta}(I^{n-k}_{1}, F_m|E_m))=1].\]
Recalling that $(p,q)$ and $(p^\theta, q^\theta)$ induce the same conditional distributions over $\Omega_{-1}$, note that $\eta_{p^\theta}(I^{k}_1, E_m|F_m)= \eta_{p}(I^{k}_1, E_m|F_m)$ and $\eta_{q^\theta}(I^{1}_{n-k}, F_m|E_m))= \eta_{q}(I^{n-k}_{1}, F_m|E_m))$. Hence, Claim 4 holds true.

\bigskip
\medskip\noindent
\emph{Claim 5}:  For any $(p,q)\in (\mathcal{P}(\hat{\Omega}))^k\times (\mathcal{P}(\hat{\Omega}))^{n-k}$ and $m=2,\ldots,M$, we have:
\[[\gamma_{\Omega_m}(p,q)=E_m] \mbox{ \textbf{iff} }
[H_m(\eta_{p}(I^{k}_1, E_m|F_m),  \eta_{q}(I^{n-k}_{1}, F_m|E_m))=1].\]

\noindent Fix $(p,q)\in (\mathcal{P}(\hat{\Omega}))^k\times (\mathcal{P}(\hat{\Omega}))^{n-k}$ and assume without loss that $\gamma_{\Omega_1}(p,q)=E_1$ ---the argument is symmetric if one assumes instead $\gamma_{\Omega_1}(p,q)=F_1$.  It is not difficult to see that we can pick belief profile
$(p',q')\in (\mathcal{P}(\hat{\Omega}))^k\times (\mathcal{P}(\hat{\Omega}))^{n-k}$ such that
 \begin{align}\label{EqClaim51}
 E_1=& \argmax\{p'_i(E_1),p'_i(F_1)\}  \\ \label{EqClaim511}
 =& \argmin\{q'_j(E_1), q'_j(F_1)\}, ~\forall (i,j)\in I_1^k\times I_1^{n-k}; \\ \label{EqClaim501}
  p'_i=&p_i \mbox{ if }  E_1= \argmax\{p_i(E_1),p_i(F_1)\},  ~\forall i\in  I_1^{k}\\  \label{EqClaim502}
    q'_i=&q_i \mbox{ if }  E_1= \argmax\{p_i(E_1),p_i(F_1)\}, ~\forall j\in  I_1^{n-k};
 \\  \label{EqClaim52} (A) ~& p_{/\Omega_{-1}}=p'_{/\Omega_{-1}}; ~(B) ~q_{/\Omega_{-1}}=q'_{/\Omega_{-1}}.
  \end{align}
  Equations (\ref{EqClaim51})-(\ref{EqClaim511}) simply say that, under $(p',q')$, all agents unanimously prefer $E_1$ to $F_1$ (in the cell $\Omega_1$), that is,  $(\eta_{p'}(I^{k}_{1}, E_1|F_1), \eta_{q'}(I^{n-k}_{1}, F_1|E_1))=(k,n-k)$. Equation (\ref{EqClaim501}) requires that $p'_i=p_i$ whenever this equality does not contradict the  condition (\ref{EqClaim51}); and  (\ref{EqClaim502}) likewise guarantees that $q'_j=q_j$ whenever possible. Finally,
    (\ref{EqClaim52}) states that  $p'_i$ ($q'_j$) has the same conditional distribution over $\Omega_{-1}$ as $p_i$ ($q_j$), for each $i\in I_1^k$ ($j\in I_1^{n-k}$).
  
  Let then $p^0,p^1,\ldots,p^k$  be the sequence of  belief profiles defined by $p^0=p$ and $p^{i}=(p'_{i}, p^{i-1}_{-i})$ for all $i=1,\ldots, k$ (so that $p^k=p'$).
Likewise, define  $q^0,q^1,\ldots,q^{n-k}$ as follows: $q^0=q$ and $q^{i}=(q'_{i}, q^{i-1}_{-i})$ for all $i=1,\ldots, n-k$ (so that $q^{n-k}=q'$).

 We remark first that $\gamma((p'_1,p_{-1}),q)=\gamma(p,q)$ [\textit{i.e.}, $\gamma(p^1,q)=\gamma(p^0,q)$]. Indeed, one can discuss two cases to show this equality. If $p'_1=p_1$ then the desired equality trivially obtains. In the other case, where $p'_1\neq p_1$, note from (\ref{EqClaim501}) that we must have
 \begin{equation}\label{EqClaim053}
p_1(F_1)>p_1(E_1)= p_1(\gamma_{\Omega_{1}}(p,q)).
 \end{equation}
The inequality in (\ref{EqClaim053}) means that we cannot have $\gamma_{\Omega_1}((p'_1,p_{-1}),q)=F_1$: this equality would imply $p'_1(\gamma_{\Omega_1}((p'_1,p_{-1}),q))=p'_1(F_1)\underbrace{<}_{by ~(\ref{EqClaim511})}p'_1(E_1)=p'_1(\gamma_{\Omega_{1}}(p,q))$, which, combined with assumption (I), yields $p'_1(\gamma_{\Omega_{-1}}((p'_1,p_{-1}),q))>p'_1(\gamma_{\Omega_{-1}}(p,q)$. Using (\ref{EqClaim52})-A and the fact that $\gamma_{\Omega_{-1}}((p'_1,p_{-1}),q)), \gamma_{\Omega_{-1}}(p,q)\subset \Omega_{-1}$, note that  $p'_1(\gamma_{\Omega_{-1}}((p'_1,p_{-1}),q))>p'_1(\gamma_{\Omega_{-1}}(p,q))$ in turn implies
   \begin{equation}\label{EqClaim53}
   p_1(\gamma_{\Omega_{-1}}((p'_1,p_{-1}),q))>p_1(\gamma_{\Omega_{-1}}(p,q)).
 \end{equation}
Summing up (\ref{EqClaim053}) and (\ref{EqClaim53}), we would get $p_1(\gamma((p'_1,p_{-1}),q))>p_1(\gamma(p,q))$, which is a violation of assumption (I) at the belied profile $(p,q)$ for $i=1$. 

Hence, we must have  $\gamma_{\Omega_1}((p'_1,p_{-1}),q)=E_1=\gamma_{\Omega_1}(p,q)$; and since $p_1$ and $p'_1$ have the same conditional distribution over $\Omega_{-1}$ [by (\ref{EqClaim52})-A],
assumption (I) yields $\gamma_{\Omega_{-1}}((p'_1,p_{-1}),q)=\gamma_{\Omega_{-1}}(p,q)$; and it follows that $\gamma((p'_1,p_{-1}),q)=\gamma(p,q)$.

We have thus shown that $\gamma(p^0,q)=\gamma(p^1,q)$. Repeating the same argument $k$ times, one can write: $\gamma(p^k,q)=\ldots=\gamma(p^1,q)=\gamma(p^0,q)=\gamma(p,q)$.
The previous sequence of equalities implies that $\gamma_{\Omega_1}(p^k,q)=E_1$. Combining this observation with (\ref{EqClaim511}), (\ref{EqClaim502}) and (\ref{EqClaim52}), one can prove (using a similar procedure as the one described in the previous paragraphs) that $\gamma(p^k,q^0)=\gamma(p^k,q^1)$. Repeating the same argument $n-k$ times, it thus comes that 
\begin{equation}\label{EqClaim5Una0}
\gamma(p',q')=\gamma(p^k,q^{n-k})= \ldots=\gamma(p^k,q^1)=\gamma(p^k,q^0)=\gamma(p^k,q)=\gamma(p,q). 
\end{equation}
 Since $(\eta_{p'}(I^{k}_{1}, E_1|F_1), \eta_{q'}(I^{n-k}_{1}, F_1|E_1))=(k,n-k)$, Claim 4 yields: for all $m=2,\ldots,M$ 
 \begin{equation}\label{EqClaim5Una1}
[\gamma_{\Omega_m}(p',q')=E_m] \mbox{ \textbf{iff} }
[H_m(\eta_{p'}(I^{k}_1, E_m|F_m),  \eta_{q'}(I^{n-k}_{1}, F_m|E_m))=1].
\end{equation}

Recalling from (\ref{EqClaim52}) that $p_i$ and  $p'_i$ [$q_j$ and $q'_j$] have the same conditional distribution on $\Omega_{-1}$ for each agent $i\in I_1^k$ [$j\in I_1^{n-k}$], we  conclude that $(\eta_{p}(I^{k}_{m}, E_m|F_m), \eta_{q}(I^{n-k}_{1}, F_m|E_m))=(\eta_{p'}(I^{k}_{m}, E_m|F_m), \eta_{q'}(I^{n-k}_{1}, F_m|E_m))$ for each $m=2,\ldots,M$; and it hence follows from (\ref{EqClaim5Una0})-(\ref{EqClaim5Una1}) that: for all $m=2,\ldots,M$
\begin{equation}\label{EqClaim5Una}
[\gamma_{\Omega_m}(p,q)=E_m] \mbox{ \textbf{iff} }
[H_m(\eta_{p}(I^{k}_1, E_m|F_m),  \eta_{q}(I^{n-k}_{1}, F_m|E_m))=1].
\end{equation}
This completes the proof of Claim 5.

\bigskip
\medskip\noindent
\emph{Claim 6}:  For any $(p,q)\in (\mathcal{P}(\hat{\Omega}))^k\times (\mathcal{P}(\hat{\Omega}))^{n-k}$ and $m=1,\ldots,M$, we have:
\[[\gamma_{\Omega_m}(p,q)=E_m] \mbox{ \textbf{iff} }
[H_m(\eta_{p}(I^{k}_1, E_m|F_m),  \eta_{q}(I^{n-k}_{1}, F_m|E_m))=1].\]

\noindent
Note that, by Claim 5, the equivalence stated in Claim 6 is satisfied at any profile $(p,q)$ whenever $m\in \{2,\ldots,M\}$. It remains to prove it in the case $m=1$.

Fix then an arbitrary $(p,q)\in (\mathcal{P}(\hat{\Omega}))^k\times (\mathcal{P}(\hat{\Omega}))^{n-k}$. 
Recalling (\ref{EqDyadcon}), pick $\theta$ (close enough to 1) so as to guarantee that $p_i^\theta$ and $q_j^\theta$ are both $\Omega_1$-dominant \textbf{for all } $(i,j)\in I_1^k\times I_1^{n-k}$. Note that the combination of Claim 2 and Claim 5 yields: for all $m=1,\ldots,M$,
 \begin{equation}\label{Eq0Claim6}
[\gamma_{\Omega_m}(p^\theta,q^\theta)=E_m] \mbox{ \textbf{iff} }
[H_m(\eta_{p^\theta}(I^{k}_1, E_m|F_m),  \eta_{q^\theta}(I^{n-k}_{1}, F_m|E_m))=1].
\end{equation}

Let  $p^0,p^1,\ldots,p^k$  be the sequence of  belief profiles defined by $p^0=p$ and $p^{i}=(p^\theta_{i}, p^{i-1}_{-i})$ for all $i=1,\ldots, k$ (so that $p^k=p^\theta$).
Likewise, define  $q^0,q^1,\ldots,q^{n-k}$ as follows: $q^0=q$ and $q^{i}=(q^\theta_{i}, q^{i-1}_{-i})$ for all $i=1,\ldots, n-k$ (so that $q^{n-k}=q^\theta$). We show first that $\gamma(p^0,q)=\gamma(p^1,q)$. Indeed, note from Claim 5 that $\gamma_{\Omega_{-1}}(p^0,q)=\gamma_{\Omega_{-1}}(p^1,q)$, since (\ref{EqDyadcon}) guarantees that, for each agent, the  conditional distribution over $\Omega_{-1}$ does not change when the belief profile varies from   $(p^0,q)$ to $(p^1,q)$.
Recalling that $p_1$ and $p^\theta_1$ generate the same conditional distribution over $\Omega_{-1}$ [by (\ref{EqDyadcon})], writing $\gamma_{\Omega_{1}}(p^0,q)=\gamma_{\Omega_{1}}(p,q)\neq\gamma_{\Omega_{1}}(p^1,q)=\gamma_{\Omega_{1}}((p^\theta_1,p_{-1}),q)$ would lead to a violation of assumption (I) ---since $\gamma_{\Omega_{-1}}(p^0,q)=\gamma_{\Omega_{-1}}(p^1,q)$.

We have thus shown that $\gamma(p^0,q)=\gamma(p^1,q)$. Repeating the argument of the previous paragraph $k$ times, one can write 
\begin{equation}\label{Eq1Claim6}
  \gamma(p^k,q)=\ldots=\gamma(p^1,q)=\gamma(p^0,q)=\gamma(p,q)
\end{equation}
Combining Claim 5, assumption (I) and the fact that $q_1$ and $q^\theta_1$ generate the same conditional distribution over $\Omega_1$ ($\Omega_{-1}$), one can prove in a similar way that $\gamma(p^k,q^0)=\gamma(p^k,q^1)$. Repeating this argument $n-k$ times then gives 
\begin{equation}\label{Eq2Claim6}
  \gamma(p^\theta,q^\theta)= \gamma(p^k,q^{n-k})=\ldots=\gamma(p^k,q^1)=\gamma(p^k,q^0).
\end{equation}
Remark that the combination of (\ref{Eq1Claim6}) and  (\ref{Eq2Claim6}) yields $\gamma(p,q)= \gamma(p^\theta,q^\theta)$. Since  $p_i$ and  $p^\theta_i$ [$q_j$ and $q^\theta_j$] have the same conditional distribution on $\Omega_{-1}$ for each agent $i\in I_1^k$ [$j\in I_1^{n-k}$], we  conclude that $(\eta_{p}(I^{k}_{m}, E_m|F_m), \eta_{q}(I^{n-k}_{1}, F_m|E_m))=(\eta_{p^\theta}(I^{k}_{m}, E_m|F_m), \eta_{q^\theta}(I^{n-k}_{1}, F_m|E_m))$ for each $m=2,\ldots,M$. Hence, substituting  $(\eta_{p}(I^{k}_{m}, E_m|F_m), \eta_{q}(I^{n-k}_{1}, F_m|E_m))$ and $\gamma(p,q)$ in (\ref{Eq0Claim6}), one obtains the desired conclusion: for all $m=1,\ldots,M$, we have $[\gamma_{\Omega_m}(p^\theta,q^\theta)=E_m] \mbox{ \textbf{iff} }
[H_m(\eta_{p^\theta}(I^{k}_1, E_m|F_m),  \eta_{q^\theta}(I^{n-k}_{1}, F_m|E_m))=1].$ \endproof

Some additional notation is needed  to state the next lemma. Recalling (\ref{EqdefsectRange}) and (\ref{Eqdecomp1})-(\ref{Eqdecomp3}), notice that $\mathcal{E}^k$ admits a maximal decomposition for all $k\in \{1,\ldots,n\}$ (potentially with rich, dyadic, and trivial components). For any $\omega\in \tilde{\Omega}$, write $\mathbf{C}^k_{\omega}$ to denote the unique component (of the decomposition of $\mathcal{E}^k$) containing (at least) one event $E$ such that $\omega\in E$. We will also write $A^k_w:=\bigcup\limits_{E\in \mathbf{C}^k_{\omega}} E$.

Moreover, since the considered decomposition of $\mathcal{E}^k$ is maximal, remark that  whenever  $\mathbf{C}^k_{\omega}$ is a dyadic component there exists a unique event $E\in \mathbf{C}^k_{\omega}$   such that $\omega\in E$; and in that case we denote this unique event by $\hat{F}^k_{\omega}$.
In other words, whenever $|\mathbf{C}^k_{\omega}|=2$, we may write $\mathbf{C}^k_{\omega}=\{\hat{F}^k_{\omega}, \hat{G}^k_\omega \}$, where $\omega\in \hat{F}^k_{\omega}$ and $\omega\notin \hat{G}^k_\omega=A^k_{\omega}\setminus \hat{F}^k_{\omega}$. 

Finally, if the maximal decomposition of $\mathcal{E}^k$ has no rich component, note from Lemma \ref{LemmaDyad} that, for all $\omega\in \Tilde{\Omega}$ satisfying $|\mathbf{C}^k_{\omega}|=2$, there exists a function $H_{\omega}^k: \SN\times \SN \rightarrow \{0,1\}$ such that,  for any $(p,q)\in \mathcal{P}^k\times\mathcal{P}^{n-k}$, we have $H_{\omega}^k(0,0)=0$, $H_{\omega}^k(k,n-k)=1$, and 
\begin{equation}\label{EqdyadHSol}
    \alpha_{A^k_\omega}(p,q)=\hat{F}^k_{\omega}\mbox{ \textbf{iff} } H^k_\omega (\eta_p(I_1^k, \hat{F}^k_{\omega}| \hat{G}^k_\omega), \eta_q(I_{k+1}^n,\hat{G}^k_\omega|\hat{F}^k_{\omega}))=1.
\end{equation}
We are now ready to state the next lemma.

\begin{lemma}\label{DyadSolidarity}  \textbf{Dyadic Nesting}\\
Suppose that $n\geq 3$ and, for any $k'\in \{0,1,\ldots,n\} $, the maximal decomposition of $\mathcal{E}^{k'}$ has no rich component. Given any $k\in \{1,\ldots,n-2\}$ and any distinct $\omega, \omega' \in \tilde{\Omega}$, we have the following properties:
\begin{align*}
   \mathbf{(I)}~& \left[\mathbf{C}^k_{\omega}=\{\hat{F}^k_{\omega}, \hat{F}^k_{\omega'}\} \mbox{ and } \mathbf{C}^{k+1}_\omega=\{\hat{F}^{k+1}_{\omega}, \hat{F}^{k+1}_{\omega'}\} \right]\Rightarrow [H_{\omega}^k(m_a,m_b)=H_{\omega}^{k+1}(m_a,m_b), \\ &\forall (m_a, m_b) \in \{0,\ldots,k\}\times\{0,\ldots,n-k-1\} ];\\
 \mathbf{(II)}~& \left[\mathbf{C}^k_{\omega}=\{\hat{F}^k_{\omega}, \hat{F}^k_{\omega'}\} \mbox{ and } \mathbf{C}^{k+1}_\omega=\{\hat{F}^{k+1}_{\omega}, \hat{F}^{k+1}_{\omega'}\} \right]\Rightarrow [H_{\omega}^{k}(m_a,m_b)=H_{\omega}^k(m_a+1,m_b-1), \\ &\forall (m_a, m_b) \in \{0,\ldots,k-1\}\times\{1,\ldots,n-k\} ];\\
   \mathbf{(III)}~& \left[ |\mathbf{C}^k_{\omega}|=|\mathbf{C}^{k+1}_{\omega}|=|\mathbf{C}^{k+1}_{\omega'}|=2  \mbox{ and }  \omega'\in \hat{F}^k_{\omega} \right]\Rightarrow [\omega'\in \hat{F}^{k+1}_{\omega} \mbox{ or }  \hat{G}^{k+1}_{\omega'}\cap A^k_\omega=\emptyset  ];\\
    \mathbf{(IV)}~& \left[ |\mathbf{C}^k_{\omega'}|=
|\mathbf{C}^{k+1}_{\omega'}|=2  \mbox{ and }  \hat{G}^{k+1}_{\omega'} \cap A^{k}_{\omega'}=\emptyset \right]\Rightarrow [H_{\omega'}^{k+1}(1,0)=1];\\
    \mathbf{(V)}~& \left[ |\mathbf{C}^k_{\omega}|=|\mathbf{C}^{k+1}_{\omega}|=2  \mbox{ and }  \hat{G}^k_\omega\neq   \hat{G}^{k+1}_{\omega} \right]\Rightarrow [ |\mathbf{C}^k_{\omega''}|=1, \forall \omega''\in \hat{G}^{k+1}_{\omega}].\end{align*}
\end{lemma}
\proof 

Let $n\geq 3$ and assume that, for any $k'\in \{0,1,\ldots,n\} $, the maximal decomposition of $\mathcal{E}^{k'}$ has no rich component. Moreover,  fix two distinct states $\omega, \omega' \in \tilde{\Omega}$.

(I) Fix $k\in \{1,\ldots, n-2\}$ and suppose that $\mathbf{C}^k_{\omega}=\{\hat{F}^k_{\omega}, \hat{F}^k_{\omega'}\} $ and $ \mathbf{C}^{k+1}_\omega=\{\hat{F}^{k+1}_{\omega}, \hat{F}^{k+1}_{\omega'}\}$. Let $ (m_a, m_b) \in \{0,\ldots,k\}\times\{0,\ldots,n-k-1\}$. 
Consider  first the case  $H^k_\omega(m_a,m_b)=1$; and  suppose by contradiction that $H_{\omega}^{k+1}(m_a,m_b)=0$. 

Pick $(v,p)\in \mathcal{\tilde{D}}^N$ such that $N_a^v=\{1,\ldots,k\}=I_1^k, N_b^v=\{k+1,\ldots,n\}=I^n_{k+1}$ and 
\begin{eqnarray}
\label{Eqmamb}
    \eta_{p}(I_1^k, \hat{F}^k_{\omega}|\hat{F}^k_{\omega'})=m_a, ~
    \eta_{p}(I_{k+1}^n, \hat{F}^k_{\omega'}|\hat{F}^k_{\omega})=m_b;\\ \label{EqLexOmDyadic1}
     p_n(\omega)>p_n(\omega')>p_n(\omega)-p_n(\omega')> p_n(\tilde{\Omega}\setminus \omega\omega');\label{EqLexOmDyadic2}
\end{eqnarray}
Combining Lemma \ref{LemmaSections}, Lemma \ref{LemmaDyad} and (\ref{Eqmamb}), observe that $ \varphi_{A^k_{\omega}}(v,p)= a \hat{F}^k_{\omega} \oplus b \hat{F}^k_{\omega'}$ [since $H^k_\omega(m_a,m_b)=1$].
Then note from (\ref{EqLexOmDyadic1}) that agent $n$ (who is in $N^v_b$) will manipulate $\varphi$ at $(v,p)$ by misreporting $(v'_n,p'_n)$ such that $v'_n(a)>v'_n(b)$ and $ p_n(\omega')>p_n(\omega)>p_n(\omega')-p_n(\omega)> p_n(\tilde{\Omega}\setminus \omega\omega')$. This manipulation will lead to the preference profile $(v',p'):=((v'_n,v_{-n}),(p'_n,p_{-n}))$, which satisfies $n_a^{v'}=k+1$, $\eta_{p'}(N_a^{v'}, \hat{F}^k_{\omega}|\hat{F}^k_{\omega'})=m_a$ and $\eta_{p'}(N_b^{v'}, \hat{F}^k_{\omega}|\hat{F}^k_{\omega'})=m_b$.  Given that $H_{\omega}^{k+1}(m_a,m_b)=0$, combining Lemma \ref{LemmaSections}, Lemma \ref{LemmaDyad} and (\ref{Eqmamb}) gives  $ \varphi_{A^{k+1}_{\omega}}(v',p')= a \hat{F}^{k+1}_{\omega'} \oplus b \hat{F}^{k+1}_{\omega}$. Hence, $ \varphi(v',p')$  respectively yields $b$ in $\omega$ and $a$ in $\omega'$, whereas $\varphi(v,p)$ yields $b$ in $\omega'$ and $a$ in $\omega$. Recalling that $v_n(b)>v_n(a)$ and  noting from (\ref{EqLexOmDyadic2}) that $(v_n,p_n)$ is $\{\omega,\omega'\}$-lexicographic, we conclude that agent $n$ prefers   $ \varphi(v',p')$ to $\varphi(v,p)$ at $(v,p)$; and this contradicts the strategy-proofness of $\varphi$.

In the case where $H^k_\omega(m_a,m_b)=0$, one can mimic the previous paragraph [assuming by contradiction that $H_{\omega}^{k+1}(m_a,m_b)=1$]; and he profitable deviation  will now  happen in the other direction: agent $n$ can manipulate $\varphi$ at $(v',p')$ by misreporting $(v_n,p_n)$ ---we omit the explicit argument. 

\bigskip
(II)  Fix $k\in \{1,\ldots, n-2\}$ and let $ (m_a, m_b) \in \{0,\ldots,k-1\}\times\{1,\ldots,n-k\} $.
Pick $(v,p)\in \mathcal{\tilde{D}}^N$ such that $N_a^v=\{1,\ldots,k\}=I_1^k, N_b^v=\{k+1,\ldots,n\}=I^n_{k+1}$ and 
\begin{eqnarray}
\label{Eq1LemSolII}
    \eta_{p}(I_1^k, \hat{F}^k_{\omega}|\hat{F}^k_{\omega'})=m_a, ~
    \eta_{p}(I_{k+1}^n ,\hat{F}^k_{\omega'}|\hat{F}^k_{\omega})=m_b;\\ \label{Eq2LemSolII}
     p_n(\omega')>p_n(\omega)>p_n(\omega')-p_n(\omega)> p_n(\tilde{\Omega}\setminus \omega\omega').
\end{eqnarray}

Combining Lemma \ref{LemmaSections}, Lemma \ref{LemmaDyad} and (\ref{Eq1LemSolII}) gives \[\varphi_{A^k_{\omega}}(v,p)= \left\{\begin{array}{cc}
     a \hat{F}^k_{\omega} \oplus b \hat{F}^k_{\omega'} & \mbox{ if } H^k_\omega(m_a,m_b)=1;  \\
    a \hat{F}^k_{\omega'} \oplus b ^k_{\omega} & \mbox{ if } H^k_\omega(m_a,m_b)=0.
\end{array}\right.\]

Let then $(v',p'):=((v'_n,v_{-n}),(p'_n,p_{-n}))$ be such that $v'_n(a)>v'_n(b)$ and $ p_n(\omega)>p_n(\omega')>p_n(\omega)-p_n(\omega')> p_n(\tilde{\Omega}\setminus \omega\omega')$. Combining the last four inequalities with (\ref{Eq2LemSolII}), note that the  preferences $(v_n,p_n)$ and $(v'_n,p'_n)$ are both $\{\omega,\omega'\}$-lexicographic and they both imply that the sub-act $a \omega \oplus b\omega'$ is strictly preferred to $a \omega' \oplus b\omega$. Combining this observation with $\mathbf{C}^k_{\omega}=\{\hat{F}^k_{\omega}, \hat{F}^k_{\omega'}\} \mbox{ and } \mathbf{C}^{k+1}_\omega=\{\hat{F}^{k+1}_{\omega}, \hat{F}^{k+1}_{\omega'}\}$, one can see that strategy-proofness of $\varphi$ requires $\varphi_{A^k_{\omega}}(v,p)=  \hat{F}^k_{\omega}$ \textbf{iff} $\varphi_{A^{k+1}_{\omega}}(v',p')= \hat{F}^{k+1}_{\omega}$. Since Lemma \ref{LemmaSections}, Lemma \ref{LemmaDyad} and $p_n(\omega)>p_n(\omega')>p_n(\omega)-p_n(\omega')> p_n(\tilde{\Omega}\setminus \omega\omega')$ imply together $\varphi_{A^{k+1}_{\omega}}(v',p')= a \hat{F}^{k+1}_{\omega} \oplus b \hat{F}^{k+1}_{\omega'}$ \textbf{if} $H^{k+1}_\omega(m_a+1,m_b-1)=1$ [and $\varphi_{A^{k+1}_{\omega}}(v',p')= a \hat{F}^{k+1}_{\omega'} \oplus b \hat{F}^{k+1}_{\omega}$ \textbf{if} $H^{k+1}_\omega(m_a+1,m_b-1)=0$], we  conclude that $H_{\omega}^{k}(m_a,m_b)=H_{\omega}^{k+1}(m_a+1,m_b-1)$, which is the desired result since we have shown in (II) that $H_{\omega}^{k+1}(m_a+1,m_b-1)=H_{\omega}^{k}(m_a+1,m_b-1)$.

\bigskip
(III) Suppose that $|\mathbf{C}^k_{\omega}|=|\mathbf{C}^{k+1}_{\omega}|=|\mathbf{C}^{k+1}_{\omega'}|=2$ and $\omega'\in \hat{F}^k_{\omega}$. Assuming that $\omega'\notin \hat{F}^{k+1}_{\omega} $ (i.e., $\hat{F}^{k+1}_{\omega} \neq \hat{F}^{k+1}_{\omega'}$), we must show that $\hat{G}^{k+1}_{\omega'}\cap A^k_\omega=\emptyset$
Since $\hat{F}^{k+1}_{\omega} \neq \hat{F}^{k+1}_{\omega'}$, we have two possibilities: either $\hat{F}^{k+1}_{\omega'}= \hat{G}^{k+1}_{\omega}$ or $A^{k+1}_{\omega'}\neq A^{k+1}_{\omega}$. These possible cases are discussed below.

\smallskip\noindent
\underline{Case 1}: $\hat{F}^{k+1}_{\omega'}= \hat{G}^{k+1}_{\omega}$. 

 \smallskip\noindent Note in this case that $\mathbf{C}^{k+1}_{\omega}= \mathbf{C}^{k+1}_{\omega'}=\{\hat{F}^{k+1}_{\omega}, \hat{F}^{k+1}_{\omega'}\}$. 
Pick $(v,p)\in \mathcal{\tilde{D}}^N$ such that $N_a^v=\{1,\ldots,k+1\}, N_b^v=\{k+2,\ldots,n\}$ and 
\begin{align}
\label{EqLexNaSol}
      p_i(\omega)>p_i(\omega')>p_i(\omega)-p_i(\omega')> p_i(\tilde{\Omega}\setminus \omega\omega'),& ~\forall i\in N^v_a. \\ \label{EqLexNbSol}
       p_j(\omega'')> p_j(\hat{F}^k_{\omega}),&  ~\forall j\in N^v_b, \forall \omega''\in \Tilde{\Omega}\setminus \hat{F}^k_{\omega}.
\end{align}
Assume without loss of generality that $\varphi_{A^{k+1}_{\omega}}(v,p)=a\hat{F}^{k+1}_{\omega}\oplus b \hat{F}^{k+1}_{\omega'}$ [the argument is identical, swapping the roles  of $\omega$ and $\omega'$, if we instead write $\varphi_{A^{k+1}_{\omega}}(v,p)=a \hat{F}^{k+1}_{\omega'}\oplus b\hat{F}^{k+1}_{\omega} $].
Then remark that agent 1 (who is in $N^v_a$) can manipulate $\varphi$ at $(v,p)$, by misreporting $(v'_1,p'_1)$ such that $v'_1(a)<v'_1(b)$ and $p'_1(\hat{F}^k_{\omega})<p'_1(\hat{G}^k_\omega)$. This manipulation will lead to the preference profile $(v',p')=((v'_1,v_{-1}),(p'_1,p_{-1}))$, which satisfies  $N_a^{v'}=\{2,\ldots,k+1\}$, $\eta_{p'}(N_a^{v'}, \hat{F}^k_{\omega}|\hat{G}^k_\omega)=k$ [by the third inequality of (\ref{EqLexNaSol})] and $\eta_{p'}(N_b^{v'}, \hat{G}^k_\omega|\hat{F}^k_{\omega})=n-k$  [by  (\ref{EqLexNbSol}) and $p'_1(\hat{F}^k_{\omega})<p'_1(\hat{G}^k_\omega)$]. Hence,  combining Lemma \ref{LemmaSections}, Lemma \ref{LemmaDyad} and  $H_\omega^k(k,n-k)=1$, we get $\varphi_{A^{k}_{\omega}}(v',p')=a\hat{F}^k_{\omega}\oplus b\hat{G}^k_\omega$, which yields $a$ in both states $\omega$ and $\omega'$ (recall that $\omega'\in \hat{F}^k_{\omega}$). Noting from (\ref{EqLexNaSol}) that $(v_n,p_n)$ is $\{\omega,\omega'\}$-lexicographic, one can see that agent $n$  prefers  $ \varphi(v',p')$ to $\varphi(v,p)$ at $(v,p)$. This contradicts the strategy-proofness of $\varphi$ (Case 1 is hence impossible).

\medskip\noindent
\underline{Case 2}: $A^{k+1}_{\omega'}\neq A^{k+1}_{\omega}$.

\smallskip\noindent Note in this case that we must have $\{\hat{F}^{k+1}_{\omega}, \hat{G}^{k+1}_{\omega} \}=\mathbf{C}^{k+1}_{\omega}\neq \mathbf{C}^{k+1}_{\omega'}=\{\hat{F}^{k+1}_{\omega'}, \hat{G}^{k+1}_{\omega'} \}$. There are four possible subcases: (i) $\hat{G}^{k+1}_{\omega}\cap A^k_\omega\neq \emptyset$ and  $\hat{G}^{k+1}_{\omega'}\cap A^k_\omega\neq \emptyset$;  (ii) $\hat{G}^{k+1}_{\omega}\cap A^k_\omega\neq \emptyset$ and  $\hat{G}^{k+1}_{\omega'}\cap A^k_\omega= \emptyset$; (iii) $\hat{G}^{k+1}_{\omega}\cap A^k_\omega= \emptyset$ and  $\hat{G}^{k+1}_{\omega'}\cap A^k_\omega\neq \emptyset$; (iv) $\hat{G}^{k+1}_{\omega}\cap A^k_\omega= \emptyset$ and  $\hat{G}^{k+1}_{\omega'}\cap A^k_\omega= \emptyset$. Since  subcase (iv) gives the desired conclusion, it suffices to rule out the first three subcases.

\medskip
First, suppose $\hat{G}^{k+1}_{\omega}\cap A^k_\omega\neq \emptyset$ and  $\hat{G}^{k+1}_{\omega'}\cap A^k_\omega\neq \emptyset$. Then there exist $\hat{\omega}\in \hat{G}^{k+1}_{\omega}\cap A^k_\omega$ and  $\tilde{\omega}\in \hat{G}^{k+1}_{\omega'}\cap A^k_\omega$. An argument identical to the discussion of Case 1 allows to find a contradiction if
$\hat{\omega}\in \hat{G}^{k+1}_{\omega}\cap \hat{F}^k_{\omega}$ or  $\tilde{\omega}\in \hat{G}^{k+1}_{\omega'}\cap \hat{F}^k_{\omega}$. We must thus have $\hat{\omega}\in \hat{G}^{k+1}_{\omega}\cap \hat{G}^k_\omega$ and  $\tilde{\omega}\in \hat{G}^{k+1}_{\omega'}\cap \hat{G}^k_\omega$; and  combining Lemma \ref{DyadSolidarity}-(I) and Lemma \ref{DyadSolidarity}-(II)  allows to write $H_\omega^{k+1}(m_a+1,m_b-1)=H_\omega^{k}(m_a,m_b)=H_\omega^{k+1}(m_a,m_b)$,  $\forall m_a\in \{0,\ldots,k\}, m_b\in \{1,\ldots, n-k\}$. Note from the previous equalities and (\ref{EqdyadHSol}) that either $H_{\omega}^{k+1}(1,0)=0$ or $H_{\hat{\omega}}^{k+1}(1,0)=0$.  Assume without loss of generality that $H_{\omega}^{k+1}(1,0)=0$. Then picking an $\{\omega,\omega',\hat{\omega}, \Tilde{\omega}\}$-lexicographic profile $(v,p)\in \tilde{\mathcal{D}}^N$ such that (A) $N_a^v=\{1,\ldots,k+1\}$,  $N_b^v=\{k+2,\ldots,n\}$, (B) $p_1(\omega)>p_1(\omega'\hat{\omega}\Tilde{\omega})$, (C) $p_i(\omega)v_i(a)+p_i(\hat{\omega})v_i(b)<p_i(\hat{\omega})v_i(a)+p_i(\omega)v_i(b)$  and $p_i(\omega \omega')v_i(a)+p_i(\hat{\omega}\tilde{\omega})v_i(b)>p_i(\hat{\omega}\tilde{\omega})v_i(a)+p_i(\omega\omega')v_i(b)$ for all $i\neq 1$, it is not difficult to check that agent 1 will manipulate $\varphi$ at $(v,p)$ by misreporting $(v_1',p_1')$ such that $v'_1(a)<v'_1(b)$ and $p_1(\omega \omega')v'_1(a)+p_1(\hat{\omega}\tilde{\omega})v'_1(b)>p_1(\hat{\omega}\tilde{\omega})v'_1(a)+p_i(\omega\omega')v'_1(b)$ ---thus inducing the unanimous choice of outcome $a$ in $\omega$ [by (\ref{EqdyadHSol}), $a$ is not chosen in $\omega$ at $(v,p)$ since $H_{\omega}^{k+1}(1,0)=0$].

\medskip
Next, suppose $\hat{G}^{k+1}_{\omega}\cap A^k_\omega\neq \emptyset$ and  $\hat{G}^{k+1}_{\omega'}\cap A^k_\omega= \emptyset$.
Then there must exist $\hat{\omega}\in \hat{G}^{k+1}_{\omega}\cap \hat{G}^k_\omega$ (since an argument identical to the discussion of Case 1 allows to find a contradiction if
$ \hat{G}^{k+1}_{\omega}\cap \hat{F}^k_{\omega}\neq \emptyset$). Combining Lemma \ref{DyadSolidarity}-(I) and Lemma \ref{DyadSolidarity}-(II) thus allows to write $H_\omega^{k+1}(m_a+1,m_b-1)=H_\omega^{k}(m_a,m_b)=H_\omega^{k+1}(m_a,m_b)$,  $\forall m_a\in \{0,\ldots,k\}, m_b\in \{1,\ldots, n-k\}$. Note from the previous equalities and (\ref{EqdyadHSol}) that either $H_{\omega}^{k+1}(1,0)=0$ or $H_{\hat{\omega}}^{k+1}(1,0)=0$. If $H_{\omega}^{k+1}(1,0)=0$, remark that an argument identical to the previous paragraph yields a manipulation of $\varphi$. Thus, we may assume  $H_{\hat{\omega}}^{k+1}(1,0)=0=H_{\hat{\omega}}^{k}(1,0)$, where the last equality holds by Lemma \ref{DyadSolidarity}-(I). Since  $\hat{G}^{k+1}_{\omega'}\cap A^k_\omega= \emptyset$, one can pick  $(v,p)\in \tilde{\mathcal{D}}^N$ such that (A) $N_a^v=\{1,\ldots,k\}$,  $N_b^v=\{k+1,\ldots,n\}$, (B)  $p_n(\omega')>p_n(\tilde{\Omega}\setminus\omega')$, (C) $p_i(\hat{G}^k_{\omega'})v_i(a)+p_i(\hat{F}^k_{\omega'})v_i(b)<p_i(\hat{F}^k_{\omega'})v_i(a)+p_i(\hat{G}^k_{\omega'})v_i(b)$  and $p_i(\hat{G}^{k+1}_{\omega'})v_i(a)+p_i(\hat{F}^{k+1}_{\omega'})v_i(b)>p_i(\hat{F}^{k+1}_{\omega'})v_i(a)+p_i(\hat{G}^{k+1}_{\omega'})v_i(b)$ for all $i\neq n$. Observe then that agent $n$ will manipulate $\varphi$ at $(v,p)$ by misreporting $(v_n',p_n')$ such that $v'_n(a)>v'_n(b)$ and $p_n(\hat{G}^{k+1}_{\omega'})v_n(a)+p_i(\hat{F}^{k+1}_{\omega'})v_n(b)>p_n(\hat{F}^{k+1}_{\omega'})v_n(a)+p_n(\hat{G}^{k+1}_{\omega'})v_n(b)$  ---thus inducing the unanimous choice of outcome $b$ in $\omega'$ [by (\ref{EqdyadHSol}), $b$ is not chosen in $\omega'$ at $(v,p)$ since $H_{\hat{\omega}}^{k}(1,0)=0$ and $\hat{\omega}\in \hat{G}^k_{\omega'}=\hat{G}^k_\omega$]. Hence, the second  subcase is ruled out.

\medskip
The subcase where $\hat{G}^{k+1}_{\omega}\cap A^k_\omega= \emptyset$ and  $\hat{G}^{k+1}_{\omega'}\cap A^k_\omega\neq \emptyset$ is identical to the discussion of the previous paragraph (just swapping the roles of $\omega$ and $\omega'$). We can thus rule it out.

\bigskip
(IV)  Suppose that $|\mathbf{C}^k_{\omega'}|=|\mathbf{C}^{k+1}_{\omega'}|=2 $  and $\hat{G}^{k+1}_{\omega'} \cap A^{k}_{\omega'}=\emptyset$. Assume by contradiction that $H_{\omega'}^{k+1}(1,0)=0$. Since $\hat{G}^{k+1}_{\omega'} \cap A^{k}_{\omega'}=\emptyset$, one can choose $(v,p)\in \tilde{\mathcal{D}}^N$ such that (A) $N_a^v=\{1,\ldots,k+1\}$,  $N_b^v=\{k+2,\ldots,n\}$, (B)  $p_1(\omega')>p_1(\tilde{\Omega}\setminus\omega')$, (C) $p_i(\hat{G}^{k+1}_{\omega'})v_i(a)+p_i(\hat{F}^{k+1}_{\omega'})v_i(b)>p_i(\hat{F}^{k+1}_{\omega'})v_i(a)+p_i(\hat{G}^{k+1}_{\omega'})v_i(b)$ and  $p_i(\hat{G}^k_{\omega'})v_i(a)+p_i(\hat{F}^k_{\omega'})v_i(b)<p_i(\hat{F}^k_{\omega'})v_i(a)+p_i(\hat{G}^k_{\omega'})v_i(b)$   for all $i\neq 1$. Note that $\eta_p(N_a^v, \hat{F}^{k+1}_{\omega'}|\hat{G}^{k+1}_{\omega'})=1$ and $\eta_p(N_a^v, \hat{G}^{k+1}_{\omega'}|\hat{F}^{k+1}_{\omega'})=0$. 
It is not difficult to check that agent $1$ will manipulate $\varphi$ at $(v,p)$ by misreporting $(v_1',p_1')$ such that $v'_1(a)<v'_1(b)$ and $p_n(\hat{G}^{k+1}_{\omega'})v_n(a)+p_i(\hat{F}^{k+1}_{\omega'})v_n(b)>p_n(\hat{F}^{k+1}_{\omega'})v_n(a)+p_n(\hat{G}^{k+1}_{\omega'})v_n(b)$. This manipulation  leads to the preference profile $(v',p')=((v'_1,v_{-1}),(p'_1,p_{-1}))$, under which $n_a^{v'}=k$ and agents unanimously prefer  selecting the outcome $a$ in $\omega'$ [by (\ref{EqdyadHSol}), $a$ is not chosen in $\omega'$ at $(v,p)$ since $H_{\omega'}^{k}(1,0)=0$]. This contradicts the strategy-proofness of $\varphi$.

\bigskip
(V) Suppose that $ |\mathbf{C}^k_{\omega}|=|\mathbf{C}^{k+1}_{\omega}|=2$
and $\hat{G}^{k+1}_{\omega}\neq \hat{G}^k_\omega$. 
Assume without loss of generality  that $\hat{G}^{k+1}_{\omega} \setminus \hat{G}^k_\omega\neq \emptyset$ (swapping the roles of $a$ and $b$, a similar argument allows to conclude if one instead writes $\hat{G}^k_\omega \setminus \hat{G}^{k+1}_{\omega}\neq \emptyset$). 
Fix then  $\omega''\in \hat{G}^{k+1}_{\omega}\setminus \hat{G}^k_\omega$.

We show first that  $\hat{G}^{k+1}_{\omega}\cap A^{k}_\omega= \emptyset$. 
Suppose by contradiction  that there exists $\hat{\omega}\in \hat{G}^{k+1}_{\omega}\cap A^{k}_\omega$. Then note that $\hat{\omega}\in \hat{G}^{k+1}_{\omega}\cap \hat{G}^k_\omega$ [assuming $ \hat{G}^{k+1}_{\omega}\cap \hat{F}^k_{\omega}\neq \emptyset$ contradicts  Lemma \ref{DyadSolidarity}-(III) if one respectively uses $a$ as $b$ and vice-versa,  $\hat{\omega}$ as $\omega$, and $\omega''$ as $\omega'$].
Combining $\hat{\omega}\in \hat{G}^{k+1}_{\omega}\cap \hat{G}^k_\omega$ and $\omega''\in \hat{G}^{k+1}_{\omega} \setminus \hat{G}^k_\omega$, note that the premise of Lemma \ref{DyadSolidarity}-(III) applies [if one respectively uses $a$ as $b$ and vice-versa, $\hat{\omega}$ as $\omega$, and $\omega''$ as $\omega'$].
Hence, remark that an argument identical to the first subcase of Case 2 [in the proof of of Lemma \ref{DyadSolidarity}-(III)] yields a manipulation of $\varphi$ (contradicting its strategy-proofness). 

Thus, we have $\hat{G}^{k+1}_{\omega}\cap A^{k}_\omega=\emptyset$. We 
must show that $|\mathbf{C}^k_{\omega''}|=1$. 
By contradiction, assume that $|\mathbf{C}^k_{\omega''}|=2$.  
Observe then that, since $|\mathbf{C}^k_{\omega''}|=|\mathbf{C}^{k+1}_{\omega''}|=2$  and $\hat{G}^{k+1}_{\omega}\neq \hat{G}^k_\omega$, an  argument identical to the previous  paragraph (just swapping the roles of $\omega$ and $\omega''$) allows  to write $\hat{G}^{k+1}_{\omega''}\cap A^{k}_{\omega''}=\emptyset$. Hence, applying  Lemma \ref{DyadSolidarity}-(IV)   yields both $H^{k+1}_{\omega}(1,0)=1$ and $H^{k+1}_{\omega''}(1,0)=1$. But this is impossible, since $\omega''\in \hat{G}^{k+1}_{\omega}$ requires that  $H^{k+1}_{\omega}(1,0)+H^{k+1}_{\omega''}(1,0)=1$.

We thus have $|\mathbf{C}^k_{\omega''}|=1,  \forall \omega''\in  \hat{G}^{k+1}_{\omega} \setminus \hat{G}^k_\omega=\hat{G}^{k+1}_{\omega}$ (recall that $\hat{G}^{k+1}_{\omega}\cap A^{k}_\omega=\emptyset$).
\endproof

\medskip
To prove the next two lemmas, we introduce additional notation and terminology.
Consider an integer $m\geq 2$ and an event $\hat{\Omega}\in 2^\Omega$. Call $m$-\textit{agent event selector} on $\hat{\Omega}$ any mapping $f:(\mathcal{P}(\hat{\Omega}))^m\rightarrow 2^{\hat{\Omega}}$, that is to say, any mapping which associates an event in $2^{\hat{\Omega}}$ with each belief profile $p\in  (\mathcal{P}(\hat{\Omega}))^m$. 

An $m$-agent event selector on $\hat{\Omega}$ will be called
strategy-proof if no agent can ever induce the choice of an event she finds more likely by misrepresenting her belief, \textit{i.e.}, $p_i(f(p))\geq p_i(f(p'_i,p_{-i}))$, for all $p\in (\mathcal{P}(\hat{\Omega}))^m$ and all $p'_i\in \mathcal{P}(\hat{\Omega})$. 

Let $\Pi_m$ denote the set of permutations of $\{1,\ldots,m\}$. An $m$-agent event selector on $\hat{\Omega}$ will be called
anonymous if the names/labels of the $m$ agents play no role in the selection of the social event,  \textit{i.e.}, $f(\sigma p)= f(p)$, for all $p\in (\mathcal{P}(\hat{\Omega}))^m$ and all $\sigma\in \Pi_m$.

Let $\mathcal{E}=\{E_1,\ldots,E_K\}$ be a collection of non-nested subsets of $\Omega$ (with $k\geq 3$); and use the shorthand $\Omega_{\mathcal{E}}:=\bigcup\limits_{k=1}^K E_k$.
We say that $(E_1,E_2,E_3)\in \mathcal{E}^3$ is a \textit{top triple} of $\mathcal{E}$ if there exist six beliefs $p_i^{123}, p_i^{132}, p_i^{213},p_i^{231}, p_i^{312}, p_i^{321}\in \Omega_{\mathcal{E}}$ such that, for any distinct $j,k,\ell\in \{1,2,3\}$, we have
\begin{align*}
    p_i^{jkl}(E_j)> p_i^{jkl}(E_k)> &p_i^{jkl}(E_{\ell}); \\
    E\in \mathcal{E}\setminus \{E_1,E_2,E_3\}& \Rightarrow   p_i^{jkl}(E_{\ell})>p_i^{jkl}(E).
    \end{align*}

We may now state the following result.

\begin{lemma}\label{LemEventChoice1}
Let $\mathcal{E}=\{E_1,\ldots,E_K\} \subsetneq 2^{\tilde{\Omega}}$ be a richly decomposable collection of $K$ non-nested events, with $K\geq 3$. Then  $\mathcal{E}$ admits a top triple.
\end{lemma}

\proof We proceed in two steps.

\medskip\noindent
\textbf{Step 1}. Let $\mathcal{E}$ be as in the statement of Lemma \ref{LemEventChoice1} and suppose  that $K=3$, that is to say, $\mathcal{E}=\{E_1,E_2,E_3\}$, where the events $E_1,E_2,E_3$ are non-nested. We describe six beliefs that result in distinct rankings of these 3 events. First, let us construct a belief $p_i^{123}\in \mathcal{P}(\Omega_{\mathcal{E}})$ such that $p_i^{123}(E_1)>p_i^{123}(E_2)>p_i^{123}(E_3)$. We distinguish two cases.

\medskip
\textit{Case 1:} $E_1\setminus (E_2\cup E_3)=\emptyset$.

 Note in this case that we must have $(E_1\cap E_2)\setminus E_3\neq \emptyset$, since $E_1$ and $E_3$ are non-nested. In the same way, since $E_1$ and $E_2$ are non-nested, we must have  $(E_1\cap E_3)\setminus E_3\neq \emptyset$. Given any $\varepsilon,\delta\in (0,1)$, we may thus select a belief $p^{123}_i\in\mathcal{P}(\Omega_{\mathcal{E}})$ satisfying the following properties
\begin{align}
  \label{EqStep1case1}  
  p^{123}_i((E_1\cap E_2)\setminus E_3)&=1-\varepsilon;\\
 \nonumber
  p^{123}_i((E_1\cap E_3)\setminus E_2)&=\varepsilon (1-\delta);\\
  \nonumber 
  p^{123}_i((E_2\cup E_3)\setminus E_1)&=\varepsilon \delta.
\end{align}
One can then see from (\ref{EqStep1case1}) that, for $\varepsilon$ and $\delta$ sufficiently small, the desired property $p_i^{123}(E_1)>p_i^{123}(E_2)>p_i^{123}(E_3)$ is satisfied.

\medskip
\textit{Case 2:} $E_1\setminus (E_2\cup E_3)\neq\emptyset$.

 Note that we have both $E_1\setminus E_2\neq \emptyset$ and  $E_2\setminus E_1\neq \emptyset$, since these three events are non-nested. We consider two subcases below.

\medskip
\textit{Subcase 2.1:} Suppose $(E_1\cap E_2)\setminus E_3 \neq \emptyset$.

 Since $E_2$ and $E_3$ are non-nested, remark that $E_3\setminus (E_2\setminus E_1)\neq \emptyset$.  In this subcase we may then select $p^{123}_i\in\mathcal{P}(\Omega_{\mathcal{E}})$  such that
\begin{align}
  \label{EqStep1case211}  
   p_i^{123}(E_1\setminus E_3)=& 1-\varepsilon \mbox{ and }   p^{123}_i(\omega)\simeq\frac{1-\varepsilon}{|E_1\setminus E_3|} , \forall \omega\in E_1\setminus E_3 ;\\
 \label{EqStep1case212}  
   p_i^{123}(E_2\setminus E_1)=& \varepsilon (1-\delta) ;\\
\label{EqStep1case213}  
 p_i^{123}(E_3\setminus (E_2\setminus E_1))=& \varepsilon \delta.
\end{align}
Observe now that, for $\varepsilon$ and $\delta$ sufficiently small, we have 
\[
p_i^{123}(E_1)  \underbrace{\geq}_{\tiny \mbox{by }(\ref{EqStep1case211})} 1-\varepsilon   
\underbrace{>}_{\tiny  \varepsilon, \delta \rightarrow 0 } (1-\varepsilon)\frac{|E_1\setminus E_3|-1}{|E_1\setminus E_3|}+ \varepsilon +\varepsilon \delta
\underbrace{\geq}_{\tiny(\ref{EqStep1case211})-(\ref{EqStep1case213})}  p_i^{123}(E_2) \underbrace{>}_{\tiny (E_1\cap E_2)\setminus E_3\neq \emptyset} p_i^{123}(E_3).
\]
    
    \medskip
    \textit{Subcase 2.2:} Suppose $(E_1\cap E_2)\setminus E_3=\emptyset$.

Note in this subcase that $E_2\setminus (E_1 \cup E_3)=[E_2\setminus \underbrace{((E_1\cap E_2)\setminus E_3)}_{=\emptyset}]\setminus E_3= E_2\setminus E_3$, since $E_1$ and $E_3$ are non-nested. Hence, the subsets three $E_1\setminus E_3$, $E_2\setminus (E_1 \cup E_3)$, and $E_3$ form a partition of $\Omega_{\mathcal{E}}$ into non-trivial cells. Selecting then 
$p^{123}_i\in\mathcal{P}(\Omega_{\mathcal{E}})$  such that 
\begin{align}
  \label{EqStep1case221}  
   p_i^{123}(E_1\setminus E_3)=& 1-\varepsilon \mbox{ and }   p^{123}_i(\omega)\simeq\frac{1-\varepsilon}{|E_1\setminus E_3|} , \forall \omega\in E_1\setminus E_3 ;\\
 \label{EqStep1case222}  
   p_i^{123}(E_2\setminus (E_1 \cup E_3))=& \varepsilon (1-\delta) ;\\
\label{EqStep1case223}  
 p_i^{123}(E_3)=& \varepsilon \delta,
\end{align}
One can then see that 
\[
p_i^{123}(E_1)  \underbrace{\geq}_{\tiny \mbox{by }(\ref{EqStep1case221})} 1-\varepsilon   
\underbrace{>}_{\tiny  \varepsilon \rightarrow 0 } \varepsilon
\underbrace{\geq}_{\tiny(\ref{EqStep1case222})-(\ref{EqStep1case223})}  p_i^{123}(E_2) \underbrace{\geq}_{\tiny(\ref{EqStep1case222})} \varepsilon(1-\delta) \underbrace{>}_{\tiny  \delta \rightarrow 0} \varepsilon \delta =p_i^{123}(E_3). \]

Of course, for any  distinct $j,k,l\in \{1,2,3\}$ the procedure described above can be replicated to construct a belief $p_i^{jkl} \in\mathcal{P}(\Omega_{\mathcal{E}}) $ satisfying $p_i^{jkl}(E_j)>p_i^{jkl}(E_k)>p_i^{jkl}(E_l)$. 
Hence, we conclude that $\mathcal{E}=\{E_1,E_2,E_3\}$ is a top triple of itself.

\medskip\noindent
\textbf{Step 2}. Fix now  $\mathcal{E}=\{E_1,E_2,\ldots, E_K\}$, a richly decomposable collection of $K\geq 4$ non-nested events. 

\noindent
Given the result of Step 1, we may formulate the induction hypothesis that \textit{any richly decomposable collection $\mathcal{E}'$ containing at most $K-1$ non-nested events has a top triple}.

Define the function $\tilde{m}:\Omega_{\mathcal{E}} \rightarrow \{0,1,2,\ldots\}$ as follows: $\forall \omega \in \Omega_{\mathcal{E}} $, 
\begin{equation}\label{Eqdefmw}
    \tilde{m}(\omega)=|\{E\in \mathcal{E}: \omega\in E\}|.\end{equation}
    In words,   $\tilde{m}(\omega)$ counts the number of events (in the collection $\mathcal{E}$) that the state $\omega$ belongs to. Moreover, define the integer 
\begin{equation}\label{EqdefM}
    M:=\max\limits_{\omega\in\Omega_{\mathcal{E}} } \tilde{m}(\omega).\end{equation}
It is easy to see from (\ref{Eqdefmw})-(\ref{EqdefM}) that $1\leq M\leq K$.

Note that we may assume without loss of generality that $1\leq M\leq K-1$.\footnote{Indeed, if $M=K$, then it comes  from (\ref{Eqdefmw})-(\ref{EqdefM}) that $\bigcap_{k=1}^K E_k\neq \emptyset$; and hence letting $\mathcal{\bar{E}}=\{E_1\setminus\bigcap_{k=1}^K E_k, \ldots, E_K\setminus \bigcap_{k=1}^K E_k\}$, one can see that (i) $\mathcal{\bar{E}}$ is richly decomposable as well, (ii) the $K$ events $\bar{E}_k=E_k\setminus \bigcap_{\ell=1}^K E_{\ell}$ (for $k=1,\ldots,K$) are non-nested, (iii) no state belongs to all events $\bar{E}_1, \ldots, \bar{E}_K$ (\textit{i.e.}, $\bar{M}\leq K-1$), and (iv) any top triple of $\bar{\mathcal{E}}$ leads to a top triple of $\mathcal{E}$ and vice-versa.}
Remark also that the desired result is not difficult to obtain if there exist $\omega_1\in \Omega_{\mathcal{E}}$ such that $\tilde{m}(\omega_1)=1$.\footnote{To see this, suppose that $\omega_1\in E_1 $ and $\tilde{m}(\omega_1)=1$.  Pick  a  belief $q_i\in \mathcal{P}(\Omega)$ such that $q_i(E_1)<q_i(\omega), \forall \omega \in \Omega_{\mathcal{E}}\setminus E_1$; and remark that this implies $q_i(E_1)=\min\limits_{E\in \mathcal{E}}q_i(E)$ (since the events in $\mathcal{E}$ are non-nested).  Assume without loss that $q_i(E_2)>q_i(E_3)>q_i(E), \forall E\in \mathcal{E}\setminus\{E_2,E_3\}$. For any given $\lambda\in (0,1)$, one can then construct a belief $_{\lambda}p_i \in \mathcal{P}(\Omega_{\mathcal{E}})$ such that $_{\lambda}p_i(\omega_1)= \lambda$ and $_{\lambda}p_i(\omega)= \frac{1-\lambda}{q_i(\Omega_{\mathcal{E}}\setminus \omega_1)} q_i(\omega)$, for all $\omega\in \Omega_{\mathcal{E}}\setminus \omega_1$. Since $\tilde{m}(\omega_1)=1$,  it is not difficult  to see from repeated application of the intermediate value theorem (to $_\lambda p_i$, which is a continuous function of $\lambda$) that there exist distinct $\underline{\lambda},\bar{\lambda} \in (0,1)$ (I) $_{\lambda}p_i(E_1)>~_{\lambda}p_i(E_2)> ~_{\lambda}p_i(E_3)>~_{\lambda}p_i(E), \forall E\in \mathcal{E}\setminus \{E_1,E_2,E_3\}$ if $\lambda >\bar{\lambda}$; (II)  $_{\lambda}p_i(E_2)>~_{\lambda}p_i(E_1> ~_{\lambda}p_i(E_3)>~_{\lambda}p_i(E), \forall E\in \mathcal{E}\setminus \{E_1,E_2,E_3\}$ if  $\lambda \in \left(\underline{\lambda}, \bar{\lambda}\right)$; and (III) $_{\lambda}p_i(E_2)>~_{\lambda}p_i(E_3)> ~_{\lambda}p_i(E_1)>~_{\lambda}p_i(E), \forall E\in \mathcal{E}\setminus \{E_1,E_2,E_3\}$ if $\lambda $ is lower than (but sufficiently close to) to $\underline{\lambda}$. Moreover, letting $\delta=q_i(E_2)-q_i(E_3)$ and  $\omega_3\in E_3\setminus E_2$,  choose a belief $q'_i\in \mathcal{P}(\Omega)$ such that  $q'_i(\omega_3)\gtrapprox q_i(\omega_3)+\delta/2$, $q'_i(E_2\setminus E_3)\lessapprox q_i(E_2\setminus E_3)-\delta/2$, and $q'_i(\omega)=q_i(\omega)$ if $\omega\in \Omega_{\mathcal{E}}\setminus ((E_2\setminus E_3)\cup \omega_3)$. Note that we have $q'_i(E_3)>q_i(E_2)>q_i(E), \forall E\in \mathcal{E}\setminus\{E_2,E_3\}$. In the same spirit, defining the continuous function $_{\lambda}p'_i$ (by replacing $q_i$ with $q'_i$ in the definition of $_{\lambda}p_i$) and  applying the intermediate value theorem,  one obtains the three  beliefs generating the  remaining three permutations of $E_1,E_2,E_3$ at the top of the collection $\mathcal{E}$.} Hence, exploiting the last two observations, we will  assume in the following lines that $\tilde{m}(\omega)\geq 2$ for all $\omega\in \Omega_{\mathcal{E}}$ and $2\leq M\leq K-1$. This means that no state belongs to all $K$ events of $\mathcal{E}$, and each state in $\Omega_{\mathcal{E}} $ belongs to at least two distinct events of $\mathcal{E}$.

Since $M:=\max\limits_{\omega\in\Omega_{\mathcal{E}} } \tilde{m}(\omega)$ and $2\leq M\leq K-1$, notice that there must exist a non-empty event $A_1\in 2^{\Omega_{\mathcal{E}}}$, and $M$ events $F_1,\ldots,F_M\in 2^{\Omega_{\mathcal{E}}}$ such that 
\begin{align}
 \label{EqA1def1}
 A_1\cup F_m \in \mathcal{E},  &\mbox{ for } m=1,\ldots,M ;
 \\  \label{EqA1def2}
A_1\cap F_m=\emptyset, & \mbox{ for } m=1,\ldots,M ;
 \\  \label{EqA1def3}
A_1\cap E=\emptyset, & ~\forall E\in \mathcal{E}\setminus\{A_1\cup F_1, A_1\cup F_2,\ldots, A_1\cup F_M \}.
\end{align}
We may assume without loss that $A_1$ is \textit{maximal}, \textit{i.e.}, there exists no event $A'_1$ that contains $A_1$  and satisfies the version of (\ref{EqA1def1})-(\ref{EqA1def3}) where $A_1$ is replaced with $A'_1$. Moreover, since the events in $\mathcal{E}$ are non-nested, it follows from (\ref{EqA1def1}) that  $\mathcal{F}=\{F_1,F_2,\ldots,F_M\}$ is a collection of non-nested events. 
Finally, remark from the combination of (\ref{EqA1def1}) and (\ref{EqA1def3}) that, for any  $p_i\in \mathcal{P}(\Omega_{\mathcal{E}})$,  $m\in \{1,\ldots,M\}$ and $E\in \mathcal{E}\setminus\{A_1\cup F_1, A_1\cup F_2,\ldots, A_1\cup F_M \}$, we have 
\begin{equation}\label{EqA1top}
    [p_i(A_1)>1/2] \Rightarrow [p(E)< p_i(A_1\cup F_m)].
\end{equation}
In other words, if the probability of  $A_1$ is sufficiently high, then the events $A_1\cup F_1, \ldots, A_1\cup F_M$ are the $M$ most likely events of the collection $\mathcal{E}$.

We discuss a number of possible cases below; and we show in each case that a top triple of $\mathcal{E}$ exists (since $\mathcal{E}$ is richly decomposable).

\medskip\noindent
\textbf{Case I:} The collection $\mathcal{F}=\{F_1,F_2,\ldots,F_M\}$ is richly decomposable.

\noindent
Since $|\mathcal{F}|=M\leq K-1$ and  $\mathcal{F}$ is richly decomposable, note from our induction hypothesis that the collection $\mathcal{F}$ has a top triple, say, $(F_1,F_2,F_3)$. Next, letting $E_m=A_1\cup F_m$ (for $m=1,2,3$), we argue that $(E_1,E_2,E_3)$ is a top triple of $\mathcal{E}$.

Since $(F_1,F_2,F_3)$ is a top triple of $\mathcal{F}$, remark that, for any distinct $j,k,l\in \{1,2,3\}$, there exists a belief $q^{jkl}_i\in \mathcal{P}(\Omega_{\mathcal{F}})$ such that 
\begin{equation}\label{EqTripleFjkl}
    q^{jkl}_i(F_j)>q^{jkl}_i(F_k)>q^{jkl}_i(F_l)>q^{jkl}_i(F), ~\forall F\in \mathcal{F}\setminus\{F_1,F_2,F_3\}.
\end{equation}

Pick then a belief $p^{jkl}_i\in  \mathcal{P}_i(\Omega_{\mathcal{E}})$ such that $p^{jkl}_i(A_1)>1/2$ and $p^{jkl}_{i/\Omega_{\mathcal{F}}}=q_i$. Then note from the combination of (\ref{EqA1top}) and (\ref{EqTripleFjkl}) that $p^{jkl}_i(E_j)>p^{jkl}_i(E_k)>p^{jkl}_i(E_l)>p^{jkl}_i(E), ~\forall E\in \mathcal{E}\setminus\{E_1,E_2,E_3\}.$ We thus conclude that $(E_1,E_2,E_3)$ is a top triple of $\mathcal{E}$.

\medskip\noindent
\textbf{Case II:} Every component of $\mathcal{F}$ is dyadic.\footnote{Note that a trivial component of $\mathcal{F}$ is ruled out by the maximality of $A_1$ and (\ref{EqA1def1}). Hence, assuming that $\mathcal{F}$ is not richly decomposable leaves only the possibility of dyadic components.}

\noindent Note in this case that there exist \textbf{disjoint events} $B_1,C_1, B_2,C_2,\ldots, B_M,C_M\in  $ such that 
\begin{equation}\label{EqFdyad}
\mathcal{F}=\{\cup_{\ell=1}^{L} E_{\ell}, E_{\ell}\in \{B_{\ell}, C_{\ell}\} \mbox{ for } \ell=1,\ldots, L\}.
\end{equation}
We discuss two subcases.

\medskip
\textit{Subcase II.1:} There exist $\bar{E}\in \Omega_{\mathcal{E}}$ and $\ell\in \{1,\ldots,L\}$ s.t. $\bar{E}\cap B_{\ell}\neq \emptyset \neq \bar{E}\cap C_{\ell}$.

To fix ideas, we prove the desired result for $\ell=1$. Assume  that there exist two states $\bar{\omega},\bar{\omega}'$ such that $\bar{\omega}\in \bar{E}\cap B_{1}$ and $\bar{\omega}'\in \bar{E}\cap C_{1}$.
Defining $\bar{B}_{-1}:=\cup_{\ell=2}^M B_{\ell}$, $E_1:=A_1\cup B_1\cup \bar{B}_{-1}$ and  $E_2:=A_1\cup C_1\cup \bar{B}_{-1}$, observe from (\ref{EqA1def1}) and (\ref{EqFdyad}) that  $E_1,E_2\in \mathcal{E}$.
 
Fix then $q_i\in \mathcal{P}(\Omega_{\mathcal{E}}\setminus (A_1\cup\bar{\omega}\bar{\omega}'))$ such that
\begin{equation}\label{EqargmaxB-1}
q_i(\bar{B}_{-1})=\max\{p_i(\cup_{\ell=2}^M E_{\ell}), E_{\ell}\in \{B_{\ell}, C_{\ell}\} \mbox{ for } \ell=2,\ldots, L\}.
\end{equation}

Define $E_3:=
\argmax\limits_{\tiny \begin{tabular}{c}
       $E\in \mathcal{E}$ s.t.\\
        $\bar{\omega},\bar{\omega}'\in E$  \\
    \end{tabular}}q _i(E\setminus \bar{\omega}\bar{\omega}')$ ---recall that $\bar{E}\in \mathcal{E}$ and $\bar{\omega},\bar{\omega}'\in \bar{E}$ (hence $E_3$ indeed exists).
    We show next that $(E_1,E_2,E_3)$ is a top triple of $\mathcal{E}$.

For any $\lambda\in (0,1)$, select  a belief  $_{\lambda}p_i\in \mathcal{P}(\Omega_{\mathcal{E}})$ such that 
\begin{align}
  \label{EqStep2case211}  
   _{\lambda}p_i(\bar{\omega})=& \frac{4\lambda}{7}, ~  _{\lambda}p_i(\bar{\omega}')= \frac{2\lambda}{7};\\
\nonumber
  _{\lambda}p_i(\omega)= &\frac{\lambda q_i(\omega)}{7}, ~\forall \omega\in \mathcal{P}(\Omega_{\mathcal{E}}\setminus (A_1\cup\bar{\omega}\bar{\omega}'));\\ \nonumber
 _{\lambda}p_i(A_1)=&1-\lambda.
\end{align}
Then, combining (\ref{EqFdyad})-(\ref{EqStep2case211}),  one can see that  
    \begin{equation}\label{EqE1E2w1w2}
          _{\lambda}p_i(E_1)> ~ _{\lambda}p_i(E_2)>   ~_{\lambda}p_i(A_1\cup F), ~\forall \lambda\in (0,1),\forall F\in \mathcal{F}\setminus\{B_1\cup \bar{B}_{-1}, C_1\cup \bar{B}_{-1}\}.
    \end{equation}

Since $E_1:=A_1\cup B_1\cup \bar{B}_{-1}$ and  $E_2:=A_1\cup C_1\cup \bar{B}_{-1}$, remark from (\ref{EqA1top}),  (\ref{EqStep2case211})-(\ref{EqE1E2w1w2}), and $E_3:=
\argmax\limits_{\tiny \begin{tabular}{c}
       $E\in \mathcal{E}$ s.t.\\
        $\bar{\omega},\bar{\omega}'\in E$  
    \end{tabular}}q _i(E\setminus \bar{\omega}\bar{\omega}')$  that  
\begin{equation}\label{EqStep2case212}
  [\lambda <1/2]\Rightarrow  [ ~_{\lambda}p_i(E_1)> ~ _{\lambda}p_i(E_2)> ~_{\lambda}p_i(E_3)> ~ _{\lambda}p_i(E), ~\forall E\in \mathcal{E}\setminus\{E_1,E_2,E_3\}]. 
     \end{equation}
On the other hand, it comes from  (\ref{EqStep2case211}) and $E_3:=
\argmax\limits_{\tiny \begin{tabular}{c}
       $E\in \mathcal{E}$ s.t.\\
        $\bar{\omega},\bar{\omega}'\in E$  
    \end{tabular}}q _i(E\setminus \bar{\omega}\bar{\omega}')$ that 
$$
 _{\lambda}p_i(E_3)> ~ _{\lambda}p_i(E_1)> ~_{\lambda}p_i(E_2),  \mbox{ for } \lambda \mbox{ close enough to } 1.
    $$
Since the function $g(\lambda)=~_{\lambda}p_i(E_3)-~_{\lambda}p_i(E_2)$ is continuous in $\lambda$, combining the intermediate value theorem and (\ref{EqStep2case212}) thus allows to claim that, for any $\varepsilon>0$, there exists $\lambda_{\varepsilon}\in (0,1)$ such that 
\begin{equation}\label{EqStep2case213}  
  ~_{\lambda_{\varepsilon}}p_i(E_2)+\varepsilon> ~ _{\lambda_{\varepsilon}}p_i(E_3)> ~_{\lambda_{\varepsilon}}p_i(E_2).
     \end{equation}

   Write $\bar{\varepsilon}=\min\limits_{\tiny \begin{tabular}{c}
       $E,E'\in  \mathcal{P}(\Omega_{\mathcal{E}}\setminus (A_1\cup\bar{\omega}\bar{\omega}'))$\\
            $E\neq E'$
    \end{tabular}}|q _i(E)-q_i(E')|\}$.  Taking then $\varepsilon$ such that $0<\varepsilon< \min\{\frac{2\lambda}{7},\bar{\varepsilon}\}$, observe that, together,  (\ref{EqStep2case211}),  (\ref{EqStep2case212}) and  (\ref{EqStep2case213}) allow to write
        \begin{equation}\label{EqStep2case214}
 ~_{\lambda_{\varepsilon}}p_i(E_1)> ~ _{\lambda_{\varepsilon}}p_i(E_3)> ~_{\lambda_{\varepsilon}}p_i(E_2)> ~ _{\lambda_{\varepsilon}}p_i(E), ~\forall E\in \mathcal{E}\setminus\{E_1,E_2,E_3\}. 
     \end{equation}

Next, 
for any $\lambda\in (0,1)$ and $\delta\in (0,\frac{3\lambda }{7})$, select  a belief  $_{\lambda \delta}\hat{p}_i\in \mathcal{P}(\Omega_{\mathcal{E}})$ such that 
\begin{align}
  \label{EqStep2case2142}  
   _{\lambda,\delta}\hat{p}_i(\bar{\omega})=&\frac{3\lambda }{7}+\delta, ~  _{\lambda,\delta}\hat{p}_i(\bar{\omega}')= \frac{3\lambda }{7}-\delta;\\
\nonumber
  _{\lambda,\delta}\hat{p}_i(\omega)= &\frac{\lambda q_i(\omega)}{7}, ~\forall \omega\in \mathcal{P}(\Omega_{\mathcal{E}}\setminus (A_1\cup\bar{\omega}\bar{\omega}'));\\ \nonumber
 _{\lambda,\delta}\hat{p}_i(A_1)=&1-\lambda.
\end{align}

A reasoning similar to the one used to derive (\ref{EqStep2case214}) allows to write the following: for all $\varepsilon, \delta>0$ satisfying $0<\varepsilon+2\delta< \bar{\varepsilon}$, there exists $\lambda_{\varepsilon}^\delta$  such that 
   \begin{equation}\label{EqStep2case215}
 ~_{\lambda_{\varepsilon}^\delta, \delta}p_i(E_3)>  ~_{\lambda_{\varepsilon}^\delta, \delta} p_i(E_1)> ~_{\lambda_{\varepsilon}^\delta, \delta}p_i(E_2)> ~ _{\lambda_{\varepsilon}^\delta, \delta}p_i(E), ~\forall E\in \mathcal{E}\setminus\{E_1,E_2,E_3\}. 
     \end{equation}

We have thus shown the existence of three beliefs $p_i^{123},p_i^{132}, p_i^{312} \in \mathcal{P}(\Omega_{\mathcal{E}})$ such that for all $(j,k,l)\in \{(1,2,3),(1,3,2), (3,1,2)\}$, we have $p^{jkl}_i(E_j)>p^{jkl}_i(E_k)>p_i^{jkl}(E_l)>p_i^{jkl}(E)$, for all $E\in \mathcal{E}\setminus\{E_1,E_2,E_3\}$.

\medskip
Swapping the roles of $\bar{\omega}$ and $\bar{\omega'}$ in both (\ref{EqStep2case211}) and (\ref{EqStep2case214}), remark that we obtain the beliefs $ _{\lambda}p'_i\in \mathcal{P}(\Omega_{\mathcal{E}})$ and $ _{\lambda,\delta}\hat{p}'_i\in \mathcal{P}(\Omega_{\mathcal{E}})$ such that
\begin{align*}
    _{\lambda}p'_i(\bar{\omega})=& \frac{2\lambda}{7}, ~  _{\lambda}p'_i(\bar{\omega}')= \frac{4\lambda}{7};\\
\nonumber
  _{\lambda}p'_i(\omega)= &\frac{\lambda q_i(\omega)}{7}, ~\forall \omega\in \mathcal{P}(\Omega_{\mathcal{E}}\setminus (A_1\cup\bar{\omega}\bar{\omega}'));\\ \nonumber
 _{\lambda}p'_i(A_1)=&1-\lambda
\end{align*}
and 
\begin{align*}
     _{\lambda\delta}\hat{p}'_i(\bar{\omega})=&\frac{3\lambda }{7}-\delta, ~  _{\lambda\delta}\hat{p}'_i(\bar{\omega}')= \frac{3\lambda }{7}+\delta;\\
\nonumber
  _{\lambda\delta}\hat{p}'_i(\omega)= &\frac{\lambda q_i(\omega)}{7}, ~\forall \omega\in \mathcal{P}(\Omega_{\mathcal{E}}\setminus (A_1\cup\bar{\omega}\bar{\omega}'));\\ \nonumber
 _{\lambda\delta}\hat{p}'_i(A_1)=&1-\lambda.
\end{align*}

Exploiting  these two new beliefs and replicating  the procedure used to derive (\ref{EqStep2case212}), (\ref{EqStep2case214}) and (\ref{EqStep2case215}), we find that there also exist $p_i^{213},p_i^{231}$ and  $p_i^{321}\in \mathcal{P}(\Omega_{\mathcal{E}})$  such that , for all  $(j,k,l)\in \{(2,1,3),(2,3,1), (3,2,1)\}$, we have $p^{jkl}_i(E_j)>p^{jkl}_i(E_k)>p_i^{jkl}(E_l)>p_i^{jkl}(E)$, for all $E\in \mathcal{E}\setminus\{E_1,E_2,E_3\}$. 

Together, these properties of the six beliefs $p_i^{123},p_i^{132}, p_i^{312}, p_i^{213},p_i^{231}, p_i^{321}\in \mathcal{P}(\Omega_{\mathcal{E}})$ allow to conclude that $(E_1,E_2,E_3)$ is a top triple of $\mathcal{E}$.

\medskip
\textit{Subcase II.2:} For all $E\in \Omega_{\mathcal{E}}$ and $\ell\in \{1,\ldots,L\}$, we have $E\cap B_{\ell}=\emptyset$ or $  E\cap C_{\ell}\neq \emptyset$.

 Note in this subcase that, for all $E\in \mathcal{E}\setminus \{A_1\cup F_1,\ldots, A_1\cup F_M\}$, there must exist $\omega\in E\setminus \Omega_{\mathcal{F}}\neq \emptyset$ (otherwise $E$ would be nested in some event $A_1\cup F$, where $F\in \mathcal{F}$; and this would contradict the fact that $\mathcal{E}$ is a collection of non-nested events.
Combining this observation with (\ref{EqA1def1})-(\ref{EqA1def3}) (and $|\mathcal{F}|=K-1<K=|\mathcal{E}|$), remark that there must exist $T\geq 2$, and nonempty events $A_2,F_1^2,\ldots, F^2_{M_2},\ldots,A_T, F_T^2\ldots F^2_{M_T}$ (with $M_t\geq 1$ for $t=2,\ldots,T$) such that the collection $\mathcal{E}$ can be written as 
\begin{equation} \label{EqUnionETriple}
    \mathcal{E}= \underbrace{\{A_1\cup F_1,\ldots,A_1\cup F_M\}}_{:=\mathcal{E}^1}\cup \ldots \cup \underbrace{\{A_T\cup F_1^T,\ldots,A_1\cup F_{M_T}^T\}}_{:=\mathcal{E}^T},
\end{equation}
where 
\begin{align}\label{EqCondF1}
    A_t \cap (F_1\cup \ldots\cup F_M)=&\emptyset, \mbox{ for } t=1,\ldots T;\\ \label{EqCondF2}
    A_t \cap (A_{t'}\cup F_m^{t'})=&\emptyset, \mbox{ for } t=1,\ldots T-1, t'=t,\ldots,T, m=1,\ldots,M_{t'}.
\end{align}

Note first that we cannot have $M_2=1$ in (\ref{EqUnionETriple}). Indeed, combining $M_2=1$ with (\ref{EqCondF1})-(\ref{EqCondF2}) would yield $\tilde{m}(\omega)=1$ for all $\omega\in A_2\neq \emptyset$; and this would contradict the fact that each state in $\Omega_{\mathcal{E}}$ belongs to at least two events of $\mathcal{E}$). We will thus have $M_2\geq 2$; and, as done for $A_1$, we may also assume without loss that  $A_2$  is maximal in the sense that $\cap_{m=1}^{M_2}F^2_{m}=\emptyset$.

Remark also that in the case where the collection $\mathcal{F}^2=\{ F_1^2,\ldots, F_{M_2}^2\}$ is richly decomposable, it comes from our induction hypothesis that $\mathcal{F}^2$ has a top triple $(F'_1,F'_2,F'_3)$ (since $M_2\leq K-1$).  Noting from (\ref{EqCondF1})-(\ref{EqCondF2}) that $A_2\cap E=\emptyset, \forall E\in \mathcal{E}\setminus \mathcal{E}_2$, it is then easy to see that $(A_2\cup F'_1, A_2\cup F'_2, A_2\cup F'_3)$ is a top triple of $\mathcal{E}$.

We are thus left with  the case where the collection $\mathcal{F}^2=\{F_1^2,\ldots,F_{M_2}^2\}$   exhibits only dyadic components in its maximal decomposition, that is to say, there exist an integer $L_2\geq 2$ and disjoint events $B_1^2, C_1^2,\ldots, B_{M_2}^2, C_{M_2}^2$ such that 
\begin{equation}\label{EqDecdyadF2}
\{F_1^2,\ldots,\cup F_{M_2}^2\}=\{\cup_{\ell=1}^{L_2} F_{\ell}, F_{\ell}\in \{B^2_{\ell}, C_{\ell}^2\} \mbox{ for } \ell=1,\ldots, L_2 \}.    
\end{equation}

We examine three configurations below.

\medskip
\underline{Configuration 1}: $( F_1^2\cup \ldots \cup F_{M_2}^2)\setminus (F_1 \cup \ldots\cup F_M)\neq \emptyset$.

 In this configuration there exists $m\in \{1,\ldots,M_2\}$ and $\omega^* \in F^2_m\setminus (F_1\cup \ldots\cup F_M)\neq \emptyset$. Recalling (\ref{EqDecdyadF2}), assume  without loss that $\omega^*\in C_{1}^2\setminus (F_1\cup\ldots \cup F_{M})$. Fix also $\omega_A\in A_2$.

Write $E_1=A_2\cup B^2_1\cup(\cup_{\ell=2}^{L_2} B^2_{\ell})$ and $E_2=A_2\cup C^2_1\cup(\cup_{\ell=2}^{L_2} B^2_{\ell})$ without loss of generality (note from what precedes that $E_1,E_2\in \mathcal{E}$).
Next, for any $\varepsilon \in (0,1/4)$, pick a belief $^{\varepsilon}p_i\in \mathcal{P}(\Omega_{\mathcal{E}})$ such that  \begin{eqnarray} \label{EqBlocksConfig1}
 ^{\varepsilon}p_i(\omega_A)>\frac{1}{2}+\varepsilon; ~^{\varepsilon}p_i(A_1)=\frac{1}{2}-2\varepsilon;\\
   \label{EqF2Config1}  0< ~^{\varepsilon}p_{i}(B^2_{1}) - ~^{\varepsilon}p_{i}(C^2_{1})< ~^{\varepsilon}p_i(B_{\ell}^2)-~^{\varepsilon} p_i(C^2_{\ell}), \mbox{ for  } ~\ell=2,\ldots, L_2.
\end{eqnarray} 

Let then $~^{\varepsilon}\hat{E}_3=A_1 \cup~\argmax\limits_{F\in \mathcal{F}} ~^{\varepsilon}p_i(F)$ ---note from (\ref{EqBlocksConfig1}) that $^{\varepsilon}\hat{E}_3\in \mathcal{E}^1$ is the most likely event in $\mathcal{E}\setminus \mathcal{E}^2$ for $\varepsilon$ small enough. Combining the properties of $^{\varepsilon}p_i$ given in  (\ref{EqBlocksConfig1})-(\ref{EqF2Config1}) with
 (\ref{EqUnionETriple})-(\ref{EqDecdyadF2}), remark that  there exists $\bar{\varepsilon}\in (0,1/4)$  such that
 \begin{align}
     \label{Eq1config1}
    ^{\varepsilon}p_i(E_1)>  ~^{\varepsilon}p_i(E_2)> ~^{\varepsilon}p_i(^{\varepsilon}\hat{E}_3) >~^{\varepsilon}p_i(E), \forall E\in \mathcal{E}\setminus\{E_1,E_2, ~^{\varepsilon}\hat{E}_3\},  \forall \varepsilon\in (0,\bar{\varepsilon}).
\end{align}
Letting $\varepsilon^*=\bar{\varepsilon}/2$  and $E_3= ~^{\varepsilon^*}\hat{E}_3$ in (\ref{Eq1config1}), one obtains  the belief $p^{123}_i =~^{\varepsilon^*}p_i$ such that 
\begin{equation}\label{Eqp123Config1}
    p_i^{123}(E_1)>  p^{123}_i(E_2)> p^{123}_i(E_3) >p^{123}_i(E), \forall E\in \mathcal{E}\setminus\{E_1,E_2, E_3\}.\end{equation}

For all $\lambda\in (0, p^{123}_i (\omega_A))$, define then the belief  $_\lambda p_i\in \mathcal{P}(\Omega_{\mathcal{E}})$ as follows:
\begin{align}\label{Eqdefplambda}
    _\lambda p_i(\omega)= &p^{123}_i(\omega)+ \frac{\lambda}{|A_1|}, ~\forall \omega\in A_1;\\ \nonumber
   _\lambda p_i(\omega_A)= &p^{123}_i(\omega_A)-\lambda;\\ \nonumber
    _\lambda p_i(\omega)= & _\lambda p^{123}_i(\omega), ~\forall \omega\in \Omega_{\mathcal{E}}\setminus (\omega_A\cup A_1).
\end{align}
Remark from (\ref{Eqdefplambda})  and $(A_1\cup A_2)\cap \Omega _{\mathcal{F}}=\emptyset$ that $\argmax\limits_{E\in \mathcal{E}_1} ~ _\lambda p_i(E)= \argmax\limits_{E\in \mathcal{E}_1}   p^{123}_i(E)=E_3$.
Since $g_1(\lambda)=~_\lambda p_i(E_3)-_\lambda p_i(E_1)$  and $g_2(\lambda)=~_\lambda p_i(E_3)-_\lambda p_i(E_2)$ are both continuous functions of $\lambda$, and $\lim\limits_{1-p^{123}_i (A_1)}g_2(\lambda)>\lim\limits_{1-p^{123}_i (A_1)}g_1(\lambda)>0$ by (\ref{EqA1top}) [while $\lim\limits_{0}g_1(\lambda)<\lim\limits_{0}g_2(\lambda)<0$ by (\ref{Eq1config1})], repeated application of the intermediate value theorem [with the properties that, for all  $\lambda\in (0,  p^{123}_i (\omega_A))$, $ _\lambda p_i(E_1)> ~_\lambda p_i(E_2)$, $ _\lambda p_i(A_1\cup A_2)\geq1-\varepsilon^*$, and $| _\lambda p_i(A_1)- ~ _\lambda p_i(A_2)|<4\varepsilon^*$] guarantees the existence of $\lambda_1$ and $\lambda_2$ such that 
\begin{eqnarray}
    \label{Eqlambda2config1}
    _{\lambda_2} p_i(E_1)> ~  _{\lambda_2} p_i(E_3)> ~_{\lambda_2} p_i(E_1)> ~_{\lambda_1} p_i(E_3)-\delta> ~_{\lambda_2} p_i(E),   \\ \label{Eqlambda1config1}
    _{\lambda_1} p_i(E_3)> ~  _{\lambda_1} p_i(E_1)>  ~_{\lambda_1} p_i(E_3)-\delta>~_{\lambda_1} p_i(E_2)> ~_{\lambda_1} p_i(E), 
\end{eqnarray}
for all $E\in \mathcal{E}\setminus\{E_1,E_2, E_3\}$ and all positive and small enough $\delta$.

Combining (\ref{Eqlambda2config1})-(\ref{Eqlambda1config1}) with (\ref{Eqp123Config1}), We have thus found 
three beliefs $p_i^{123},p_i^{132}, p_i^{312} \in \mathcal{P}(\Omega_{\mathcal{E}})$ such that for all $(j,k,l)\in \{(1,2,3),(1,3,2), (3,1,2)\}$, we have $p^{jkl}_i(E_j)>p^{jkl}_i(E_k)>p_i^{jkl}(E_l)>p_i^{jkl}(E)$, for all $E\in \mathcal{E}\setminus\{E_1,E_2,E_3\}$.

Next, define the belief $p^{213}\in \mathcal{P}(\Omega_{\mathcal{E}})$ as follows: 
\begin{align}\label{Eqdefp213}
  p^{213}(\omega^*)= &p^{123}_i(\omega^*)+ \varepsilon;\\ \nonumber
  p^{213}(\omega_A)= &p^{123}_i(\omega_A)- \varepsilon;\\ \nonumber
   p^{213}(\omega)= & _\lambda p^{123}_i(\omega), ~\forall \omega\in \Omega_{\mathcal{E}}\setminus (\omega_1\cup A_2).
\end{align}
Remark from (\ref{Eqdefp213})  and $(A_2\cup \omega^*)\cap \Omega _{\mathcal{F}}=\emptyset$ that $\argmax\limits_{E\in \mathcal{E}_1} ~ _\lambda  p^{213}(E)= \argmax\limits_{E\in \mathcal{E}_1}   p^{123}_i(E)=E_3$. Remark also from (\ref{EqBlocksConfig1})-(\ref{EqF2Config1}) that $p^{213}_i(C^2_1)>p^{213}_i(B^2_1)$.
Since $ p^{123}=~^{\varepsilon^*}p_i$ and $ p_i^{213}(A_2)>1/2$, it comes from these remarks, (\ref{Eqdefp213}) and (\ref{Eqp123Config1}) that 
\begin{equation}\label{Eqdefp213Conf1}
    p_i^{213}(E_2)>  p^{213}_i(E_1)> p^{213}_i(E_3) >p^{213}_i(E), \forall E\in \mathcal{E}\setminus\{E_1,E_2, E_3\}.\end{equation}
Finally, starting from $p_i^{231}$ above, one can replicate the procedure described between (\ref{Eqp123Config1}) and (\ref{Eqlambda1config1}) to derive the beliefs $P_i^{231}$ and $P_i^{321}$ corresponding to the remaining two permutations of $\{E_1,E_2,E_3\}$ at the top of the collection $\mathcal{E}$. We thus conclude that $(E_1,E_2,E_3)$ is a top triple of $\mathcal{E}$.

\medskip
\underline{Configuration 2}: $( F_1\cup \ldots \cup F_{M})\setminus ( F_1^2\cup \ldots \cup F_{M_2}^2)\neq  \emptyset$.

The construction of a top triple in this configuration is similar to that described in Configuration 1 (the explicit argument is thus omitted)).

The last possible configuration is discussed in the following lines.

\medskip
\underline{Configuration 3}: $ F_1^2\cup \ldots \cup F_{M_2}^2= F_1 \cup \ldots\cup F_M$ (\textit{i.e.}, $\Omega_{\mathcal{F}^2}=\Omega_{\mathcal{F}}$).

Recall from (\ref{EqFdyad}) and (\ref{EqDecdyadF2}) that both $\mathcal{F}$ and $\mathcal{F}^2$ have maximal decompositions with only dyadic components. We first show that a top triple of $\mathcal{E}$ exists if $\mathcal{F}\neq \mathcal{F}^2$, i.e., these two decompositions do not coincide. Observe that the two decompositions coincide if, for all $\ell\in \{1,\ldots,L\}$, there exists $\ell_2\in \{1,\ldots, L_2\}$ such that $\{B_\ell,C_\ell\}=\{B_{\ell_2}, C_{\ell_2}\}$, and vice-versa.

Assuming that the maximal decompositions of $\mathcal{F}$ and $\mathcal{F}^2$ do not coincide, suppose without loss of generality  that there exist  $\bar{\omega},\bar{\omega}'\in B^2_{1}$ such that  $\bar{\omega}\in B_{1}$, \textbf{but} $\bar{\omega}'\notin B_{1}$.

Recalling in Subcase II.2 that $E\cap B_1=\emptyset$ or $E\cap C_1=\emptyset$ (for all $E\in \Omega_{\mathcal{E}}$), remark that  $\bar{\omega}'\notin C_{1}$. Given that $\bar{\omega}' \in \Omega_{\mathcal{F}^2}=\Omega_{\mathcal{F}}$ and $\{B_1,C_1,\ldots, B_{L},C_{L}\}$ is a partition of $\Omega_{\mathcal{F}}$,  there must then exist $\bar{\ell}\in \{2,\ldots,L\}$ such that $\bar{\omega}'\in B_{\bar{\ell}}\cup C_{\bar{\ell}}$. To fix ideas, assume  that $\bar{\ell}=2$  and $\bar{\omega}'\in B_{\bar{\ell}}$ (symmetric argument if one instead assumes $\bar{\ell}\neq 2$ or $\bar{\omega}'\in C_{\bar{\ell}}$).

Pick then a conditional belief $q_i\in \mathcal{P}(\Omega_{\mathcal{F}})$ such that 
\begin{align}
    \label{Eq1Config3}
      q_i(\bar{\omega})> ~ 1/2 ~~[\mbox{hence, } q_i(B_{1})>  q_i(C_{1})];
    \\ \nonumber
   0< q_i(C_{2})-  q_i(B_{2})<q_i(B_{\ell})-  q_i(C_{\ell}) \mbox{ for } \ell\in \{1,\ldots,L_2\}\setminus \{2\} 
\end{align}
Moreover, for any $\varepsilon\in (0,1/4)$, select a belief $^\varepsilon p_i\in \mathcal{P}(\Omega_{\mathcal{E}})$ such that 
\begin{align}
\label{Eq2Config3}
   ^\varepsilon p_i(A_1)=&\frac{1}{2}+\varepsilon , ~  ^\varepsilon p_i(A_2)=\frac{1}{2}-2\varepsilon ;
    \\\nonumber
     ^\varepsilon p_i(\Omega_{\mathcal{F}})=&~q_i.
\end{align}
Defining the three events $E_1:=A_1\cup C_2\cup (\cup_{\ell\neq 2}B_{\ell})$, $E_2:=A_1\cup B_2\cup (\cup_{\ell\neq 2}B_{\ell})$, and $E_3:=A_2 \cup (\argmax\limits_{F\in \mathcal{F}} q_i(F))$, remark from (\ref{Eq1Config3})-(\ref{Eq2Config3}) and (\ref{EqUnionETriple}) that,  there exists $\varepsilon^*$ (small enough) such that
\begin{equation}\label{Eqp123Config3}
     ^{\varepsilon^*} p_i(E_1)> ~  ^{\varepsilon^*}p_i(E_2)>~   ^{\varepsilon^*} p_i(E_3)>~  ^{\varepsilon^*} p_i(E), ~\forall E\in \mathcal{E}\setminus\{E_1,E_2,E_3\}.
\end{equation}

Note from $ q_i(\bar{\omega})> ~ 1/2$ [see (\ref{Eq1Config3})] and $\bar{\omega}\in B_1^2$ that we must have $B_1^2\subseteq E_3$ (recall that $\mathcal{F}^2$ has a dyadic maximal decomposition); and this means that $E_3$ will remain at the top of $\mathcal{E}^2$ if one transfers weight from $A_1$ to $A_2$ (while preserving $q_i$ as the conditional distribution on $\Omega_{\mathcal{F}}$). 
Hence, starting from $ ^{\varepsilon^*} p_i$ and following a procedure similar to those applied in the previous subcases and  configurations (i.e., defining a continuous function allowing to transfer weight from $A_1$ to $A_2$, and making use of the intermediate value theorem), one can construct all six beliefs leading to distinct permutations of $E_1,E_2,E_3$ at the top of the collection $\mathcal{E}$. We thus conclude that $(E_1,E_2,E_3)$ is a top triple of $\mathcal{E}$.

\bigskip It now remains to examine only the case where $\mathcal{F}= \mathcal{F}^2$. Using (\ref{EqUnionETriple}), remark in this case that we must have $T\geq 3$ (indeed, since $\{F_1,\ldots,F_M\}=\{F^2_1,\ldots,F^2_{M_2}\}$ has $L$ dyadic components, writing $T=2$  would mean that $\mathcal{E}$ has a maximal decomposition into $L+1$ dyadic components, thus contradicting the fact that $\mathcal{E}$ is richly decomposable).

Fix then $t\in \{3,\ldots,T\}$; and assume by induction that $\{F^{t'}_1,\ldots,F^{t'}_{M_{t'}}\}=\{F_1,\ldots,F_M\}$, for all $t'=2,\ldots,t-1$. Combining these equalities with (\ref{EqCondF1})-(\ref{EqCondF2}), remark that we have 
\begin{equation}\label{EqindFt}
    A_t \cap E=\emptyset, ~\forall E\in \mathcal{E}\setminus\mathcal{E}^t.
\end{equation}
Therefore, exploiting (\ref{EqindFt}) and applying to $\mathcal{F}^t$ the same procedure (used in the previous paragraphs for $\mathcal{F}^2$), one can prove the existence of a top triple of $\mathcal{E}$, except maybe in the case where $\{F_1,\ldots,F_M\}=\{F^2_1,\ldots,F^2_{M_2}\}=\ldots=\{F_1^{t},\ldots,F^t_{M^t}\}$. At the end of this induction argument, we can thus claim the existence of a top triple of $\mathcal{E}$ in all configurations but the one where $\{F_1,\ldots,F_M\}=\{F^2_1,\ldots,F^2_{M_2}\}=\ldots=\{F_1^{T},\ldots,F^T_{M^T}\}$.

\medskip
Finally, suppose then that  $\{F_1,\ldots,F_M\}=\{F^2_1,\ldots,F^2_{M_2}\}=\ldots=\{F_1^{T},\ldots,F^T_{M^T}\}$. 
Since $\mathcal{F}=\{F_1,\ldots,F_M\}$ admits a maximal decomposition into $L$ dyadic components, observe that $\mathcal{E}$ has a maximal decomposition with $L+1$  components: the rich component $\mathcal{A}=\{A_1,\ldots,A_T\}$ (where $3\leq T\leq K-1$) and the $L$ dyadic components of $\mathcal{F}$.\footnote{Note from (\ref{EqUnionETriple})-(\ref{EqCondF2}) that $\mathcal{A}=\{A_1,\ldots,A_T\}$ is indecomposable; and hence it is indeed a component in the maximal decomposition of $\mathcal{E}$.} From our induction hypothesis, note that there exist three events of $\mathcal{A}$, say,  $A_1,A_2,A_3$, and six conditional beliefs $q_i^{jkl}\in \mathcal{P}(\Omega_{\mathcal{A}})$ (for any distinct $j,k,l\in \{1,2,3\}$) such that 
\begin{equation}
 \label{EqConf3ATriple}  
 q_i^{jkl}(A_j)>q_i^{jkl}(A_k)>q_i^{jkl}(A_l)> q_i^{jkl}(A), ~\forall A\in \mathcal{A}\setminus \{A_1,A_2,A_3\}.
\end{equation}

Construct the events $E_1:= A_1\cap F_1$, $E_2:= A_1\cap F_2$, and $E_3:= A_1\cap F_3$; and pick a  belief $p_i^{jkl}$ (for any distinct $j,k,l\in \{1,2,3\}$) such that 
\begin{align}
    \label{Eq1AFmax}  
   p_i^{jkl}(\omega)> p_i^{jkl}&(\Omega_{\mathcal{F}}\setminus F_1), ~\forall \omega \in F_1;
   \\    \label{Eq2AFmax}  
   p_{i/\Omega_{\mathcal{A}}}^{jkl}=q_i^{jkl}&.
\end{align}
Note from (\ref{Eq1AFmax}) that $F_1=\argmax\limits_{F\in \mathcal{F}} p^{jkl}_i(F)$. Thus, combining $(\ref{EqConf3ATriple})-(\ref{Eq2AFmax})$ with the fact (obtained from (\ref{EqindFt})) that $A_t\cap \Omega_{\mathcal{F}}=\emptyset$ for $t=1,\ldots,T$, one can write
\begin{equation*}
    p_i^{jkl}(E_j)>p_i^{jkl}(E_k)>p_i^{jkl}(E_l)> p_i^{jkl}(E), ~\forall E\in \mathcal{E}\setminus \{E_1,E_2,E_3\},
\end{equation*}
for any distinct $j,k,l\in \{1,2,3\}$.
This shows that  $(E_1,E_2,E_3)$ is a top triple of $\mathcal{E}$; and the proof of Lemma \ref{LemEventChoice1} is now complete.
\endproof

\begin{lemma}\label{LemEventChoice2}
Let  $m\in \{2,3,4,\ldots,\}$ and  $\hat{\Omega}\in 2^{\tilde{\Omega}}$.   Suppose that $f$ is a strategy-proof $m$-agent event selector on $\hat{\Omega}$. Moreover, assume that the range of $f$, $\mathcal{E}=\{E_1,\ldots,E_K\}$, is indecomposable and has three or more events (\textit{i.e.,} $K\geq 3$). Then $f$ is not anonymous.
\end{lemma}
\proof
Suppose that $m$, $\hat{\Omega}$ and $f$ are as depicted in the statement of Lemma \ref{LemEventChoice2}. 
Note first that repeated application of strategy-proofness of $f$ allows to see that the range of $f$ ( which is $\mathcal{E}=\{E_1,\ldots,E_K\}$) consists of non-nested events. Since $K\geq 3$ and $\mathcal{E}$ is indecomposable, it follows that $\mathcal{E}$ is richly decomposable. One can then use Lemma \ref{LemEventChoice1}  to claim that there exist three events $E_1,E_2,E_3$ and six beliefs $p^{jkl}\in \mathcal{P}(\Omega_{\mathcal{E}})$ such that 
\begin{equation}\label{Eq1Lem14sixp}
     p_i^{jkl}(E_j)>p_i^{jkl}(E_k)>p_i^{jkm}(E_l)> p_i^{jkl}(E), ~\forall E\in \mathcal{E}\setminus \{E_1,E_2,E_3\},
\end{equation}
for any distinct $j,k,l\in \{1,2,3\}$.

Combining the strategy-proofness of $f$ with the fact that $E_1,E_2,E_3\in \mathcal{E}=Range(f)$, observe that we must have: for all $p_i\in \mathcal{P}(\Omega_{\mathcal{E}})$,
\begin{equation}\label{Eq2Lem14sixp}
    [p_i\in \{p_i^{123}, p_i^{132}, p_i^{213}, p_i^{231}, p_i^{312}, p_i^{321} \}] \Rightarrow  [f(p_i)\in \{E_1,E_2,E_3\}].
\end{equation}

Let then $\succ^{jkl}$ be the linear order (defined over $\{E_1,E_2,E_3\}$) such that, for all $E,E'\in \{E_1,E_2,E_3\}$, we have $E$ is $\succ^{jkl}$-preferred to $E'$ if and only if $p_i^{jkl}(E)>p_i^{jkl}(E')$. For instance, taking $(j,k,l)=(1,2,3)$, remark from (\ref{Eq1Lem14sixp}) that $E_1$ is $\succ^{123}$-preferred to $E_2$ and $E_2$ is $\succ^{123}$-preferred to $E_3$. Moreover, write $\mathcal{L}=\{\succ^{123}, \succ^{132}, \succ^{213}, \succ^{231}, \succ^{312}, \succ^{321}\}$ to denote the set of linear orderings of  $\{E_1,E_2,E_3\}$.

Construct the deterministic social choice function $\rho_{GS}:\mathcal{L}^m\rightarrow  \{E_1,E_2,E_3\}$ as follows: for any $L=(\succ^{j_1k_1l_1},\ldots,\succ^{j_mk_ml_m})$ (with distinct $j_i,k_i,l_i\in \{1,2,3\}$ for  $i=1,\ldots,m$),
\begin{equation}\label{Eq3Lem14}
    \rho_{GS}(L)= f(p_1^{j_1k_1l_1},\ldots,p_m^{j_mk_ml_m}).
\end{equation}

Observe from (\ref{Eq2Lem14sixp}) and (\ref{Eq3Lem14}) that $\rho_{GS}$ is well-defined, it has full range $\{E_1,E_2,E_3\}$, and it is strategy-proof [by the combination of (\ref{Eq1Lem14sixp}), (\ref{Eq3Lem14})  and the strategy-proofness of $f$]. It thus follows from the Gibbard-Satterthwaite theorem that $\rho_{GS}$ is dictatorial. Assuming without loss that the dictator is agent 1, note then  from  (\ref{Eq2Lem14sixp})-(\ref{Eq3Lem14})  that
\[f(p)=\argmax_{E\in \{E_1,E_2,E_3\}} p_1(E), ~\forall p_1\in \{p_1^{123}, p_1^{132}, p_1^{213}, p_1^{231}, p_1^{312}, p_1^{321}\};\]
and this means that $f$ is not anonymous.\endproof

\bigskip
We are now ready to prove the  result stated in Proposition \ref{PropTriadic} for the class of filtering factors (this is the final step of the proof of Proposition \ref{LemmaDec}). 

\bigskip
\noindent \textbf{Proof of Proposition 
\ref{PropTriadic}.}
\proof
 Suppose that $n\geq 3$ and $|\Tilde{\Omega}|\geq 3$. 
 Looking at   (\ref{EqTriadFilter2})-(\ref{EqTriadFilter0}), it is not difficult to check that an iso-filtering factor is strategy-proof. Conversely, we will now argue  that every strategy-proof filtering factor must   satisfy (\ref{EqTriadFilter2})-(\ref{EqTriadFilter0}).  
 Consider then a strategy-proof filtering factor  $\varphi$, defined by  the filter $\mathcal{C}^k=\{(E_1^k,F_1^k), \ldots,(E_{M_k}^k,F_{M_k}^k), (G^k_a, G^k_b)\}$ (with $k= \underline{k},\ldots, \Bar{k}$) and the voting quotas   $\hat{t}^k_m\in \{1,\ldots,k+1\}$ and $ \tilde{t}^k_m\in   \{1,\ldots, n-k\}$ for any $k=\underline{k},\ldots,\Bar{k}$ and $m=1,\ldots,M_k$. 
 
 Noting from (\ref{EqTriad}) that $\varphi$ is anonymous and range-unanimous, we conclude that this BSCF satisfies our three axioms; and hence Lemma \ref{DyadSolidarity} applies to $\varphi$. Indeed, observe from  (\ref{EqTriad})  that  the maximal decomposition of every sectional range $\mathcal{E}^k=\{E\in \tilde{\Omega}: \varphi^a(v,p)=E, \mbox{ for some } (v,p)\in \mathcal{D}^N \mbox{ s.t. } n_a^v=k\}$ has no rich component, thus  
 satisfying the assumptions of Lemma \ref{DyadSolidarity}. 

Fix now $k\in\{\underline{k},\ldots,\Bar{k}\}$, $m\in\{1,\ldots,M_k\}$, and $m'\in\{1,\ldots,M_{k+1}\}$. We must prove (\ref{EqTriadFilter2}), that is, $[ E^k_m\cap F^{k+1}_{m'}\neq \emptyset]\Rightarrow [\hat{t}^{k+1}_{m'}=\Tilde{t}^k_m=1]$.
Suppose first  that $\emptyset\neq E^k_m\cap F^{k+1}_{m'}\neq E^k_m$. Letting then $\omega\in  E^k_m\cap F^{k+1}_{m'}$ and $\omega'\in E^k_m\setminus F^{k+1}_{m'}$, remark first from Lemma \ref{DyadSolidarity}  that  $\hat{G}^{k+1}_{\omega}\cap A_\omega^k=\emptyset$. Indeed, writing $\hat{G}^{k+1}_{\omega} \cap \hat{G}^k_\omega\neq \emptyset$ would lead to  the contradiction $|C^k_\omega|=1$ [by Lemma \ref{DyadSolidarity}-(V), applied to any $\Bar{\omega}\in \hat{G}^{k+1}_{\omega}\cap \hat{G}^k_\omega$, since  $\omega\in F^{k+1}_{m'}=\hat{G}^{k+1}_{\Bar{\omega}}$]; and writing $(\hat{F}^{k+1}_{\omega}\setminus \omega) \cap \hat{G}^k_\omega \neq \emptyset$ would contradict the homologue of Lemma \ref{DyadSolidarity}-(III) where the roles of $a$ and $b$ are swapped. Using  $\hat{G}^{k+1}_{\omega}\cap A_\omega^k=\emptyset$, remark now from Lemma \ref{DyadSolidarity}-(IV) that $H^{k+1}_{\omega}(1,0)=1$. Combining this last equality with (\ref{EqTriad}) then allows to write $\tilde{t}^{k+1}_{m'}=1$ (since $\omega\in F^{k+1}_{m'}$).  
Next, suppose instead that $\emptyset\neq E^k_m\cap F^{k+1}_{m'}= E^k_m$. Then  note from Lemma \ref{DyadSolidarity}-(V) that $\hat{G}^{k+1}_\omega\cap A^{k+1}_\omega=\emptyset$, for any $\omega\in F^{k+1}_{m'}$. It thus comes from Lemma \ref{DyadSolidarity}-(IV) that $H^{k+1}_{\omega}(1,0)=1$; and combining  with (\ref{EqTriad}) then gives $\tilde{t}^{k+1}_{m'}=1$.

Repeating the analysis of the previous paragraph while reversing the roles of $a$ and $b$ in the definition of $\varphi$ and in all previous results, one obtains in a similar way $\hat{t}^k_m=1$.

\medskip
It remains to prove (\ref{EqTriadFilter0}), \textit{i.e.}, $[(E^{k+1}_{m'}, F^{k+1}_{m'})\in \mathcal{C}^k]\Rightarrow [\hat{t}^{k}_{m'}=\tilde{t}^{k}_{m}= 1=\hat{t}^{k+1}_{m'}=\tilde{t}^{k+1}_{m'}]$. Assume  then that $ (E^{k+1}_{m'}, F^{k+1}_{m'})\in \mathcal{C}^k$. Then recalling conditions (\ref{EqFilterdef1})-(\ref{EqFilterdef2}) in the definition of a filter, and the fact that the iso-filtering factor $\varphi$ is prime (and not a dyadic factor), note that exactly one of the following two statements must hold:
\begin{abet}
    \item[(A)]    $\emptyset\neq \overbrace{E^{k'}_{m_0}}^{=E^{k+1}_{m'}}\cap F^{k'+1}_{m_1}\neq E^{k'}_{m_0}$, for some $ k'\in \{k+1, \ldots,\Bar{k}\}$, $m_0\in \{1,\ldots,M_{k'}\}$, $m_1\in \{1,\ldots,M_{k'+1}\}$; and $(E^{k+1}_{m'}, F^{k+1}_{m'})\in \mathcal{C}^{k''}$  for all $k''=\underline{k},\ldots,k'$.
    \item[(B)] $\emptyset\neq E^{k'}_{m_0}\cap \overbrace{F^{k'+1}_{m_1}}^{=F^{k}_{m}} \neq E^{k'}_{m_0}$, for some $ k'\in \{\underline{k}, \ldots,k-1\}$, $m_0\in \{1,\ldots,M_{k'}\}$, $m_1\in \{1,\ldots,M_{k'+1}\}$; and $(E^{k+1}_{m'}, F^{k+1}_{m'})\in \mathcal{C}^{k''}$  for all $k''=k'+1,\ldots,k+1$.
\end{abet}
Assume without loss that (A) holds (the argument is similar in the other case). Then combining $\emptyset\neq \overbrace{E^{k'}_{m_0}}^{=E^{k+1}_{m'}}\cap F^{k'+1}_{m_1}\neq E^{k'}_{m_0}$ and  (\ref{EqTriadFilter2}) [which we have already established] yields $\hat{t}^{k'}_{m_0}=\tilde{t}^{k'+1}_{m_1}=1$. Moreover, since $(E^{k+1}_{m'}, F^{k+1}_{m'})\in \mathcal{C}^{k''}$  for all $k''=\underline{k},\ldots,k'$, combining $\hat{t}^{k'}_{m_0}=\tilde{t}^{k'+1}_{m_1}=1$ with the repeated application of Lemma \ref{DyadSolidarity}-(I) and Lemma \ref{DyadSolidarity}-(II) finally gives $\hat{t}^{k'}_{m_0}=\tilde{t}^{k'+1}_{m_1}=1=\hat{t}^{k}_{m'}=\tilde{t}^{k}_{m}=\hat{t}^{k+1}_{m'}=\tilde{t}^{k+1}_{m'}$, which is the desired result. \endproof

\bigskip
To proceed with the proof of Proposition \ref{LemmaDec}, we discuss the three cases below (which are exhaustive).

\bigskip\noindent
\textbf{Case I:} Suppose that there exists \emph{a simple state} $\omega_1\in \tilde{\Omega}$.

\smallskip\noindent Fix any belief profile $\bar{p}\in \mathcal{\tilde{P}}^N$ that is  $\{\omega_1\}$-dominant. By range-unanimity of $\varphi$, we have
\begin{align}\label{Eq0CaseI}
  \varphi(v, \bar{p}; \omega_1)= b, & ~ \forall (v,p)\in \mathcal{\tilde{D}}^N \mbox{ s.t. }  v\in  \mathcal{V}_ 0^N;\\ \nonumber
 \varphi(v, \bar{p}; \omega_1)= a, & ~\forall (v,p)\in \mathcal{\tilde{D}}^N \mbox{ s.t. }  v\in  \mathcal{V}_ n^N.
 \end{align}
 Recalling $\bar{u}^k\in \mathcal{V}^N_k$ [which was introduced in (\ref{EqNaproof})], note  from  (\ref{Eq0CaseI}) that we may define
 \begin{equation}\label{Eq00CaseI}
   \bar{k}:=\min \{k\in \{1,\ldots,n\}: \varphi(\bar{u}^k, \bar{p}; \omega_1)= a\}.
 \end{equation}
 Since $\varphi$ is binary, it thus comes from (\ref{Eq0CaseI})-(\ref{Eq00CaseI}) that
\begin{equation*}
\varphi(\bar{u}^k, \bar{p}; \omega_1)= b \mbox{ if } 0\leq k\leq \bar{k}-1.
\end{equation*}
We show next that
\begin{equation}\label{Eq11CaseI}
\varphi(\bar{u}^k, \bar{p}; \omega_1)= a \mbox{ if } \bar{k}\leq k\leq n.
\end{equation}
Remark  that, if $\bar{k}\geq n-1$, then (\ref{Eq11CaseI}) easily comes from the combination of (\ref{Eq0CaseI}) and (\ref{Eq00CaseI}). Suppose now that $1\leq \bar{k}\leq n-2 $; and notice that we must have $\varphi(\bar{u}^{\bar{k}+1}, \bar{p}; \omega_1)= a$. Indeed, writing $\varphi(\bar{u}^{\bar{k}+1}, \bar{p}; \omega_1)= b$  leads to
\begin{equation*}\bar{p}_{\bar{k}+1}(\omega_1) \underbrace{\bar{u}^{\bar{k}+1}_{\bar{k}+1}(\varphi(\bar{u}^{\bar{k}+1}, \bar{p}; \omega_1))}_{=0}+\sum\limits_{\omega\neq \omega_1}\bar{p}_{\bar{k}+1}(\omega) \underbrace{\bar{u}^{\bar{k}+1}_{\bar{k}+1}(\varphi(\bar{u}^{\bar{k}+1}, \bar{p}; \omega))}_{\in \{0,1\}}\leq \bar{p}_{\bar{k}+1}(\omega_1) \underbrace{\bar{u}^{\bar{k}+1}_{\bar{k}+1}(\varphi(\bar{u}^{\bar{k}}, \bar{p}; \omega_1))}_{=1},
 \end{equation*}
 where the inequality holds because $\sum\limits_{\omega\neq \omega_1}\bar{p}_{\bar{k}+1}(\omega)\leq \bar{p}_{\bar{k}+1}(\omega_1) $ (since $\bar{p}$ is $\{\omega_1\}$-dominant).
Hence, writing $\varphi(\bar{u}^{\bar{k}+1}, \bar{p}; \omega_1)= b$ implies
\begin{equation}\label{Eq2CaseI}
\sum\limits_{\omega\in\tilde{\Omega}}\bar{p}_{\bar{k}+1}(\omega) \bar{u}^{\bar{k}+1}_{\bar{k}+1}(\varphi(\bar{u}^{\bar{k}+1}, \bar{p}; \omega_1))\leq
\sum\limits_{\omega\in\tilde{\Omega}}\bar{p}_{\bar{k}+1}(\omega) \bar{u}^{\bar{k}+1}_{\bar{k}+1}(\varphi(\bar{u}^{\bar{k}}, \bar{p}; \omega)),
\end{equation}
which contradicts the strategy-proofness of $\varphi$.
We have thus shown that $\varphi(\bar{u}^{\bar{k}+1}, \bar{p}; \omega_1)= \varphi(\bar{u}^{\bar{k}}, \bar{p}; \omega_1)= a$. Repeating the same argument as needed, one can show that $\varphi(\bar{u}^{\bar{k}}, \bar{p}; \omega_1)=\varphi(\bar{u}^{\bar{k}+1}, \bar{p}; \omega_1)=\ldots= \varphi(\bar{u}^{n-1}, \bar{p}; \omega_1)=\varphi(\bar{u}^{n}, \omega_1)= a$.
We can therefore write
\begin{align}\label{Eq22CaseI}
 \varphi(\bar{u}^k, \bar{p}; \omega_1)= b &~ \mbox{ if } 0\leq k\leq \bar{k}-1\\ \nonumber
 \varphi(\bar{u}^k, \bar{p}; \omega_1)= a & ~\mbox{ if } \bar{k}\leq k\leq n.
\end{align}

Recalling that $\omega_1$ is a simple state, note from  (\ref{Eq22CaseI}) that: for all $p\in \mathcal{\tilde{P}}^N$
\begin{align}\label{Eq23CaseI}
 \varphi(\bar{u}^k, p; \omega_1)= b, &~ \forall k\in \{0, \ldots, \bar{k}-1\} \mbox{ s.t. } (\bar{u}^k, p)\in \mathcal{D}^N\\ \nonumber
 \varphi(\bar{u}^k, p; \omega_1)= a, &~  \forall k\in \{k+1, \ldots, n\} \mbox{ s.t. } (\bar{u}^k, p)\in \mathcal{D}^N.
\end{align}
Finally applying Lemma \ref{LemmaOrdinal} to (\ref{Eq23CaseI}), one may write
\begin{align}\label{Eq3CaseI}
 \varphi(v, p; \omega_1)= b, &~ \forall (v,p)\in \mathcal{\tilde{D}}^N  \mbox{ s.t. } n^v_a\in \{0, \ldots, \bar{k}-1\} \\\ \nonumber
 \varphi(v, p; \omega_1)= a, &~  \forall (v,p)\in \mathcal{\tilde{D}}^N \mbox{ s.t. } n^v_a\in \{\bar{k}+1, \ldots, n\}
\end{align}

 Let $\varphi_1: (\mathcal{V}\times \mathcal{P}(\{\omega_1\})^N \rightarrow \{a,b\}^{\{\omega_1\}}$ be the $\{\omega_1\}$-SCF defined as follows: for all $(v,q)\in (\mathcal{V}\times \mathcal{P}(\{\omega_1\})^N$,
\begin{align}\label{Eq32CaseI}
 \varphi_{1}(v,q; \omega_1)= b &~ \mbox{ if } 0\leq k\leq \bar{k}-1\\ \nonumber
 \varphi_1(v, q; \omega_1)= a & ~\mbox{ if } \bar{k}\leq k\leq n.
\end{align}
Then it comes from (\ref{Eq3CaseI})-(\ref{Eq32CaseI}) that, for all $(v,p)\in \mathcal{\tilde{D}}^N$
\begin{equation}\label{Eq4CaseI}
  \varphi(v,p) = a \left(\varphi^a_{1}(v,p_{/\{\omega_1\}})  \cup \varphi^a_{\overline{\omega_1}}(v,p) \right) \oplus b \left(\varphi^b_{1}(v,p) \cup \varphi^b_{\overline{\omega_1}}(v,p) \right).
\end{equation}
Recalling (\ref{Eq3CaseI}), it comes from Lemma \ref{LemmaSepSimpleDict}-I that $\varphi_{\overline{\omega_1}}(v,p)$ depends only on $(v,p_{/\overline{\omega_1}})$, \emph{i.e.},  there exists a mapping $\hat{\varphi}: (\mathcal{D}({\tilde{\Omega}\setminus \{\omega_1\}}))^N \rightarrow \{a,b\}^{{\tilde{\Omega}\setminus \{\omega_1\}}}$ such that
\begin{equation}\label{EqdecompI}
  \varphi(v,p) = a \left(\varphi^a_{1}(v,p_{/\{\omega_1\}})  \cup \hat{\varphi}^a(v,p_{/\overline{\omega_1}}) \right) \oplus b \left(\varphi^b_{1}(v,p_{/\{\omega_1\}}) \cup \hat{\varphi}^b(v,p_{/\overline{\omega_1}}) \right), \forall (v,p)\in \mathcal{\tilde{D}}^N.
\end{equation}
Moreover, using (\ref{EqdecompI}) and the fact that $\varphi_1$ is trivial [which stems from (\ref{Eq32CaseI})], it is easy to see   that the binary ($\tilde{\Omega}\setminus \{\omega_1\}$)-SCF $\hat{\varphi}$ must be strategyproof, range-unanimous and anonymous (just as $\varphi$). Since $|\tilde{\Omega}\setminus \{\omega_1\}|=S-1<S$, it comes from our induction hypothesis that there exist (i) a partition $\{\Omega_2, \ldots, \Omega_T\}$  of $\tilde{\Omega}\setminus \{\omega_1\}$ (with $T\geq 2$)  and (ii) sub-SCFs $\varphi_t: (\mathcal{D}(\Omega_t))^N\rightarrow \{a,b\}^{\Omega_t}$ (for $t=2,\ldots, T$) such that:
\begin{equation}\label{EqsubdecompI}\hat{\varphi}(v,q)= a\left(\bigcup_{t=2}^T \varphi_t^a(v,q_{/\Omega^t}) \right) \oplus b\left(\bigcup_{t=2}^T \varphi_t^b(v,q_{/\Omega^t}) \right),  ~\forall (v,q)\in (\mathcal{D}(\overline{\omega_1}))^N , \end{equation}
with the property that each $\varphi_t$ (for $t=2, \ldots, T$) is  either simple, quasi-dictatorial,  dyadic, or iso-filtering.
Combining (\ref{Eq32CaseI}), (\ref{EqdecompI}) and (\ref{EqsubdecompI}) then yields the desired result.

\bigskip
\bigskip\noindent
\textbf{Case II:}  Suppose that the maximal decomposition of $\mathcal{E}^k$ has at least \emph{one rich component}, for some $k\in \{1,\ldots n-1\}$.

\medskip\noindent Precisely, assume   that  the maximal decomposition of $\mathcal{E}^k$ has a component $\{E_1,E_2,\ldots, E_L\}$ such that $L\geq 3$; and define $\Omega_1=\bigcup_{l=1,\ldots,L} E_{\ell}$. Note from Lemma \ref{LemmaClut} that the events $E_1,E_2,\ldots, E_L$ must be non-nested. Recalling  Lemma \ref{LemmaSections}, define the mapping $s=(s_1,s_2): \mathcal{\tilde{P}}\times \mathcal{\tilde{P}} \rightarrow 2^{\tilde{\Omega}}\times 2^{\tilde{\Omega}}$ as follows: for all $(p_1,p_2)\in \mathcal{\tilde{P}}\times \mathcal{\tilde{P}}$,
\begin{equation}\label{Eqassign}
  s_1(p_1,p_2)= \alpha^k((\underbrace{p_1,\ldots,p_1}_{k~ times}), (\underbrace{p_2,\ldots,p_2}_{n-k~ times})); ~~s_2(p_1,p_2)=  \overline{ s_1(p_1,p_2)}.
\end{equation}
Remark from (\ref{Eqassign}) that $s(p_1,p_2)$ is a \emph{bona fide}  bilateral assignment rule in the sense of \cite{bahel2020strategyproof}. Moreover, recalling Lemma \ref{LemmaSemiSP}, it is easy to see that $s(p_1,p_2)$ is strategyproof (in the sense that none of the groups $k=1,2$ can increase the probability of its share $s_k(p_1,p_2)$ by lying about its belief $p_k$).

 Given that $\{E\in 2^{\tilde{\Omega}}: s_1(p_1,p_2)=E, \mbox{ for some }  (p_1,p_2)\in \mathcal{\tilde{P}}\times \mathcal{\tilde{P}}\}=\mathcal{E}^k$ by Lemma \ref{LemmaEUnanimous}, we conclude that $\{E_1,\ldots, E_L\}$ is  a rich component in the decomposition of the range of $s_1$. It thus follows from the characterization of strategyproof binary assignment rules of Bahel and Sprumont (2020) that exactly one belief (either $p_1$ or $p_2$) must dictate the sub-event chosen in $\Omega_1$.  In the case where $p_1$ dictates, we  thus write
 \begin{equation}\label{Eq1CaseII}
 \alpha^k_{ \Omega_1}((\underbrace{p_1,\ldots,p_1}_{k~ times}), (\underbrace{p_2,\ldots,p_2}_{n-k~ times}))= \argmax_{\ell=1,\ldots,L} p_1(E_{\ell}), ~\forall (p_1,p_2)\in \mathcal{\tilde{P}}\times \mathcal{\tilde{P}}.
 \end{equation}
 If instead $p_2$ dictates, it comes that
 \begin{equation}\label{Eq2CaseII}
  \alpha^k_{ \Omega_1}((\underbrace{p_1,\ldots,p_1}_{k~ times}), (\underbrace{p_2,\ldots,p_2}_{n-k~ times}))= \argmin_{\ell=1,\ldots,L} p_2(E_{\ell}), ~\forall (p_1,p_2)\in \mathcal{\tilde{P}}\times \mathcal{\tilde{P}}.
 \end{equation}
Next, we prove the following claims.

\bigskip\noindent
\emph{Claim 1}: Suppose that $p_1$ dictates in (\ref{Eqassign}), so that (\ref{Eq1CaseII}) holds. Then we have
\begin{equation}\label{Eqclaim1}
\alpha^k_{ \Omega_1}((\underbrace{p_1,\ldots,p_1}_{k~ times}), (p_{k+1},\ldots,p_n))= \argmax\limits_{\ell=1,\ldots,L} p_1(E_{\ell}), ~\forall p_1,
 p_{k+1},\ldots,p_n\in \mathcal{\tilde{P}}.
\end{equation}

\medskip\noindent Fix  $p_1, p_{k+1},\ldots,p_n\in \mathcal{\tilde{P}}$; and assume without loss of generality that $\argmax\limits_{\ell=1,\ldots,L} p_1(E_{\ell})=E_1$.
Choose a belief $p'_{2}$ that is $ E_1$-dominant; and note from (\ref{Eq1CaseII}) that
\begin{equation}\label{Eq3CaseII}
  \alpha^k_{ \Omega_1}((\underbrace{p_1,\ldots,p_1}_{k~ times}), (\underbrace{p'_{2},\ldots,p'_{2}}_{n-k~ times}))= \argmax_{\ell=1,\ldots,L} p_1(E_{\ell})=E_1.
\end{equation}
Since $p'_2$ is $ E_1$-dominant, observe that, for any $\ell=2,\ldots,L$, we have $p'_2(E_{\ell})<p'_2(E_1)$ (since $E_1$ and $E_{\ell}$ are non-nested), and even $p'_2(E)<p'_2(F)$ for all $E,F\in \mathcal{E}^k$ such that $E\cap \Omega_1=E_{\ell}$ and $F\cap \Omega_1=E_1$. Combining this last observation with (\ref{Eq3CaseII}) and Lemma \ref{LemmaSemiSP}-II, we then conclude that $ \alpha^k_{ \Omega_1}((\underbrace{p_1,\ldots,p_1}_{k~ times}), (p_{k+1},p'_{2},\ldots,p'_{2}))=E_1$.

Repeating this argument $n-k$ times, we get the desired result, that is to say,  
\begin{align*}
E_1=\alpha^k_{ \Omega_1}((\underbrace{p_1,\ldots,p_1}_{k~ times}), (p_{k+1},p'_{2},\ldots,p'_{2}))=&\alpha^k_{ \Omega_1}((\underbrace{p_1,\ldots,p_1}_{k~ times}), (p_{k+1},p_{k+2},\ldots,p'_{2})) \\
\vdots\\ 
=&\ldots= \alpha^k_{ \Omega_1}((\underbrace{p_1,\ldots,p_1}_{k~ times}), (p_{k+1},p_{k+2},\ldots,p_{n})).
\end{align*}

Of course, the analogue of Claim 1 obtains in a similar way if one assumes that $p_2$ dictates in (\ref{Eqassign}), so that (\ref{Eq2CaseII}) holds.

\bigskip\noindent
\emph{Claim 2}:  We have either ($k=1$ and $\alpha^k_{ \Omega_1}$ is $1$-dictatorial) or ($k=n-1$ and $\alpha^k_{ \Omega_1}$ is $n$-dictatorial).

\smallskip\noindent  Without loss of generality, assume that $p_1$ dictates in (\ref{Eqassign}), so that Claim 1 is valid.
We will show that $k=1$ must hold.
 Fix $\bar{q}=(\bar{q}_{k+1}, \ldots, \bar{q}_{k+1})\in \mathcal{\tilde{P}}^{\{k+1,\ldots,n\}}$; and  define the mapping $\gamma: (\mathcal{P}(\Omega_1))^{\{1,\ldots,k\}}\rightarrow 2^{\Omega_1}$ as follows: for all $(p_1,\ldots, p_k)\in \tilde{P}^{\{1,\ldots, k\}}$,
\begin{equation}\label{Eq4CaseII}
  \gamma(p_1,\ldots,p_k)=  \alpha^k_{\Omega_1}((p_1,\ldots, p_k), \bar{q}).
\end{equation}


First, note from Lemma \ref{LemmaDomalloc}-(I) that $\gamma(p)=\gamma(p')$, for all $p,p'\in  \tilde{P}^{\{1,\ldots, k\}}$ such that [$p_{/\Omega_1}=p'_{/\Omega_1}$ and  $p,p'$ are $\Omega_1$-dominant].
 Thus, only the conditional beliefs $p_{/\Omega_1}$ matter for the restriction of the function $\gamma(p)$ to the set of $\Omega_1$-dominant belief profiles, \emph{i.e.}, there exists $\gamma': (\mathcal{P}(\Omega_1))^{\{1,\ldots,k\}}\rightarrow 2^{\Omega_1}$ such that
 \begin{equation}\label{Eq5CaseII}
   \gamma(p)=\gamma'(p_{/\Omega_1}), \mbox{ for any } \Omega_1\mbox{-dominant } p\in\tilde{P}^{\{1,\ldots, k\}}.
 \end{equation}

Next, note that $\{E\in 2^{\Omega_1}:  \exists p\in  \mbox{ s.t. } \gamma'(p_1,\ldots, p_k)=E \}=\{E_1,\ldots, E_L\}$, \emph{i.e.}, the range of $\gamma'$ is precisely $\{E_1,\ldots,E_L\}$. To see why, fix any $E_{\ell}\in \{E_1,\ldots, E_L\}$ and let    $p=(p_1,\ldots,p_1)\in \mathcal{\tilde{P}}^{\{1,\ldots,k\}}$  be an $\Omega_1$-dominant belief satisfying $\argmax\limits_{\ell'=1,\ldots,L}p_1(E_{\ell'})=E_{\ell}$, for all $i=1,\ldots, k$. Then it comes from Claim 1 and (\ref{Eq4CaseII}) that $\gamma(p)=\argmax\limits_{\ell' =1,\ldots,L}p_1(E_{\ell'})=E_{\ell}$.
Hence, recalling (\ref{Eq5CaseII}), one may write $\gamma(p)=\gamma'(p_{/\Omega_1})=E_{\ell}$, which is the desired result.

It remains to see that we necessarily have $q_i(\gamma'(q))=\max\limits_{q'_i\in\mathcal{P}(\Omega_1)}q_i(\gamma'(q'_i,q_{-i}))$, for all $i\in \{1,\ldots,k\}$ and $q\in (\mathcal{P}(\Omega_1))^{\{1,\ldots,k\}}$. Indeed, assuming that there exists $i\in \{1,\ldots,k\}$, $q\in \mathcal{\tilde{P}}^{\{1,\ldots,k\}}$, $q'_i\in\mathcal{P}(\Omega_1)$
such that $q_i(\gamma'(q))<q_i(\gamma'(q'_i,q_{-i}))$ would lead to a violation of Lemma \ref{LemmaSemiSP}-I at any $\Omega_1$-dominant belief profile $p$ such that $p_{/\Omega_1}=q$ ---agent $i$ would have $p_i(\gamma(p))<p_i(\gamma(p'_i,p_{-i}))$ for some $\Omega_1$-dominant  $p_i\in \mathcal{\tilde{P}}$ such that $p_i{/\Omega_1}=q_i$.

We have thus shown that $\gamma'$ is a strategy-proof $k$-agent event selector, with an indecomposable range $\{E_1,\ldots, E_L\}$ (where $L\geq 3$). 
Assuming by contradiction that $k\geq 2$,  it hence follows from Lemma \ref{LemEventChoice2} that $\gamma'$ is \textbf{not} anonymous.
But this is violation of Lemma \ref{LemmaSemiAnon}-I: swapping the beliefs of any two agents in $\{1,\ldots,k\}$) should never change the event selected by $\alpha^k$; and hence [recalling  (\ref{Eq4CaseII})-(\ref{Eq5CaseII})] it should never change the event chosen by the $k$-agent event selector $\gamma'$.

Thus, we necessarily have $k=1$; and using this fact in Claim 1 yields
\[
\alpha^k_{\Omega_1}(p_1, (q_{2},\ldots,q_n))= \argmax\limits_{\ell=1,\ldots,L} p_1(E_{\ell}), ~\forall (p_1,q)\in \mathcal{\tilde{P}}\times \mathcal{\tilde{P}}^{n-1}.\]
 In other words, the mapping $\alpha^k_{\Omega_1}$ is 1-dictatorial.

Of course, had we instead assumed from the start that $p_2$ dictates in (\ref{Eqassign}), we would have obtained in a similar way that $k=n-1$ and
the mapping $ \alpha^k_{\Omega_1}$ is $n$-dictatorial.
This concludes the proof of Claim 2.

\bigskip\noindent
\emph{Claim 3}: Suppose that  $k=1$ (\emph{i.e.}, $\alpha^k_{ \Omega_1}$ is 1-dictatorial). Then we have $\alpha^{k'}_{ \Omega_1}(p,q)=\Omega_1$, for all $k'=2,\ldots,n$ and $(p,q)\in \mathcal{\tilde{P}}^{k'}\times \mathcal{\tilde{P}}^{n-k'}$.

\smallskip\noindent
Recalling Lemma \ref{LemmaInvariancePres}, it suffices to prove Claim 3 in the case where $k'=2$. Note that the desired result trivially holds by range-unanimity if $n=k'=2$.  Assume then  $n\geq3> k'=2$.
Combining Lemma \ref{LemmaDomalloc}-(I) and Lemma \ref{LemmaDomalloc}-(II), we get $\alpha^2_{\Omega_1}(p,q)=\alpha^2_{\Omega_1}(p',q')$, for all $\Omega_1$-dominant $(p,q),(p',q')\in \mathcal{\tilde{P}}^2\times \mathcal{\tilde{P}}^{n-2}$ s.t. $p_{/\Omega_1}=p'_{/\Omega_1}$ and $q_{/\Omega_1}=q'_{/\Omega_1}$. In other words, there exists $\gamma'': (\mathcal{P}(\Omega_1))^2\times (\mathcal{P}(\Omega_1))^{n-2} \rightarrow 2^{\Omega_1}$ s.t.
\begin{equation}\label{Eqphi2}
  \alpha^2_{\Omega_1}(p,q)=\gamma''(p_{/\Omega_1},q_{/\Omega_1}), \mbox{ for any } \Omega_1\mbox{-dominant } (p,q)\in \mathcal{\tilde{P}}^2\times \mathcal{\tilde{P}}^{n-2}.
\end{equation}

Moreover, recalling Lemma \ref{LemmaSemiSP}, one can check that, for all $i\in \{1,\ldots,k\}$, $j\in \{k+1,\ldots,n\}$ and  $(p,q)\in \mathcal{\tilde{P}}^2\times \mathcal{\tilde{P}}^{n-2}$, we have $p_i(\gamma''(p,q))=\max\limits_{p'_i\in \mathcal{P}(\Omega_1)}p_i(\gamma''((p'_i,p_{-i}),q))$ and $q_i(\gamma''(p,q))=\min\limits_{q'_j\in \mathcal{P}(\Omega_1)}q_j(\gamma''(p,(q_j,q_{-j}))$ ---similar to what was done for $\gamma'$ in the proof of Claim 2.
Then letting
\[\mathcal{R}_2^{\Omega_1}:=\{E\in 2^{\Omega_1}\vert E=\gamma''(p,q) \mbox{ for some } (p,q) \in (\mathcal{P}(\Omega_1))^2\times (\mathcal{P}(\Omega_1))^{n-2} \} \] be the range of $\gamma''$, note from Lemma \ref{LemmaClut} that $\mathcal{R}_2^{\Omega_1}$ is a collection of non-nested events. Observe that the maximal decomposition of $\mathcal{R}_2^{\Omega_1}$ cannot have a rich component. Indeed, recall first from the proof of Claim 2 that a rich component requires either $k'=1$ or $k'=n-1$. Since we have $k'=2$ in this case, the only possibility for a rich component $\{F_1,\ldots,F_M\}$ (with $M\geq 3$ and $F_1,\ldots,F_M\subseteq \Omega_1$) to exist is hence $k'=2=n-1$ (\textit{i.e.}, $n=3$). Therefore, mirroring  the proof of Claim 2,  we know that agent 3 (the only supporter of $b$) would dictate the choice on $\{F_1,\ldots,F_M\}$. Writing $\Omega_2=F_1\cup\ldots\cup F_M$, we would thus have  
\begin{equation}\label{Eqdictat}
    \alpha^2_{\Omega_2}((p_1,p_2), q_{3}))= \argmin\limits_{m=1,\ldots,M} q_3(F_{m}), ~\forall ((p_1,p_2), q_{3})\in \mathcal{\tilde{P}}^2\times \mathcal{\tilde{P}}.
\end{equation}

Second, observe that agent 2 would manipulate $\varphi$. Fix $\omega\in \Omega_2\subseteq \Omega_1$. Given that $\{F_1,\ldots,F_M\}$ is a collection of non-nested events, note that \textbf{(i)} there exists an event  $F\in \{F_1,\ldots,F_M\}$ such that $\omega\notin F$.
Next, pick  $(v,\pi)\in \mathcal{\tilde{D}}^3$ such that \textbf{(ii)} $N_a^v=\{1\}$ (and hence $N_b^v=\{2,3\}$); \textbf{(iii)} $\pi_1(\omega)=\pi_2(\omega)=\pi_3(\omega)>1/2$.

Combining (ii)-(iii) with 
$\alpha^1_{\Omega_1}(\pi_1, (\pi_2,\pi_3))= \argmax\limits_{\ell=1,\ldots,L} p_1(E_{\ell})$ (which stems from Claim 1 and $n=3$), one can see that $\omega\in \alpha^1_{\Omega_2}(\pi_1, (\pi_2,\pi_3))\subseteq \alpha^1_{\Omega_1}(\pi_1, (\pi_2,\pi_3))$. Recalling Lemma \ref{LemmaSections} thus gives $\varphi(v,\pi; \omega)=a$; and it follows from (iii) that $\pi_2(\varphi^a(v,\pi))>1/2$. But this selection of $a$ in state $\omega$ is not good for agent 2 (who would rather have her top $b$ selected in $\omega$). In fact, note that agent 2 can profitably deviate from $\varphi(v,\pi)$ by misreporting $(v'_2,\pi'_2)$ such that \textbf{(iv)} $v'_2(a)>v'_2(b)$ and \textbf{(v)} $\pi'_2=\pi_2$. Indeed, note from (ii) and (iv) that $N_a^{(v_1,v'_2,v_3)}=\{1,2\}$; and hence, combining  (\ref{Eqdictat}) and Lemma \ref{LemmaSections}, we get $\varphi_{\Omega_3}^a((v_1,v'_2,v_3),(\pi_1,\pi'_2,\pi_3))=\argmin\limits_{m=1,\ldots,M} \pi_3(F_{m})$. Finally, combining the last equality with (i) and (iii), we get $\omega\notin \varphi_{\Omega_3}^a((v_1,v'_2,v_3),(\pi_1,\pi'_2,\pi_3))$ and hence $\pi_2(\varphi^a((v_1,v'_2,v_3),(\pi_1,\pi'_2,\pi_3)))<1/2<\pi_2(\varphi^a(v,\pi))$, in contradiction to the strategy-proofness of $\varphi$ ---since agent 2 prefers $b$ to $a$ at the preference profile $(v,\pi)$.

\medskip
 We have thus shown that the maximal decomposition of $\mathcal{R}_2^{\Omega_1}$ may have only trivial or dyadic components (no rich component is allowed).  We will  rule out as well the possibility of a dyadic component. Suppose by contradiction that the maximal decomposition of $\mathcal{R}_2^{\Omega_1}$ (the range of $\gamma''$) has a dyadic component $\{F,F'\}$ (with $F\cap F'=\emptyset\neq F,F'$). Thus, letting  $\Omega_3= F \cup F'$, it comes from (\ref{Eqphi2}) that
\begin{equation}\label{Eqdyad}
  \alpha^2_{\Omega_3}(p,q)\in \{F,F'\}, \mbox{ for any } \Omega_1\mbox{-dominant } (p,q)\in \mathcal{\tilde{P}}^2\times \mathcal{\tilde{P}}^{n-2}.
\end{equation}

Next, define $E'_1=E_1\cap \Omega_3,\ldots, E'_L=E_L\cap \Omega_3$; and remark that we may have $E_{\ell}=E_{\ell'}$ for some distinct $\ell, \ell'\in \{1,\ldots,L\}$. Letting  then $L^*$ be the number of distinct events from the list $E'_1,\ldots, E'_L$ (with $1\leq L^*\leq L$), we may assume without loss that $E'_1, \ldots, E'_{L^*}$ are distinct. It thus follows that $\{E'_1,\ldots, E'_{L^*}\}$ is the range of $\alpha^1_{\Omega_3}$  (recall that $\Omega_3\subseteq \Omega_1$).

Let $\omega\in F$, $\omega'\in F'$; and consider an $\Omega_1$-dominant profile $(p,q)\in \mathcal{\tilde{P}}^2\times \mathcal{\tilde{P}}^{n-2}$ satisfying \textbf{(i) }   $p_1(\omega)>1/2$ and  \textbf{(ii)} $p_2(\omega')>1/2$. Recalling (\ref{Eqdyad}), remark that we have either $\alpha^2_{\Omega_3}(p,q)=F$ or   $\alpha^2_{\Omega_3}(p,q)=F'$.

Assume without loss that  $\alpha^2_{\Omega_3}(p,q)=F'$  holds. Then picking  $v\in \mathcal{V}^N$ such that $N^v_a=\{1\}$ and $(v,(p,q))\in \mathcal{\tilde{D}}^N$, we claim that agent 2 can manipulate $\varphi$ at $(v,((p_1,p_1),q))$ by reporting $(v'_2,p_2)$, where $v'_2$ satisfies $v'_2(a)>v'_2(b)$. Indeed, note from the previous paragraph  (and Lemma \ref{LemmaSections}) that $\varphi^a_{\Omega_3}(v,(p,q))=\alpha^1_{\Omega_3}(p,q)\in \{E'_1,\ldots, E'_{L^*}\}$. Write  $\alpha^1_{\Omega_3}(p,q)= E'_1$ without loss of generality. Since $E'_1\cup\ldots\cup E'_{L^*}=\Omega_3$, there exists $\ell\in \{1,\ldots, L^*\}$ such that $\omega\in E'_{\ell}$; and writing $\omega\notin E'_1$ would imply $\omega\notin \alpha^1_{\Omega_1}(p,q)$: it would  thus follow from (i) above that $p_1(\alpha^1_{\Omega_1}(p,q))<1/2<p_1(E'_\ell)$. But note that $p_1(\alpha^1_{\Omega_1}(p,q))<p_1(E'_\ell)$ contradicts the fact that agent 1 is a dictator for $\alpha^1_{\Omega_1}$, since $\{E_1,\ldots, E_L\}$ is the range of $\alpha^1_{\Omega_1}$ and $E'_\ell=E_l\cap \Omega_3$ for some $l=1,\ldots,L$. Therefore, we must have $\omega\in E'_1$.

Recalling  Lemma \ref{LemmaSections} and  $\omega\in E'_1\subseteq \Omega_1$, note that agent 2 [who seeks to minimize the probability of $a$ at $(v,((p_1,p_1),q))$] receives $a$ with probability $p_1(\alpha^1((p_1,p_1),q)))\geq p_1(\omega)>1/2$ ---where the last inequality comes from (i).
On the other hand,  since $\omega\notin F'$, agent 2's  deviation to $(v'_2,p_2)$, yields the outcome $a$ with probability $p_1(\alpha^2(p,q))\leq 1-p_1(\omega)<1/2$.
But this contradicts the strategy-proofness of $\varphi$ at $(v,((p_1,p_1),q))$.

We thus conclude that the maximal decomposition of $\mathcal{R}_2^{\Omega_1}$  has only a trivial component, i.e., there exists $\bar{F}\in 2^{\tilde{\Omega}_1}$ such that
\begin{equation}\label{EqFixOm1}
  \alpha^2_{\Omega_1}(p,q)=\bar{F}, \mbox{ for any } \Omega_1\mbox{-dominant }  (p,q)\in \mathcal{\tilde{P}}^2\times \mathcal{\tilde{P}}^{n-2}.
    \end{equation}
 It is not difficult to see that we must have $\bar{F}=\Omega_1$ in (\ref{EqFixOm1}). Suppose by contradiction that $\Omega_1\setminus \bar{F}\neq \emptyset$; and let $\omega\in \Omega_1\setminus \bar{F}$. Consider a profile $(v, (p,q))\in \mathcal{D}^N$ such that (i) $(p,q)$ is $\Omega_1$-dominant, (ii) $p_1=p_2$  and  $p_1(\omega)>1/2$,  (iii) 
 $N^v_a=\{1,2\}$.
Then note from (ii) and (\ref{EqFixOm1}) that $p_2(\alpha^2(p,q))<p_2(\overline{\omega})<1/2$; and agent 2 will manipulate $\varphi$ at any $(v,(p,q))$ by reporting $(v'_2,p_2)$ such that $v'_2(a)<v'_2(b)$ so as to let agent 1 select (as the dictator found in Claim 2 for $\alpha^1_{\Omega_1}$) her best event in $\{E_1,\ldots,E_L\}$, whose probability is $\max\limits_{l=1,\ldots,L}p_1(E_l)\geq p_1(\omega)>1/2$ since $\omega\in \Omega_3\subseteq \Omega_1=E_1\cup \ldots \cup E_L$. Given that $p_2=p_1$, this deviation at  $(v,(p,q))$ results in a higher (lower) probability of $a$  ($b$) for agent 2, in contradiction with the strategyproofness of $\varphi$.

 We have thus shown that
\begin{equation}\label{EqOm1Dom}
  \alpha^2_{\Omega_1}(p,q)=\Omega_1, \mbox{ for any } \Omega_1\mbox{-dominant }  (p,q)\in \mathcal{\tilde{P}}^2\times \mathcal{\tilde{P}}^{n-2}.
\end{equation}
Using (\ref{EqOm1Dom}) we will now show that  $\alpha^2(p,q)=\Omega_1$, for any  $(p,q)\in \mathcal{\tilde{P}}^2\times \mathcal{\tilde{P}}^{n-2}$ such that $p$ (but not necessarily $q$) is $\Omega_1$-dominant.

Fix then $(p,q)\in \mathcal{\tilde{P}}^2\times \mathcal{\tilde{P}}^{n-2} $ such that $p$ is  $\Omega_1$-dominant; and pick any $\bar{q}\in \mathcal{\tilde{P}}^{n-2} $ that is $\Omega_1$-dominant. Define $q^2=\bar{q}$ and $q^i=(q_i,q^{i-1}_{-i})$ for any $i=3,\ldots,n$ (so as to have $q^n=q$).
Since $(p,\bar{q})$ is $\Omega_1$-dominant, it comes from (\ref{EqOm1Dom}) that  $ \alpha^2(p,\bar{q})= \alpha^2(p,q^2)=\Omega_1$. Next, note that  $\alpha^2_{\Omega_1}(p,q^2)=\Omega_1$ implies  $ \alpha^2_{\Omega_1}(p,q^3)=\Omega_1$. Indeed, assuming (by contradiction) that  $ \alpha^2_{\Omega_1}(p,q^3)\subset \Omega_1=\alpha^2_{\Omega_1}(p,q^2) $ yields $q^2_3(\alpha^2_{\Omega_1}(p,q^3))<q^2_3(\alpha^2_{\Omega_1}(p,q^2))$. That is to say, $\bar{q}_2(\alpha^2_{\Omega_1}(p,q^3))<\bar{q}_2(\alpha^2_{\Omega_1}(p,q^2))$. Since $\bar{q}_2$ is $\Omega_1$-dominant, the previous inequality implies  $\bar{q}_2(\alpha^2(p,q^3))<\bar{q}_2(\alpha^2(p,q^2))$, which contradicts Lemma \ref{LemmaSemiSP}-II. Hence, we must have $ \alpha^2_{\Omega_1}(p,q^3)=\alpha^2_{\Omega_1}(p,q^2)=\Omega_1$.
Repeating the same argument $n-2$ times, it comes that $\alpha^2_{\Omega_1}(p,q)=\alpha^2_{\Omega_1}(p,q^{n-2})= \ldots= \alpha^2_{\Omega_1}(p,q^3)=\alpha^2_{\Omega_1}(p,q^2)=\Omega_1$.

Therefore, we have\begin{equation}\label{EqOm1Dom12}
  \alpha^2_{\Omega_1}(p,q)=\Omega_1, \mbox{ for any }  (p,q)\in \mathcal{\tilde{P}}^2\times \mathcal{\tilde{P}}^{n-2} \mbox{ s.t. } p \mbox{ is } \Omega_1\mbox{-dominant}.
\end{equation}

Let us continue our proof that $ \alpha^2_{\Omega_1}(p,q)=\Omega_1$, for any   $(p,q)\in \mathcal{\tilde{P}}^2\times \mathcal{\tilde{P}}^{n-2}$. Fix an arbitrary $\bar{q}\in \mathcal{\tilde{P}}^{n-2}$; and consider the mapping $\psi': \mathcal{\tilde{P}}^{2}\rightarrow 2^{\tilde{\Omega}}$ defined as follows:
\begin{equation}\label{Eqpsiprime}
  \psi'(p)=\alpha^2(p,\bar{q}), ~\forall p=(p_1,p_2) \in \mathcal{\tilde{P}}^{2}.
\end{equation}
It is easy to see from (\ref{Eqpsiprime}) and Lemma \ref{LemmaSemiSP}-I that, for all $ p=(p_1,p_2) \in \mathcal{\tilde{P}}^{2}$,
\begin{eqnarray*}
  p_1(\psi'(p_1,p_2))=\max\limits_{p_1'\in \mathcal{\tilde{P}}}p_1(\psi'(p'_1,p_2))),\\  \nonumber
  p_2(\psi'(p_1,p_2))=\max\limits_{p_2'\in \mathcal{\tilde{P}}}p_2(\psi'(p_1,p'_2))).
  \end{eqnarray*}

In other words, $\psi'$ is a strategy-proof 2-agent event selector.   Moreover, remark from
 (\ref{Eqpsiprime}) and Lemma \ref{LemmaSemiAnon}-I that $\psi'(p_1,p_2)=\psi'(p_2,p_1)$, for all $p=(p_1,p_2)\in  \mathcal{\tilde{P}}^{2}$, \textit{i.e.}, $\psi'$ is anonymous. Since $\psi'$ is strategy-proof and anonymous, remark from Lemma \ref{LemEventChoice2}  that  the maximal decomposition of $\mathcal{F}$ \emph{may have only trivial or dyadic components} (no rich component is allowed, since it would imply that the restriction of $\psi'$ to that component is \textbf{not} anonymous, a contradiction!).

Let then $\tilde{p}=(\tilde{p}_1, \tilde{p}_2)\in \mathcal{\tilde{P}}^2$ be an $\Omega_1$-dominant belief subprofile, [i.e., $\omega_1\in  \psi'(\tilde{p}_1,\tilde{p}_1)=\alpha^2((\tilde{p}_1,\tilde{p}_1),\bar{q})=\Omega_1$ [by (\ref{EqOm1Dom12})-(\ref{Eqpsiprime})].
Consider now $p=(p_1,p_2)\in \mathcal{\tilde{P}}^2$; and suppose by contradiction that $\alpha^2_{\Omega_1}(p,\bar{q})\subset \Omega_1$. This strict inclusion means that there exists $\omega_1\in \Omega_1\setminus \alpha^2_{\Omega_1}(p,\bar{q})$.

Since $\omega_1\notin \psi'(p_1,p_2)=\alpha^2(p,\bar{q})$ and $\omega_1\in  \psi'(\tilde{p}_1,\tilde{p}_1)=\alpha^2((\tilde{p}_1,\tilde{p}_1),\bar{q})=\Omega_1$, the state $\omega_1$ belongs to a non-trivial component of the maximal decomposition of $\mathcal{F}$ (the range of $\psi'$); and given that there exists no rich component, there must exist a dyadic component $\{F,F'\}$ containing $\omega_1$ and satisfying the following properties [where $\Omega_4=F\cup F'$]: \textbf{(i)} $\omega_1\in F$ and $F\cap F'=\emptyset$, \textbf{(ii)}  $\psi'(p')\cap \Omega_4\in \{F,F'\}$,  for all $ p'=(p'_1,p'_2)\in \mathcal{\tilde{P}}^2$, and \textbf{(iii)} $\exists \kappa\in \{1,2\}$ s.t. $\psi'(p')\cap \Omega_4=F'$ if and only if  $\eta_{p'}(\{1,2\}, F'|F))\geq \kappa$.\footnote{Note that condition (iii) obtains from the anonymity and strategy-proofness of $\psi'$, and the fact that the two agents $1,2$ are offered only  two options ($F$ and $F'$) on the cell $\Omega_4$. In other words, the strategy-proof and anonymous restriction of $\psi'$ can be fully implemented just by counting the number of agents who prefer $F'$ to F, and by then selecting $F'$ if  that number is at least equal to the exogenous quota $\kappa$ (otherwise, $F$ is selected).}

The desired result can now be established in two steps. In the first step, we show that $F'\cap \Omega_1=\emptyset$. Suppose by contradiction that there exists $\omega_2\in F'\cap \Omega_1\neq \emptyset$. Then one can pick an $\Omega_1$-dominant subprofile $\bar{p}\in \mathcal{\tilde{P}}^2$ such that $\bar{p}_i(\omega_2)>1/2$ for $i=1,2$; and it thus comes that $\bar{p}_i(F')\geq \bar{p}_i(\omega_2)>1/2> \bar{p}_i(F)$, where the last inequality holds because $\omega_2\notin F$ [by (i) above]. Given that $\bar{p}_i(F')>\bar{p}_i(F)$ (for $i=1,2$), property (iii) above yields $\psi'(\bar{p})\cap \Omega_4=\alpha^2_{\Omega_4}(\bar{p},\bar{q})=F'$.
Thus,  we get $\omega_1\notin \alpha^2_{\Omega_4}(\bar{p},\bar{q})$ ---because $\omega_1\notin F'$ by (i).
Since $\omega_1\in F\subset \Omega_4$ (and hence $\omega_1\notin \overline{\Omega_4}$), it thus comes that $\omega_1\notin \alpha^2(\bar{p},\bar{q})= \alpha^2_{\Omega_4}(\bar{p},\bar{q})\cup \alpha^2_{\overline{\Omega_4}}(\bar{p},\bar{q})$. But note that  $\omega_1\notin \alpha^2(\bar{p},\bar{q})$ contradicts $\omega_1\in \Omega_1=\alpha^2_{\Omega_1}(\bar{p},\bar{q})$ [which comes from (\ref{EqOm1Dom12}) since  $\bar{p}$ is $\Omega_1$-dominant].

In the second step, we show that agent 2 can manipulate $\varphi$ at some preference profile. Pick $\omega'\in F'$; and note from the previous paragraph that $\omega'\notin \Omega_1$. Consider a  subprofile $\hat{p}\in \mathcal{\tilde{P}}^2$ such that \textbf{(iv)} $\hat{p}_1(\omega')>1/2>\hat{p}_1(\omega_1)>\hat{p}_1(\overline{\{\omega_1,\omega'\}})$ and \textbf{(v)} $\hat{p}_2(\omega_1)>1/2$. Moreover, let $v\in \mathcal{V}^N$ be such that $N^v_a=\{1\}$ and $(v,(\hat{p},\bar{q}))\in \mathcal{\tilde{D}}^N$ [note in particular that $v_1(a)>v_1(b)$ and $v_2(a)<v_2(b)$)]. Then recalling Claim 2 (and  $k=1$), one can write \textbf{(vi)} $\varphi^a_{\Omega_1}(v,(\hat{p},\bar{q}))= \alpha^1_{\Omega_1}(\hat{p},\bar{q}) =\argmax\limits_{\ell=1,\ldots,L}\hat{p}_1(E_\ell)$.

Since $\omega_1\in \Omega_1=E_1\cup\ldots\cup E_L$, there exists $l\in \{1,\ldots,L\}$ such that $\omega_1\in E_l$. Observing from (iv) that $\hat{p}_1(\omega_1)>\hat{p}_1(E_{\ell})$, for any $\ell\in \{1,\ldots,L\}$ such that $\omega_\ell\notin E_{\ell}$, we conclude that $\max\limits_{\ell=1,\ldots,L}\hat{p}_1(E_\ell)\geq \hat{p}_1(E_l)\geq \hat{p}_1(\omega_1)>\hat{p}_1(E_{\ell})$ for any $\ell \in \{1,\ldots, L\}$ such that $\omega_1 \notin E_l$. Hence, we must have $\omega_1\in \argmax\limits_{\ell=1,\ldots,L}\hat{p}_1(E_\ell)$; and it thus comes from (v)-(vi) that $\hat{p}_2(\varphi^a(v,(\hat{p},\bar{q})))=\hat{p_2}(\alpha^1(\hat{p},\bar{q}))\geq \hat{p_2}(\alpha^1_{\Omega_1}(\hat{p},\bar{q}))\geq \hat{p}_2(\omega_1)>1/2$.

By reporting $(v'_2,p'_2)\in \mathcal{\tilde{D}}$ such that $v'_2(a)>v'_2(b)$ and $p'_2=\hat{p}_1$, agent 2 induces the choice of $\varphi((v'_{2},v_{-2})((\hat{p}_1,\hat{p}_1),\bar{q}))$. Noting from (iv) that $\hat{p}_1(F')>\hat{p}_1(F)$, one can invoke (iii) to claim that $\omega_1\notin \varphi^a_{\Omega_4}((v'_{2},v_{-2})((\hat{p}_1,\hat{p}_1),\bar{q}))=F'$. Since $\omega_1\notin \overline{\Omega_4}$ (because $\omega_1\in \Omega_4$) we also  have $\omega_1\notin\varphi^a_{\overline{\Omega_4}}((v'_{2},v_{-2})((\hat{p}_1,\hat{p}_1),\bar{q}))$, and hence \textbf{(vii)} $\omega_1\notin \varphi^a((v'_{2},v_{-2})((\hat{p}_1,\hat{p}_1),\bar{q}))= \varphi^a_{\Omega_4}((v'_{2},v_{-2})((\hat{p}_1,\hat{p}_1),\bar{q}))\cup \varphi^a_{\overline{\Omega_4}}((v'_{2},v_{-2})((\hat{p}_1,\hat{p}_1),\bar{q}))$. Combining (v) and (vi), it thus comes that $\hat{p}_2(\varphi^a((v'_{2},v_{-2})((\hat{p}_1,\hat{p}_1),\bar{q})))<1-\hat{p}_2(\omega_1)<1/2$. Since $\hat{p}_2(\varphi^a((v'_{2},v_{-2})((\hat{p}_1,\hat{p}_1),\bar{q})))<1/2< \hat{p}_2(\varphi^a(v,(\hat{p},\bar{q})))$, we conclude that $(v'_2,\hat{p}_1)$ is a profitable deviation for agent 2 at
$(v,(\hat{p},\bar{q})))$; which contradicts the strategy-proofness of $\varphi$. This completes the proof of Claim 3.

\bigskip

We are now ready to conclude the discussion of Case II. Recalling Claim 2, assume without loss of generality that $\alpha^k_{\Omega_1}$ is 1-dictatorial (the argument is symmetric if we instead assume that agent $n$ is the dictator for  $\alpha^k_{\Omega_1}$).

Define the  $\Omega_1$-SCF $\varphi_1$ as follows: for all $(v,q)\in (\mathcal{D}(\Omega_1))^N$,
\begin{equation}\label{Defquasidicta}
  \varphi_1 (v,q)=\left \{ \begin{array}{cc}
                             b\Omega_1 & \mbox{ if } n^v_a=0; \\
                             a\left(\argmax\limits_{\ell=1,\ldots,L}q_i(E_{\ell})\right)\oplus b \left(\Omega_1\setminus \argmax\limits_{\ell=1,\ldots,L}q_i(E_{\ell})\right) &\mbox{ if } N^v_a=\{i\};
                              \\a\Omega_1 & \mbox{ if } n^v_a\geq 2;
                             \end{array} \right.
  \end{equation}
 Combining the range-unanimity of $\varphi$ with Claims 2-3 and (\ref{Defquasidicta}) gives  
 \begin{equation}\label{EqQuasidicta2}
 \varphi_{\Omega_1}(v,p)= \varphi_1 (v,p_{/\Omega_1}), ~\forall (v,p)\in \mathcal{\tilde{D}}^N.
   \end{equation}
   Note that (\ref{EqQuasidicta2}) gives the desired result if $\Omega_1=\tilde{\Omega}$. Suppose then  that $\Omega_1\subset\tilde{\Omega}$.
 In this non-trivial case, it comes from (\ref{EqQuasidicta2}) that
\begin{equation}\label{EqDecII}
\varphi(v,p)= a \left(\varphi^a_{1}(v,p_{/\Omega_1})  \cup \varphi^a_{\overline{\Omega_1}}(v,p) \right) \oplus b \left(\varphi^b_{1}(v,p_{/\Omega_1}) \cup \varphi^b_{\overline{\Omega_1}}(v,p) \right), ~\forall (v,p)\in \mathcal{\tilde{D}}^N.
\end{equation}
Given that the sub-act chosen by $\varphi_1$ in every section $\mathcal{V}^N_k$ (for all possible beliefs) is either constant (whenever $k\neq 1$) or dictated by the only agent in $N^v_a$ (if $k=1$), applying Lemma \ref{LemmaSepSimpleDict} (I and II) to (\ref{EqDecII}) allows to see that $\varphi_{\overline{\Omega_1}}(v,p)$ only depends on $(v,p_{/\overline{\Omega_1}})$, \emph{i.e.}, there exists a map $\hat{\varphi}: (\mathcal{D}(\overline{\Omega_1}))^N \rightarrow \{a,b\}^{\overline{\Omega_1}}$ such that
$\varphi^a_{\overline{\Omega_1}}(v,p)=\hat{\varphi}(v,p_{/\overline{\Omega_1}}) $, for all $(v,p)\in \mathcal{\tilde{D}}^N$. Substituting this equality in (\ref{EqDecII}) thus gives
\begin{equation}\label{EqdecompII}
  \varphi(v,p) = a \left(\varphi^a_{1}(v,p_{/\Omega_1})  \cup \hat{\varphi}^a(v,p_{/\overline{\Omega_1}}) \right) \oplus b \left(\varphi^b_{1}(v,p_{/\Omega_1}) \cup \hat{\varphi}^b(v,p_{/\overline{\Omega_1}}) \right), \forall (v,p)\in \mathcal{\tilde{D}}^N.
\end{equation}

Combining (\ref{EqdecompII}) with the fact that $\varphi_1$ is dictatorial and $\varphi$ is strategy-proof, range-unanimous and anonymous,  it comes that $ \hat{\varphi}$ is  strategy-proof, range-unanimous and anonymous.  Moreover, note that $\hat{\varphi}$ is binary.
 Hence, given that $|\overline{\Omega_1}|=S-|\Omega_1|<S$, it comes from our induction hypothesis that there exist (i) a partition $\{\Omega_2, \ldots, \Omega_T\}$  of $\overline{\Omega_1}$ (with $T\geq 2$)  and (ii) sub-SCFs $\varphi_t: (\mathcal{D}(\Omega_t))^N\rightarrow \{a,b\}^{\Omega_t}$ (for $t=2,\ldots, T$) such that:
\begin{equation}\label{EqsubdecompII}\hat{\varphi}(v,q)= a\left(\bigcup_{t=2}^T \varphi_t^a(v,q_{/\Omega^t}) \right) \oplus b\left(\bigcup_{t=2}^T \varphi_t^b(v,q_{/\Omega^t}) \right),  ~\forall (v,q)\in (\mathcal{D}(\overline{\Omega_1}))^N , \end{equation}
with the property that each sub-SCF $\varphi_t$ (for $t=2, \ldots, T$) is   simple, quasi-dictatorial,  dyadic, or iso-filtering.
Combining (\ref{EqdecompII}) and (\ref{EqsubdecompII}) then yields the desired result.

\bigskip\noindent
\textbf{Case III:}  Suppose that (A) the maximal decomposition of $\mathcal{E}^k$ has (at least) \emph{one dyadic component} for some $k\in \{1,\ldots n-1\}$ and (B) the maximal decomposition of $\mathcal{E}^{k'}$ has \emph{no rich component} for all $k'\in \{1,\ldots n-1\}$.

\smallskip\noindent
Fix then $k\in \{1,\ldots, n-1\}$  such that the maximal decomposition of $\mathcal{E}_k$ has  exactly $M$ dyadic components $\{\hat{B}_1,\tilde{B}_1\}, \dots, \{\hat{B}_M,\tilde{B}_M\}$, with $M\geq 1$. A trivial component may or may not exist; and  we denote it by $\{B_0\}$ (with the convention that $B_0=\emptyset$ if it does not exist).  Moreover, assume without loss that $k$ is \textit{minimal}, that is, if $k'\in \{0,\ldots,k-1\}$ then the decomposition of $\mathcal{E}_{k'}$  only has  a trivial component: $ \alpha^{k'}(p,q)=\alpha^{k'}(\hat{p},\hat{q})$, for all $(p,q),(\hat{p},\hat{q})\in \mathcal{\tilde{P}}^k\times \mathcal{\tilde{P}}^{n-k}$. Combining Lemma \ref{LemmaInvariancePres} and unanimity, note that the minimality of $k$ implies
\begin{equation}\label{EqlowUnan}
 \varphi(v,p)= [\emptyset]_a^{\Tilde{\Omega}}, \mbox{ for all }  (v,p)\in \mathcal{\tilde{D}}^N \mbox{ s.t. } n_a^v\leq k-1.
\end{equation}

Let then $B_m=\hat{B}_m\cup \tilde{B}_m$, for all $m=1,\ldots,M$. From what precedes,  we must have $\Tilde{\Omega}=B_0\cup B_1\cup \ldots\cup B_M$. Pick $\hat{\omega}\in \hat{B}_1$; and define $ \hat{k}:=\max\{k'\in \{k,\ldots, n-1\}: |\mathbf{C}^{k'}_{\hat{\omega}}|=2\}$. Note that $ \hat{k}$ is well-defined and $ \hat{k}\geq k$, since $k\in \{k'\in \{k,\ldots,n-1\}: |\mathbf{C}^{k'}_{\hat{\omega}}|=2\}\neq \emptyset$.
We discuss two  subcases covering all remaining possibilities.

\medskip
 \textbf{Subcase III-1:}  Suppose that  $\mathbf{C}^{k'}_{\hat{\omega}}=\mathbf{C}^{k}_{\hat{\omega}}$, for all $k'=k,\ldots,\hat{k}$. 

 \smallskip Note in this  subcase that $\mathbf{C}^{k'}_{\hat{\omega}}=\{\hat{B}_1,\tilde{B}_1\}$, for all $k'=k,\ldots,\hat{k}$. 
 Combining this with the definition of  $ \hat{k}$, range-unanimity of $\varphi$, and Lemma \ref{LemmaSections}, we get 
\begin{equation}\label{EqhighUnan}
 \varphi_{B_1}(v,p)= [B_1]_a^{B_1}, \mbox{ for all }  (v,p)\in \mathcal{\tilde{D}}^N \mbox{ s.t. } n_a^v\geq \hat{k}+1.
\end{equation}
 
 In addition, if $\hat{k}>k$,  Lemma  \ref{DyadSolidarity}-(II)  allows to write  $H^{k'}_{\hat{\omega}}(m_a,m_b)=H^{k'}_{\hat{\omega}}(m_a+1,m_b-1)$, for all $k'\in \{k,\ldots,  \hat{k}\}$, $m_a \in \{0,\ldots, k'-1\}$, $ m_b\in\{1,\ldots,n-k'\}$. Recalling then  Lemma  \ref{DyadSolidarity}-(I) guarantees the existence of a non-decreasing function $h:\{1,\ldots, n\}\rightarrow \{0,1\}$ satisfying $H^{k'}_{\hat{\omega}}(m_a+1,m_b-1)=h(m_a+m_b)$, for all $k'\in \{k, \ldots, \hat{k}\}$,  $m_a \in \{0,\ldots, k'\}$, $ m_b\in\{1,\ldots,n-k'\}$. 
  Hence, whether $ \hat{k}=k$ or $ \hat{k}>k$, there exists a non-decreasing function $H: \SN\times \SN \rightarrow \{0,1\}$ such that $H^{k'}_{\hat{\omega}}(m_a,m_b)=H(m_a,m_b)$, for all $k'\in \{k, \ldots, \hat{k}\}$,  $m_a \in \{0,\ldots, k'\}$, $ m_b\in\{1,\ldots,n-k'\}$. 

 Write   then  $\Omega_1=B_1$ and  define $\Omega_1$-SCF $\varphi_1$ as follows: for all $(v,q)\in (\mathcal{D}(\Omega_1))^N$,
\
\begin{equation}\label{EqThmDyad}
  \varphi_1 (v,q)=\left \{ \begin{array}{cl}
                             [\emptyset]^{\Omega_1}_a & \mbox{ if } n^v_a\leq  k-1; \\ \\
                             \left[\hat{B}_1\right]_a^{\Omega_1} &\mbox{ if } k\leq  n^v_a\leq \hat{k}  \mbox{ and } H(\eta_q(N^v_a, \hat{B}_1|\tilde{B}_1), \eta_q(N^v_b, F|E) )=1;\\ \\
                            \left[\tilde{B}_1\right]_a^{\Omega_1} &\mbox{ if } k\leq  n^v_a\leq \hat{k}  \mbox{ and } H(\eta_q(N^v_a, E|F), \eta_q(N^v_b, F|E) )=0;\\ \\
                             \left[\Omega_1\right]_a^{\Omega_1} & \mbox{ if } n^v_a\geq  \hat{k}  +1.
                             \end{array} \right.
  \end{equation}
The reader should recognize a dyadic binary factor in (\ref{EqThmDyad}). The previous paragraph and (\ref{EqlowUnan})-(\ref{EqThmDyad}) guarantee that 
\begin{equation}\label{EqThmDyad2}
    \varphi_{\Omega_1}(v,p)=\varphi_1(v,p_{/\Omega_1}).
\end{equation}
 The equality in (\ref{EqThmDyad2}) is  the desired result if $\tilde{\Omega}=\Omega_1$. Suppose then  that  $\overline{\Omega_1}=\tilde{\Omega}\setminus \Omega_1\neq \emptyset$. Then combining (\ref{EqThmDyad2}) and the fact that the maximal decomposition of $\mathcal{E}^{k'}$ has \emph{no rich component} (for all $k'\in \{1,\ldots n-1\}$), one can use the separability property stated in  Lemma \ref{LemmaDyad} to write $\varphi(v,p)=\varphi_1(v,p_{/\Omega_1}) \oplus \varphi_{\overline{\Omega_1}}(v,p_{/\overline{\Omega_1}})$, for all  $(v,p)\in (v,p)\in \mathcal{\tilde{D}}^N$.
 Hence, there exists a binary $\overline{\Omega_1}$-SCF $\hat{\varphi}: (\mathcal{D}(\overline{\Omega_1}))^N \rightarrow \{a,b\}$ such that 
\begin{equation}\label{EqThmDyad3} 
\varphi(v,p)=\varphi_1(v,p_{/\Omega_1}) \oplus \hat{\varphi}(v,p_{/\overline{\Omega_1}}), ~\forall (v,p)\in \mathcal{\tilde{D}}^N.
\end{equation}
Moreover, since $\varphi$ is strategy-proof, range-unanimous and anonymous, it is not difficult to see from (\ref{EqThmDyad3}) that $\hat{\varphi}$ must be strategy-proof, range-unanimous and anonymous. Given that $|\overline{\Omega_1}|=S-|\Omega_1|<S$, our induction hypothesis guarantees the existence of  (i) a partition $\{\Omega_2, \ldots, \Omega_T\}$  of $\overline{\Omega_1}$ (with $T\geq 2$)  and (ii) sub-SCFs $\varphi_t: (\mathcal{D}(\Omega_t))^N\rightarrow \{a,b\}^{\Omega_t}$ (for $t=2,\ldots, T$) such that:
\begin{equation}\label{EqsubdecompIII}\hat{\varphi}(v,q)= \varphi_2 (v,q_{/\Omega_2}) \oplus \ldots \oplus  \varphi_T (v,q_{/\Omega_T}),  ~\forall (v,q)\in (\mathcal{D}(\overline{\Omega_1}))^N , \end{equation}
with the property that each sub-SCF $\varphi_t$ (for $t=2, \ldots, T$) is  simple, quasi-dictatorial,  dyadic, or iso-filtering.
Combining (\ref{EqThmDyad3}) and (\ref{EqsubdecompIII}) then yields the desired result.

\bigskip
 \textbf{Subcase III-2:} Suppose now  that $\mathbf{C}^{\hat{k}}_{\hat{\omega}}\neq\mathbf{C}^{k}_{\hat{\omega}}$ (which requires $\hat{k}>k$). 

\medskip\noindent Define the following binary relations on $\tilde{\Omega}$: $\forall \omega_1, \omega_2\in \tilde{\Omega}$, $\forall \ell \in \{1,\ldots,n-1\} $,
\begin{equation}\label{Eqdefrelfilter}
    \omega_1~ r_{\ell} ~\omega_2 \mbox{ \textbf{iff} }  [ \mathbf{C}^{\ell}_{\omega_1}= \mathbf{C}^{\ell}_{\omega_2} \mbox{ and }   |\mathbf{C}^{\ell}_{\omega_1}|= |\mathbf{C}^{\ell}_{\omega_2}|=2  ].
\end{equation}
 Let then $\bar{r}_{\ell}$ be the \textit{transitive closure of the binary relation} $\bigcup\limits_{\ell'=1}^{\ell}r_{\ell'}$. For all $\omega\in \tilde{\Omega}$, write 
\begin{equation}\label{Eqdefclassr}
    \bar{s}_{\ell}(\omega):=\{\omega'\in \tilde{\Omega}: \omega~ \bar{r}_{\ell} ~\omega_1~ \bar{r}_{\ell}\ldots \bar{r}_{\ell} ~\omega_{\ell}~ \bar{r}_{\ell}~ \omega', \mbox{ for some } \omega_1,\ldots,\omega_{\ell} \in \tilde{\Omega}\}.
\end{equation}
Write
\begin{equation}\label{Eqktilde}\tilde{k}:=\max\{\ell\in \{k,\ldots,n-1\}:   \bar{s}_{\ell}(\hat{\omega}) \neq \emptyset\};
\end{equation}
and remark from (\ref{Eqdefrelfilter})-(\ref{Eqktilde}) that any state $\omega$ that is  $\bar{r}_{\tilde{k}}$-related to $\hat{\omega}$ must be in a trivial component (\textit{i.e.}, $|\mathbf{C}^k_{\omega}|=1$) as soon as the number $k$ of supporters of $a$ reaches $\tilde{k}+1$. Combining this observation with Lemma \ref{LemmaSections} and Lemma \ref{LemmaInvariancePres}, and defining $\Omega_1:=\bar{s}_{\tilde{k}}(\hat{\omega})$, it thus comes that 
\begin{equation}\label{EqPhioverktilde}
\varphi_{\Omega_1}(v,p)= [\Omega_1]_a^{\Omega_1} \mbox{ if } n_a^v \geq \tilde{k}+1.
\end{equation}

 \bigskip
 Moreover,  recalling  (\ref{Eqktilde}) and the fact that the
 component of $\hat{\omega}$ (in the maximal decomposition of $\mathcal{E}^k$) is dyadic, notice that, for all $k'=k,\ldots, \tilde{k}$, the  maximal decomposition of $\mathcal{E}^{k'}$ has components [the last two being trivial and the other(s) being dyadic] which can be written as follows:\footnote{We  slightly abuse notation by allowing $B^{k'}_a$ and $B^{k'}_b$ to be empty as part of a maximal decomposition: this slight abuse is convenient and  no possible confusion can arise from it.}
$$ \{\hat{B}^{k'}_1,  \tilde{B}^{k'}_{1}\}, \ldots,\{\hat{B}^{k'}_{M_{k'}}, \tilde{B}^{k'}_{M_{k'}}\}, \{B_a^{k'}\}, \{B_b^{k'}\},$$
with the constraints $B_a^{k}=\emptyset=B_b^{\bar{k}}$, $\hat{B}^{k'}_1\cup \tilde{B}^{k'}_{1}\cup \ldots \cup\hat{B}^{k'}_{M_{k'}}\cup \tilde{B}^{k'}_{M_{k'}}\cup B_a^{k'}\cup B_b^{k'}= \Omega_1 $.


One can therefore define the following  dipartition of $\Omega_1$
$$\mathcal{C}^{k'}=\{(\hat{B}^{k'}_1,  \tilde{B}^{k'}_{1}), \ldots,(\hat{B}^{k'}_{M_{k'}}, \tilde{B}^{k'}_{M_{k'}}), (B_a^{k'}, B_b^{k'})\}, \forall k'\in \{k,\ldots,\tilde{k}\};$$ and, recalling  (\ref{Eqdefrelfilter})-(\ref{Eqktilde}) then allows to see that  the transitive closure of   $\bigcup\limits_{k'=k}^{\tilde{k}}R_{\mathcal{C}^{k'}}$ coincides with $\bar{r}_{\tilde{k}}$ and has full graph $\Omega_1\times \Omega_1$. 

Recalling Lemma \ref{DyadSolidarity}-(II), one can check that, for all $\omega,\omega'\in \Omega_1$,
$$
  [ \mathbf{C}^{k'}_\omega=\{\hat{F}^{k'}_\omega, \hat{F}^{k'}_{\omega'}\} \mbox{ and } \mathbf{C}^{k'+1}_\omega=\{\hat{F}^{k'+1}_\omega, \hat{F}^{k'+1}_{\omega'}\}] \Rightarrow \mathbf{C}^{k'}_\omega=\mathbf{C}^{k'+1}_\omega.$$ Combining this implication with Lemma \ref{DyadSolidarity}-(V) allows to see that the dipartition  $\mathcal{C}^{k'}$  satisfies the properties (\ref{EqFilterdef1})-(\ref{EqFilterdef2}).\footnote{Strictly speaking, it may be necessary to swap the roles of  $\hat{B}^{k'}_{m}$ and $\tilde{B}^{k'}_{m}$ in the definition of $\mathcal{C}^{k'}$  (for some $k'=k,\ldots,\tilde{k}$ and $m=1,\ldots, M_{k'}$) to ensure that (\ref{EqFilterdef1})-(\ref{EqFilterdef2}) follow from Lemma \ref{DyadSolidarity}-(V), since a dipartition involves ordered pairs (whereas a dyadic component does not specify the order of the two sets $\hat{B}^{k'}_{M_{k'}},\tilde{B}^{k'}_{M_{k'}}$); but this is easily done without loss of generality. In particular, if $\{\hat{B}^{k'}_1,  \tilde{B}^{k'}_{1}\}=\{\hat{B}^{k'+1}_1,  \tilde{B}^{k'+1}_{1}\}$ then we write $\hat{B}^{k'}_1=\hat{B}^{k'+1}_1$ and $\tilde{B}^{k'}_1=\tilde{B}^{k'+1}_1$ (instead of $\hat{B}^{k'}_1=\tilde{B}^{k'+1}_1$ and $\tilde{B}^{k'}_1=\hat{B}^{k'+1}_1$).} 
We thus conclude that the sequence $\mathcal{C}^{k'}$ (with $k'=k,\ldots,\tilde{k}$) is a filter of $\Omega_1$.

Fix now $k'\in \{k,\ldots,\tilde{k}\}$ and $m'\in \{1,\ldots,M_{k'}\}$. Since the binary relation $\bar{r}$ (the transitive closure of $\bigcup\limits_{k'=k}^{\tilde{k}}R_{\mathcal{C}^{k'}}$) has full graph $\Omega_1\times \Omega_1$, remark from the combination of (\ref{Eqdefrelfilter})-(\ref{Eqktilde}) and $\mathbf{C}^{\tilde{k}}_{\hat{\omega}}\neq \mathbf{C}^{k}_{\hat{\omega}}$ that there must exist integers  $k''\in \{k,\ldots,\tilde{k}\}\setminus \{k'\}$ and $m''\in \{1,\ldots,M_{k''}\}$ such that
\begin{align}
\label{EqcomponentFilter1}
    (\hat{B}^{k''}_{m''}\cup \tilde{B}^{k''}_{m''})\cap &(\hat{B}^{k'}_{m'}\cup \tilde{B}^{k'}_{m'})\neq \emptyset,
    \\ \label{EqcomponentFilter2}
    (\hat{B}^{k''}_{m''}, \tilde{B}^{k''}_{m''})\neq & (\hat{B}^{k'}_{m'},\tilde{B}^{k'}_{m'}),\\ \label{EqcomponentFilter3}
     (\hat{B}^{k'}_{m'},\tilde{B}^{k'}_{m'})\in \mathcal{C}^{\ell},& ~\forall \ell\in \{\min\{k',k''\}, \ldots, \max\{k',k''\}\}\setminus\{k''\}.
    \end{align} 

To fix ideas, assume that $k'>k''$ (similar argument if one instead assumes  $k'<k''$). Then note that  (\ref{EqcomponentFilter3}) becomes 
\begin{align}
\label{EqcomponentFilter4}
     (\hat{B}^{k'}_{m'},\tilde{B}^{k'}_{m'})\in \mathcal{C}^{\ell},& \mbox{ for }  \ell=k''+1, \ldots, k'.
    \end{align} 
Substituting (\ref{EqcomponentFilter4}) in  (\ref{EqcomponentFilter1})-(\ref{EqcomponentFilter2}) then yields  
\[(\hat{B}^{k''}_{m''}\cup \tilde{B}^{k''}_{m''})\cap (\hat{B}^{k''+1}_{m^*}\cup \tilde{B}^{k''+1}_{m^*})\neq \emptyset \mbox{ and } (\hat{B}^{k''}_{m''}, \tilde{B}^{k''}_{m''})\neq  (\hat{B}^{k''+1}_{m^*},\tilde{B}^{k''+1}_{m^*}),\] for some $m^*\in \{1,\ldots, M_{k''+1}\}$; and combining these two properties with (\ref{EqFilterdef1})-(\ref{EqFilterdef2}), we conclude that 
\begin{equation}\label{Eqfiltquotas}
    \tilde{B}^{k''+1}_{m^*}\cap \hat{B}^{k''}_{m''}\neq \emptyset \mbox{ and }   \hat{B}^{k''+1}_{m^*}\subseteq G^{k''}_b  \mbox{ and }  \tilde{B}^{k''}_{m''}\subseteq G^{k''+1}_a.
\end{equation}

Using (\ref{Eqfiltquotas})  and applying  Lemma \ref{DyadSolidarity}-(IV), notice that the sub-act $a\tilde{B}^{k''+1}_{m^*}\oplus b\hat{B}^{k''+1}_{m^*}$ must be selected on $\tilde{B}^{k''+1}_{m^*}\cup \hat{B}^{k''+1}_{m^*}$ by $\varphi$ (for valuation profiles in $\mathcal{V}_{k''+1}$) on  as soon as one of the $k''+1$ supporters of $a$ prefers it to $a\hat{B}^{k''+1}_{m^*}\oplus b\tilde{B}^{k''+1}_{m^*}$. Combining this last observation with (\ref{EqcomponentFilter4}) and Lemma \ref{DyadSolidarity}-(I), and letting $\tilde{t}^{k''+1}_{m^*}=1$, we conclude that $b$ must be selected  as soon as (at least) $\tilde{t}^{k''+1}_{m^*}$ of the $k''+1$ supporters of $a$ prefer it.

In the case where no supporters of $a$ prefers the sub-act $a\tilde{B}^{k''+1}_{m^*}\oplus b\hat{B}^{k''+1}_{m^*}$, remark from the monotonicity of $H^{k''+1}_{\omega}$ (for any $\omega\in \tilde{B}^{k''+1}_{m^*}$) that there must exist a threshold $\hat{t}^{k''+1}_{m^*}\in \{1,\ldots,n-k''\}$ such that $a\tilde{B}^{k''+1}_{m^*}\oplus b\hat{B}^{k''+1}_{m^*}$ is selected by $\varphi$ if and only the number of supporters of $b$ who prefer it is at least $\hat{t}^{k''+1}_{m^*}$. Recalling then (\ref{EqcomponentFilter4}) and Lemma \ref{DyadSolidarity}-(I), we conclude that  there exists a  threshold $\hat{t}^{k'}_{m'}$ ($\tilde{t}^{k'}_{m'}$) that must be reached by the number  of  supporters of the outcome $b$ ($a$) to trigger the selection of  $a\tilde{B}^{k'}_{m'}\oplus b\hat{B}^{k'}_{m'}$ by $\varphi$ in every dyadic cell $\{\hat{B}^{k'}_{m'},\tilde{B}^{k'}_{m'}\}$ of $\Omega_1$.

Combining this last result with (\ref{EqlowUnan}) and (\ref{EqPhioverktilde}), and letting $\varphi_1$ be the filtering factor defined on $\Omega_1$ by the filter  $\{\mathcal{C}^{k'}\}_{k'=k,\ldots,\tilde{k}}$ and the quotas $\hat{t}^{k'}_{m'},\tilde{t}^{k'}_{m'}$ ($k'=k, \ldots,\tilde{k}$, $m=1,\ldots, M_{k'}$),  one can  claim that
\begin{equation}\label{EqExtrFilterIII2}
    \varphi_{\Omega_1}(v,p)=\varphi_1(v,p_{/\Omega_1}), ~\forall (v,p)\in \tilde{\mathcal{D}}^N.
\end{equation}
Since $\varphi_1$ inherits the strategy-proofness property from $\varphi$, remark from Proposition \ref{PropTriadic} that $\varphi_1$ must be an iso-filtering factor. Hence, the equality in (\ref{EqExtrFilterIII2}) is  the desired result if $\tilde{\Omega}=\Omega_1$. Suppose then  that  $\overline{\Omega_1}=\tilde{\Omega}\setminus \Omega_1\neq \emptyset$. Then combining (\ref{EqExtrFilterIII2}) with the fact that the maximal decomposition of $\mathcal{E}^{k'}$ has \emph{no rich component} (for all $k'\in \{1,\ldots n-1\}$), one can use the separability property stated in  Lemma \ref{LemmaDyad} to write $\varphi(v,p)=\varphi_1(v,p_{/\Omega_1}) \oplus \varphi_{\overline{\Omega_1}}(v,p_{/\overline{\Omega_1}})$, for all  $(v,p)\in (v,p)\in \mathcal{\tilde{D}}^N$.
 Hence, there exists a binary $\overline{\Omega_1}$-SCF $\hat{\varphi}: (\mathcal{D}(\overline{\Omega_1}))^N \rightarrow \{a,b\}$ such that 
\begin{equation}\label{EqThmDyad32} 
\varphi(v,p)=\varphi_1(v,p_{/\Omega_1}) \oplus \hat{\varphi}(v,p_{/\overline{\Omega_1}}), ~\forall (v,p)\in \mathcal{\tilde{D}}^N.
\end{equation}
Moreover, since $\varphi$ is strategy-proof, range-unanimous and anonymous, it is not difficult to see from (\ref{EqThmDyad32}) that $\hat{\varphi}$ must be strategy-proof, range-unanimous and anonymous. Given that $|\overline{\Omega_1}|=S-|\Omega_1|<S$ (recall that $\hat{\omega}\in \Omega_1$), our induction hypothesis guarantees the existence of  (i) a partition $\{\Omega_2, \ldots, \Omega_T\}$  of $\overline{\Omega_1}$ (with $T\geq 2$)  and (ii) sub-SCFs $\varphi_t: (\mathcal{D}(\Omega_t))^N\rightarrow \{a,b\}^{\Omega_t}$ (for $t=2,\ldots, T$) such that:
\begin{equation}\label{EqsubdecompIII2}\hat{\varphi}(v,q)= \varphi_2 (v,q_{/\Omega_2}) \oplus \ldots \oplus  \varphi_T (v,q_{/\Omega_T}),  ~\forall (v,q)\in (\mathcal{D}(\overline{\Omega_1}))^N , \end{equation}
with the property that each sub-SCF $\varphi_t$ (for $t=2, \ldots, T$) is  simple, quasi-dictatorial,  dyadic, or iso-filtering.
Combining (\ref{EqThmDyad32}) and (\ref{EqsubdecompIII2}) then yields the desired result.\endproof


The proof of Theorem \ref{ThmChar} then easily follows from the combination of Proposition \ref{Propbinarydecomp} and Proposition \ref{LemmaDec}.
\end{document}